\documentclass[utf8]{aa}  
\usepackage{caption}
\def\arraystretch{0.85}
\def\l{$\lambda$}
\def\mbh{$M_{\rm BH}$\/}
\def\nh{$n_{\mathrm{H}}$\/}
\def\lledd{$L/L_{\rm Edd}$}

\def\rfe{$R_\mathrm{FeII}$}

\def\feiiq{\rm Fe{\sc ii}$\lambda$4570\/}

\def\ltsima{$\; \buildrel < \over \sim \;$}
\def\ltsim{\lower.5ex\hbox{\ltsima}}  
\def\gtsima{$\; \buildrel > \over \sim \;$}

\def\gtsim{\lower.1ex\hbox{\gtsima}} 
\def\h{$H_{0}$}
\def\lya{{ Ly}$\alpha$}
\def\civfull{{\sc{Civ}}$\lambda$1549\/}
\def\civ{{\sc Civ}}

\def\civbc{{\sc{Civ}}$\lambda$1549$_{\rm BC}$\/}

\def\cmc{cm$^{-3}$\/}
\def\hb{{\sc{H}}$\beta$\/}

\def\mgii{{Mg\sc{ii}}$\lambda$2800\/}

\def\nivfull{{\sc{Niv]}}$\lambda$1486\/}
\def\ciii{{\sc{Ciii]}}\/}
\def\cii{{C{\sc ii}}}
\def\ciiifull{{\sc{Ciii]}}$\lambda$1909\/}

\def\oi{{\sc{[Oi]}}$\lambda$6300\/}
\def\oiuv{{\sc{[Oi]}}$\lambda$1304\/}

\def\oiiiuv{{\sc{Oiii]}}$\lambda$1663\/}

\def\siiii{Si{\sc iii]}\/}
\def\siiiifull{Si{\sc iii]}$\lambda$1892\/}
\def\aliii{Al{\sc  iii}}
\def\heiiuv{He{\sc{ii}}}

\def\aliiifull{Al{\sc iii}$\lambda$1860\/}
\def\heiiuvfull{He{\sc{ii}}$\lambda$1640}
\def\nvfull{{N\sc{v}}$\lambda$1240}

\def\feii{{Fe\sc{ii}}\/}
\def\siii{{Si\sc{ii}}$\lambda$1814\/}
\def\feiii{{Fe\sc{iii}}\/}
\def\fe{{\sc{Fe}}\/}

\def\fe76087{{\sc [Fe vii]}$\lambda$6087\/}

\def\b{$\beta$}
\def\kms{km~s$^{-1}$}

\def\ergss{erg s$^{-1}$\/}

\def\heii{{{\sc H}e{\sc ii}}$\lambda$1640\/}

\def\siiv{Si{\sc iv}\/}
\def\oiv{O{\sc iv]}\/}
\def\siivfull{Si{\sc iv}$\lambda$1397\/}
\def\oivfull{O{\sc iv]}$\lambda$1402\/}

\usepackage{float}
\usepackage{graphicx}
\usepackage{subfig}
\usepackage{amsmath}
\usepackage{txfonts}
\usepackage{xcolor}
\usepackage{csvsimple}
\usepackage{soul}
\definecolor{seagreen}{rgb}{0.190, 0.525, 0.361}
\definecolor{darksalmon}{rgb}{0.914, 0.588, 0.478}
\definecolor{steelblue}{rgb}{0.274 0.510 0.706}
\definecolor{purple}{rgb}{0.5 0.0 0.5}
\definecolor{darkred}{rgb}{0.8 0.0 0.0}
\definecolor{darkgold}{rgb}{0.8 0.543 0.0}
\definecolor{disagreeablegray}{rgb}{0.00 0.00 0.00}
\definecolor{darkgreen}{rgb}{0,0.6,0}

\def\nh{$n_\mathrm{H}$\/}
\usepackage{rotating}
 \date{ }
 \makeatletter
\def\blfootnote{\xdef\@thefnmark{}\@footnotetext}
\makeatother
\begin{document}
\titlerunning{Metallicity in highly accreting quasars — II}
%\begin{tikzpicture}
%\node[inner sep=0pt] (russell) at (1,-1)
%    {\includegraphics[width=.25\textwidth]{stamp.png}};
%\end{tikzpicture}
\title{High metal content of highly accreting quasars: analysis of an extended sample}
\author{ K. Garnica\inst{1}   \and C. A. Negrete\inst{2} \and P. Marziani\inst{3} \and D. Dultzin\inst{1}  \and M. \'Sniegowska\inst{4,5} \and S. Panda\inst{5,6,4,\thanks{CNPq Fellow}}}
\institute{{Instituto de Astronom\'{\i}a, UNAM, 04510, Mexico}
\and{CONACyT Research Fellow - Instituto de Astronom\'{\i}a, UNAM, 04510, Mexico}
\and
{National Institute for Astrophysics (INAF), Padova Astronomical Observatory, IT 35122, Padova, Italy}
\and
{Nicolaus Copernicus Astronomical Center, Polish Academy of Sciences, ul. Bartycka 18, 00-716 Warsaw, Poland}
\and {Center for Theoretical Physics, Polish Academy of Sciences, Al. Lotnik\' ow 32/46, 02-668 Warsaw, Poland}
\and {Laborat\'orio Nacional de Astrof\'isica - MCTIC, R. dos Estados Unidos, 154 - Na\c{c}\~oes, Itajub\'a - MG, 37504-364, Brazil}
}

\abstract
{We present an analysis of UV spectra of quasars at intermediate redshifts  believed to belong to the extreme Population A (xA), aimed to estimate  the chemical abundances of the broad line emitting gas. We follow  the approach described in a previous work %\citet{sniegowskaetal21} 
extending the %ir 
sample to 42 sources.}
{Our aim is to test the robustness of the analysis carried out previously, %by \citet{sniegowskaetal21}, 
as well as to confirm the two most intriguing results of this investigation: evidence of very high solar metallicities, and deviation of the relative abundance of elements with respect to solar {values}.} {The basis of our analysis are multi-component fits in three regions of the spectra centered at 1900, 1550 and 1400  \AA \ in order to deblend the broad components of \aliiifull, \ciiifull, \civfull, \heiiuvfull, and \siivfull{} + \oivfull{} and their blue excess.} {By comparing the observed flux ratios of these components  with the same ratios predicted by photoionization code {\tt CLOUDY}  we found that the virialized gas (broad components) presents a metallicity {($Z$)} higher than 10Z$_\odot$. For non-virialized clouds we derive a lower limit to the metallicity around $\sim$ 5Z$_\odot$ under the assumption of chemical composition proportional to the solar one, confirming the previous results. %obtained by \citet{sniegowskaetal21}. 
We especially {rely}  on the ratios between metal lines and \heiiuvfull. This allowed us to  confirm  systematic differences in the solar-scaled metallicity derived from the lines of Aluminium and Silicon, and of Carbon, with the first being a factor $\approx$ 2 higher. } { For luminous quasars accreting at high rates, high $Z$\ values are likely, but that $Z$\ scaled-values are affected by the possible pollution due to highly-enriched gas associated with the circumnuclear star formation. The high-$Z$ values suggest a complex process  involving nuclear and circumnuclear star formation, interaction between nuclear compact objects and accretion disk, possibly with the formation of accretion-modified stars.}
%\end{abstract}
\maketitle

\defcitealias{sniegowskaetal21}{S21}
\defcitealias{marzianisulentic14}{MS14}

%%%%%%%%%%%%%%%%%%%%%%%%%%%%%%%%%%%%%%%%%%%%%%%%%%%%%%%
\begin{figure}[h] %fig 1
     \centering
     %\hspace{-1cm}
     %trim=left bottom right top
%     \includegraphics[trim= 0.5 0.5 0.5 0.5, clip, width=0.5\textwidth]{Lyke.png}
     \includegraphics[trim= 0.5 0.5 0.0 0.5, clip,width=0.52\textwidth]{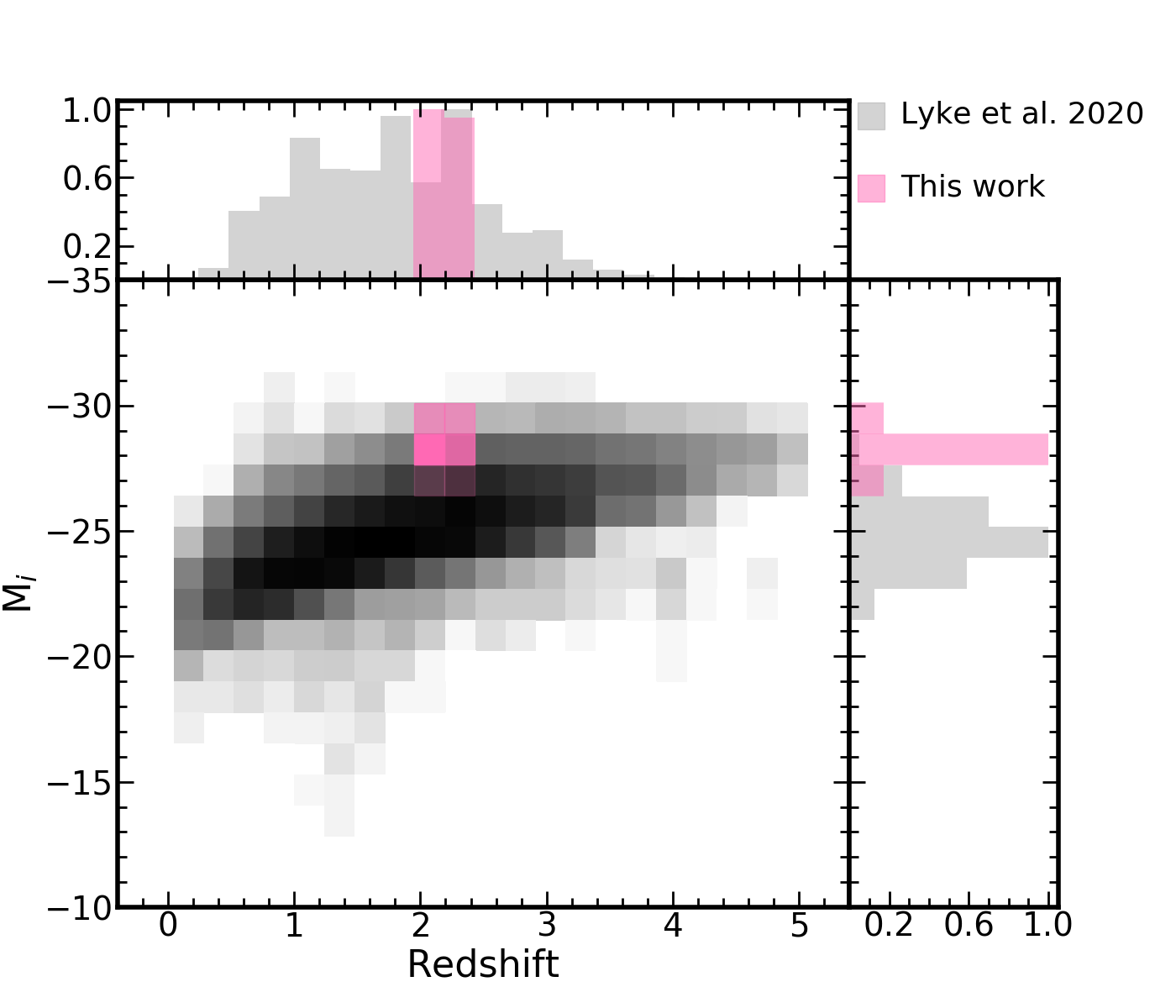}
    \caption{Redshift distribution of the sample as a function of the absolute $i-$band magnitude $M_\mathrm{i}$. The gray scale squares show the \citet{Lyke_2020ApJS250} data, binned over small bins in $z$ and $M_\mathrm{i}$. The pink squares identify the sample used in this work. \textit{Upper panel}: normalized redshift distribution. The pink histogram shows the corrected redshift adopted in this paper (see Sect. \ref{redshift}). \textit{Right panel}: normalized absolute $i-$band magnitude distributions for both \citet{Lyke_2020ApJS250} and our sample.}   
    \label{fig:Lyke}
\end{figure}

%%%%%%%%%%%%%%%%%%%%%%%%%%%%%%%%%%%%%%%%%%%%%%%%%%%%%
\section{Introduction} 
\label{sec:intro}

The diversity of broad-line quasars has been efficiently organized through the parameter space of the Eigenvector 1 (E1) in several studies  during the last three decades \citep{borosongreen92,sulenticetal00a,shenho14}. \citet{sulenticetal00a,sulenticetal00c,sulenticetal07} proposed an E1 space that  is represented by four dimensions (4DE1) with two spectroscopical parameters defining an optical plane:

\begin{itemize}
    \item FWHM(\hb), the full width at half maximum of the \hb\ broad component;
    \rfe\ = F(\feiiq)/F(\hb), the ratio of the optical FeII emission between 4434  \AA\ and 4684   \AA\ to the \hb\ intensity,
\end{itemize}

\noindent and two remaining parameters involving spectral measurements in the UV and X-ray observations:

\begin{itemize}
    \item \civ\ c(1/2), the centroid line shift of \civ;
     \item $\Gamma_\mathrm{soft}$, the soft X-ray photon index.
\end{itemize}
The observed pattern in the  optical plane of the 4DE1 (FWHM(\hb) vs. \rfe) has been identified  as the quasar main sequence (MS), since it is associated with trends of physical properties  such as the mass of the supermassive black hole (\mbh), the broad-line region (BLR) gas density (\nh), the ionization parameter (U), and the  Eddington ratio  \citep[\lledd, ][]{borosongreen92,boroson02,pandaetal18,pandaetal19a,wildyetal19}. However, the {principal driver of the MS} is probably the Eddington ratio \citep{borosongreen92,marzianietal01,sunshen15}, a basic parameter that expresses the balance between radiation and gravitational forces,  and that is related to the structure of the accretion flow \citep{giustiniproga19,donofrioetal21}. 

Two Populations, A and B, have been defined from this parameter space. They are delimited at 4000 \kms, Pop. B objects {being} the ones with the broadest widths \citep{sulenticetal00a,sulenticetal09,sulenticmarziani15}. However, although this condition subdivides quasars of different  properties, spectral differences of objects in the same population can still be found, particularly in Population A (Figure 2 of \citealt{sulenticetal02}; cf. \citealt{shenho14}). This is why a subdivision in bins of $\Delta$FWHM(\hb) = 4000 \kms\ and $\Delta$\rfe\ = 0.5 was introduced. Bins A1, A2, A3, and A4 are defined in increasing \rfe, while bins B1, B1+, and B1++ are defined in increasing FWHM(\hb) (see Figure 1 of \citealt{sulenticetal02}; cf. \citealt{shenho14}). Sources of the same spectral type show similar spectroscopic measures (e.g., line profiles and UV line ratios, \citealt{bachevetal04,sulenticetal07}) and similar physical parameters .

Most interesting along the MS {are the} spectral types at the extremes \citep{marzianietal21}. At the high \rfe\ end, quasars are found to be radiating at high Eddington ratio from a variety of theoretical and observational analyses \citep[e.g.,][]{marzianietal01,sunshen15,duetal16a,pandaetal19}. High Eddington ratio  quasars can be  selected   on the basis of  empirical criteria based on the MS spectral types \citep[e.g.,][]{wangetal13,marzianisulentic14,wangetal14b,duetal16a}. According to \citet[][hereafter \citetalias{marzianisulentic14}]{marzianisulentic14}, they are defined by having 
\begin{equation}
R_\mathrm{FeII} = \frac{F(\mathrm{FeII}\lambda 4570)}{F(\mathrm H\beta)} \ge 1.0    
\label{eq:optcrit}
\end{equation}
i.e.,  the flux of the \feiiq\ blend on the blue side of \hb\ (as defined by \citealt{borosongreen92}) exceeds the flux of \hb, and the spectral type is An with $n \ge 3$.  In the optical diagram of the quasar MS \citep{sulenticetal00c,shenho14,marzianietal18} they are at the extreme tip in terms of \feii\ prominence, and identified as extreme Population A (hereafter xA). 

At relatively modest redshifts ($z \gtrsim 1$), the \hb\ line is already shifted into the NIR. Since NIR spectroscopic observations are still difficult, it is expedient to resort to the quasar-rest frame UV spectrum shifted into the visual range. For z $\gtrsim$ 1, the  criteria for selecting xA objects based on two UV line intensity ratios are believed to be satisfied:  
 \begin{eqnarray}
 \frac{F(\mathrm{AlIII}\lambda 1860)}{F(\mathrm{SiIII]}\lambda1892)} &\ge& 0.5\\ \nonumber
 \frac{F(\mathrm{SiIII]}\lambda1892)}{F(\mathrm{CIII]}\lambda 1909)} &\ge& 1.0\\ \nonumber \label{eq:uvcrit}
  \end{eqnarray}
following \citetalias{marzianisulentic14}.  These criteria are believed to be met by the sources satisfying the condition of Eq. \ref{eq:optcrit}. 

There have been previous claim of very high metallicity needed to account for the extreme \feii\ emission \citep{pandaetal18,pandaetal19,pandaetal20} as well as to account for the metallicity-sensitive UV emission line ratios  \citep{baldwinetal03,warneretal04,shinetal13,sulenticetal14}.  
 The aim of this work is to investigate the metallicity-sensitive diagnostic ratios of the UV spectral range for an extended sample of extreme Population A quasars i.e., {for highly accreting quasars}, capitalizing on the recent developments by \citet[][hereafter \citetalias{sniegowskaetal21}]{sniegowskaetal21}. At variance with previous work, \citetalias{sniegowskaetal21} relied on (1) the decomposition of the emission line profiles in a broad component, symmetric at rest frame,  and a BLUE component associated with the prominent {blue-shifted} excess observed in high-ionization lines of these sources \citep[e.g.,][]{sulenticetal17,martinez-aldamaetal18}, {and} (2) the use of diagnostic intensity ratios involving metal lines and \heii. 
 
 In the present work, we extend the method of \citetalias{sniegowskaetal21} to a larger sample of xA quasars (Sect. \ref{sec:sample}). In Sect. \ref{sec:methods}, we describe the methodology used. Results on {the} 42 objects are provided in Section \ref{sec:results}. They include the measurements on the fit of the emission blends (Section \ref{sec:immediate}), along with the estimates of the metal content for fixed physical conditions and for ionization and density free to vary. The discussion of the results (Section \ref{Discussion}) is mainly focused on the  test of  the stability of the method with two fitting techniques and of one of the main results of \citetalias{sniegowskaetal21}, namely that metallicity values derived from the diagnostic ratios are not consistent with scaled solar abundance ratios.

%%%%%%%%%%%%%%%%%%%%%%%%%%%%%%%%%%%%%%%%%%%%%%%%%%%%%%%%%%%%%%%%%%%%%%%%%%%%%%%%%%

\section{Sample}
\label{sec:sample}

\subsection{Sample definition}

The xA sources considered in this study have rest-frame UV spectra from SDSS DR12  with   coverage of 1000-3000  \AA \ where we can find: (1) the strongest high ionization lines (HILs) associated with resonance transitions (i.e. \siivfull{}, \civfull{}), (2) usually weak low ionization lines   (LILs; e.g. \siii{}) and (3) several inter-mediate ionization lines (IILs) from transitions leading to the ground state (e.g. \aliiifull{}, and the intercombination lines \siiiifull{}, \ciiifull{}). For some of the spectra the blend \lya\ + \nvfull{}\  is present. As in the case of \citetalias{sniegowskaetal21}, measurements were discarded because of the strong absorptions alongside the emissions that make it difficult to interpret the blend. 

The initial sample  consisted of 526 objects with a redshift range centered at ~\mbox{$z = 1.5 - 2.2$} from SDSS-DR12 to ensure moderate-to-high S/N in the continua. Based on a  semi-automatic selection using the \aliii/ \siiii\ flux ratio as a high accretion rate indicator \citepalias{marzianisulentic14}, the original sample was reduced to 120 objects. Subsequently we restricted the redshift range, centering it at $z \approx 2.2$, so it would be possible to observe these objects in the H band and to obtain observations of H$\beta$\ in the infrared. This will give us the possibility of an eventual comparison of the indicators of accretion at different regions of the electromagnetic spectrum  using different indicators. Finally, this {restricts} our sample to the 42 objects that are used in this work,  all type-1 quasars, with 6 of them classified as Broad Absorption Line Quasars (BAL QSOs).

In summary, the 42 SDSS sources in this study are bright quasars ($r<19$) with S/N  from 10 up to 50 around 1400 \AA. They are located between $-11^{\circ}   < \delta< 10^{\circ}$ under a redshift coverage of $2.13<z<2.42$. Each spectrum is similar to {the other}, except for a few outliers which will be analyzed in Section \ref{Discussion}.

%%%%%%%%%%%%%%%%%%%%%%%%%%%%%%%%%%%%%%%%%%%%%%%%%%%%%%%%%%%%%%%%%%%%%%%%%%%%%%%%%%
\subsection{Sample properties}
\begin{table*}[htpb] %26-ago
\centering
\caption{Source identification and basic properties}
\begin{tabular}{rccccccccc}
\hline\hline
\multicolumn{1}{c}{SDSS FULL JCODE}  & \multicolumn{1}{c}{$z_\mathrm{SDSS}$} & \multicolumn{1}{c}{$\delta z$} &  \multicolumn{1}{c}{${g}$} & \multicolumn{1}{c}{$g-r$}   &   \multicolumn{1}{c}{$f_{1700}$} & \multicolumn{1}{c}{$f_{1350}$}     & \multicolumn{1}{c}{S/N } & \multicolumn{1}{c}{Notes}  \\
\multicolumn{1}{c}{(1)}& \multicolumn{1}{c}{(2)}& \multicolumn{1}{c}{(3)}&\multicolumn{1}{c}{(4)}& \multicolumn{1}{c}{(5)}& \multicolumn{1}{c}{(6)}&\multicolumn{1}{c}{(7)}& \multicolumn{1}{c}{(8)}& \multicolumn{1}{c}{(9)} 
\\
\hline \\
J002023.12+074041.1	&	2.4172	&	0.0047		&	19.01	&	0.19		&	293	&	424	&	21	&		\\
J003411.37$-$032618.2	&	2.1771	&	0.0049		&	19.01	&	0.19		&	206	&	412	&	25	&		\\
J003751.90$-$023845.2	&	2.2734	&	-		&	18.92	&	0.21		&	146	&	351	&	16	&		\\
J010328.71$-$110414.4	&	2.1912	&	-		&	18.31	&	0.31		&	689	&	1016	&	20	&		\\
J010657.94$-$085500.1	&	2.3550	&	0.0054		&	18.18	&	0.11		&	665	&	972	&	20	&	\citetalias{sniegowskaetal21}	\\
J012303.46+032900.1	&	2.3328	&	0.0032		&	18.35	&	0.25		&	666	&	969	&	29	&		\\
J021039.51$-$082303.6	&	2.1898	&	0.0043		&	19.04	&	0.34		&	393	&	551	&	12	&		\\
J021606.40+011509.5	&	2.2294	&	-0.0019	\ 	&	18.82	&	0.52		&	369	&	516	&	28	&	BALQ	\\
J025255.65$-$042022.8	&	2.1309	&	-0.0033	\	&	19.09	&	0.44		&	269	&	229	&	15	&	BALQ	\\
J082714.60+030616.1	&	2.2106	&	-		&	19.10	&	0.32		&	296	&	526	&	14	&		\\
J082936.30+080140.6	&	2.1891	&	0.0070		&	18.37	&	0.31		&	687	&	957	&	36	&	\citetalias{sniegowskaetal21}	\\
J083611.10+054806.1	&	2.2575	&	0.0035		&	18.87	&	0.22		&	357	&	549	&	22	&		\\
J084525.84+072222.3	&	2.2687	&	0.0196		&	18.20	&	0.33		&	647	&	975	&	26	&	\citetalias{sniegowskaetal21}	\\
J084719.12+094323.4	&	2.2947	&	0.0043		&	18.87	&	0.15		&	388	&	504	&	21	&	\citetalias{sniegowskaetal21}	\\
J085856.00+015219.4	&	2.1598	&	0.0044		&	17.92	&	0.26		&	701	&	1188	&	45	&	\citetalias{sniegowskaetal21}	\\
J090312.22+070832.4	&	2.2225	&	0.0085		&	18.93	&	0.26		&	387	&	503	&	30	&		\\
J091500.43$-$020228.5	&	2.2806	&	-		&	18.95	&	0.13		&	96	&	230	&	15	&		\\
J092641.41+013506.6	&	2.1809	&	0.0059		&	18.59	&	0.34		&	385	&	625	&	28	&	\citetalias{sniegowskaetal21}	\\
J092919.45+033303.4	&	2.2275	&	0.0048		&	18.90	&	0.20		&	266	&	395	&	19	&		\\
J093251.98+023727.0	&	2.1529	&	0.0049		&	18.32	&	0.29		&	537	&	684	&	36	&	BALQ	\\
J094637.83$-$012411.5	&	2.2125	&	0.0017		&	18.56	&	0.18		&	400	&	612	&	40	&	\citetalias{sniegowskaetal21}	\\
J101341.85+085126.1	&	2.2526	&	0.0100		&	17.95	&	0.63		&	1272	&	1330	&	40	&	BALQ	\\
J102421.32+024520.2	&	2.3195	&	0.0081		&	18.49	&	0.18		&	460	&	695	&	34	&	\citetalias{sniegowskaetal21}	\\
J102606.67+011459.0	&	2.2596	&	0.0071		&	18.98	&	0.20		&	440	&	543	&	24	&	\citetalias{sniegowskaetal21}	\\
J114557.84+080029.0	&	2.3389	&	0.0092		&	18.55	&	0.37		&	250	&	357	&	13	&	\citetalias{sniegowskaetal21}	\\
J120550.19+020131.5	&	2.1548	&	0.0022		&	17.46	&	0.41		&	1635	&	1840	&	45	&	BALQ	\\
J121423.01+024252.8	&	2.1894	&	0.0129		&	17.64	&	0.23		&	1100	&	1450	&	35	&		\\
J121506.00+032642.6	&	2.2240	&	0.0029		&	18.49	&	0.24		&	683	&	993	&	31	&		\\
J121906.02+025433.5	&	2.2976	&	-0.0066	\	&	18.46	&	0.14		&	615	&	869	&	30	&		\\
J123120.55+072552.6	&	2.3872	&	-		&	18.01	&	-0.08	\	&	658	&	1039	&	45	&		\\
J124409.88+082155.2	&	2.3569	&	0.0065		&	18.58	&	0.18		&	614	&	877	&	34	&		\\
J125934.29+075200.7	&	2.4188	&	0.0017		&	17.92	&	0.16		&	862	&	1354	&	29	&		\\
J131452.64+092735.3	&	2.2367	&	0.0098		&	18.85	&	0.27		&	439	&	624	&	18	&		\\
J141925.48+074953.5	&	2.3837	&	0.0030		&	17.51	&	0.15		&	997	&	1495	&	50	&		\\
J150959.16+074450.1	&	2.2529	&	0.0099		&	18.94	&	0.28		&	253	&	342	&	17	&	\citetalias{sniegowskaetal21}	\\
J151636.79+002940.4	&	2.2659	&	-0.0057	\	&	18.77	&	0.96		&	320	&	238	&	25	&	BALQ \\
J151929.45+072328.7	&	2.3942	&	0.0079		&	18.66	&	0.17		&	384	&	502	&	21	&	\citetalias{sniegowskaetal21}	\\
J154503.23+015614.7	&	2.1607	&	0.0122		&	18.25	&	0.35		&	813	&	1107	&	24	&		\\
J160955.41+065401.9	&	2.1413	&	0.0074		&	18.44	&	0.33		&	697	&	925	&	28	&		\\
J161801.70+070450.3	&	2.2324	&	0.0032		&	18.48	&	0.26		&	490	&	674	&	26	&		\\
J211651.48+044123.7	&	2.3517	&	-		&	18.82	&	0.22		&	410	&	565	&	43	&	\citetalias{sniegowskaetal21}	\\
J214502.56$-$075805.6	&	2.1455	&	0.0025		&	18.58	&	0.22		&	540	&	744	&	17	&		\\
\\ \hline																			
$\mu$	&	2.2552	&	0.0052		&	18.53	&	0.27		&	542	&	756	&	27	&		\\
$\sigma$	&	0.0623	&	0.0025		&	0.29	&	0.07		&	159	&	236	&	7	&		\\
SIQR	&	0.0817	&	0.0050		&	0.43	&	0.16		&	301	&	373	&	10	&		\\

 \hline
 \end{tabular} 
\tablefoot{(1) SDSS coordinate name;   (2) SDSS redshift; (3) correction to {the} redshift estimated in the present work ($\delta z $ = $z - z_\mathrm{SDSS}$); (4) $g$-band magnitude from SDSS Sky Server; (5) color index $g-r$; (6) continuum flux measured at 1700  \AA\ in units of 10$^{-17}$ erg s$^{-1}$ cm$^{-2}$  \AA$^{-1}$; (7) continuum flux measured at 1350  \AA\ in the same units; (8) S/N measured at continuum level at 1450  \AA; (9) additional notes per source. 
Last three rows shows the mean values ($\mu$), the standard deviation ($\sigma$), and the semi inter-quartile range (SIQR) {for each column}.}
\label{tab:general}
 \end{table*}
 
 \begin{table*}[h]%\tabletypesize{\scriptsize}
\centering
{\fontsize{9.5}{10}\selectfont %\tabcolsep=3pt
\caption{Measurements in the 1900 \AA\ blend region.} 
\begin{tabular}{cccccccccc}
\hline\hline
SDSS JCODE   & \aliii & BC & \aliii  & \siiii & \multicolumn{2}{c}{\ciii}  & \multicolumn{2}{c}{\feiii}   \\ 
\cline{6-7}\cline{8-9} 
 & $W$ & FWHM & Flux  &  Flux  &  FWHM  &  Flux  &  FWHM  &  Flux\\
(1) & (2) & (3)  & (4)  & (5) & (6) & (7) & (8) & (9)   \\
\hline \\
J0020+0740	&	6	&	$	3340	\pm	1040	$	&	$	4.7	\pm	1.4	$	&	$	5.6	\pm	1.7	$	&	$	2920	\pm	1080	$	&	$	3.0	\pm	0.9	$	&	$				$	&	$				$	\\
J0034$-$0326	&	4	&	$ 2930	\pm	690 	$ \ \	&	$	3.0	\pm	0.7	$	&	$	2.6	\pm	0.7	$	&	$	2070	\pm	830	$ \ \	&	$	1.0	\pm	0.3	$	&	$				$	&	$				$	\\
J0037$-$0238	&	6	&	$	3060	\pm	960	$ \ \	&	$	2.7	\pm	0.8	$	&	$	2.3	\pm	0.7	$	&	$	3210	\pm	1180	$	&	$	2.1	\pm	0.6	$	&	$				$	&	$				$	\\
J0103$-$1104	&	7	&	$	2600	\pm	930	$ \ \	&	$	2.2	\pm	0.8	$	&	$	2.5	\pm	0.7	$	&	$	2920	\pm	1240	$	&	$	4.2	\pm	1.6	$	&	$	2950	\pm	1200	$	&	$	2.1	\pm	0.8	$	\\
J0106$-$0855	&	7	&	$	3210	\pm	910	$ \ \	&	$	3.3	\pm	1.0	$	&	$	4.8	\pm	1.0	$	&	$	2270	\pm	950	$ \ \	&	$	1.1	\pm	0.5	$	&	$	2660	\pm	1060	$	&	$	1.3	\pm	0.5	$	\\
J0123+0329	&	7	&	$	3120	\pm	1180	$	&	$	3.1	\pm	1.1	$	&	$	2.5	\pm	0.7	$	&	$	3060	\pm	1190	$	&	$	1.3	\pm	0.4	$	&	$				$	&	$				$	\\
J0210$-$0823	&	10	&	$	3080	\pm	1340	$	&	$	3.6	\pm	1.5	$	&	$	2.4	\pm	1.0	$	&	$	2980	\pm	1240	$	&	$	2.0	\pm	0.8	$	&	$				$	&	$				$	\\
\ \ J0216+0115*	&	8	&	$	3020	\pm	760	$ \ \	&	$	5.2	\pm	1.5	$	&	$	3.3	\pm	0.8	$	&	$	2820	\pm	1190	$	&	$	1.3	\pm	0.4	$	&	$	2560	\pm	1100	$	&	$	2.4	\pm	0.9	$	\\
\ \ J0252$-$0420*	&	5	&	$	3160	\pm	1370	$	&	$	6.5	\pm	2.6	$	&	$	6.5	\pm	2.6	$	&	$	2270	\pm	960	$ \ \	&	$	5.3	\pm	2.1	$	&	$				$	&	$				$	\\
J0827+0306	&	6	&	$	2720	\pm	1180	$	&	$	2.2	\pm	0.9	$	&	$	2.0	\pm	0.8	$	&	$	2220	\pm	940	$ \ \	&	$	0.9	\pm	0.4	$	&	$				$	&	$				$	\\
J0829+0801	&	8	&	$	3340	\pm	900	$	\ \ &	$	4.3	\pm	1.3	$	&	$	3.3	\pm	0.9	$	&	$	2750	\pm	1080	$	&	$	1.2	\pm	0.4	$	&	$				$	&	$				$	\\
J0836+0548	&	12	&	$	3900	\pm	1220	$	&	$	7.2	\pm	2.2	$	&	$	5.2	\pm	1.6	$	&	$	3150	\pm	1160	$	&	$	2.7	\pm	0.8	$	&	$				$	&	$				$	\\
J0845+0722	&	5	&	$	3340	\pm	1460	$	&	$	3.9	\pm	1.6	$	&	$	3.3	\pm	1.4	$	&	$	2430	\pm	1010	$	&	$	1.3	\pm	0.5	$	&	$				$	&	$				$	\\
J0847+0943	&	9	&	$	3440	\pm	1490	$	&	$	5.5	\pm	2.2	$	&	$	4.6	\pm	1.8	$	&	$	2330	\pm	990	$ \ \	&	$	1.9	\pm	0.7	$	&	$				$	&	$				$	\\
J0858+0152	&	6	&	$	3000	\pm	350	$	\ \ &	$	2.9	\pm	0.4	$	&	$	3.5	\pm	0.5	$	&	$	2330	\pm	780	$ \ \	&	$	1.5	\pm	0.4	$	&	$	2690	\pm	1090	$	&	$	0.9	\pm	0.2	$	\\
J0903+0708	&	11	&	$	3710	\pm	1460	$	&	$	3.0	\pm	1.0	$	&	$	2.2	\pm	0.8	$	&	$	2310	\pm	990	$ \ \	&	$	0.7	\pm	0.3	$	&	$				$	&	$				$	\\
J0915$-$0202	&	8	&	$	2190	\pm	920	$ \ \	&	$	1.9	\pm	0.7	$	&	$	1.6	\pm	0.6	$	&	$	2200	\pm	920	$\ \	&	$	1.0	\pm	0.3	$	&	$	2120	\pm	910	$ \ \	&	$	0.9	\pm	0.3	$	\\
J0926+0135	&	9	&	$	3770	\pm	880	$	\ \ &	$	3.1	\pm	0.7	$	&	$	4.9	\pm	1.2	$	&	$	3060	\pm	1220	$	&	$	1.9	\pm	0.5	$	&	$				$	&	$				$	\\
J0929+0333	&	7	&	$	3240	\pm	1010	$	&	$	6.0	\pm	1.8	$	&	$	6.0	\pm	1.9	$	&	$	2760	\pm	1010	$	&	$	4.1	\pm	1.2	$	&	$				$	&	$				$	\\
\ \ J0932+0237*	&	7	&	$	3900	\pm	710	$ \ \	&	$	7.3	\pm	1.4	$	&	$	7.4	\pm	1.5	$	&	$	3700	\pm	1390	$	&	$	5.5	\pm	1.3	$	&	$				$	&	$				$	\\
J0946$-$0124	&	11	&	$	1970	\pm	600	$ \ \	&	$	2.7	\pm	0.8	$	&	$	2.9	\pm	0.7	$	&	$	2210	\pm	680	$ \ \	&	$	3.2	\pm	0.7	$	&	$	2470	\pm	910	$ \ \	&	$	4.2	\pm	1.3	$	\\
\ \ J1013+0851*	&	10	&	$	3860	\pm	1090	$	&	$	9.2	\pm	2.7	$	&	$	7.6	\pm	1.6	$	&	$	2750	\pm	1150	$	&	$	2.3	\pm	0.9	$	&	$	3400	\pm	1350	$	&	$	4.2	\pm	1.5	$	\\
J1024+0245	&	6	&	$	3150	\pm	850	$	\ \ &	$	3.9	\pm	1.1	$	&	$	4.0	\pm	1.1	$	&	$	2820	\pm	1100	$	&	$	2.2	\pm	0.6	$	&	$				$	&	$				$	\\
J1026+0114	&	6	&	$	3000	\pm	850	$	\ \ &	$	5.2	\pm	1.5	$	&	$	3.9	\pm	0.8	$	&	$	2290	\pm	960	$ \ \	&	$	1.6	\pm	0.7	$	&	$				$	&	$				$	\\
J1145+0800	&	9	&	$	2990	\pm	1300	$	&	$	5.2	\pm	2.2	$	&	$	4.5	\pm	1.9	$	&	$	2480	\pm	1030	$	&	$	1.8	\pm	0.7	$	&	$				$	&	$				$	\\
\ \ J1205+0201*	&	8	&	$	4270	\pm	1160	$	&	$	5.3	\pm	1.1	$	&	$	4.4	\pm	1.1	$	&	$	3690	\pm	1470	$	&	$	2.4	\pm	0.7	$	&	$				$	&	$				$	\\
J1214+0242	&	6	&	$	3900	\pm	1050	$	&	$	4.5	\pm	1.3	$	&	$	4.7	\pm	1.3	$	&	$	3670	\pm	1440	$	&	$	2.9	\pm	0.8	$	&	$				$	&	$				$	\\
J1215+0326	&	5	&	$	2910	\pm	730	$	\ \ &	$	3.3	\pm	0.9	$	&	$	4.4	\pm	1.0	$	&	$	2930	\pm	1240	$	&	$	2.2	\pm	0.7	$	&	$	2560	\pm	1100	$	&	$	2.5	\pm	0.9	$	\\
J1219+0254	&	8	&	$	3880	\pm	970	$	\ \ &	$	4.9	\pm	1.4	$	&	$	7.0	\pm	1.6	$	&	$	3220	\pm	1360	$	&	$	2.7	\pm	0.9	$	&	$	2780	\pm	1190	$	&	$	2.6	\pm	0.9	$	\\
J1231+0725	&	5	&	$	2000	\pm	440	$	\ \ &	$	2.3	\pm	0.5	$	&	$	2.7	\pm	0.6	$	&	$	2110	\pm	670	$ \ \	&	$	3.9	\pm	0.9	$	&	$	2150	\pm	830	$ \ \	&	$	3.2	\pm	1.0	$	\\
J1244+0821	&	8	&	$	3530	\pm	950	$	\ \ &	$	4.0	\pm	1.2	$	&	$	4.3	\pm	1.2	$	&	$	2590	\pm	1020	$	&	$	2.0	\pm	0.6	$	&	$				$	&	$				$	\\
J1259+0752	&	6	&	$	3190	\pm	1320	$	&	$	1.7	\pm	0.7	$	&	$	1.3	\pm	0.6	$	&	$	2550	\pm	1040	$	&	$	0.4	\pm	0.2	$	&	$				$	&	$				$	\\
J1314+0927	&	5	&	$	3020	\pm	1310	$	&	$	2.4	\pm	1.0	$	&	$	2.7	\pm	1.1	$	&	$	2930	\pm	1190	$	&	$	1.7	\pm	0.7	$	&	$				$	&	$				$	\\
J1419+0749	&	6	&	$	3340	\pm	860	$	\ \ &	$	2.7	\pm	0.8	$	&	$	3.8	\pm	0.9	$	&	$	2980	\pm	1060	$	&	$	2.3	\pm	0.6	$	&	$				$	&	$				$	\\
J1509+0744	&	9	&	$	3300	\pm	1430	$	&	$	3.8	\pm	1.6	$	&	$	5.0	\pm	2.0	$	&	$	2210	\pm	940	$ \ \	&	$	1.3	\pm	0.5	$	&	$				$	&	$				$	\\
\ \ J1516+0029*	&	8	&	$	2790	\pm	700	$ \ \ &	$	12.9	\pm	3.7	$ \ \	&	$	6.4	\pm	1.5	$	&	$	2730	\pm	1150	$	&	$	3.5	\pm	1.2	$	&	$	3100	\pm	1330	$	&	$	4.5	\pm	1.6	$	\\
J1519+0723	&	5	&	$	3230	\pm	1010	$	&	$	5.2	\pm	1.6	$	&	$	6.2	\pm	1.9	$	&	$	2130	\pm	780	$ \ \	&	$	3.0	\pm	0.9	$	&	$				$	&	$				$	\\
J1545+0156	&	7	&	$	3500	\pm	1330	$	&	$	4.5	\pm	1.6	$	&	$	3.7	\pm	1.1	$	&	$	3090	\pm	1200	$	&	$	1.6	\pm	0.5	$	&	$				$	&	$				$	\\
J1609+0654	&	3	&	$	3750	\pm	1550	$	&	$	4.1	\pm	1.6	$	&	$	3.3	\pm	1.4	$	&	$	2470	\pm	1000	$	&	$	0.9	\pm	0.4	$	&	$				$	&	$				$	\\
J1618+0704	&	6	&	$	3390	\pm	1190	$	&	$	3.0	\pm	0.9	$	&	$	1.6	\pm	0.5	$	&	$	2330	\pm	970	$ \ \	&	$	1.0	\pm	0.4	$	&	$				$	&	$				$	\\
J2116+0441	&	6	&	$	3160	\pm	810	$	\ \ &	$	3.0	\pm	0.8	$	&	$	3.6	\pm	0.9	$	&	$	2420	\pm	860	$ \ \	&	$	2.5	\pm	0.6	$	&	$				$	&	$				$	\\
J2145$-$0758	&	6	&	$	2910	\pm	1040	$	&	$	3.8	\pm	1.4	$	&	$	6.2	\pm	1.8	$	&	$	2300	\pm	970	$ \ \	&	$	2.5	\pm	0.9	$	&	$	2660	\pm	1080	$	&	$	2.5	\pm	1.0	$	\\
\\ \hline																																													
$\mu$	&	7	&	$	3200	\pm	1010	$	&	$	3.85	\pm	1.3	$	&	$	3.9	\pm	1.1	$	&	$	2660	\pm	1035	$	&	$	2.0	\pm	0.7	$	&	$	2660	\pm	1095	$	&	$	2.5	\pm	0.9	$	\\
$\sigma$	&	2	&	$	493	\pm	288	$	&	$	2.10	\pm	0.7	$	&	$	1.7	\pm	0.5	$	&	$	443	\pm	186	$	&	$	1.2	\pm	0.4	$	&	$	363	\pm	160	$	&	$	1.2	\pm	0.4	$	\\
SIQR	&	1	&	$	243	\pm	214	$	&	$	1.10	\pm	0.4	$	&	$	1.1	\pm	0.4	$	&	$	333	\pm	115	$	&	$	0.7	\pm	0.2	$	&	$	143	\pm	85	$ \ \	&	$	0.8	\pm	0.2	$	\\

\hline  
\end{tabular}
    \tablefoot{Equivalent widths {($W$)} are in units of \AA. 
    Rest-frame FWHM are in units of \kms. 
    Rest-frame line flux normalized to continuum flux at 1350 \AA, in units of  \ergss\ cm$^{-2}$. 
    BALQ are identified with an asterisk (*), their spectra {are shown in} Appendix \ref{app:spec}.}
\label{tab:aliii}}
\end{table*}

%%%%%%%%%%%%%%%%%%%%%%%%%%%%%%%%%%%%%%%%%%%%%%%%%%%%%%%%%%%%%%%%%%%%%%%%%%%%%%%%%
 \begin{figure*}
    \centering
     \includegraphics[trim= 0.0 0. 0. 0., clip, width=\textwidth]{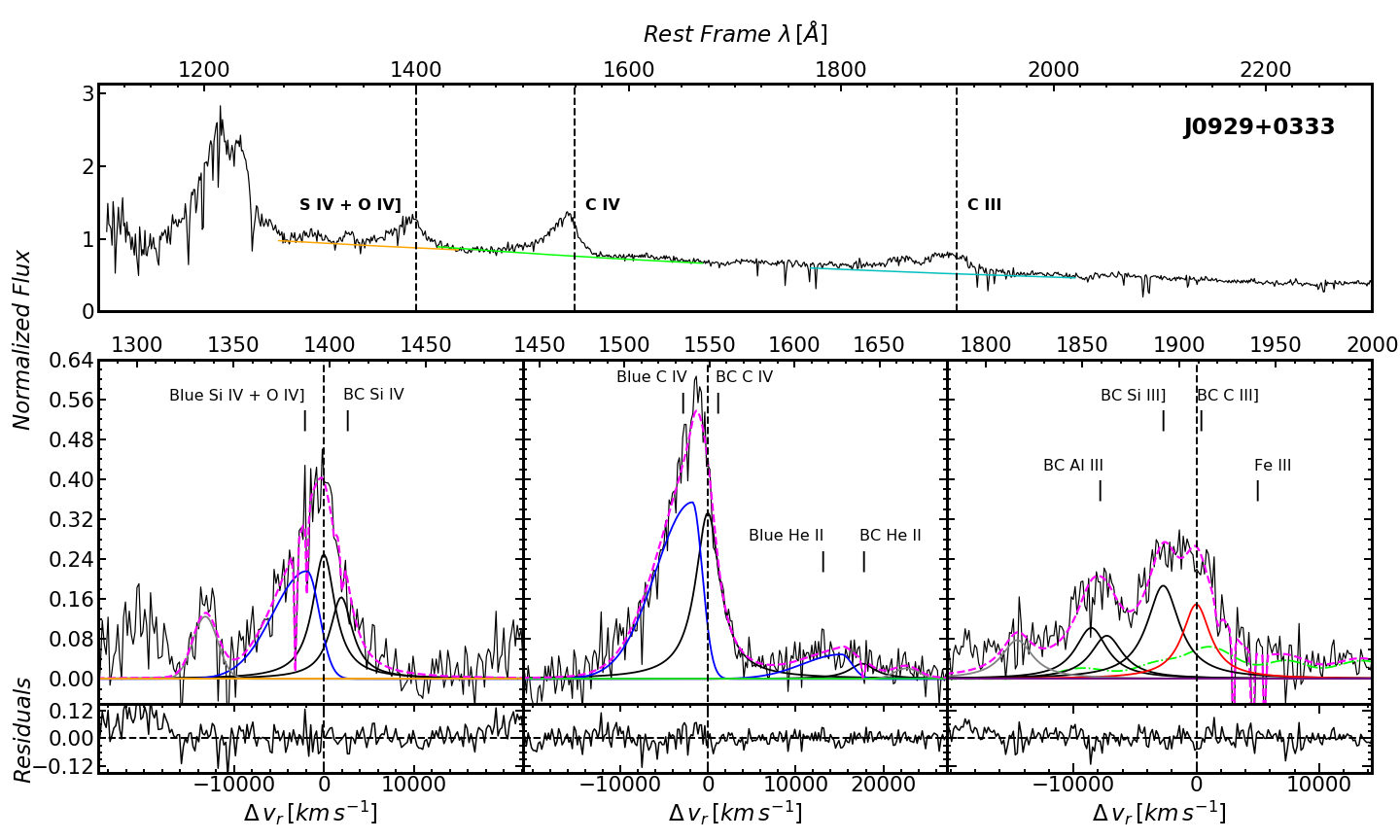}
    \caption{\object{SDSSJ092919.45+033303.4, with {\tt specfit} analysis results. For each spectrum, the top panel shows the full  spectrum  as a function of rest-frame  wavelength, after normalization at 1350 \AA. The actual continuum ranges employed in the fitting of the three spectral regions discussed in the paper are marked. The bottom panels show the decomposition of the blends. Black lines: observed spectrum under continuum subtraction. Magenta dashed lines: full model of the observed spectrum.  Thick black lines: BC of \siiv, \civ, \heii, \aliii\ and \siiii. Blue lines: {blue-shifted} components of \siiv+\oiv, \civ\ and \heii. Grey lines: faint lines affecting the blends or \aliii\ BLUE component.}}  
    \label{fig:example_fit}
 \end{figure*}

\begin{figure*}[t]%3
     \centering
     %trim=left bottom right top
     \includegraphics[trim= 0.5 0.5 0.5 0.5, clip, width=\textwidth]{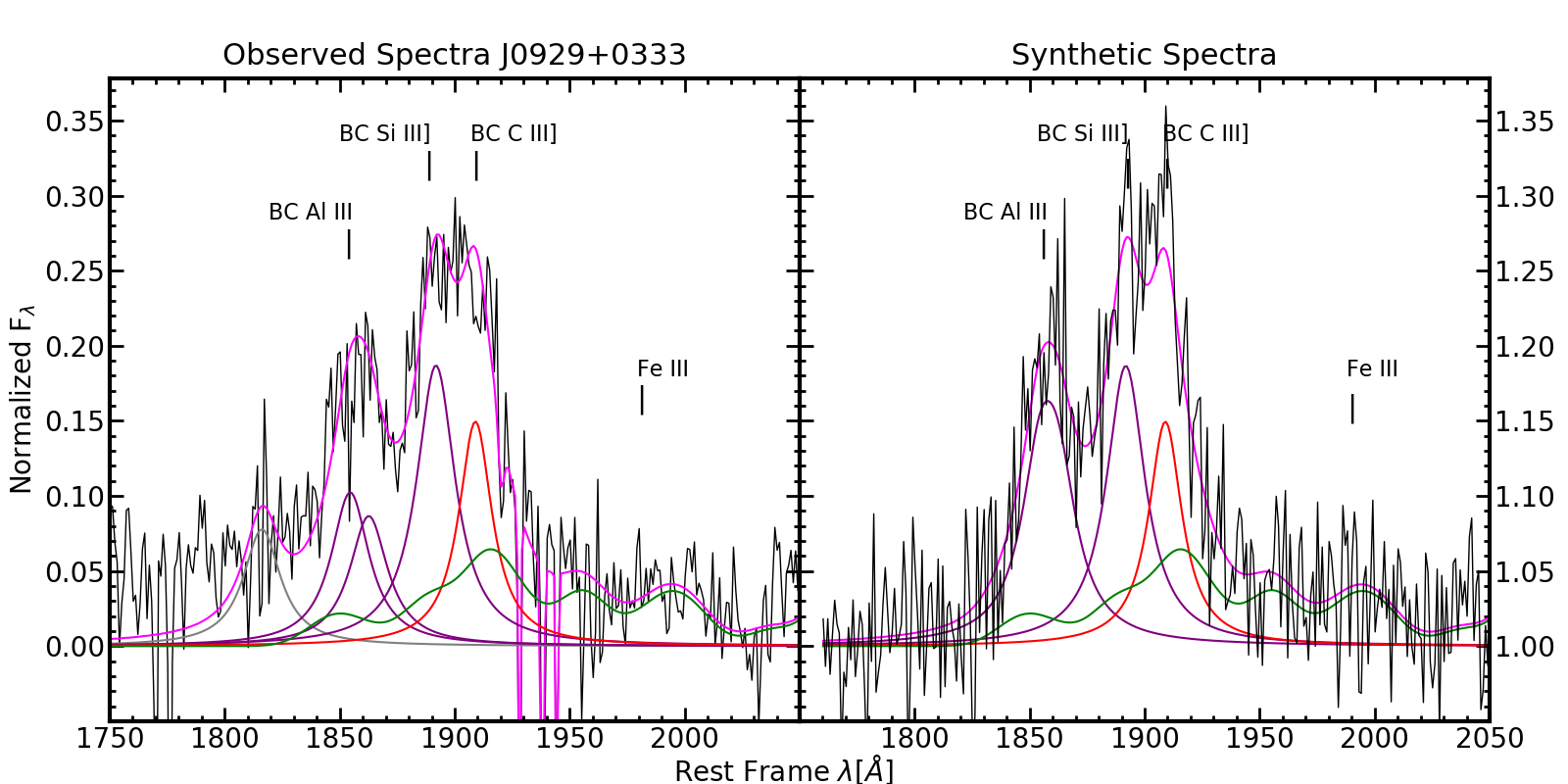}
    \caption{Observed and synthetic spectra of J0929+0333. The abscissa corresponds to the rest frame wavelength in \AA; the ordinate is {the flux normalized to the continuum at 1350 \AA}. \textit{Left:}  Black line:  observed spectra minus the continuum; pink: {model fit using  {\tt specfit}}. The BC emission of \aliii \ and \siiii\  are shown in purple, \ciii\ emission is {shown} in red, and the \feiii{} template {is shown} in green. \textit{Right:} Black line: the synthetic spectrum composed {of} the sum of the emission line components plus noise. The BC emission of \aliii\ (the sum of the doublet) and \siiii\  are again shown in purple, the \ciii\ emission {and the} \feiii{} template {are shown in red and green, respectively.}}   
    \label{fig:error}
 \end{figure*}
 %%%%%%%%%%%%%%%%%%%%%%%%%%%%%%%%%%%%%%%%%%%%%%%%%%%%%%%%%%%%%%%%%%%%%%%%%%%%%%%%%%%%
\begin{table}[h]
\hspace{-2cm}
\centering
{\fontsize{9.5}{10}\selectfont %\tabcolsep=3pt
\caption{Additional measurements in the $\lambda1900$\ region.}
\begin{tabular}{ccccc}\hline\hline
{SDSS JCODE} & FWHM & BLUE \aliii & $\Delta \lambda$ & skew\\
  &   &  Flux &  &    \\
        (1) & {(2)} & (3) & (4)  & (5)   \\
\hline \\
J0106$-$0855	&	$	2640	\pm	940	$ \ \	&	$	0.60	\pm	0.20	$	&	$	14.8	\pm	4.5	$	&	$	0.80	$	\\
J0903+0708	&	$	2680	\pm	990	$ \ \	&	$	0.80	\pm	0.30	$	&	$	11.1	\pm	5.5	$	&	$	0.70	$	\\
\ \ J1013+0851*	&	$	3280	\pm	1160	$	&	$	1.70	\pm	0.70	$	&	$	12.9	\pm	4.5	$	&	$	0.80	$	\\
J1026+0114	&	$	3040	\pm	1080	$	&	$	0.80	\pm	0.30	$	& \ \	$	8.9	\pm	4.6	$	&	$	0.30	$	\\
\ \ J1205+0201*	&	$	4290	\pm	1520	$	&	$	1.00	\pm	0.30	$	&	$	11.8	\pm	3.6	$	&	$	0.40	$	\\
J1618+0704	&	$	2880	\pm	610	$ \ \	&	$	0.50	\pm	0.20	$	&	\ \ $	9.2	\pm	5.5	$	&	$	0.20	$	\\
\\ \hline																							
$\mu$	&	$	2960	\pm	1035	$	&	$	0.80	\pm	0.30	$	&	$	11.5	\pm	4.6	$ \ \	&	$	0.55	$	\\
$\sigma$	&	$	613	\pm	298	$	&	$	0.43	\pm	0.19	$	&	$	2.2	\pm	0.7	$	&	$	0.27	$	\\
SIQR	&	$	245	\pm	94	$ \ \	&	$	0.15	\pm	0.04	$	&	$	1.5	\pm	0.4	$	&	$	0.23	$	\\
\hline
\end{tabular}
\tablefoot{(1) SDSS short name; (2)   {FWHM of }\aliii \ asymmetric emission in  km s$^{-1}$; (3) normalized flux at 1350 \AA ; (4) separation of the peak of the BLUE from the rest-frame \aliii \ emission in  \AA; (5){asymmetry reported by {\tt specfit}}. {BALQs are marked with an asterisk}(*).}
\label{tab:BlueAliii}}
\end{table}

The basic properties of our sample are presented in Table \ref{tab:general}: (1) SDSS name, %(2) SDSS short name, 
(2) redshift from the SDSS, (3) the difference between our redshift estimation using Mg II and the SDSS redshift $\delta z = z - z_{SDSS}$, (4) the $g$-band magnitude provided by SDSS Sky Server, (5) the $g - r$\ color index, (6) and (7) are the rest-frame specific continuum flux at 1700 \AA\ and 1350 \AA\ respectively, (8) the ~\mbox{S/N $\sim$ 25} at 1450 \AA, and (9) indicates the BAL QSOs and the objects from the S21 sample. Fig. \ref{fig:Lyke} and Table \ref{tab:general} indicate that the sources of our sample are very luminous, {i.e., they are at} the high-end of the luminosity distribution of BOSS sources at redshift $z \approx 2$, and of rather homogeneous properties. As a consequence, we may not expect strong correlations with physical properties such as $L$, \mbh, \lledd. The sample is best suited to confirm the main results seen by \citetalias{sniegowskaetal21}, and to improve their statistical significance. 

In addition, the scatter in the color index SIQR (semi interquartile range) $g-r \approx 0.07$\ with a median $g-r \approx 0.27$ suggests that the sample is not suffering strong extinction save \object{J151636.79+002940.4} ($g-r \approx 1$) and other sources with $g-r \approx 0.4 - 0.5$ that are all BAL QSOs {(BALQ)} and were not considered {in} the analysis.

 %%%%%%%%%%%%%%%%%%%%%%%%%%%%%%%%%%%%%%% SECTION 3 %%%%%%%%%%%%%%%%%%%%%%%%%%%%%%%%%%%

\section{Methods} 
\label{sec:methods}

This is {the second paper in the series (the first being \citetalias{sniegowskaetal21})} with an extended sample, and therefore we will follow the same basic {methodology}: (1) multi-component analysis of the emission blends in the UV spectra, (2)  {\tt CLOUDY} photoionization modelling and (3) metallicity estimates of the sample by comparing observed fluxes and {\tt CLOUDY} synthetic fluxes at different $U$\ and \nh\ values. The 13 sources from \citetalias{sniegowskaetal21} are {also} presented in this article (identified in Column 9 of Table \ref{tab:general}) but they were re-processed and re-fitted. A systematic comparison between the new measurements and the ones of \citetalias{sniegowskaetal21} is carried out in Section \ref{s21}. 

%%%%%%%%%%%%%%%%%%%%%%%%%%%%%%%%%%%%%%%%%%%%%%%%%%%%%%%%%%%%%%%%%%%%%%%%%%%%%%%%%%
\subsection{Additional redshift determination}
\label{redshift}

xA sources are characterized by a strong asymmetric blue-shifted component mainly in the regions  centered at 1400 and 1550  \AA \ \citep[e.g.,][]{sulenticetal07,sulenticetal17,vietrietal18}.  These asymmetric emissions are frequently even stronger than the expected broad component emission.  The peak of the line profile might therefore appear shifted by several thousands \kms\ \citep[e.g.,][]{hewettwild10}. Even if the \aliiifull\ line is in most cases unshifted (within $\pm$ 200 \kms) with respect to the rest frame \citep{buendiarios2022}, an excess component is occasionally present on the blue wing of the \aliiifull\ doublet causing a wrong measurement of the rest frame \citep{martinez-aldamaetal17,marinelloetal20,marzianietal22}. {Hence we estimate the redshift from the   \mgii\ low-ionization emission line usually characterized by a symmetric line profile.}  Other spectral lines such as \siii\ or \oiuv\ are weak and are affected in some cases by low S/N or contaminated by absorptions lines.  \ciiifull\ and \siiiifull\ are heavily blended together. To improve the precision of the redshift estimate, we fitted a Gaussian to top half of the \mgii\ line using the IRAF \texttt{splot} task to measure its central wavelength.  Afterward we added a redshift correction to the original redshift from the Sloan when necessary using the IRAF \texttt{dopcor} task. These $z$ corrections were up to 0.019  with respect to the SDSS redshift  and affected more than two thirds of our sample (as shown in Column 3 of Table \ref{tab:general}) 
that guarantee us a better analysis of the data specially for the strong {blue-shifted} components at $\lambda$1400 and $\lambda$1550.  The \mgii\ emission may however present a blue displacement less than 200 -- 300 km s$^{-1}$\ \citep{marzianietal13} that  does not lead to any drastic misinterpretation of the line blends. We cross-checked the reshift obtained from \mgii\ with the one from  the \aliii\ doublet broad component, and found consistency with the two $z$ estimates.

\subsection{Line interpretation and diagnostic ratios}

As mentioned in Section \ref{sec:sample}, our spectral coverage is characterized by  {HILs, IILs and LILs}. These emissions need a favorable environment for their associated transitions to occur, and in turn they provide insight on the gas surrounding the ionization source. Population A sources are characterized by broad components described with symmetric profiles plus strong asymmetric {blue-shifted} components associated with high-ionization emission \citep{leighlymoore04,marzianietal10}, hence we decomposed our regions of interest mainly using two emission profiles:

\begin{itemize}
    \item The broad component (BC), a rest-frame emission modeled with a Lorentzian profile. This symmetric profile is believed to be associated with a virialized system {of clouds in the vicinity of the central supermassive black hole}.
    \item The {blue-shifted} component (BLUE),  a strong flux excess mainly in the blue wing of HILs that are resonance transitions, ultimately ascribed to high velocity winds. We modeled this emission with a very broad {skewed} Gaussian  profile as an interpretation of the high velocity winds in which these emissions occur. 
\end{itemize}

The profiles  of both BC {and} BLUE are constrained by the reduction to minimum $\chi^2$ on the {full extent} of the line.

%%%%%%%%%%%%%%%%%%%%%%%%%%%%%%%%%%%%%%%%%%%%%%%%%%%%%%%%%%%%%%%%%%%%%%%%%%%%%%%%%%
\subsubsection{Broad component}

The analysis {follows} the same procedure of \citetalias{sniegowskaetal21}, and relies mainly on the ratios \aliii, \civ, \siiv\ +\oiv\  over \heiiuv.  These ratios give us  more direct information about the metal content even if  \heiiuv\ is a weak emission mixed in the red wing of \civ. It might be underestimated and {thus} set an unrealistic upper limit for the metallicity. The issue will be discussed in Section \ref{specpr}. 

In addition, we considered the line ratios (\siiv+\oiv)/\civ, \aliii/\civ, and \aliii/\siiii. The first ratio is frequently computed in metallicity studies of the BLR \citep{nagaoetal06,shinetal13}. The second one is more related to the possibility of \aliii\ overabundance, and the third one is associated with the selection of xA sources and is strongly dependent, for a fixed {spectral energy distribution (SED)}, on both $Z$\ and \nh\ \citep{marzianietal20}. 
%%%%%%%%%%%%%%%%%%%%%%%%%%%%%%%%%%%%%%%%%%%%%%%%%%%%%%%%%%%%%%%%%%%%%%%%%%%%%%%%%%
\subsubsection{BLUE component}

\begin{table*}[h]
\centering
%\tabletypesize{\scriptsize\tabcolsep=2pt}
{\fontsize{9.5}{10}\selectfont %\tabcolsep=3pt
\caption{Measurements in the $\lambda1550$\ region. }
\begin{tabular}{ccccc | cccccccccccccc}\hline\hline
{SDSS JCODE} & 1550 & BC & \civ & \heiiuv{} &  BLUE &  \civ &  \heiiuv \\
&  W  & FWHM &  Flux  &  Flux & FWHM &  Flux & Flux & $\Delta \lambda$ & skew   \\
        (1) & (2) & (3) & (4)  & (5) &  (6)  & (7) & (8) & (9) & (10)    \\ 
\hline
& & & & & & & \\
J0020+0740	&	12	&	$	3528	\pm	742	$	&	$	\ \	9.7	\pm	1.8	$	&	$	1.5	\pm	0.4	$	&	$		13117	\pm	1514		$	&	$		12.4	\pm	1.5	$	&	$	2.9	\pm	0.8	$	&	$		13.6	\pm	1.8	$	&	0.28	\\
J0034$-$0326	&	7	&	$	2842	\pm	650	$	&	$	\ \	3.0	\pm	0.8	$	&	$	1.4	\pm	0.4	$	&	$	\ \ 	8091	\pm	1510		$	&	$	\ \	8.2	\pm	1.6	$	&	$	1.7	\pm	0.5	$	&	$		14.9	\pm	4.6	$	&	0.56	\\
J0037$-$0238	&	18	&	$	3800	\pm	800	$	&	$		12.8	\pm	2.4	$	&	$	2.1	\pm	0.6	$	&	$		10207	\pm	1178		$	&	$	\ \	6.9	\pm	0.8	$	&	$	1.8	\pm	0.5	$	&	$		14.5	\pm	1.8	$	&	0.28	\\
J0103$-$1104	&	7	&	$	3800	\pm	848	$	&	$	\ \	6.6	\pm	1.3	$	&	$	1.2	\pm	0.4	$	&	$		10841	\pm	1470		$	&	$	\ \	5.3	\pm	0.7	$	&	$	1.0	\pm	0.3	$	&	$		13.8	\pm	3.0	$	&	0.25	\\
J0106$-$0855	&	7	&	$	3021	\pm	674	$	&	$	\ \	3.3	\pm	0.7	$	&	$	1.1	\pm	0.3	$	&	$		10505	\pm	1425		$	&	$		10.9	\pm	1.5	$	&	$	1.9	\pm	0.5	$	&	$		17.3	\pm	3.0	$	&	0.45	\\
J0123+0329	&	4	&	$	3254	\pm	670	$	&	$	\ \	2.5	\pm	0.4	$	&	$	0.8	\pm	0.2	$	&	$		11305	\pm	1294		$	&	$	\ \	8.1	\pm	0.9	$	&	$	1.1	\pm	0.3	$	&	$		20.4	\pm	3.0	$	&	0.43	\\
J0210$-$0823	&	3	&	$	3388	\pm	760	$	&	$	\ \	2.7	\pm	0.7	$	&	$	0.8	\pm	0.2	$	&	$		12133	\pm	2168		$	&	$	\ \	6.4	\pm	1.7	$	&	$	1.6	\pm	0.5	$	&	$		26.5	\pm	5.1	$	&	0.48	\\
\ \ J0216+0115*	&	9	&	$	3260	\pm	746	$	&	$	\ \	7.2	\pm	2.0	$	&	$	1.1	\pm	0.3	$	&	$	\ \ 	6303	\pm	1176		$	&	$	\ \	5.3	\pm	1.0	$	&	$	0.9	\pm	0.3	$	&	$	\ \	9.0	\pm	4.6	$	&	0.45	\\
J0827+0306	&	6	&	$	3282	\pm	736	$	&	$	\ \	3.7	\pm	1.0	$	&	$	1.1	\pm	0.3	$	&	$		13751	\pm	2457		$	&	$	\ \	7.3	\pm	1.9	$	&	$	1.3	\pm	0.4	$	&	$		13.2	\pm	5.1	$	&	0.22	\\
J0829+0801	&	6	&	$	3191	\pm	578	$	&	$	\ \	4.2	\pm	0.5	$	&	$	0.9	\pm	0.2	$	&	$		11802	\pm	860	\ \ 	$	&	$		10.4	\pm	0.8	$	&	$	1.7	\pm	0.5	$	&	$		20.2	\pm	1.8	$	&	0.39	\\
J0836+0548	&	8	&	$	3401	\pm	759	$	&	$	\ \	7.1	\pm	1.4	$	&	$	1.0	\pm	0.3	$	&	$		11974	\pm	1624		$	&	$	\ \	6.6	\pm	0.9	$	&	$	1.6	\pm	0.5	$	&	$		12.2	\pm	3.0	$	&	0.15	\\
J0845+0722	&	5	&	$	3388	\pm	697	$	&	$	\ \	4.9	\pm	0.8	$	&	$	1.1	\pm	0.3	$	&	$		13961	\pm	1598		$	&	$	\ \	9.9	\pm	1.1	$	&	$	1.7	\pm	0.5	$	&	$		16.3	\pm	3.0	$	&	0.26	\\
J0847+0943	&	9	&	$	3399	\pm	715	$	&	$	\ \	7.6	\pm	1.4	$	&	$	1.3	\pm	0.4	$	&	$		12944	\pm	1494		$	&	$		12.6	\pm	1.5	$	&	$	2.7	\pm	0.7	$	&	$		10.9	\pm	1.8	$	&	0.17	\\
J0858+0152	&	8	&	$	3089	\pm	524	$	&	$	\ \	4.6	\pm	0.5	$	&	$	1.2	\pm	0.3	$	&	$		15923	\pm	1084		$	&	$		15.0	\pm	1.1	$	&	$	2.7	\pm	0.7	$	&	$		15.2	\pm	1.1	$	&	0.25	\\
J0903+0708	&	2	&	$	3266	\pm	748	$	&	$	\ \	2.3	\pm	0.6	$	&	$	1.1	\pm	0.3	$	&	$		14584	\pm	2721		$	&	$	\ \	5.7	\pm	1.1	$	&	$	1.1	\pm	0.3	$	&	$		19.2	\pm	4.6	$	&	0.37	\\
J0915$-$0202	&	10	&	$	3093	\pm	690	$	&	$	\ \	4.9	\pm	1.0	$	&	$	1.4	\pm	0.4	$	&	$		11479	\pm	1557		$	&	$	\ \	9.3	\pm	1.3	$	&	$	2.2	\pm	0.6	$	&	$		12.9	\pm	3.0	$	&	0.32	\\
J0926+0135	&	8	&	$	3400	\pm	602	$	&	$	\ \	7.7	\pm	1.1	$	&	$	1.4	\pm	0.4	$	&	$		11100	\pm	910	\ \ 	$	&	$	\ \	8.6	\pm	0.7	$	&	$	1.6	\pm	0.4	$	&	$		16.6	\pm	1.4	$	&	0.26	\\
J0929+0333	&	14	&	$	3395	\pm	714	$	&	$	\ \	9.2	\pm	2.0	$	&	$	0.9	\pm	0.3	$	&	$	\ \ 	9984	\pm	1374		$	&	$		12.2	\pm	1.8	$	&	$	1.8	\pm	0.5	$	&	$	\ \	9.0	\pm	2.8	$	&	0.26	\\
\ \ J0932+0237*	&	17	&	$	3356	\pm	266	$	&	$		11.8	\pm	0.7	$	&	$	1.9	\pm	0.5	$	&	$	\ \ 	8979	\pm	1246		$	&	$		18.1	\pm	2.4	$	&	$	3.0	\pm	0.8	$	&	$		10.3	\pm	5.1	$	&	0.35	\\
J0946$-$0124	&	14	&	$	3400	\pm	686	$	&	$		11.7	\pm	1.2	$	&	$	1.8	\pm	0.5	$	&	$	\ \ 	8928	\pm	607	\ \ 	$	&	$	\ \	5.6	\pm	0.5	$	&	$	1.7	\pm	0.5	$	&	$		12.4	\pm	1.6	$	&	0.40	\\
\ \ J1013+0851*	&	6	&	$	3399	\pm	576	$	&	$	\ \	6.3	\pm	0.7	$	&	$	2.4	\pm	0.7	$	&	$		10824	\pm	737	\ \ 	$	&	$		10.3	\pm	0.7	$	&	$	2.3	\pm	0.6	$	&	$		13.1	\pm	1.1	$	&	0.24	\\
J1024+0245	&	6	&	$	3331	\pm	686	$	&	$	\ \	5.2	\pm	0.8	$	&	$	0.8	\pm	0.2	$	&	$		11969	\pm	1370		$	&	$		10.6	\pm	1.2	$	&	$	2.2	\pm	0.6	$	&	$		17.5	\pm	3.0	$	&	0.38	\\
J1026+0114	&	6	&	$	3113	\pm	655	$	&	$	\ \	4.0	\pm	0.7	$	&	$	1.0	\pm	0.3	$	&	$		15812	\pm	1825		$	&	$		12.6	\pm	1.5	$	&	$	1.9	\pm	0.5	$	&	$		13.6	\pm	1.8	$	&	0.19	\\
J1145+0800	&	4	&	$	3011	\pm	675	$	&	$	\ \	4.4	\pm	1.2	$	&	$	0.9	\pm	0.3	$	&	$		11432	\pm	2043		$	&	$		11.2	\pm	2.9	$	&	$	1.9	\pm	0.5	$	&	$		15.9	\pm	5.1	$	&	0.36	\\
\ \ J1205+0201*	&	5	&	$	3155	\pm	589	$	&	$	\ \	3.6	\pm	0.4	$	&	$	0.9	\pm	0.3	$	&	$		13004	\pm	921	\ \ 	$	&	$	\ \	8.1	\pm	0.6	$	&	$	0.8	\pm	0.2	$	&	$		17.0	\pm	1.5	$	&	0.31	\\
J1214+0242	&	9	&	$	3400	\pm	616	$	&	$	\ \	6.9	\pm	0.9	$	&	$	1.1	\pm	0.3	$	&	$		12643	\pm	921	\ \ 	$	&	$		10.8	\pm	0.9	$	&	$	1.8	\pm	0.5	$	&	$		12.5	\pm	1.8	$	&	0.21	\\
J1215+0326	&	12	&	$	3414	\pm	604	$	&	$		10.1	\pm	1.4	$	&	$	1.3	\pm	0.4	$	&	$		10520	\pm	863	\ \ 	$	&	$		11.2	\pm	0.9	$	&	$	2.7	\pm	0.7	$	&	$		11.4	\pm	1.4	$	&	0.28	\\
J1219+0254	&	9	&	$	3500	\pm	672	$	&	$		15.7	\pm	2.1	$	&	$	2.1	\pm	0.6	$	&	$	\ \ 	9008	\pm	863	\ \ 	$	&	$	\ \	9.1	\pm	1.0	$	&	$	2.9	\pm	0.8	$	&	$		13.0	\pm	2.1	$	&	0.34	\\
J1231+0725	&	21	&	$	3496	\pm	277	$	&	$		18.2	\pm	1.1	$	&	$	2.6	\pm	0.8	$	&	$		12467	\pm	1730		$	&	$	\ \	8.0	\pm	1.0	$	&	$	2.6	\pm	0.7	$	&	$		12.0	\pm	5.1	$	&	0.66	\\
J1244+0821	&	8	&	$	3600	\pm	637	$	&	$	\ \	5.6	\pm	0.8	$	&	$	0.8	\pm	0.2	$	&	$		12083	\pm	991	\ \ 	$	&	$	\ \	9.5	\pm	0.8	$	&	$	2.1	\pm	0.6	$	&	$		16.1	\pm	1.4	$	&	0.32	\\
J1259+0752	&	3	&	$	3128	\pm	716	$	&	$	\ \	2.2	\pm	0.6	$	&	$	0.8	\pm	0.2	$	&	$		12427	\pm	2319		$	&	$	\ \	6.8	\pm	1.3	$	&	$	1.4	\pm	0.4	$	&	$		20.1	\pm	4.6	$	&	0.46	\\
J1314+0927	&	2	&	$	2925	\pm	675	$	&	$	\ \	2.0	\pm	0.6	$	&	$	0.8	\pm	0.2	$	&	$		13244	\pm	2806		$	&	$	\ \	5.3	\pm	1.5	$	&	$	0.8	\pm	0.2	$	&	$		20.8	\pm	5.2	$	&	0.32	\\
J1419+0749	&	7	&	$	3439	\pm	642	$	&	$	\ \	5.2	\pm	0.5	$	&	$	0.9	\pm	0.3	$	&	$		11735	\pm	831	\ \ 	$	&	$	\ \	7.2	\pm	0.6	$	&	$	1.7	\pm	0.5	$	&	$		16.0	\pm	1.5	$	&	0.34	\\
J1509+0744	&	6	&	$	3264	\pm	729	$	&	$	\ \	4.6	\pm	0.9	$	&	$	1.2	\pm	0.3	$	&	$		13001	\pm	1763		$	&	$	\ \	9.9	\pm	1.4	$	&	$	1.6	\pm	0.4	$	&	$		17.1	\pm	3.0	$	&	0.28	\\
J1519+0723	&	7	&	$	3122	\pm	657	$	&	$	\ \	5.8	\pm	1.1	$	&	$	1.2	\pm	0.3	$	&	$		12118	\pm	1399		$	&	$		12.9	\pm	1.6	$	&	$	2.4	\pm	0.6	$	&	$		15.1	\pm	1.8	$	&	0.27	\\
J1545+0156	&	5	&	$	3117	\pm	696	$	&	$	\ \	3.4	\pm	0.7	$	&	$	1.2	\pm	0.3	$	&	$		13128	\pm	1780		$	&	$	\ \	7.9	\pm	1.1	$	&	$	1.2	\pm	0.3	$	&	$		18.3	\pm	3.0	$	&	0.35	\\
J1609+0654	&	5	&	$	3098	\pm	638	$	&	$	\ \	3.7	\pm	0.6	$	&	$	0.3	\pm	0.1	$	&	$		12761	\pm	1460		$	&	$		10.6	\pm	1.2	$	&	$	1.6	\pm	0.5	$	&	$		18.0	\pm	3.0	$	&	0.32	\\
J1618+0704	&	3	&	$	3352	\pm	767	$	&	$	\ \	3.0	\pm	0.8	$	&	$	0.8	\pm	0.2	$	&	$		13050	\pm	2435		$	&	$	\ \	4.7	\pm	0.9	$	&	$	0.5	\pm	0.1	$	&	$		16.4	\pm	4.6	$	&	0.25	\\
J2116+0441	&	9	&	$	3601	\pm	653	$	&	$	\ \	8.8	\pm	1.1	$	&	$	0.8	\pm	0.2	$	&	$		12940	\pm	943	\ \ 	$	&	$	\ \	5.0	\pm	0.4	$	&	$	2.5	\pm	0.7	$	&	$		13.6	\pm	1.8	$	&	0.63	\\
J2145$-$0758	&	11	&	$	3297	\pm	694	$	&	$	\ \	9.1	\pm	2.0	$	&	$	1.2	\pm	0.4	$	&	$	\ \ 	9937	\pm	1368		$	&	$	\ \	8.5	\pm	1.3	$	&	$	3.0	\pm	0.9	$	&	$	\ \	9.0	\pm	2.8	$	&	0.20	\\
	&		&						&							&						&								&							&						&							&		\\
\hline
$\mu$	&	8	&	$	3308	\pm	662	$	&	$	\ \	6.4	\pm	1.0	$	&	$	1.2	\pm	0.3	$	&	$		11800	\pm	1466		$	&	$	\ \	9.1	\pm	1.2	$	&	$	1.8	\pm	0.5	$	&	$		15.1	\pm	2.9	$	&	0.33	\\
$\sigma$	&	4	&	$	3314	\pm	111	$	&	$	\ \	3.8	\pm	0.5	$	&	$	0.5	\pm	0.1	$	&	$		1950	\pm	552		$	&	$	\ \	2.9	\pm	0.5	$	&	$	0.6	\pm	0.2	$	&	$		3.7	\pm	1.4	$	&	0.12	\\
SIQR	&	2	&	$	3302	\pm	41	$	&	$	\ \	2.2	\pm	0.3	$	&	$	0.2	\pm	0.1	$	&	$		1127	\pm	380		$	&	$	\ \	2.0	\pm	0.3	$	&	$	0.4	\pm	0.1	$	&	$		2.1	\pm	1.4	$	&	0.06	\\
\hline
\end{tabular}
\tablefoot{(1) SDSS short name; (2) Equivalent width of the \civ \ BC emission in \AA; (3) FWHM of the \civ \ and \heiiuv\ broad components; (4,5) normalized flux of \civ \ and \heiiuv\ broad components; (6) FWHM of the \civ \ and \heiiuv\ BLUE components; (7,8) normalized flux of the \civ \ and \heiiuv\ BLUE components; (9) separation {of the peak of the BLUE} from the rest-frame  \civ\ emission in  \AA; (10) {asymmetry reported by {\tt specfit}}. {BALQ are marked with an asterisk}(*). The measurements of J0252$-$0420 and J1516+0029 BALQ are excluded from this table. %due to its BALQ nature, their spectra can be seen in appendix \ref{app:spec}
}
\label{tab:civ}}
\end{table*}

We   measured    BLUE components for three blends: the one at $\lambda 1400$, mostly due to \siiv\ and \oiv\ emission, the ones  for the \civ \  and \heiiuv\ at the 1550  \AA\ region. We will base our analysis on three ratios, with only two of them being independent: \civ/\heii, (\siiv+\oiv)/\heii\  and (\siiv+\oiv)/\civ. We also sporadically measure a  component on the blue side of  \aliii, although we would not consider  diagnostic ratios involving BLUE \aliii\ because this component shows a profile narrower than those of \civ, \siiv+\oiv\ and \heii. 
Also, the measurement of \aliii\ BLUE intensity is uncertain, as it is affected by the \siii\ emission. %at $\lambda 1816$  \AA. 
Specific to the BLUE intensity ratios, we note that the (\siiv+\oiv) is often affected by heavy absorptions (even more than \civ) that may systematically lower the $Z$\ estimates from the ratio (\siiv+\oiv)/\heii. %\\

%%%%%%%%%%%%%%%%%%%%%%%%%%%%%%%%%%%%%%%%%%%%%%%%%%%%%%%%%%%%%%%%%%%%%%%%%%%%%%%%%
\subsection{Analysis via multicomponent fits}
\label{fitting}

\begin{table*}[h]
\centering
%\tabletypesize{\scriptsize\tabcolsep=2pt}
{\fontsize{9.5}{10}\selectfont %\tabcolsep=3pt
\caption{Measurements in the $\lambda1400$\ region. }
\begin{tabular}{cccc|cccc}
 \hline\hline
{SDSS JCODE} & 1400 & BC & \siiv  &  BLUE &  \siiv $+$\oiv{} & \\
&  W  & FWHM &  Flux  &  FWHM &  Flux & $\Delta \lambda$ & skew   \\
        (1) & (2) & (3) & (4)  & (5) &  (6) & (7) & (8)   \\
 \hline
J0020+0740	&	10	&	$	3275	\pm	689	$	&	$		10.1	\pm	1.8	$	&	$	9204	\pm	1062		$	&	$	4.2	\pm	0.5	$	&	$		10.4	\pm	1.6	$	&	0.31	\\
J0034$-$0326	&	4	&	$	3236	\pm	741	$	&	$	\ \	4.1	\pm	1.2	$	&	$	8891	\pm	1659		$	&	$	2.1	\pm	0.4	$	&	$	\ \	8.8	\pm	4.2	$	&	0.35	\\
J0037$-$0238	&	10	&	$	3184	\pm	670	$	&	$		10.3	\pm	1.9	$	&	$	9099	\pm	1050		$	&	$	3.0	\pm	0.4	$	&	$	\ \	8.0	\pm	1.6	$	&	0.22	\\
J0103$-$1104	&	4	&	$	2844	\pm	635	$	&	$	\ \	5.2	\pm	1.0	$	&	$	8014	\pm	1087		$	&	$	1.4	\pm	0.2	$	&	$	\ \	6.8	\pm	2.7	$	&	0.19	\\
J0106$-$0855	&	6	&	$	3298	\pm	736	$	&	$	\ \	6.4	\pm	1.3	$	&	$	9908	\pm	1344		$	&	$	5.3	\pm	0.7	$	&	$		10.4	\pm	2.7	$	&	0.35	\\
J0123+0329	&	6	&	$	2791	\pm	575	$	&	$	\ \	5.1	\pm	0.8	$	&	$	9047	\pm	1035		$	&	$	4.2	\pm	0.5	$	&	$		10.9	\pm	2.8	$	&	0.25	\\
J0210$-$0823	&	1	&	$	2302	\pm	516	$	&	$	\ \	2.4	\pm	0.7	$	&	$	9203	\pm	1645		$	&	$	2.5	\pm	0.7	$	&	$		11.7	\pm	4.6	$	&	0.21	\\
J0827+0306	&	5	&	$	3014	\pm	676	$	&	$	\ \	4.3	\pm	1.2	$	&	$	7999	\pm	1429		$	&	$	1.8	\pm	0.5	$	&	$	\ \	7.6	\pm	4.7	$	&	0.29	\\
J0829+0801	&	5	&	$	2614	\pm	474	$	&	$	\ \	4.5	\pm	0.6	$	&	$	7971	\pm	581	\ \	$	&	$	4.1	\pm	0.3	$	&	$		12.1	\pm	1.6	$	&	0.35	\\
J0836+0548	&	6	&	$	2904	\pm	648	$	&	$	\ \	7.5	\pm	1.5	$	&	$	8010	\pm	1086		$	&	$	3.4	\pm	0.5	$	&	$	\ \	7.7	\pm	2.7	$	&	0.23	\\
J0845+0722	&	6	&	$	3307	\pm	681	$	&	$	\ \	6.3	\pm	1.0	$	&	$	10612	\pm	1214		$	&	$	4.3	\pm	0.5	$	&	$		14.0	\pm	2.7	$	&	0.33	\\
J0847+0943	&	9	&	$	3006	\pm	633	$	&	$	\ \	6.1	\pm	1.1	$	&	$	7914	\pm	913	\ \	$	&	$	6.5	\pm	0.8	$	&	$	\ \	8.5	\pm	1.6	$	&	0.45	\\
J0858+0152	&	5	&	$	2999	\pm	508	$	&	$	\ \	5.4	\pm	0.6	$	&	$	12065	\pm	821		$	&	$	6.8	\pm	0.5	$	&	$		13.6	\pm	1.0	$	&	0.35	\\
J0903+0708	&	2	&	$	2601	\pm	595	$	&	$	\ \	2.0	\pm	0.6	$	&	$	7500	\pm	1399		$	&	$	2.0	\pm	0.4	$	&	$	\ \	9.7	\pm	4.2	$	&	0.31	\\
J0915$-$0202	&	9	&	$	3038	\pm	678	$	&	$	\ \	7.4	\pm	1.5	$	&	$	8016	\pm	1087		$	&	$	3.2	\pm	0.5	$	&	$	\ \	8.5	\pm	2.7	$	&	0.28	\\
J0926+0135	&	7	&	$	2876	\pm	509	$	&	$	\ \	6.6	\pm	0.9	$	&	$	9357	\pm	767	\ \	$	&	$	4.4	\pm	0.4	$	&	$	\ \	8.4	\pm	1.3	$	&	0.20	\\
J0929+0333	&	11	&	$	3016	\pm	635	$	&	$	\ \	9.2	\pm	2.0	$	&	$	9007	\pm	1240		$	&	$	6.5	\pm	1.0	$	&	$	\ \	8.7	\pm	2.5	$	&	0.36	\\
J0946$-$0124	&	10	&	$	2762	\pm	558	$	&	$		10.3	\pm	1.1	$	&	$	9111	\pm	620	\ \	$	&	$	3.2	\pm	0.3	$	&	$	\ \	8.6	\pm	1.5	$	&	0.19	\\
J1024+0245	&	6	&	$	3014	\pm	620	$	&	$	\ \	4.8	\pm	0.8	$	&	$	9004	\pm	1030		$	&	$	4.5	\pm	0.5	$	&	$	\ \	9.7	\pm	2.8	$	&	0.42	\\
J1026+0114	&	6	&	$	3160	\pm	665	$	&	$	\ \	6.3	\pm	1.2	$	&	$	9041	\pm	1044		$	&	$	4.4	\pm	0.5	$	&	$	\ \	9.6	\pm	1.6	$	&	0.24	\\
J1145+0800	&	7	&	$	3264	\pm	732	$	&	$	\ \	7.8	\pm	2.1	$	&	$	9085	\pm	1623		$	&	$	4.8	\pm	1.3	$	&	$		12.0	\pm	4.6	$	&	0.30	\\
J1214+0242	&	6	&	$	3122	\pm	566	$	&	$	\ \	5.8	\pm	0.7	$	&	$	8021	\pm	584	\ \	$	&	$	3.4	\pm	0.3	$	&	$	\ \	9.4	\pm	1.6	$	&	0.34	\\
J1215+0326	&	7	&	$	2405	\pm	426	$	&	$	\ \	8.8	\pm	1.2	$	&	$	7993	\pm	655	\ \	$	&	$	2.6	\pm	0.2	$	&	$		10.1	\pm	1.3	$	&	0.34	\\
J1219+0254	&	11	&	$	3236	\pm	622	$	&	$		11.2	\pm	1.5	$	&	$	7012	\pm	672	\ \	$	&	$	2.8	\pm	0.3	$	&	$		10.2	\pm	1.9	$	&	0.35	\\
J1231+0725	&	8	&	$	2421	\pm	192	$	&	$	\ \	7.4	\pm	0.4	$	&	$	9046	\pm	1255		$	&	$	3.2	\pm	0.4	$	&	$		11.1	\pm	4.6	$	&	0.50	\\
J1244+0821	&	6	&	$	3157	\pm	559	$	&	$	\ \	5.6	\pm	0.8	$	&	$	8999	\pm	738	\ \	$	&	$	3.7	\pm	0.3	$	&	$	\ \	8.6	\pm	1.3	$	&	0.28	\\
J1259+0752	&	4	&	$	3421	\pm	783	$	&	$	\ \	4.7	\pm	1.3	$	&	$	8034	\pm	1499		$	&	$	2.0	\pm	0.4	$	&	$		10.0	\pm	4.2	$	&	0.19	\\
J1314+0927	&	3	&	$	3628	\pm	837	$	&	$	\ \	2.6	\pm	0.8	$	&	$	8573	\pm	1816		$	&	$	2.2	\pm	0.6	$	&	$		12.3	\pm	4.7	$	&	0.28	\\
J1419+0749	&	5	&	$	2977	\pm	556	$	&	$	\ \	3.7	\pm	0.4	$	&	$	8143	\pm	577	\ \	$	&	$	2.9	\pm	0.2	$	&	$	\ \	9.9	\pm	1.4	$	&	0.42	\\
J1509+0744	&	6	&	$	3129	\pm	699	$	&	$	\ \	4.2	\pm	0.8	$	&	$	8245	\pm	1118		$	&	$	4.2	\pm	0.6	$	&	$	\ \	8.3	\pm	2.7	$	&	0.30	\\
J1519+0723	&	8	&	$	3124	\pm	657	$	&	$	\ \	6.5	\pm	1.2	$	&	$	8305	\pm	959	\ \	$	&	$	6.3	\pm	0.8	$	&	$		10.3	\pm	1.6	$	&	0.42	\\
J1545+0156	&	5	&	$	3004	\pm	671	$	&	$	\ \	3.7	\pm	0.7	$	&	$	8290	\pm	1124		$	&	$	4.3	\pm	0.6	$	&	$		11.0	\pm	2.7	$	&	0.38	\\
J1609+0654	&	4	&	$	3037	\pm	625	$	&	$	\ \	4.4	\pm	0.7	$	&	$	9305	\pm	1065		$	&	$	3.8	\pm	0.4	$	&	$		11.8	\pm	2.7	$	&	0.30	\\
J1618+0704	&	3	&	$	3048	\pm	698	$	&	$	\ \	3.6	\pm	1.0	$	&	$	8264	\pm	1542		$	&	$	2.2	\pm	0.4	$	&	$		11.4	\pm	4.1	$	&	0.35	\\
J2116+0441	&	4	&	$	3112	\pm	564	$	&	$	\ \	6.6	\pm	0.8	$	&	$	8699	\pm	634	\ \	$	&	$	1.5	\pm	0.1	$	&	$	\ \	8.6	\pm	1.6	$	&	0.17	\\
J2145$-$0758	&	6	&	$	3000	\pm	631	$	&	$	\ \	6.8	\pm	1.5	$	&	$	8921	\pm	1228		$	&	$	2.8	\pm	0.4	$	&	$		10.7	\pm	2.5	$	&	0.27	\\
	&		&						&							&							&						&							&		\\
\hline													
$\mu$	&	6	&	$	3009	\pm	618	$	&	$	\ \	6.0	\pm	1.1	$	&	$	8720	\pm	1089		$	&	$	3.6	\pm	0.5	$	&	$		10.0	\pm	2.6	$	&	0.31	\\
$\sigma$	&	2	&	$	282	\pm	113	$	&	$	\ \	2.3	\pm	0.4	$	&	$	906	\pm	343		$	&	$	1.4	\pm	0.2	$	&	$		1.7	\pm	1.2	$	&	0.08	\\
SIQR	&	1	&	$	134	\pm	58	$	&	$	\ \	1.5	\pm	0.3	$	&	$	534	\pm	235		$	&	$	0.9	\pm	0.1	$	&	$		1.2	\pm	0.7	$	&	0.05	\\
\hline
\end{tabular}
\tablefoot{(1) SDSS short name; (2) \siiv\ equivalent width {(in \AA~)} between 1340-1450 \AA\; (3) FWHM of the \siiv\  broad component {in km s$^{-1}$}; (4) BC \siiv\  flux {normalized to the continuum at 1350 \AA}; (5,6) FWHM  and flux of the \siiv\ + \oiv\ asymmetric emission; (7) separation  {of the peak of the BLUE?} from the BC \siiv\ rest-frame in \AA; (8) {asymmetry reported by {\tt specfit}}. The measurements of the six BALQ were excluded from this table. % J0216+0115, J0252$-$0420, J0932+0237, J1013+0851 and J1205+0201 were excluded from this table due to its BALQ nature, their spectra can be seen in appendix \ref{app:spec}. 
}
\label{tab:siiv}}
\end{table*}

\begin{table*}[h]
%\tabletypesize{\scriptsize\tabcolsep=2pt}
\centering
\tabcolsep=4pt
\caption{\texttt{specfit} Flux ratios dependent on metallicity}
\begin{tabular}{ccccccccccccccccccc}
\hline\hline
 & {BC} & {BC} & {BC} & {BC} & {BC} & {BC} & {BLUE} & {BLUE} & {BLUE} \\
{SDSS JCODE}  & {\aliii/\heiiuv} & {\civ{}/\heiiuv} & {\siiv{}/\heiiuv} &
{\civ{}/\aliii} & {\siiv/\civ{}} & 
{\siiv/\aliii} & {\civ{}/\heiiuv} & {(\siiv{}+\oiv)/} & {\civ{}/}\\
 &  &  &  &  &  &  &  & {\heiiuv} & {(\siiv{}+\oiv)}\\
(1) & (2) & (3) & (4)  & (5)  & (6) & (7) & (8) & (9) & (10)  \\
\hline
\\
J0020+0740	&	$	3.1	\pm	1.2		$	&	$	6.4	\pm	2.2		$	&	$	6.6	\pm	2.3		$	&	$	2.0	\pm	0.7	$	&	$	1.0	\pm	0.2	$	&	$	2.1	\pm	0.7	$	&	$	4.3	\pm	1.3	$	&	$	1.5	\pm	0.4	$	&	$	2.9	\pm	0.5	$	\\
J0034$-$0326	&	$	2.1	\pm	0.8		$	&	$	2.1	\pm	0.8		$	&	$	2.9	\pm	1.2		$	&	$	1.0	\pm	0.4	$	&	$	0.7	\pm	0.3	$	&	$	1.4	\pm	0.5	$	&	$	5.0	\pm	1.7	$	&	$	1.3	\pm	0.5	$	&	$	3.8	\pm	1.1	$	\\
J0037$-$0238	&	$	1.3	\pm	0.5		$	&	$	6.0	\pm	2.0		$	&	$	4.8	\pm	1.6		$	&	$	4.7	\pm	1.5	$	&	$	1.2	\pm	0.3	$	&	$	3.8	\pm	1.2	$	&	$	3.8	\pm	1.1	$	&	$	1.7	\pm	0.5	$	&	$	2.3	\pm	0.4	$	\\
J0103$-$1104	&	$	1.8	\pm	0.7		$	&	$	5.4	\pm	1.9		$	&	$	4.3	\pm	1.5		$	&	$	2.9	\pm	1.0	$	&	$	1.2	\pm	0.4	$	&	$	2.4	\pm	0.8	$	&	$	5.4	\pm	1.7	$	&	$	1.4	\pm	0.4	$	&	$	3.9	\pm	0.8	$	\\
J0106$-$0855	&	$	3.2	\pm	1.3		$	&	$	3.2	\pm	1.1		$	&	$	6.1	\pm	2.1		$	&	$	1.0	\pm	0.3	$	&	$	0.5	\pm	0.1	$	&	$	1.9	\pm	0.7	$	&	$	5.7	\pm	1.8	$	&	$	2.8	\pm	0.9	$	&	$	2.0	\pm	0.4	$	\\
J0123+0329	&	$	3.6	\pm	1.3		$	&	$	2.9	\pm	1.0		$	&	$	6.0	\pm	2.0		$	&	$	0.8	\pm	0.2	$	&	$	0.5	\pm	0.1	$	&	$	1.7	\pm	0.5	$	&	$	7.3	\pm	2.2	$	&	$	3.8	\pm	1.2	$	&	$	1.9	\pm	0.3	$	\\
J0210$-$0823	&	$	4.5	\pm	1.9		$	&	$	3.4	\pm	1.3		$	&	$	3.0	\pm	1.2		$	&	$	0.7	\pm	0.3	$	&	$	1.1	\pm	0.4	$	&	$	0.7	\pm	0.3	$	&	$	4.0	\pm	1.6	$	&	$	1.6	\pm	0.6	$	&	$	2.6	\pm	0.9	$	\\
J0827+0306	&	$	2.1	\pm	0.9		$	&	$	3.5	\pm	1.4		$	&	$	4.0	\pm	1.6		$	&	$	1.7	\pm	0.7	$	&	$	0.9	\pm	0.3	$	&	$	1.9	\pm	0.8	$	&	$	5.6	\pm	2.2	$	&	$	1.4	\pm	0.5	$	&	$	4.0	\pm	1.5	$	\\
J0829+0801	&	$	5.1	\pm	1.9		$	&	$	4.9	\pm	1.6		$	&	$	5.3	\pm	1.7		$	&	$	1.0	\pm	0.2	$	&	$	0.9	\pm	0.2	$	&	$	1.0	\pm	0.3	$	&	$	6.0	\pm	1.7	$	&	$	2.4	\pm	0.7	$	&	$	2.5	\pm	0.3	$	\\
J0836+0548	&	$	7.2	\pm	2.9		$	&	$	7.0	\pm	2.5		$	&	$	7.5	\pm	2.6		$	&	$	1.0	\pm	0.3	$	&	$	0.9	\pm	0.3	$	&	$	1.0	\pm	0.3	$	&	$	4.1	\pm	1.3	$	&	$	2.1	\pm	0.7	$	&	$	1.9	\pm	0.4	$	\\
J0845+0722	&	$	3.5	\pm	1.3		$	&	$	4.4	\pm	1.5		$	&	$	5.7	\pm	1.9		$	&	$	1.3	\pm	0.4	$	&	$	0.8	\pm	0.2	$	&	$	1.6	\pm	0.5	$	&	$	5.7	\pm	1.7	$	&	$	2.5	\pm	0.8	$	&	$	2.3	\pm	0.4	$	\\
J0847+0943	&	$	4.4	\pm	1.7		$	&	$	6.1	\pm	2.1		$	&	$	4.9	\pm	1.7		$	&	$	1.4	\pm	0.5	$	&	$	1.2	\pm	0.3	$	&	$	1.1	\pm	0.4	$	&	$	4.6	\pm	1.4	$	&	$	2.4	\pm	0.7	$	&	$	1.9	\pm	0.3	$	\\
J0858+0152	&	$	2.5	\pm	0.8		$	&	$	3.9	\pm	1.2		$	&	$	4.6	\pm	1.4		$	&	$	1.6	\pm	0.3	$	&	$	0.8	\pm	0.1	$	&	$	1.9	\pm	0.4	$	&	$	5.6	\pm	1.4	$	&	$	2.5	\pm	0.6	$	&	$	2.2	\pm	0.2	$	\\
J0903+0708	&	$	2.7	\pm	1.1		$	&	$	2.0	\pm	0.8		$	&	$	1.8	\pm	0.7		$	&	$	0.8	\pm	0.3	$	&	$	1.1	\pm	0.5	$	&	$	0.7	\pm	0.3	$	&	$	5.0	\pm	1.7	$	&	$	1.8	\pm	0.6	$	&	$	2.8	\pm	0.8	$	\\
J0915$-$0202	&	$	1.4	\pm	0.6		$	&	$	3.5	\pm	1.2		$	&	$	5.2	\pm	1.8		$	&	$	2.5	\pm	0.9	$	&	$	0.7	\pm	0.2	$	&	$	3.8	\pm	1.3	$	&	$	4.3	\pm	1.4	$	&	$	1.5	\pm	0.5	$	&	$	2.9	\pm	0.6	$	\\
J0926+0135	&	$	2.3	\pm	0.9		$	&	$	5.6	\pm	1.8		$	&	$	4.8	\pm	1.5		$	&	$	2.5	\pm	0.7	$	&	$	1.2	\pm	0.2	$	&	$	2.1	\pm	0.6	$	&	$	5.3	\pm	1.5	$	&	$	2.7	\pm	0.8	$	&	$	1.9	\pm	0.2	$	\\
J0929+0333	&	$	6.7	\pm	2.7		$	&	$	10.3	\pm	3.7	\	$	&	$	10.3	\pm	3.7	\	$	&	$	1.5	\pm	0.5	$	&	$	1.0	\pm	0.3	$	&	$	1.5	\pm	0.5	$	&	$	6.8	\pm	2.2	$	&	$	3.6	\pm	1.2	$	&	$	1.9	\pm	0.4	$	\\
J0946$-$0124	&	$	1.5	\pm	0.5		$	&	$	6.4	\pm	2.0		$	&	$	5.6	\pm	1.8		$	&	$	4.4	\pm	1.1	$	&	$	1.1	\pm	0.2	$	&	$	3.9	\pm	1.0	$	&	$	3.2	\pm	0.9	$	&	$	1.8	\pm	0.5	$	&	$	1.8	\pm	0.2	$	\\
J1024+0245	&	$	4.9	\pm	1.9		$	&	$	6.6	\pm	2.2		$	&	$	6.1	\pm	2.0		$	&	$	1.3	\pm	0.4	$	&	$	1.1	\pm	0.2	$	&	$	1.2	\pm	0.4	$	&	$	4.9	\pm	1.5	$	&	$	2.1	\pm	0.6	$	&	$	2.3	\pm	0.4	$	\\
J1026+0114	&	$	5.5	\pm	2.2		$	&	$	4.2	\pm	1.4		$	&	$	6.6	\pm	2.3		$	&	$	0.8	\pm	0.3	$	&	$	0.6	\pm	0.2	$	&	$	1.2	\pm	0.4	$	&	$	6.7	\pm	2.0	$	&	$	2.3	\pm	0.7	$	&	$	2.9	\pm	0.5	$	\\
J1145+0800	&	$	5.8	\pm	2.4		$	&	$	4.9	\pm	2.0		$	&	$	8.7	\pm	3.5		$	&	$	0.8	\pm	0.3	$	&	$	0.6	\pm	0.2	$	&	$	1.5	\pm	0.6	$	&	$	6.0	\pm	2.3	$	&	$	2.6	\pm	1.0	$	&	$	2.3	\pm	0.9	$	\\
J1214+0242	&	$	4.0	\pm	1.5		$	&	$	6.2	\pm	2.0		$	&	$	5.1	\pm	1.6		$	&	$	1.5	\pm	0.4	$	&	$	1.2	\pm	0.2	$	&	$	1.3	\pm	0.3	$	&	$	5.9	\pm	1.7	$	&	$	1.9	\pm	0.5	$	&	$	3.2	\pm	0.4	$	\\
J1215+0326	&	$	2.6	\pm	1.0		$	&	$	8.0	\pm	2.5		$	&	$	6.9	\pm	2.2		$	&	$	3.1	\pm	0.9	$	&	$	1.1	\pm	0.2	$	&	$	2.7	\pm	0.8	$	&	$	4.2	\pm	1.2	$	&	$	1.0	\pm	0.3	$	&	$	4.4	\pm	0.5	$	\\
J1219+0254	&	$	2.4	\pm	0.9		$	&	$	7.5	\pm	2.5		$	&	$	5.4	\pm	1.7		$	&	$	3.2	\pm	0.9	$	&	$	1.4	\pm	0.3	$	&	$	2.3	\pm	0.7	$	&	$	3.2	\pm	1.0	$	&	$	1.0	\pm	0.3	$	&	$	3.2	\pm	0.5	$	\\
J1231+0725	&	$	0.9	\pm	0.3		$	&	$	7.1	\pm	2.1		$	&	$	2.9	\pm	0.9		$	&	$	7.8	\pm	1.5	$	&	$	2.5	\pm	0.2	$	&	$	3.2	\pm	0.6	$	&	$	3.1	\pm	1.0	$	&	$	1.2	\pm	0.4	$	&	$	2.5	\pm	0.5	$	\\
J1244+0821	&	$	4.7	\pm	1.8		$	&	$	6.7	\pm	2.1		$	&	$	6.7	\pm	2.1		$	&	$	1.4	\pm	0.4	$	&	$	1.0	\pm	0.2	$	&	$	1.4	\pm	0.4	$	&	$	4.6	\pm	1.3	$	&	$	1.8	\pm	0.5	$	&	$	2.6	\pm	0.3	$	\\
J1259+0752	&	$	2.1	\pm	0.9		$	&	$	2.7	\pm	1.1		$	&	$	5.9	\pm	2.4		$	&	$	1.3	\pm	0.5	$	&	$	0.5	\pm	0.2	$	&	$	2.8	\pm	1.1	$	&	$	4.9	\pm	1.7	$	&	$	1.4	\pm	0.5	$	&	$	3.5	\pm	1.0	$	\\
J1314+0927	&	$	2.9	\pm	1.2		$	&	$	2.4	\pm	1.0		$	&	$	3.1	\pm	1.3		$	&	$	0.8	\pm	0.3	$	&	$	0.8	\pm	0.3	$	&	$	1.1	\pm	0.4	$	&	$	6.5	\pm	2.6	$	&	$	2.6	\pm	1.0	$	&	$	2.5	\pm	0.9	$	\\
J1419+0749	&	$	2.9	\pm	1.0		$	&	$	5.6	\pm	1.7		$	&	$	4.0	\pm	1.3		$	&	$	1.9	\pm	0.4	$	&	$	1.4	\pm	0.2	$	&	$	1.4	\pm	0.3	$	&	$	4.2	\pm	1.2	$	&	$	1.7	\pm	0.5	$	&	$	2.5	\pm	0.3	$	\\
J1509+0744	&	$	3.2	\pm	1.3		$	&	$	3.9	\pm	1.4		$	&	$	3.5	\pm	1.2		$	&	$	1.2	\pm	0.4	$	&	$	1.1	\pm	0.3	$	&	$	1.1	\pm	0.4	$	&	$	6.4	\pm	2.0	$	&	$	2.7	\pm	0.9	$	&	$	2.3	\pm	0.5	$	\\
J1519+0723	&	$	4.4	\pm	1.7		$	&	$	4.8	\pm	1.7		$	&	$	5.5	\pm	1.9		$	&	$	1.1	\pm	0.4	$	&	$	0.9	\pm	0.2	$	&	$	1.2	\pm	0.4	$	&	$	5.4	\pm	1.6	$	&	$	2.6	\pm	0.8	$	&	$	2.1	\pm	0.4	$	\\
J1545+0156	&	$	3.9	\pm	1.6		$	&	$	2.9	\pm	1.0		$	&	$	3.1	\pm	1.1		$	&	$	0.7	\pm	0.3	$	&	$	0.9	\pm	0.3	$	&	$	0.8	\pm	0.3	$	&	$	6.7	\pm	2.1	$	&	$	3.6	\pm	1.2	$	&	$	1.9	\pm	0.4	$	\\
J1609+0654	&	$	13.8	\pm	5.6	\	$	&	$	12.3	\pm	4.1	\	$	&	$	14.6	\pm	4.8	\	$	&	$	0.9	\pm	0.3	$	&	$	0.8	\pm	0.2	$	&	$	1.1	\pm	0.4	$	&	$	6.6	\pm	2.0	$	&	$	2.4	\pm	0.7	$	&	$	2.8	\pm	0.4	$	\\
J1618+0704	&	$	3.6	\pm	1.5		$	&	$	3.6	\pm	1.4		$	&	$	4.3	\pm	1.7		$	&	$	1.0	\pm	0.4	$	&	$	0.8	\pm	0.3	$	&	$	1.2	\pm	0.5	$	&	$	10.1	\pm	3.5	$	&	$	4.7	\pm	1.6	$	&	$	2.2	\pm	0.6	$	\\
J2116+0441	&	$	3.8	\pm	1.4		$	&	$	11.4	\pm	3.6	\	$	&	$	8.5	\pm	2.7		$	&	$	3.0	\pm	0.7	$	&	$	1.3	\pm	0.2	$	&	$	2.2	\pm	0.5	$	&	$	2.0	\pm	0.6	$	&	$	0.6	\pm	0.2	$	&	$	3.4	\pm	0.4	$	\\
J2145$-$0758	&	$	3.1	\pm	1.2		$	&	$	7.5	\pm	2.7		$	&	$	5.6	\pm	2.0		$	&	$	2.4	\pm	0.8	$	&	$	1.3	\pm	0.4	$	&	$	1.8	\pm	0.6	$	&	$	2.9	\pm	0.9	$	&	$	0.9	\pm	0.3	$	&	$	3.0	\pm	0.6	$	\\
\\	\hline																																																									
$\mu$	&	$	3.7	\pm	1.4		$	&	$	5.4	\pm	1.9		$	&	$	5.5	\pm	1.9		$	&	$	1.9	\pm	0.6	$	&	$	1.0	\pm	0.3	$	&	$	1.8	\pm	0.6	$	&	$	5.2	\pm	1.6	$	&	$	2.1	\pm	0.7	$	&	$	2.6	\pm	0.5	$	\\
$\sigma$	&	$	2.3	\pm	0.9		$	&	$	2.5	\pm	0.8		$	&	$	2.4	\pm	0.8		$	&	$	1.4	\pm	0.3	$	&	$	0.4	\pm	0.1	$	&	$	0.9	\pm	0.3	$	&	$	1.5	\pm	0.6	$	&	$	0.9	\pm	0.3	$	&	$	0.7	\pm	0.3	$	\\
SIQR	&	$	1.0	\pm	0.4		$	&	$	1.6	\pm	0.4		$	&	$	1.0	\pm	0.3		$	&	$	0.7	\pm	0.2	$	&	$	0.2	\pm	0.1	$	&	$	0.5	\pm	0.1	$	&	$	0.9	\pm	0.4	$	&	$	0.6	\pm	0.1	$	&	$	0.4	\pm	0.1	$	\\
\hline
\end{tabular}
\tablefoot{(1) SDSS name, (2-7) line ratios of the broad components, (8-10) line ratios of the BLUE components. %(2) FWHM of \siiv\ in  km s$^{-1}$, (3) and (4) list fluxes of the broad components and the blue component line in units of  {10$^{-14}$} ergs$^{-1}$cm$^{-2}$. 
The measurements of the six BALQ were excluded from this table.}
\label{tab:ratios}
\end{table*}

\begin{table*}[h]
\centering
%\tabletypesize{\scriptsize\tabcolsep=4pt}
\fontsize{8.2}{10}\selectfont %\tabcolsep=3pt
\caption{Normalized intensities}
\begin{tabular}{ccccccccccccccccccccc}
\hline\hline
SDSS JCODE  & \aliii & \siiii & \civ & BLUE \civ & \heiiuv{} & BLUE \heiiuv{} & \siiv  & \siiv + \oiv   \\
& {$\lambda 1860$} & {$\lambda 1891$} & {$\lambda 1549$} & {$\lambda 1536$}  & {$\lambda 1640$} & {$\lambda 1626$} & {$\lambda 1400$} & {$\lambda 1385$} \\
(1) & (2) & (3) & (4)  & (5) & (6) & (7) & (8) & (9)   \\ \hline
\\
J0020+0740	&	$	0.141	\pm	0.024	$	&	$	0.201	\pm	0.021	$	&	$	0.34	\pm	0.09	$	&	$	0.33	\pm	0.03	$	&	$	0.054	\pm	0.018	$	&	$	0.064	\pm	0.024	$	&	$	0.31	\pm	0.03	$	&	$	0.175	\pm	0.040	$	\\
J0034$-$0326	&	$	0.089	\pm	0.027	$	&	$	0.116	\pm	0.021	$	&	$	0.15	\pm	0.11	$	&	$	0.17	\pm	0.11	$	&	$	0.040	\pm	0.016	$	&	$	0.012	\pm	0.080	$	&	$	0.13	\pm	0.03	$	&	$	0.055	\pm	0.044	$	\\
J0037$-$0238	&	$	0.099	\pm	0.017	$	&	$	0.111	\pm	0.015	$	&	$	0.38	\pm	0.04	$	&	$	0.26	\pm	0.04	$	&	$	0.050	\pm	0.016	$	&	$	0.060	\pm	0.012	$	&	$	0.32	\pm	0.03	$	&	$	0.153	\pm	0.046	$	\\
J0103$-$1104	&	$	0.081	\pm	0.015	$	&	$	0.132	\pm	0.030	$	&	$	0.20	\pm	0.03	$	&	$	0.17	\pm	0.04	$	&	$	0.046	\pm	0.024	$	&	$	0.031	\pm	0.021	$	&	$	0.18	\pm	0.03	$	&	$	0.064	\pm	0.036	$	\\
J0106$-$0855	&	$	0.102	\pm	0.018	$	&	$	0.164	\pm	0.018	$	&	$	0.18	\pm	0.05	$	&	$	0.26	\pm	0.02	$	&	$	0.028	\pm	0.012	$	&	$	0.039	\pm	0.026	$	&	$	0.23	\pm	0.04	$	&	$	0.140	\pm	0.066	$	\\
J0123+0329	&	$	0.086	\pm	0.015	$	&	$	0.111	\pm	0.023	$	&	$	0.10	\pm	0.03	$	&	$	0.19	\pm	0.02	$	&	$	0.005	\pm	0.023	$	&	$	0.030	\pm	0.016	$	&	$	0.15	\pm	0.03	$	&	$	0.089	\pm	0.151	$	\\
J0210$-$0823	&	$	0.110	\pm	0.027	$	&	$	0.147	\pm	0.041	$	&	$	0.10	\pm	0.06	$	&	$	0.11	\pm	0.05	$	&	$	0.039	\pm	0.031	$	&	$	0.042	\pm	0.020	$	&	$	0.03	\pm	0.03	$	&	$	0.088	\pm	0.045	$	\\
J0216+0115*	&	$	0.157	\pm	0.015	$	&	$	0.153	\pm	0.018	$	&	$	0.34	\pm	0.04	$	&	\ldots	&	$	0.030	\pm	0.012	$	&	$	0.039	\pm	0.020	$	&	$		\ldots		$	&	$		\ldots		$	\\
J0252$-$0420*	&	$	0.198	\pm	0.035	$	&	$	0.226	\pm	0.098	$	&	$		\ldots		$	&	$		\ldots		$	&	$		\ldots		$	&	$		\ldots		$	&	$		\ldots		$	&	$		\ldots		$	\\
J0827+0306	&	$	0.082	\pm	0.027	$	&	$	0.083	\pm	0.020	$	&	$	0.16	\pm	0.04	$	&	$	0.130	\pm	0.049	$	&	$	0.016	\pm	0.028	$	&	$	0.043	\pm	0.026	$	&	$	0.15	\pm	0.04	$	&	$	0.078	\pm	0.042	$	\\
J0829+0801	&	$	0.129	\pm	0.012	$	&	$	0.144	\pm	0.014	$	&	$	0.16	\pm	0.05	$	&	$	0.255	\pm	0.021	$	&	$	0.027	\pm	0.014	$	&	$	0.048	\pm	0.014	$	&	$	0.17	\pm	0.02	$	&	$	0.115	\pm	0.067	$	\\
J0836+0548	&	$	0.118	\pm	0.016	$	&	$	0.189	\pm	0.022	$	&	$	0.22	\pm	0.04	$	&	$	0.201	\pm	0.094	$	&	$	0.025	\pm	0.012	$	&	$	0.050	\pm	0.022	$	&	$	0.26	\pm	0.03	$	&	$	0.146	\pm	0.041	$	\\
J0845+0722	&	$	0.104	\pm	0.023	$	&	$	0.126	\pm	0.016	$	&	$	0.20	\pm	0.04	$	&	$	0.241	\pm	0.024	$	&	$	0.027	\pm	0.019	$	&	$	0.051	\pm	0.018	$	&	$	0.20	\pm	0.05	$	&	$	0.160	\pm	0.023	$	\\
J0847+0943	&	$	0.158	\pm	0.034	$	&	$	0.137	\pm	0.082	$	&	$	0.30	\pm	0.08	$	&	$	0.351	\pm	0.027	$	&	$	0.044	\pm	0.020	$	&	$	0.066	\pm	0.020	$	&	$	0.30	\pm	0.05	$	&	$	0.206	\pm	0.036	$	\\
J0858+0152	&	$	0.086	\pm	0.013	$	&	$	0.130	\pm	0.007	$	&	$	0.18	\pm	0.09	$	&	$	0.205	\pm	0.137	$	&	$	0.023	\pm	0.009	$	&	$	0.043	\pm	0.009	$	&	$	0.17	\pm	0.03	$	&	$	0.069	\pm	0.141	$	\\
J0903+0708	&	$	0.084	\pm	0.015	$	&	$	0.084	\pm	0.016	$	&	$	0.04	\pm	0.09	$	&	$	0.130	\pm	0.014	$	&	$	0.022	\pm	0.019	$	&	$	0.020	\pm	0.015	$	&	$	0.08	\pm	0.02	$	&	$	0.085	\pm	0.028	$	\\
J0915$-$0202	&	$	0.067	\pm	0.024	$	&	$	0.094	\pm	0.043	$	&	$	0.21	\pm	0.11	$	&	$	0.228	\pm	0.025	$	&	$	0.047	\pm	0.020	$	&	$	0.053	\pm	0.017	$	&	$	0.25	\pm	0.04	$	&	$	0.137	\pm	0.038	$	\\
J0926+0135	&	$	0.100	\pm	0.010	$	&	$	0.156	\pm	0.022	$	&	$	0.26	\pm	0.04	$	&	$	0.273	\pm	0.021	$	&	$	0.051	\pm	0.015	$	&	$	0.047	\pm	0.008	$	&	$	0.23	\pm	0.02	$	&	$	0.119	\pm	0.089	$	\\
J0929+0333	&	$	0.179	\pm	0.037	$	&	$	0.249	\pm	0.022	$	&	$	0.41	\pm	0.12	$	&	$	0.354	\pm	0.056	$	&	$	0.044	\pm	0.028	$	&	$	0.058	\pm	0.024	$	&	$	0.38	\pm	0.05	$	&	$	0.196	\pm	0.071	$	\\
J0932+0237*	&	$	0.177	\pm	0.063	$	&	$	0.217	\pm	0.095	$	&	$	0.59	\pm	0.15	$	&	$	0.520	\pm	0.071	$	&	$	0.069	\pm	0.023	$	&	$	0.102	\pm	0.024	$	&	$		\ldots		$	&	$		\ldots		$	\\
J0946$-$0124	&	$	0.114	\pm	0.016	$	&	$	0.161	\pm	0.053	$	&	$	0.42	\pm	0.05	$	&	$	0.220	\pm	0.047	$	&	$	0.065	\pm	0.010	$	&	$	0.054	\pm	0.011	$	&	$	0.34	\pm	0.05	$	&	$	0.086	\pm	0.030	$	\\
J1013+0851*	&	$	0.239	\pm	0.029	$	&	$	0.286	\pm	0.010	$	&	$	0.21	\pm	0.05	$	&	$	0.188	\pm	0.146	$	&	$	0.080	\pm	0.019	$	&	$	0.073	\pm	0.011	$	&	$		\ldots		$	&	$		\ldots		$	\\
J1024+0245	&	$	0.117	\pm	0.019	$	&	$	0.151	\pm	0.016	$	&	$	0.16	\pm	0.07	$	&	$	0.257	\pm	0.025	$	&	$	0.023	\pm	0.019	$	&	$	0.040	\pm	0.014	$	&	$	0.20	\pm	0.02	$	&	$	0.145	\pm	0.027	$	\\
J1026+0114	&	$	0.164	\pm	0.024	$	&	$	0.172	\pm	0.028	$	&	$	0.18	\pm	0.04	$	&	$	0.286	\pm	0.027	$	&	$	0.034	\pm	0.013	$	&	$	0.042	\pm	0.020	$	&	$	0.21	\pm	0.02	$	&	$	0.180	\pm	0.021	$	\\
J1145+0800	&	$	0.146	\pm	0.030	$	&	$	0.188	\pm	0.046	$	&	$	0.07	\pm	0.25	$	&	$	0.238	\pm	0.052	$	&	$	0.028	\pm	0.019	$	&	$	0.032	\pm	0.044	$	&	$	0.23	\pm	0.03	$	&	$	0.152	\pm	0.037	$	\\
J1205+0201*	&	$	0.104	\pm	0.026	$	&	$	0.140	\pm	0.011	$	&	$	0.15	\pm	0.03	$	&	$	0.188	\pm	0.012	$	&	$	0.018	\pm	0.010	$	&	$	0.014	\pm	0.011	$	&	$		\ldots		$	&	$		\ldots		$	\\
J1214+0242	&	$	0.120	\pm	0.011	$	&	$	0.162	\pm	0.007	$	&	$	0.25	\pm	0.05	$	&	$	0.118	\pm	0.177	$	&	$	0.028	\pm	0.008	$	&	$	0.049	\pm	0.007	$	&	$	0.21	\pm	0.02	$	&	$	0.089	\pm	0.135	$	\\
J1215+0326	&	$	0.098	\pm	0.013	$	&	$	0.168	\pm	0.029	$	&	$	0.30	\pm	0.15	$	&	$	0.312	\pm	0.089	$	&	$	0.047	\pm	0.010	$	&	$	0.062	\pm	0.012	$	&	$	0.31	\pm	0.02	$	&	$	0.107	\pm	0.054	$	\\
J1219+0254	&	$	0.131	\pm	0.018	$	&	$	0.226	\pm	0.016	$	&	$	0.52	\pm	0.04	$	&	$	0.293	\pm	0.037	$	&	$	0.084	\pm	0.017	$	&	$	0.087	\pm	0.015	$	&	$	0.33	\pm	0.02	$	&	$	0.106	\pm	0.055	$	\\
J1231+0725	&	$	0.096	\pm	0.022	$	&	$	0.161	\pm	0.040	$	&	$	0.67	\pm	0.07	$	&	$	0.282	\pm	0.057	$	&	$	0.110	\pm	0.009	$	&	$	0.067	\pm	0.020	$	&	$	0.30	\pm	0.04	$	&	$	0.003	\pm	0.080	$	\\
J1244+0821	&	$	0.098	\pm	0.024	$	&	$	0.142	\pm	0.011	$	&	$	0.21	\pm	0.03	$	&	$	0.237	\pm	0.033	$	&	$	0.027	\pm	0.010	$	&	$	0.046	\pm	0.010	$	&	$	0.19	\pm	0.06	$	&	$	0.133	\pm	0.019	$	\\
J1259+0752	&	$	0.050	\pm	0.011	$	&	$	0.052	\pm	0.013	$	&	$	0.05	\pm	0.07	$	&	$	0.139	\pm	0.025	$	&	$	0.019	\pm	0.013	$	&	$	0.027	\pm	0.008	$	&	$	0.11	\pm	0.06	$	&	$	0.109	\pm	0.015	$	\\
J1314+0927	&	$	0.092	\pm	0.020	$	&	$	0.098	\pm	0.027	$	&	$	0.03	\pm	0.05	$	&	$	0.095	\pm	0.024	$	&	$	0.026	\pm	0.026	$	&	$	0.021	\pm	0.016	$	&	$	0.05	\pm	0.05	$	&	$	0.066	\pm	0.038	$	\\
J1419+0749	&	$	0.089	\pm	0.016	$	&	$	0.137	\pm	0.011	$	&	$	0.20	\pm	0.03	$	&	$	0.195	\pm	0.020	$	&	$	0.029	\pm	0.030	$	&	$	0.037	\pm	0.008	$	&	$	0.16	\pm	0.02	$	&	$	0.106	\pm	0.019	$	\\
J1509+0744	&	$	0.109	\pm	0.017	$	&	$	0.169	\pm	0.028	$	&	$	0.16	\pm	0.05	$	&	$	0.237	\pm	0.029	$	&	$	0.032	\pm	0.015	$	&	$	0.050	\pm	0.024	$	&	$	0.18	\pm	0.02	$	&	$	0.164	\pm	0.054	$	\\
J1516+0029*	&	$	0.487	\pm	0.053	$	&	$	0.389	\pm	0.085	$	&	$		\ldots		$	&	$		\ldots		$	&	$		\ldots		$	&	$		\ldots		$	&	$		\ldots		$	&	$		\ldots		$	\\
J1519+0723	&	$	0.153	\pm	0.026	$	&	$	0.238	\pm	0.022	$	&	$	0.22	\pm	0.06	$	&	$	0.345	\pm	0.038	$	&	$	0.033	\pm	0.019	$	&	$	0.072	\pm	0.029	$	&	$	0.20	\pm	0.15	$	&	$	0.225	\pm	0.041	$	\\
J1545+0156	&	$	0.143	\pm	0.017	$	&	$	0.158	\pm	0.014	$	&	$	0.15	\pm	0.03	$	&	$	0.001	\pm	0.163	$	&	$	0.025	\pm	0.033	$	&	$	0.030	\pm	0.014	$	&	$	0.14	\pm	0.02	$	&	$	0.077	\pm	0.093	$	\\
J1609+0654	&	$	0.101	\pm	0.028	$	&	$	0.120	\pm	0.011	$	&	$	0.13	\pm	0.07	$	&	$	0.228	\pm	0.016	$	&	$	0.014	\pm	0.013	$	&	$	0.001	\pm	0.051	$	&	$	0.15	\pm	0.03	$	&	$	0.139 	\pm	0.016	$	\\
J1618+0704	&	$	0.088	\pm	0.018	$	&	$	0.092	\pm	0.017	$	&	$	0.09	\pm	0.03	$	&	$	0.131	\pm	0.021	$	&	$	0.012	\pm	0.017	$	&	$	0.022	\pm	0.022	$	&	$	0.12	\pm	0.02	$	&	$	0.083	\pm	0.040	$	\\
J2116+0441	&	$	0.092	\pm	0.016	$	&	$	0.135	\pm	0.020	$	&	$	0.34	\pm	0.03	$	&	$	0.112	\pm	0.046	$	&	$	0.031	\pm	0.011	$	&	$	0.045	\pm	0.012	$	&	$	0.19	\pm	0.03	$	&	$	0.041	\pm	0.067	$	\\
J2145$-$0758	&	$	0.124	\pm	0.035	$	&	$	0.237	\pm	0.028	$	&	$	0.35	\pm	0.09	$	&	$	0.276	\pm	0.073	$	&	$	0.021	\pm	0.033	$	&	$	0.086	\pm	0.025	$	&	$	0.23	\pm	0.05	$	&	$	0.026	\pm	0.111	$	\\
		\\	\hline									
$\mu$	&	$	0.126	\pm	0.023	$	&	$	0.161	\pm	0.028	$	&	$	0.23	\pm	0.07	$	&	$	0.21	\pm	0.06	$	&	$	0.037	\pm	0.018	$	&	$	0.046	\pm	0.035	$	&	$	0.20	\pm	0.04	$	&	$	0.114	\pm	0.054	$	\\
$\sigma$	&	$	0.068	\pm	0.011	$	&	$	0.061	\pm	0.023	$	&	$	0.14	\pm	0.04	$	&	$	0.12	\pm	0.06	$	&	$	0.021	\pm	0.007	$	&	$	0.021	\pm	0.013	$	&	$	0.08	\pm	0.02	$	&	$	0.051	\pm	0.035	$	\\
SIQR	&	$	0.025	\pm	0.006	$	&	$	0.028	\pm	0.007	$	&	$	0.08	\pm	0.02	$	&	$	0.06	\pm	0.02	$	&	$	0.011	\pm	0.005	$	&	$	0.013	\pm	0.006	$	&	$	0.05	\pm	0.01	$	&	$	0.033	\pm	0.016	$	\\

\hline
\end{tabular}
\tablefoot{%Columns are as follows: (1) SDSS short name. 
All intensities are normalized to the continuum flux at 1350 \AA\ (Col.7 of Table \ref{tab:general}). {BALQ are marked with an asterisk}(*). %The missing measurements correspond to BALQ objects or measurements that ended up negative due to absorptions. 
}
\label{tab:ProfileMeasurement}
\end{table*}

\begin{table*}[h]
\centering
\fontsize{8.2}{10}\selectfont %\tabcolsep=3pt
%\tabletypesize{\scriptsize\tabcolsep=2pt}
\tabcolsep=3pt
\caption{Flux ratios dependent on metallicity of normalized intensities measurements}
\begin{tabular}{ccccccccccccccccccc}
\hline\hline
 & {BC} & {BC} & {BC} & {BC} & {BC} & {BC} & {BLUE} & {BLUE} & {BLUE} \\
{SDSS JCODE}  & {\aliii/\heiiuv} & {\civ{}/\heiiuv} & {\siiv{}/\heiiuv} &
{\civ{}/\aliii} & {\siiv/\civ{}} & 
{\siiv/\aliii} & {\civ{}/\heiiuv} & {(\siiv{}+\oiv)/\heiiuv} & {\civ{}/(\siiv{}+\oiv)}\\
(1) & (2) & (3) & (4)  & (5)  & (6) & (7) & (8) & (9) & (10)  \\
\hline
\\
J0020+0740	&	$	2.62	\pm	1.00	$	&	$	6.25	\pm	2.72	$	&	$	5.83	\pm	2.06	$	&	$	2.38	\pm	1.54	$	&	$	1.07	\pm	0.30	$	&	$	2.22	\pm	0.43	$	&	$	5.12	\pm	2.00	$	&	$	2.72	\pm	1.20	$	&	$	1.88	\pm	0.46	$	\\
J0034$-$0326	&	$	2.24	\pm	1.13	$	&	$	3.78	\pm	3.07	$	&	$	3.29	\pm	1.53	$	&	$	1.69	\pm	2.04	$	&	$	1.15	\pm	0.86	$	&	$	1.47	\pm	0.57	$	&			\ldots			&			\ldots			&	$	3.06	\pm	3.21	$	\\
J0037$-$0238	&	$	1.96	\pm	0.70	$	&	$	7.53	\pm	2.49	$	&	$	6.29	\pm	2.08	$	&	$	3.83	\pm	1.48	$	&	$	1.20	\pm	0.17	$	&	$	3.20	\pm	0.63	$	&	$	4.29	\pm	1.09	$	&	$	2.55	\pm	0.92	$	&	$	1.68	\pm	0.56	$	\\
J0103$-$1104	&	$	1.77	\pm	0.98	$	&	$	4.43	\pm	2.41	$	&	$	3.88	\pm	2.15	$	&	$	2.50	\pm	1.01	$	&	$	1.14	\pm	0.28	$	&	$	2.19	\pm	0.58	$	&	$	5.50	\pm	4.00	$	&	$	2.09	\pm	1.86	$	&	$	2.63	\pm	1.59	$	\\
J0106$-$0855	&	$	3.59	\pm	1.61	$	&	$	6.25	\pm	3.05	$	&	$	8.14	\pm	3.64	$	&	$	1.74	\pm	0.81	$	&	$	0.77	\pm	0.24	$	&	$	2.27	\pm	0.57	$	&	$	6.66	\pm	4.49	$	&	$	3.56	\pm	2.91	$	&	$	1.87	\pm	0.90	$	\\
J0123+0329	&			\ldots			&			\ldots			&			\ldots			&	$	1.11	\pm	0.46	$	&	$	0.66	\pm	0.26	$	&	$	1.68	\pm	0.43	$	&	$	6.36	\pm	3.53	$	&	$	3.01	\pm	5.36	$	&	$	2.11	\pm	3.58	$	\\
J0210$-$0823	&	$	2.80	\pm	2.34	$	&	$	2.63	\pm	2.53	$	&	$	0.77	\pm	1.04	$	&	$	0.94	\pm	0.53	$	&	$	3.42	\pm	4.15	$	&	$	0.27	\pm	0.31	$	&	$	2.59	\pm	1.74	$	&	$	2.07	\pm	1.46	$	&	$	1.25	\pm	0.87	$	\\

J0827+0306	&	$	5.07	\pm	8.96	$	&	$		\ldots		$	&			\ldots			&	$	1.95	\pm	1.08	$	&	$	1.08	\pm	0.42	$	&	$	1.80	\pm	0.79	$	&	$	3.05	\pm	2.18	$	&	$	1.84	\pm	1.50	$	&	$	1.66	\pm	1.08	$	\\
J0829+0801	&	$	4.79	\pm	2.48	$	&	$	5.78	\pm	3.44	$	&	$	6.31	\pm	3.29	$	&	$	1.21	\pm	0.46	$	&	$	0.92	\pm	0.30	$	&	$	1.32	\pm	0.18	$	&	$	5.33	\pm	1.65	$	&	$	2.42	\pm	1.58	$	&	$	2.21	\pm	1.30	$	\\
J0836+0548	&	$	4.64	\pm	2.24	$	&	$	8.44	\pm	4.26	$	&	$	10.22	\pm	4.91	$	&	$	1.82	\pm	0.66	$	&	$	0.83	\pm	0.19	$	&	$	2.20	\pm	0.40	$	&	$	4.05	\pm	2.63	$	&	$	2.95	\pm	1.57	$	&	$	1.38	\pm	0.75	$	\\
J0845+0722	&	$	3.90	\pm	2.90	$	&	$	7.37	\pm	5.43	$	&	$	7.41	\pm	5.56	$	&	$	1.89	\pm	0.72	$	&	$	0.99	\pm	0.31	$	&	$	1.90	\pm	0.63	$	&	$	4.70	\pm	1.74	$	&	$	3.14	\pm	1.21	$	&	$	1.50	\pm	0.26	$	\\
J0847+0943	&	$	3.63	\pm	1.81	$	&	$	6.78	\pm	3.56	$	&	$	6.79	\pm	3.28	$	&	$	1.87	\pm	0.97	$	&	$	1.00	\pm	0.32	$	&	$	1.87	\pm	0.52	$	&	$	5.31	\pm	1.62	$	&	$	3.12	\pm	1.07	$	&	$	1.70	\pm	0.33	$	\\
J0858+0152	&	$	3.66	\pm	1.57	$	&	$	7.87	\pm	4.88	$	&	$	7.14	\pm	3.22	$	&	$	2.15	\pm	2.18	$	&	$	1.10	\pm	0.56	$	&	$	1.95	\pm	0.49	$	&	$	4.78	\pm	3.36	$	&	$	1.60	\pm	3.32	$	&	$	2.99	\pm	6.48	$	\\
J0903+0708	&	$	3.80	\pm	3.41	$	&	$	1.79	\pm	4.27	$	&	$	3.53	\pm	3.25	$	&	$	0.47	\pm	0.52	$	&	$	0.51	\pm	1.13	$	&	$	0.93	\pm	0.30	$	&	$	6.44	\pm	4.88	$	&	$	4.23	\pm	3.46	$	&	$	1.52	\pm	0.52	$	\\
J0915$-$0202	&	$	1.43	\pm	0.80	$	&	$	4.40	\pm	3.00	$	&	$	5.41	\pm	2.48	$	&	$	3.09	\pm	5.03	$	&	$	0.81	\pm	0.44	$	&	$	3.79	\pm	1.45	$	&	$	4.31	\pm	1.49	$	&	$	2.59	\pm	1.12	$	&	$	1.66	\pm	0.50	$	\\
J0926+0135	&	$	1.96	\pm	0.61	$	&	$	5.03	\pm	1.71	$	&	$	4.41	\pm	1.38	$	&	$	2.56	\pm	1.10	$	&	$	1.14	\pm	0.22	$	&	$	2.25	\pm	0.31	$	&	$	5.78	\pm	1.12	$	&	$	2.53	\pm	1.95	$	&	$	2.29	\pm	1.73	$	\\
J0929+0333	&	$	4.09	\pm	2.73	$	&	$	9.45	\pm	6.58	$	&	$	8.64	\pm	5.59	$	&	$	2.31	\pm	1.55	$	&	$	1.09	\pm	0.34	$	&	$	2.11	\pm	0.51	$	&	$	6.15	\pm	2.72	$	&	$	3.41	\pm	1.87	$	&	$	1.80	\pm	0.71	$	\\
J0946$-$0124	&	$	1.77	\pm	0.36	$	&	$	6.52	\pm	1.27	$	&	$	5.23	\pm	1.07	$	&	$	3.69	\pm	1.67	$	&	$	1.25	\pm	0.23	$	&	$	2.96	\pm	0.58	$	&	$	4.06	\pm	1.18	$	&	$	1.59	\pm	0.64	$	&	$	2.56	\pm	1.05	$	\\
J1024+0245	&	$	5.18	\pm	4.35	$	&	$	7.17	\pm	6.68	$	&	$	8.81	\pm	7.34	$	&	$	1.38	\pm	0.85	$	&	$	0.81	\pm	0.37	$	&	$	1.70	\pm	0.34	$	&	$	6.51	\pm	2.32	$	&	$	3.67	\pm	1.44	$	&	$	1.77	\pm	0.37	$	\\
J1026+0114	&	$	4.77	\pm	1.92	$	&	$	5.09	\pm	2.31	$	&	$	5.96	\pm	2.33	$	&	$	1.07	\pm	0.33	$	&	$	0.85	\pm	0.24	$	&	$	1.25	\pm	0.23	$	&	$	6.82	\pm	3.32	$	&	$	4.29	\pm	2.11	$	&	$	1.59	\pm	0.24	$	\\
J1145+0800	&	$	5.24	\pm	3.67	$	&	$	2.47	\pm	9.30	$	&	$	8.28	\pm	5.68	$	&	$	0.47	\pm	0.85	$	&	$	0.30	\pm	1.11	$	&	$	1.58	\pm	0.40	$	&			\ldots			&	$	4.79	\pm	6.75	$	&	$	1.57	\pm	0.51	$	\\
J1214+0242	&	$	4.22	\pm	1.23	$	&	$	8.76	\pm	2.92	$	&	$	7.20	\pm	2.16	$	&	$	2.08	\pm	0.79	$	&	$	1.22	\pm	0.26	$	&	$	1.71	\pm	0.24	$	&	$	2.43	\pm	3.66	$	&	$	1.84	\pm	2.79	$	&	$	1.32	\pm	2.82	$	\\
J1215+0326	&	$	2.11	\pm	0.52	$	&	$	6.52	\pm	3.54	$	&	$	6.70	\pm	1.47	$	&	$	3.09	\pm	4.79	$	&	$	0.97	\pm	0.49	$	&	$	3.17	\pm	0.48	$	&	$	5.06	\pm	1.72	$	&	$	1.74	\pm	0.94	$	&	$	2.90	\pm	1.68	$	\\
J1219+0254	&	$	1.56	\pm	0.37	$	&	$	6.24	\pm	1.32	$	&	$	3.90	\pm	0.81	$	&	$	4.00	\pm	1.20	$	&	$	1.60	\pm	0.16	$	&	$	2.50	\pm	0.38	$	&	$	3.36	\pm	0.72	$	&	$	1.22	\pm	0.67	$	&	$	2.76	\pm	1.47	$	\\
J1231+0725	&	$	0.87	\pm	0.21	$	&	$	6.06	\pm	0.80	$	&	$	2.71	\pm	0.40	$	&	$	6.96	\pm	5.13	$	&	$	2.23	\pm	0.36	$	&	$	3.12	\pm	0.81	$	&	$	4.19	\pm	1.49	$	&	$	0.04	\pm	1.18	$	&	$		\ldots		$	\\
J1244+0821	&	$	3.63	\pm	1.59	$	&	$	7.65	\pm	3.01	$	&	$	6.92	\pm	3.30	$	&	$	2.11	\pm	0.68	$	&	$	1.11	\pm	0.37	$	&	$	1.91	\pm	0.74	$	&	$	5.14	\pm	1.32	$	&	$	2.89	\pm	0.74	$	&	$	1.78	\pm	0.36	$	\\
J1259+0752	&	$	2.60	\pm	1.83	$	&	$	2.65	\pm	4.05	$	&	$	5.87	\pm	4.86	$	&	$	1.02	\pm	1.44	$	&	$	0.45	\pm	0.66	$	&	$	2.26	\pm	1.23	$	&	$	5.14	\pm	1.84	$	&	$	4.03	\pm	1.37	$	&	$	1.28	\pm	0.29	$	\\
J1314+0927	&	$	3.55	\pm	3.63	$	&	$	1.19	\pm	2.21	$	&	$	2.07	\pm	2.82	$	&	$	0.34	\pm	0.28	$	&	$	0.58	\pm	1.05	$	&	$	0.58	\pm	0.56	$	&	$	4.56	\pm	3.71	$	&	$	3.14	\pm	3.03	$	&	$	1.45	\pm	0.92	$	\\
J1419+0749	&	$	3.03	\pm	3.16	$	&	$	6.81	\pm	7.07	$	&	$	5.31	\pm	5.50	$	&	$	2.25	\pm	0.71	$	&	$	1.28	\pm	0.23	$	&	$	1.75	\pm	0.37	$	&	$	5.27	\pm	1.22	$	&	$	2.86	\pm	0.78	$	&	$	1.84	\pm	0.38	$	\\
J1509+0744	&	$	3.40	\pm	1.66	$	&	$	4.94	\pm	2.74	$	&	$	5.63	\pm	2.69	$	&	$	1.45	\pm	0.66	$	&	$	0.88	\pm	0.29	$	&	$	1.66	\pm	0.32	$	&	$	4.77	\pm	2.36	$	&	$	3.31	\pm	1.92	$	&	$	1.44	\pm	0.50	$	\\
J1519+0723	&	$	4.69	\pm	2.78	$	&	$	6.76	\pm	4.26	$	&	$	6.26	\pm	5.74	$	&	$	1.44	\pm	0.59	$	&	$	1.08	\pm	0.83	$	&	$	1.34	\pm	0.99	$	&	$	4.82	\pm	2.00	$	&	$	3.14	\pm	1.38	$	&	$	1.53	\pm	0.32	$	\\
J1545+0156	&	$	5.66	\pm	7.46	$	&	$	6.12	\pm	8.14	$	&	$	5.65	\pm	7.47	$	&	$	1.08	\pm	0.29	$	&	$	1.08	\pm	0.31	$	&	$	1.00	\pm	0.21	$	&	$	0.04	\pm	5.37	$	&	$	2.52	\pm	3.30	$	&	$	0.02	\pm	2.13	$	\\
J1609+0654	&	$	7.34	\pm	7.28	$	&	$		\ldots		$	&			\ldots			&	$	1.30	\pm	0.92	$	&	$	0.87	\pm	0.48	$	&	$	1.50	\pm	0.51	$	&			\ldots			&	$		\ldots		$	&	$	1.64	\pm	0.22	$	\\
J1618+0704	&			\ldots			&			\ldots			&			\ldots			&	$	0.97	\pm	0.40	$	&	$	0.72	\pm	0.29	$	&	$	1.36	\pm	0.38	$	&	$	5.90	\pm	5.79	$	&	$	3.75	\pm	4.05	$	&	$	1.57	\pm	0.79	$	\\
J2116+0441	&	$	3.01	\pm	1.22	$	&	$	11.13	\pm	4.21	$	&	$	6.33	\pm	2.48	$	&	$	3.70	\pm	1.20	$	&	$	1.76	\pm	0.28	$	&	$	2.10	\pm	0.46	$	&	$	2.52	\pm	1.24	$	&	$	0.92	\pm	1.51	$	&	$	2.75	\pm	4.60	$	\\
J2145$-$0758	&	$	5.85	\pm	9.23	$	&	$		\ldots		$	&			\ldots			&	$	2.81	\pm	2.03	$	&	$	1.50	\pm	0.51	$	&	$	1.87	\pm	0.67	$	&	$	3.20	\pm	1.26	$	&	$	0.31	\pm	1.29	$	&			\ldots			\\
		\\	\hline										
$\mu$	&	$	3.54	\pm	2.58	$	&	$	5.93	\pm	3.78	$	&	$	5.84	\pm	3.28	$	&	$	2.08	\pm	1.30	$	&	$	1.10	\pm	0.53	$	&	$	1.91	\pm	0.53	$	&	$	4.66	\pm	2.47	$	&	$	2.63	\pm	1.91	$	&	$	1.80	\pm	1.26	$	\\
$\sigma$	&	$	1.48	\pm	2.36	$	&	$	2.27	\pm	2.02	$	&	$	2.07	\pm	1.91	$	&	$	1.28	\pm	1.23	$	&	$	0.54	\pm	0.67	$	&	$	0.73	\pm	0.27	$	&	$	1.47	\pm	1.34	$	&	$	1.09	\pm	1.37	$	&	$	0.62	\pm	1.39	$	\\
SIQR	&	$	1.17	\pm	1.03	$	&	$	1.29	\pm	0.88	$	&	$	1.10	\pm	1.41	$	&	$	0.66	\pm	0.42	$	&	$	0.17	\pm	0.12	$	&	$	0.37	\pm	0.11	$	&	$	0.72	\pm	0.94	$	&	$	0.68	\pm	0.72	$	&	$	0.33	\pm	0.58	$	\\

\hline
\end{tabular}
\tablefoot{(1) SDSS short name, (2-7) line ratios of the broad components, (8-10) line ratios of the BLUE components. The measurements of the six BALQ were excluded from this table. }%\\ %Columns are as follows: (1) SDSS short name, ratios from normalized intensities measurements. The measurements of the six BALQ were excluded from this table.
%$^\ast$  The 6 BALQSO were excluded for this table.\\
%$^{\ast \ast}$ The ellipsis correspond to deficient measurements.
\label{tab:ratios_pm}
\end{table*}

We analysed our sample using the IRAF \texttt{specfit} task \citep{kriss94} to deblend the regions of interest centered at 1400, 1550 and 1900  \AA \ in the SDSS spectra. These regions are a mix of emission lines (such as {blue-shifted} and broad components)  and narrow absorption lines, making it necessary to obtain the best   fit  according to the minimum $\chi^2$ computed by a non-linear multi-component fit. %program \texttt{specfit} within {\tt IRAF}. 
The {Figures} of the fit results for the full sample are shown in Appendix \ref{app:spec}, and an example is shown in Figure \ref{fig:example_fit}. Each {Figure} shows  the continuum placement for the three regions (upper panels) and the fit of the blended lines with the appropriate components {(lower panels)}. The regions of each spectra are fit separately, however imposing an overall consistency on the continuum shape.

Each region was analyzed according to \citet{negreteetal12}. The components specific to the three spectral regions used are as follows: \paragraph{$1900$  \AA \ $region \ $:}
\begin{itemize}
    \item {The continuum}, modeled as a local power law using {spectral windows free from lines} at the edges of the emission regions to be {analyzed};
    \item {Broad components}, \aliii \ doublet and \siiii \ were modeled using symmetric Lorentzian  profiles  sharing the same FWHM, since both emissions have {an}
    ionization potential below 20 eV in most cases;
    \item{\ciii\ emission}, modeled using symmetric Lorentzian  profiles  with an independent FWHM; 
    \item \feiii\ emission was modeled using a template from \cite{bvb2008}, with {a} maximum in the red wing {of} the 1900 \AA\ blend; 
    \item {\feii\ emission} was found just for a third of the sample. A \feii\ template could not model  our data. We individually fit the  emission  at $\approx$ 1785   \AA\ due to the \feii\ multiplet UV 191 {using a symmetric Gaussian profile.}; %These emissions were usually found as a pair being the 1915  \AA \ the stronger emission and with a intensity similar to C III].
    \item{BLUE \aliii \ emission}. Although the region emitting most \aliii\ is believed to be the  virialized part of the BLR, we found evidence of outflows (i.e., excess blueshift with respect to the symmetric profile) in the blue wing of \aliii\ for 6 sources. This emission was modeled using a {blue-shifted} asymmetric Gaussian profile.
\end{itemize}

\paragraph{$1550$  \AA \ $region\ $:} 
\begin{itemize}
    \item {The continuum}, modeled as a local power law using {spectral windows free from lines} at the edges of the spectral region; 
    \item Broad components of \civ \ and \heiiuv \ modeled with Lorentzian profiles sharing the same FWHM;
    \item BLUE components of \civ \ and \heiiuv \ were modeled with asymmetric Gaussian profiles sharing the same FWHM and displacement from the broad component. {Blue-shifted} emissions tend to dominate this region with FWHM up to $\sim$ 15,000 km s$^{-1}$;
    \item Absorption lines were modeled as narrow Gaussians and included when necessary to improve the fit, especially in the blue wing of the profile.
\end{itemize}

\paragraph{$1400$  \AA \ $region\ $:} 

\begin{itemize}
    \item {The continuum}, modeled as a local power law using {spectral windows free from lines} prior and posterior to the emission regions;
    \item {Broad components.} The modeling was done as for the 1550  \AA \ region but skipping the BC emission of \oiiiuv{}. {This} emission is associated with a transition between levels with a well-defined and relatively low critical density \citep{marzianietal20}, {thus} its BC intensity is probably low;
    \item {The BLUE component} of the {blended} emission of \siiv + \oiv was modeled as a broad asymmetric Gaussian profile;
    \item {\cii\ emission}, if detected. The \cii $\lambda$1335 line was modeled with a symmetric profile; 
    %to improve the $\chi^2_\nu$;
     \item Absorption lines were modeled as narrow Gaussians and included when necessary to improve the fit, especially in the blue wing of the region.
\end{itemize}

%%%%%%%%%%%%%%%%%%%%%%%%%%%%%%%%%%%%%%%%%%%%%%%%%%%%%%%%%%%%%%%%%%%%%%%%%%%%%
\subsection{Error estimation on line fluxes}
\label{errors}

As described above, the regions  of interest in our spectra are blended into a very broad spectral emission. Each final model is affected by the  interpretation of the authors; however, the final parameters of each model component were determined by  \texttt{specfit} via an objective determination of the minimum $\chi^2$ i.e., of the lowest residuals. 

We believe the main source of uncertainty in our measurements   is due to noise and inter-line blending (once the model components have been selected, for fixed line width and equivalent width). 

To  estimate the errors in the flux, centroid and FWHM we created a Python routine based on  a \texttt{specfit} model (from Sec. \ref{fitting})  intended to cover the  components, EW and S/N range  for a selection of  sources from our sample. The goal of the routine is to create numerous random models and therefore measure the distributions of the free parameters and estimate an error to be associated to a group of spectra with similar characteristics.  The input is what we call a \textit{synthetic} spectrum: a replica of the characteristic \texttt{specfit} model plus a Gaussian white noise (as shown in Fig. 3). For each characteristic spectra we create  5 error models with  S/N = 10, 20, 30, 40, 50.

Although our sample is composed of objects from the same population, we still expect some spectral diversity, most spectra  differ in S/N and fitted components (e.g {BLUE} \aliii, \feiii\ $\lambda$1914). Therefore we chose a set of  characteristic spectra for the 1900 \AA\  region and another set for the 1550\AA \ and 1400 \AA \ region.

For the 1900 \AA \ region  we chose 7 models from  the spectra J1259+0752, J0020+0740, J1419+0749, J2145$-$0758, J1231+0726, J1618+0705 and J0107$-$0855, with EWs of 11, 23, 15, 20, 19, 11 and 17 \AA\ respectively (the input parameters for these models are described in Tables \ref{tab:aliii} and \ref{tab:BlueAliii}). Taking as an example the error models derived from J0020+0740, for the S/N=50 model the routine estimated an error of $\sim$20\% for the flux and $\sim$25\% for the FWHM, as we decrease the S/N the error percentage increases, so for the S/N=10 model the routine estimated an error of $\sim$40\% for the flux and $\sim$45\% for the FWHM. So for spectrum with a high S/N the associated error will be lower than those with a lower S/N, this behavior is shown in all of our error models and is in agreement with our measurement's main source of uncertainty.

Since the 1400 \AA \ and 1500 \AA \ region are similar in components and shape, the same uncertainties  obtained for the \civ\ line  has been used also  for the 1400 \siiv\ + \oiv\ blend. For these regions we chose 5 models from the spectra J0847+0943, J0929+0333, J1214+0242, J1231+0725 and J1259+0752 with EWs of 9, 14, 9, 21 and 3 \AA\ respectively (the input parameters for these models are described in Table \ref{tab:civ}).

The random models are generated with our routine, which consists of five basic steps:

\begin{enumerate}
    \item Create an array of possible values for each free parameter, where the free parameters are:
    \begin{itemize}
    \item intensity of the emission line ~\mbox{$-$50/+250} $\%$ in steps of 1 $\%$ from the original amplitude.
    \item   centroid position within $\pm$10 \AA \ for symmetric unblended  emissions, and $\pm$2 for heavily blended BC lines (e.g. \siiiifull, \ciiifull). For asymmetric  emissions we used an asymmetric range of variation $-$15/+2 \AA \ to avoid contamination from the BLUE component to the BC one. All variations were carry out in steps of 0.1 \AA.
    \item FWHM:  we vary the width of the emission $-$50/+250 $\%$ in steps of 1 km s$^{-1}$. 
\end{itemize}
\end{enumerate}

\begin{enumerate}
\setcounter{enumi}{1}
    \item Create a model, using our routine, with random parameters from our arrays;
    \item Iterate the routine; we decided to settled for 1,000,000 iterations which translates into 1,000,000 random models. The  random iterations yield a distribution of values for line fluxes.
    \item  Prune the distribution so   that only cases with $\chi^2$\ consistent within a  2$\sigma$ confidence level from the minimum $\chi^2$.
    \item Compute the dispersion of the distribution for each model with given $W$\ and S/N. The dispersion is the uncertainty at 1$\sigma$\ confidence level
\end{enumerate}
 
 The routine identifies the best fit  minimizing the reduced $\chi_\nu^2$ defined as:

\begin{equation}
    \chi_\nu^2 = \frac{1}{n_\mathrm{d}}\sum_\mathrm{i=1}^{n} \frac{(f^\mathrm{o}_\mathrm{i} - f^\mathrm{e}_\mathrm{i})^2}{\sigma_\mathrm{i}^2},
\end{equation}

where $f^\mathrm{o}$ is an element of the random model array, $f^\mathrm{e}$ is an element of the given input, $\sigma^2$ is the quadratic spread of the random model with the original signal, and $n\mathrm{d}$ is the number of degrees of freedom. The summation runs up to $n$, the length of the arrays which match the length of the {array for the} input spectra (i.e., the analyzed spectral region  re-sampled to 1 \AA).
 
%%%%%%%%%%%%%%%%%%%%%%%%%%%%%%%%%%%%%%%%%%%%%%%%%%%%%%%%%%%%%%%%%%%%%%%%%%%%%%%%
\subsection{Measurements of profile normalized intensities} % cambiar nombre 
\label{sec:profile}
\begin{figure}[h]
    %\vspace{-0.9cm}
     \centering
     %trim=left bottom right top
     \includegraphics[trim= 0.0 0.0 0.0 0., clip,width=0.45\textwidth]{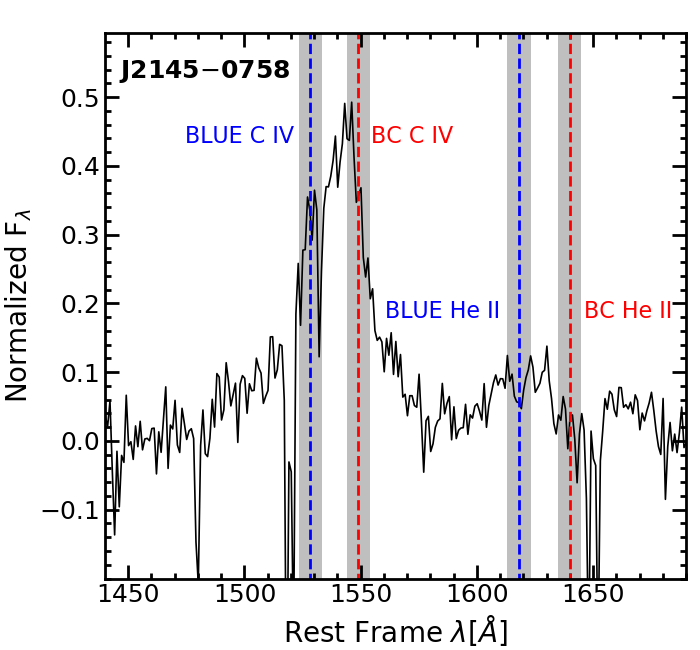}
    \caption[The 1550 \AA \ region used for profile ratio measurements.]{Example of the 1550 \AA \ region used for profile ratio measurements. The Figure shows in black  the observed spectra  minus the continuum. The red dashed lines indicate the rest frame wavelength of \civ \ and \heiiuv{}, the gray area surrounding these lines show the $\pm$1000 km s$^{-1}$ range of intensity measurement. The blue dashed lines indicate the BLUE emissions and are $-$4000 km s$^{-1}$ from their broad component companion, the gray area surrounding these lines also show the $\pm$1000 km s$^{-1}$ range of measurement. Abscissa correspond to the rest frame wavelength in \AA \ and ordinate normalized flux.}   
    \label{fig:pm_plot}
\end{figure}

The \texttt{specfit} multi-components fits are time-consuming procedures. In order to compare \texttt{specfit} results  with a simpler procedure we measure the intensity in specific radial velocity ranges of the emission lines to be used as metallicity indicators.  

We create a routine using Python which yields the amplitude in the region given an input central wavelength and around it $\pm$1000 km s$^{-1}$. The first  range is centered at 0 \kms\  and corresponds to the wavelengths where the BC emission is dominating (red dashed lines in Figure \ref{fig:pm_plot}). To represent  the BLUE component with any actual profile decomposition, we measure  the line intensity on the profile at $-$4000 km s$^{-1}$ from the broad component (blue dashed lines in Figure \ref{fig:pm_plot}) and $\pm$1000 km s$^{-1}$ around this wavelength (gray area in Figure \ref{fig:pm_plot}). {We chose the BLUE component position carefully with an offset with respect to the rest frame in order to avoid absorptions and to sample a spectral range where we expect emission from outflowing gas. In the case of an absorption line within the range of our measurement this results in negative values, which are not   physical meaningful and therefore were ignored.} To estimate errors we measured the distribution of the amplitudes within  $\pm$1000 km s$^{-1}$ for each central wavelength component (grey areas in Fig. \ref{fig:pm_plot}).

Even though the BC \civ{} \ is the main emission of the 1550 \AA \ region, Figure \ref{fig:pm_plot} shows that the peak of maximum intensity is on the blue side of the region despite of the \mgii{} \ redshift correction. This behavior is present in the majority of the sample which lead as to believe that the BLUE emissions are a constant in the process in xA sources instead of a occasionally emission.

%%%%%%%%%%%%%%%%%%%%%%%%%%%%%%%%%%%%%%%%%%%%%%%%%%%%%%%%%%%%%%%
\begin{table}[h]
\centering
\fontsize{9}{10}\selectfont
\tabcolsep=2pt
%\tabletypesize{\scriptsize\tabcolsep=2pt}
\caption{Physical Parameters}
\begin{tabular}{ccccc}
\hline\hline
{SDSS JCODE}  &  {$\log M_\mathrm{BH}$} & $\log L_\mathrm{bol}$ & {$\log L_\mathrm{vir}$} & \lledd\ \\
& [$M_\odot$] & [\ergss]  &  [\ergss] & \\
(1) & (2) & (3) & (4)  & (5)   \\  \hline
\\
J0020+0740	&	9.26	&	46.98	&	46.99	&	0.78	\\
J0034$-$0326	&	9.04	&	46.78	&	46.76	&	0.81	\\
J0037$-$0238	&	9.00	&	46.65	&	46.83	&	0.65	\\
J0103$-$1104	&	9.24	&	47.31	&	46.55	&	1.82	\\
J0106$-$0855	&	9.43	&	47.33	&	46.92	&	1.22	\\
J0123+0329	&	9.40	&	47.32	&	46.87	&	1.29	\\
J0210$-$0823	&	9.24	&	47.07	&	46.85	&	1.00	\\
\ \ J0216+0115*	&	9.21	&	47.05	&	46.81	&	1.02	\\
\ \ J0252$-$0420*	&	9.16	&	46.89	&	46.89	&	0.78	\\
J0827+0306	&	9.07	&	46.95	&	46.63	&	1.12	\\
J0829+0801	&	9.45	&	47.31	&	46.99	&	1.10	\\
J0836+0548	&	9.43	&	47.04	&	47.26	&	0.60	\\
J0845+0722	&	9.45	&	47.30	&	46.99	&	1.09	\\
J0847+0943	&	9.35	&	47.08	&	47.04	&	0.81	\\
J0858+0152	&	9.36	&	47.31	&	46.80	&	1.37	\\
J0903+0708	&	9.40	&	47.07	&	47.17	&	0.69	\\
J0915$-$0202	&	8.60	&	46.47	&	46.25	&	1.04	\\
J0926+0135	&	9.41	&	47.06	&	47.20	&	0.66	\\
J0929+0333	&	9.19	&	46.91	&	46.93	&	0.76	\\
\ \ J0932+0237*	&	9.52	&	47.19	&	47.26	&	0.71	\\
J0946$-$0124	&	8.86	&	47.08	&	46.07	&	2.46	\\
\ \ J1013+0851*	&	9.74	&	47.59	&	47.24	&	1.12	\\
J1024+0245	&	9.32	&	47.16	&	46.89	&	1.06	\\
J1026+0114	&	9.26	&	47.13	&	46.80	&	1.13	\\
J1145+0800	&	9.12	&	46.90	&	46.79	&	0.89	\\
\ \ J1205+0201*	&	9.88	&	47.68	&	47.41	&	1.00	\\
J1214+0242	&	9.71	&	47.52	&	47.26	&	1.01	\\
J1215+0326	&	9.34	&	47.31	&	46.75	&	1.46	\\
J1219+0254	&	9.57	&	47.28	&	47.25	&	0.79	\\
J1231+0725	&	9.02	&	47.33	&	46.10	&	3.14	\\
J1244+0821	&	9.49	&	47.29	&	47.08	&	0.97	\\
J1259+0752	&	9.50	&	47.45	&	46.91	&	1.41	\\
J1314+0927	&	9.26	&	47.13	&	46.81	&	1.11	\\
J1419+0749	&	9.57	&	47.51	&	46.99	&	1.37	\\
J1509+0744	&	9.20	&	46.89	&	46.97	&	0.72	\\
\ \ J1516+0029*	&	9.11	&	46.99	&	46.67	&	1.12	\\
 J1519+0723	&	9.30	&	47.10	&	46.93	&	0.94	\\
J1545+0156	&	9.53	&	47.38	&	47.07	&	1.08	\\
J1609+0654	&	9.55	&	47.31	&	47.19	&	0.87	\\
J1618+0704	&	9.39	&	47.17	&	47.02	&	0.92	\\
J2116+0441	&	9.29	&	47.12	&	46.89	&	1.00	\\
J2145$-$0758	&	9.27	&	47.20	&	46.75	&	1.29	\\
\\	\hline								
$\mu$	&	$9.33 \pm 0.13	$&	$47.15 \pm 		0.14$ &	$46.91 \pm 0.13$	&$	1.01\pm 	0.16$	\\
$\sigma$	&	0.24	&	0.24	&	0.29	&	0.47	\\
\hline
\end{tabular}
\tablefoot{(1) SDSS short name, (2) logarithm of the black hole mass in solar masses, (3,4) logarithm of the bolometric and virial luminosities in units of \ergss, and (5) logarithm of the Eddington ratio. Last two rows show the mean and the standard deviation {for each physical parameter}. {BALQ are marked with an asterisk}(*).}
\label{tab:Phys_par}
\end{table}

\subsection{ Estimation of   physical parameters: \mbh, luminosity, \lledd} 
\label{parameters}

 We determined some basic physical parameters using the continuum \texttt{splot} measurements  such as the bolometric luminosity ($L_{bol}$), the central supermassive black hole mass ($M_\mathrm{BH}$), and the Eddington ratio (\lledd);  in addition, we computed the virial luminosity ($L_\mathrm{vir}$).  The estimated parameter values are listed in Table \ref{tab:Phys_par}, and details of the computations are reported below.

\paragraph{\textit{Bolometric Luminosity:}}

We used the conventional way to derive the luminosity, \mbox{$L = 4 \pi d_\mathrm{c}^{2}f$}  where $d_\mathrm{c}$ is the co-moving distance  according to \citet{sulenticetal06}

 \begin{equation}
 \centering
 d_\mathrm{c} \approx \frac{c}{H_0}\left[ 1.500(1-e^{-\frac{z}{6.309}}) + 0.996(1-e^{-\frac{z}{1.266}})\right],
\label{dcaprox}
\end{equation}

\noindent and $c$ is the speed of light, $H_0$ the Hubble constant assumed as 70 \kms\ {Mpc$^{-1}$} and $z$\ the  source final redshift i.e., the SDSS redshift after correction. The flux $f$ = $\lambda f_\lambda$\ is in the {quasar rest frame}. A bolometric correction factor according to \citet{richardsetal06} was applied to compute the bolometric luminosity.

\paragraph{\textit{Black Hole Mass and Eddington ratio:}}

Black hole mass estimations tend to be over-estimated when using \civ-based scaling relations because of its strong blue excess \citep{marzianietal19}. In an effort to obtain a more accurate estimation we used the \aliii-based mass scaling relation proposed by \citet{marzianietal22}: 
\begin{eqnarray} 
\log  {M_\mathrm{BH}}(\mathrm{Al\ III}) & \approx & (0.579^{+0.031}_{-0.029})\log L_{1700,44} + \\ \nonumber
& &  2\log [\mathrm{FWHM (Al\ III)}]   + (0.490 ^{+0.110}_{-0.060}). 
\label{NewMass}
\end{eqnarray}

\noindent This relation requires only two parameters  $L_{1700,44}$  the continuum luminosity at 1700 \AA \ normalized by $10^{44}$ {erg s$^{-1}$} and the FWHM of \aliii \ in units of km s$^{-1}$.

Using the $M_{BH}$ {obtained from the} previous {step} we estimated the Eddington luminosity  following

\begin{equation}
L_{Edd}=1.3 \mathsf{x} 10^{38} \left(\frac{M}{M_{\odot}}\right) \mathrm{erg\, s}^{-1}. \label{Le}
\end{equation}

Given this luminosity we obtained  the Eddington ratio (\lledd), which is  an indicator of the BH accretion rate.

\paragraph{\textit{Virial Luminosity:}}

We computed the redshift-independent virial luminosity in the form proposed by  \citetalias{marzianisulentic14}:

\begin{equation}
L_\mathrm{vir} \approx 7.8 \mathsf{x} 10^{44} \gimel \frac{1}{(n_H U)_{10^{9.6}}} (\delta v)^4_{1000} 
\label{eq:virL}
\end{equation}

\noindent where the $\gimel$\ is a function of the AGN SED and of the Eddington ratio that can be assumed constant and equal to 1 for the xA population. The product $n_HU$\ has been normalized to the typical value $10^{9.6}$ cm$^{-3}$ and $\delta\nu$ to 1000 km s$^{-1}$\ \citep[more details are provided by][]{negreteetal18,dultzinetal20}. In practice, the virial luminosity has been computed by   multiplying the numerical factor and the virial estimator  $\delta v$\ (FWHM \aliii), raised to the fourth power.

%%%%%%%%%%%%%%%%%%%%%%%%%%%%%%%%%%%%%%%%%%%%%%%%%%%%%%%%%%%%%%%%%%%%%%%%%%%%%%%%%%
\subsection{Photoionization modeling}
\label{photoion}

To interpret our fitting results we compare the line intensity ratios for BC and BLUE observed components with the ones predicted by {\tt CLOUDY} simulations \citep{ferlandetal17}. An array of simulations is used as reference for comparison with the observed line intensity ratios. It was computed under the assumption that  (1) the column density is $N_\mathrm{c}$\ = 10$^{23}$ cm$^{-2}$; (2) the continuum is represented  by the model continuum of \citet{mathewsferland87} which is believed to be appropriate for Population A quasars, and (3)  {the} microturbulence is negligible. The simulation arrays cover the hydrogen  density range 7.00 $\leq \log(n_\mathrm{H}) \leq$ 14.00 and the ionization parameter $-4.5 \leq \log(U) \leq$ 1.00, each in intervals of 0.25 dex. They are repeated for values of metallicities in a range encompassing five orders of magnitude: 0.01, 0.1, 1, 2, 5, 10, 20, 50, 100, 200, 500 and 1000 $Z_{\odot}$. Extremely high metallicity $Z \gtrsim 100 Z_{\odot}$\ {are} considered  {to be} physically unrealistic ($Z \approx 100 Z_\odot$ implies that more than half of the gas mass is made up by metals!), unless the enrichment is provided {\em in situ} within the disk, a possibility briefly discussed in Section \ref{pollu}. In several cases, simulations suffered convergence problems if $Z \gtrsim 500 Z_\odot$. The behavior of diagnostic line ratios as a function of $U$ and \nh\ for selected values of $Z$ {are} shown in the Appendices of \citetalias{sniegowskaetal21}.

%%%%%%%%%%%%%%%%%%%%%%%%%%%%%%%%%%%%%%%%%%%%%%%%%%%%%%%%%%%%%%%%%%%%%%%%
\section{Results}
\label{sec:results}

We aim to estimate the metallicity {($Z$)} of the xA sources, as well as the correlations between $Z$\ and physical parameters. We first report the measurements of the emission line components, and of their ratios (Sect. \ref{sec:immediate}), and then consider the results of $Z$\ estimates. After confirming the xA nature for most sources of our sample (Sect. \ref{intruders}), we  seek for correlations in the sample (Sect. \ref{sec:correlation}).  {Thereafter we} present the $Z$\ estimates,  first assuming fixed physical conditions (i.e., for a given \nh\ and $U$, different for the regions associated with BLUE and BC, Sect. \ref{fixed}), and then considering \nh\ and $U$\ free to vary (Sect. \ref{ZfreenU}). 
 We discuss the spectral appearance of a few subgroups and of some remarkable objects in the sample in Appendix \ref{ind}.

%%%%%%%%%%%%%%%%%%%%%%%%%%%%%%%%%%%%%%%%%%%%%%%%%%%%%%%%%%%%%%%%%%%%%%%%%%%%%%%%%%
\subsection{Immediate Results}
\label{sec:immediate}
The analysis of the spectra involve the de-blending of broad  and BLUE components. The rest-frame spectra with the continuum placements, and the fits to the blends of the spectra are shown in Appendix \ref{app:spec}. We isolate the contribution of the components in two ways: (1) using the \texttt{specfit} task, and (2) measuring the amplitude of the emission $\pm$ 1000 km s$^{-1}$ around a central wavelength.  

\paragraph{{\tt Specfit} measurements.} Tables \ref{tab:aliii} and \ref{tab:BlueAliii} report the measurements of the 1900 \AA \ region. The columns in Table \ref{tab:aliii} list (1) SDSS identification short code, (2) equivalent width between 1780-2000  \AA, (3) the FWHM of \aliii\ and  \siiii, (4,5) \aliii\ and \siiii\ line fluxes, (6,7) \ciii \ FWHM and flux and (8,9) the  measurements of the   emission of iron at 1914  \AA. Table \ref{tab:BlueAliii} reports the additional measurements for the presence of an excess of \aliii\ for six sources. The Columns list (1) the SDSS identification short code, (2,3) the FWHM and flux, (4) the separation {of  the peak of the BLUE component} from the rest-frame wavelength of \aliii\ in units of angstroms, and (5) the asymmetry (the skewness parameter as reported by {\tt specfit}) of the emission line. Table \ref{tab:civ} reports the measurements of the 1500  \AA \ region for both BLUE and BC components. The columns list (1) the SDSS identification short code, (2) the equivalent width between 1450-1680  \AA, (3) broad components' FWHM, (4,5) flux of \civ \ and \heiiuv{} BC, (6) BLUE components' FWHM, (7,8) flux of \civ \ and \heiiuv{} BLUE, (9) the separation {of the peak of the BLUE component} from the rest-frame wavelength of \civbc\ in units of \AA\ and (10) the asymmetry of the BLUE component according to \texttt{specfit}. Lastly, Table \ref{tab:siiv} lists the results for the 1400 \AA\ region for both BLUE and BC components in the following order: (1) SDSS identification short code, (2) equivalent width between 1340-1450  \AA, (3) broad component FWHM, (4) flux of \siiv BC, (5) BLUE component FWHM, (6) flux of  \siiv $+$\oiv{} \ BLUE, (7) the separation {of the peak of the BLUE component} from the rest-frame wavelength of \siiv\ BC in units of \AA~, and (8) the asymmetry of the emission line. Table \ref{tab:ratios} reports the flux ratios of the metallicity indicators that will be used to estimate the metal content alongside {\tt CLOUDY} simulations.
    
\paragraph{Normalized intensities measurements.} Table \ref{tab:ProfileMeasurement} reports the measurements of the three regions. For the \civ\ and \siiv\ regions, measurements are reported for two ranges, one corresponding to the rest frame and a second meant to measure the {blue-shifted} excess.   The columns list SDSS identification short code and the normalized intensities. Table \ref{tab:ratios_pm} reports the ratios of the metallicity indicators.
 
%%%%%%%%%%%%%%%%%%%%%%%%%%%%%%%%%%%%%%%%%%%%%%%%%%%%%%%%%%

\begin{figure*}
     \centering
     %\hspace{-1cm}
     %trim=left bottom right top
          \includegraphics[trim= 0.5 0.5 0.5 300.5, clip, height=18cm]{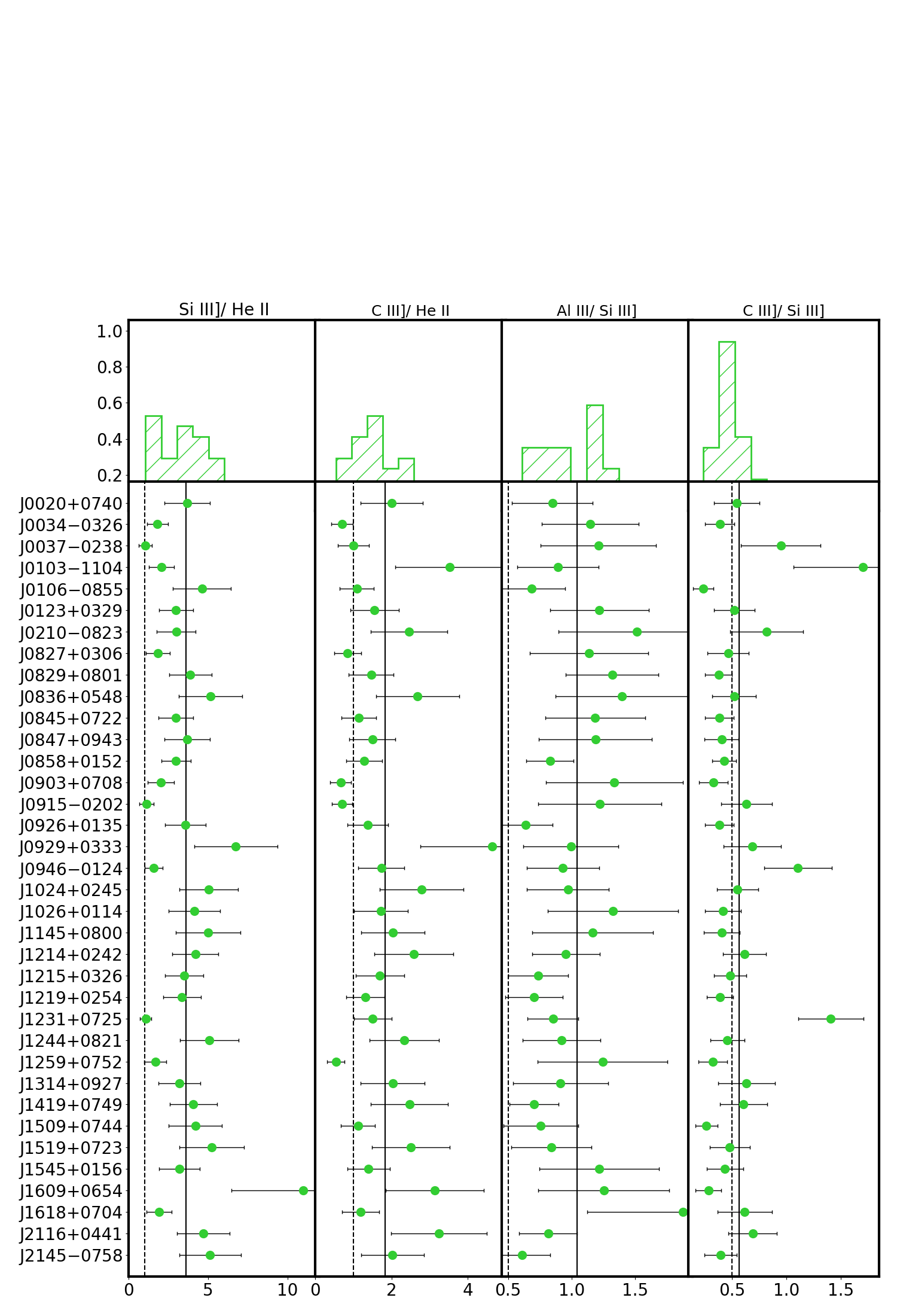}
    \caption{Distribution (top) and individual (bottom) values of intensity ratios between prominent lines not used to estimate metallicity. The two panels on the right (\aliii/\siiii\ and \ciii/\siiii) show the ratios involved in the identification of xA sources. Vertical solid lines identify the mean value of the distribution. Dashed lines are set to 1 for the \siiii/\heiiuv\ and \ciii/\heiiuv\ distributions, and 0.5 and 1.0 for the \aliii/\siiii\ and \ciii/\siiii\ distributions respectively, according to the UV selection criteria {for the xA} (Sect. \ref{intruders}).}    
    \label{fig:p_ratios}
 \end{figure*}

\subsubsection{BC intensity ratios}

Fig. \ref{fig:p_ratios} includes the individual  ratios that are   used to {distinguish} xA sources. The lower panels show the results for individual non-BAL QSOs  and their errors. The vertical continuous lines show the mean values of the distributions and the dashed vertical lines are set to 1 for   \siiii/\heiiuv\ and \ciii/\heiiuv. For \aliii/\siiii\ and \ciii/\siiii, the dashed vertical lines are set to 1 and 0.5 respectively according to xA criteria and Fig. \ref{fig:F6_MP}. The outlying sources identified were J0103$-$1104, a low EW source with a strong \ciii\ emission, and J1231+0725, a BC-dominated spectrum with {a} faint BLUE. 

\begin{figure*}[]
     \centering
     \hspace{-0.3cm}
     %trim=left bottom right to   
     \includegraphics[trim= 0.5 0.5 0.5 200.5, clip, width=0.5\textwidth]{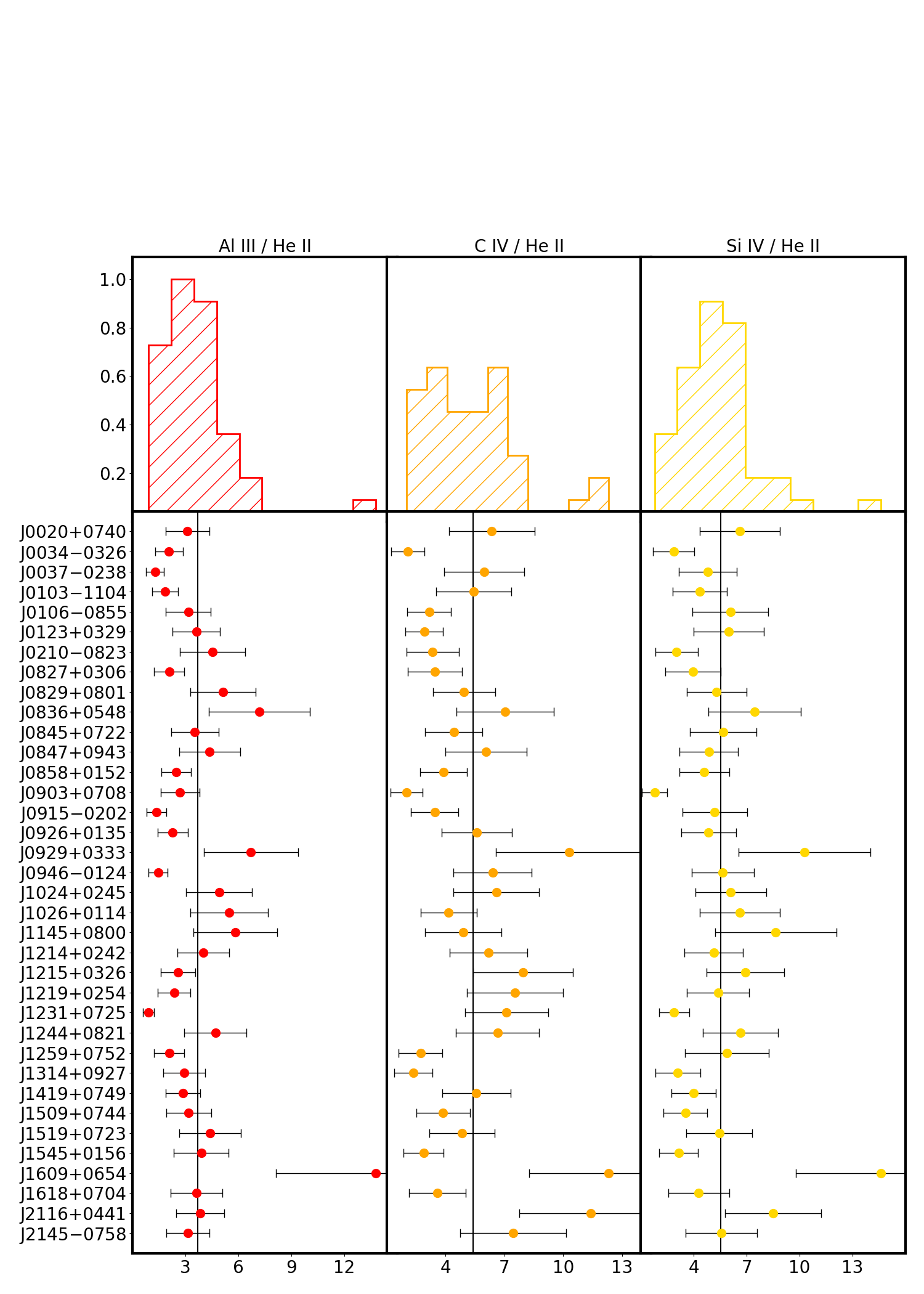}
        \includegraphics[trim= 0.5 0.5 0.5 200.5, clip, width=0.5\textwidth]{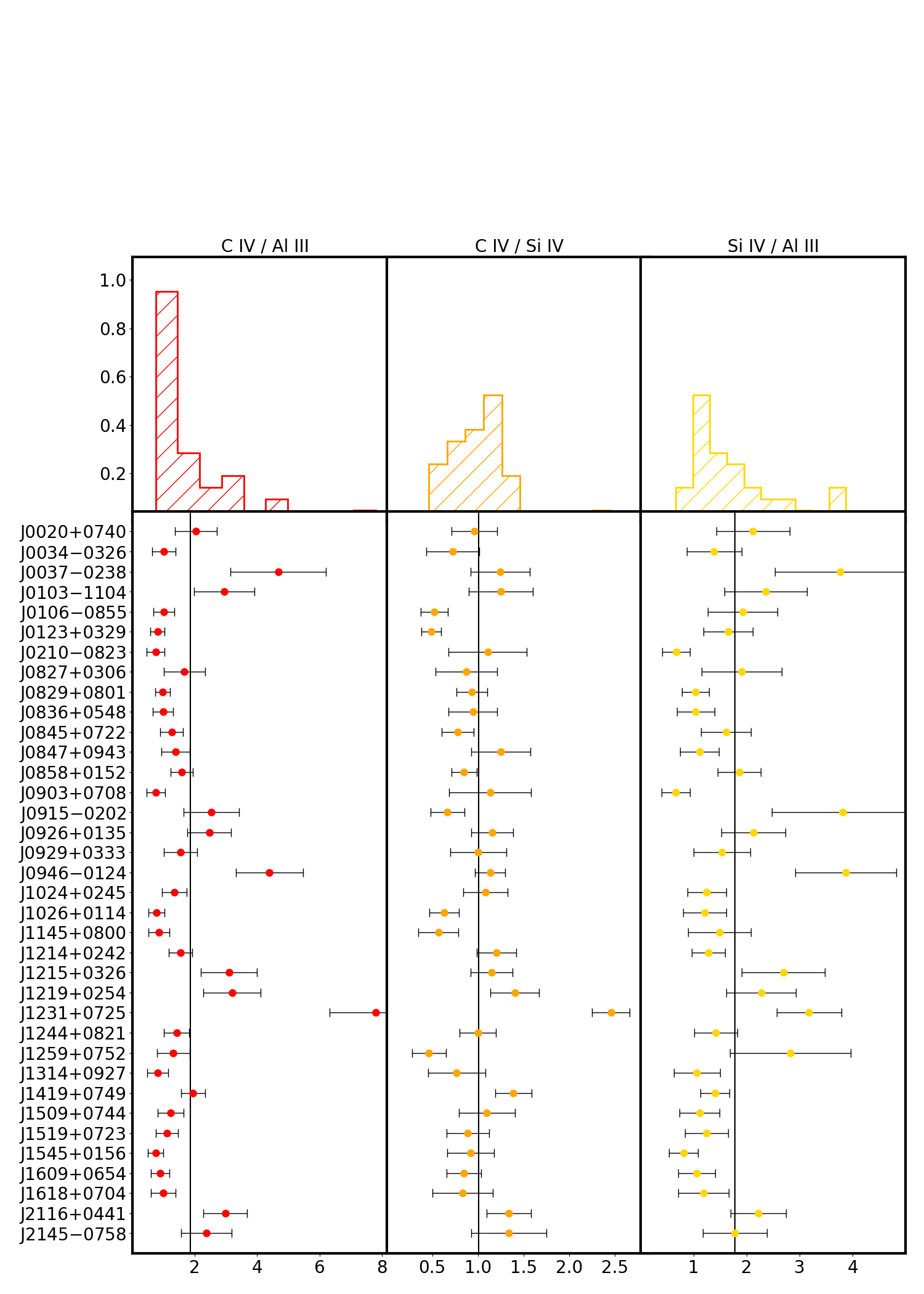}
    \caption{Distribution (top) and individual (bottom) values of intensity ratios of the broad components used as diagnostics for $Z$\ involving \heii\ (left), and not involving \heii\ (right). Values are reported in Columns from 2 to 7 of Table \ref{tab:ratios_pm}.}   
    \label{fig:redmix}
 \end{figure*}

The left panel of Fig. \ref{fig:redmix} shows the normalized distributions of the diagnostic intensity ratios \aliii/\heiiuv,  \civ/\heiiuv\ and \siiv/\heiiuv\  for the BC, {while the right panel shows} the  diagnostic intensity ratios of \civ/\aliii, \siiv/\civ\ and \siiv/\aliii\ broad components.  The vertical lines identify the mean values of the distributions:  $\mu$(\aliii/\heiiuv)=3.70, $\mu$(\civ/\heiiuv)=5.42, $\mu$(\siiv/\heiiuv)=5.55,  $\mu$(\civ/\aliii)=1.86, $\mu$(\siiv/\civ)=1.01 and $\mu$(\siiv/\aliii)=1.78.

The distribution of the data points involving \heiiuv \ (Fig. \ref{fig:redmix} left panel) shows a small dispersion from the sample mean with the exception of J1609+0654, a type-1 QSO with weak broad components especially for \heiiuv\ which for this case was more likely underestimated due to absorptions and low S/N. Fig. \ref{fig:redmix} right panel distributions show more sources outside the trend. An example is J1231+0725, a clear non-xA source. Higher values of the \civ/\heiiuv\ and \siiv/\heiiuv\ ratios are associated with higher ionization parameters and with the 3 cases of non-xA and borderline sources identified in Sect. \ref{intruders}.

Since the three \heiiuv \ ratios are, for fixed physical conditions, proportional to metallicity, we expect an overall
consistency in their behavior i.e., if {the ratio for} one object is higher than the sample mean, the other intensity ratios {for} the same object should also be higher. The lower panels are helpful to identify sources for which only one intensity ratio deviates significantly from the rest of the sample. The lack of consistency {in the distributions of line ratios not involving \heii shown in the right panel of }Fig. \ref{fig:redmix} was also found for its {\tt CLOUDY} metallicity relations for fixed physical conditions (see Sect. \ref{ZfixednU} ). {Hence,} these relations won't be considered for further analysis.

\begin{figure}
     \centering
     \hspace{-0.3cm}
     %trim=left bottom right to   
     \includegraphics[trim= 0.5 0.5 0.5 200.5, clip, width=0.5\textwidth]{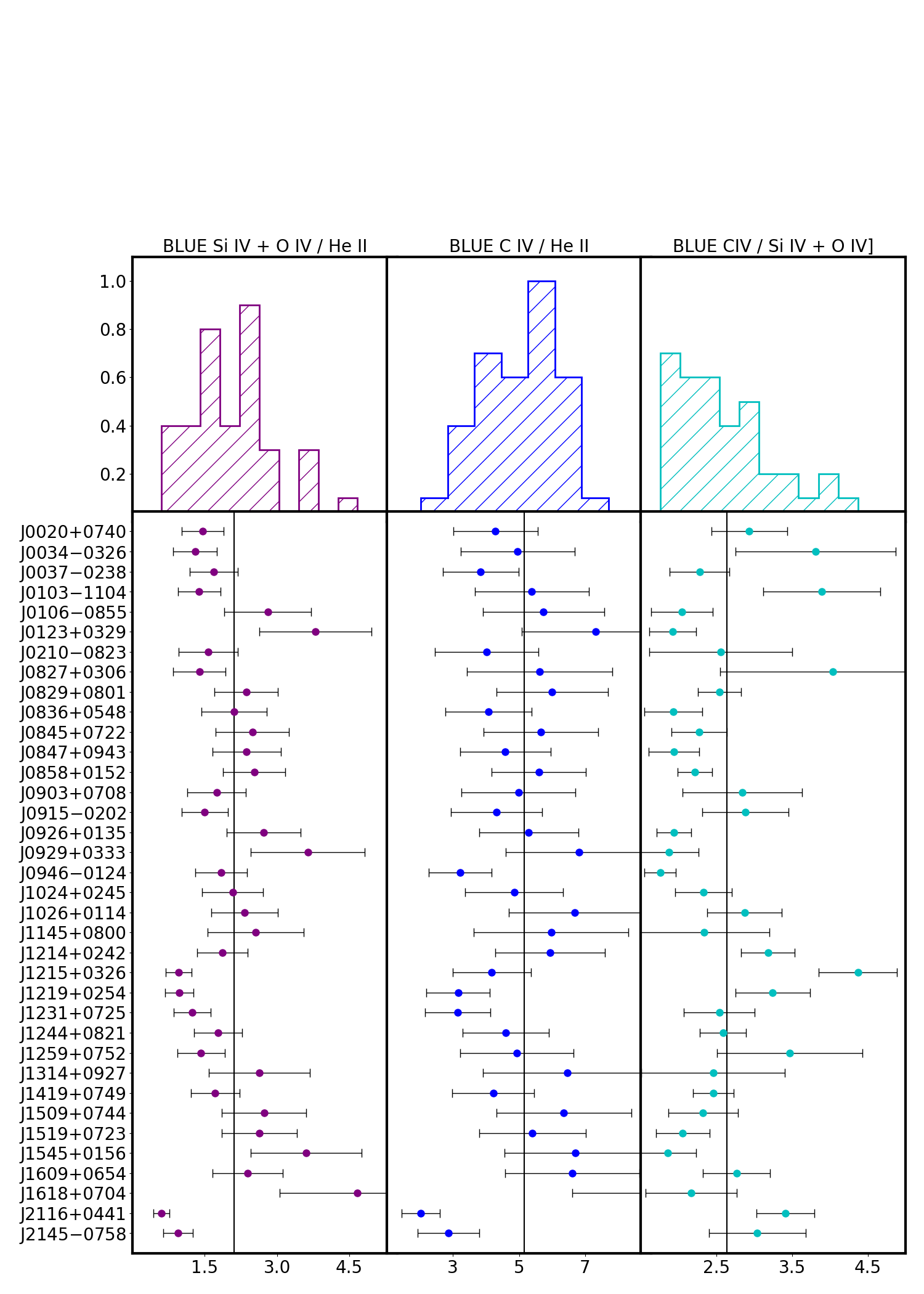}
    \caption{Intensity ratios for the BLUE components, shown in the same layout of the previous figures. Values are reported in Columns from 8 to 10 of Table \ref{tab:ratios_pm}.}   
    \label{fig:bluemix}
\end{figure}

 %%%%%%%%%%%%%%%%%%%%%%%%%%%%%%%%%%%%%%%%%%%%%%%%%%%%%%%%%%%
\subsubsection{BLUE intensity ratios}

Fig. \ref{fig:bluemix} shows the distributions of the diagnostic intensity ratios between the BLUE components of  \civ / \heiiuv{}, (\siiv + \oiv)/ \heiiuv{} and \civ /(\siiv + \oiv). The vertical lines identify the mean values of the distributions,  $\mu$(\civ / \heiiuv{})= 5.17,  $\mu$((\siiv + \oiv)/ \heiiuv{})=2.11, $\mu$(\civ /(\siiv + \oiv))=2.64.
The distributions show wide error bars as well as large {dispersions}. For most of the sources, the \siiv + \oiv \  BLUE component emission is slightly weaker than \civ \ BLUE component, resulting in {well-behaved} distributions with the exception of J2116+0441 and J2145$-$0758 that show weak \siiv + \oiv \ BLUE emission surrounded by {absorptions} and affected by  poor S/N. { BLUE \civ /(\siiv + \oiv)  ratios  show an apparently more confused distribution. However, two groups can be identified:  a majority one with median BLUE \civ /(\siiv + \oiv $\approx$ 2.5, and a minority one with larger BLUE \civ /(\siiv + \oiv) $\approx 4$. The higher ratios that can be {linked to heavy absorptions affect} the \siiv + \oiv\ profile. The source J1215+0326 shows the highest ratio due to the strong absorption in both wings of the 1400 \AA \ region. Indeed the BLUE ratios for J1215+0326 appear not to be fully consistent due to the absorption that affects also the red side of the \siiv\ profile and introduces a large source of uncertainty.}

%%%%%%%%%%%%%%%%%%%%%%%%%%%%%%%%%%%%%%%%%%%%%%%%%%%%%%%%%%%%%%%%%%%%%%%%%%%

\begin{figure}[h]%9
    \centering
     %trim=left bottom right top
     \includegraphics[trim= 0.0 0. 0. 0., clip, width=0.45\textwidth]{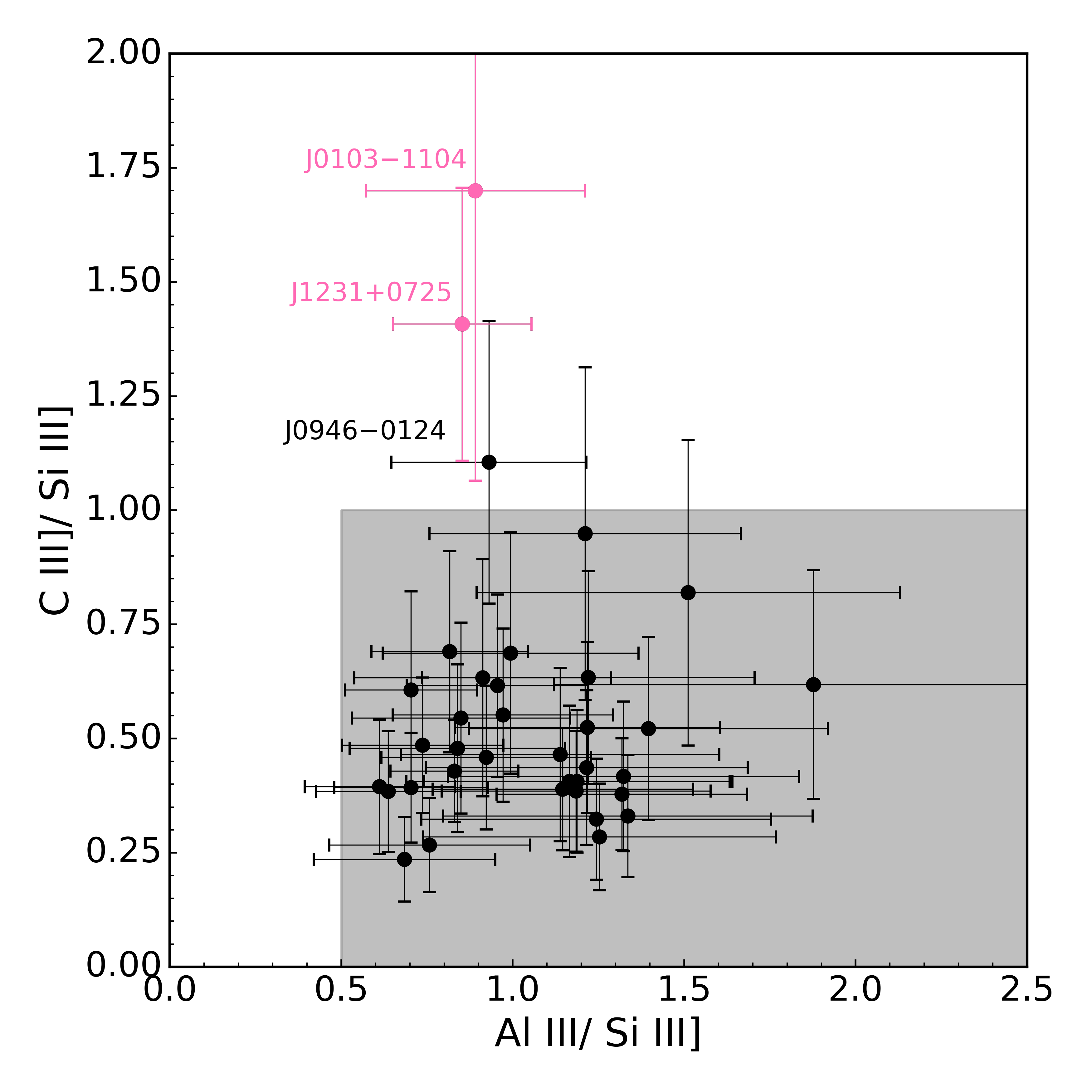}
    \caption{Relation between \aliiifull/\siiiifull\ and \ciiifull/\siiiifull\ intensity ratios. The gray area corresponds to the parameter space occupied by xA objects. Outlying sources are {shown} in pink color.}   
    \label{fig:F6_MP}
 \end{figure}

\subsection{Identification of xA sources and of ``intruders"}
\label{intruders}

\begin{figure*}[h]%10
    \centering
     %trim=left bottom right top
     \includegraphics[trim= 0.0 0. 0. 0., clip, width=\textwidth]{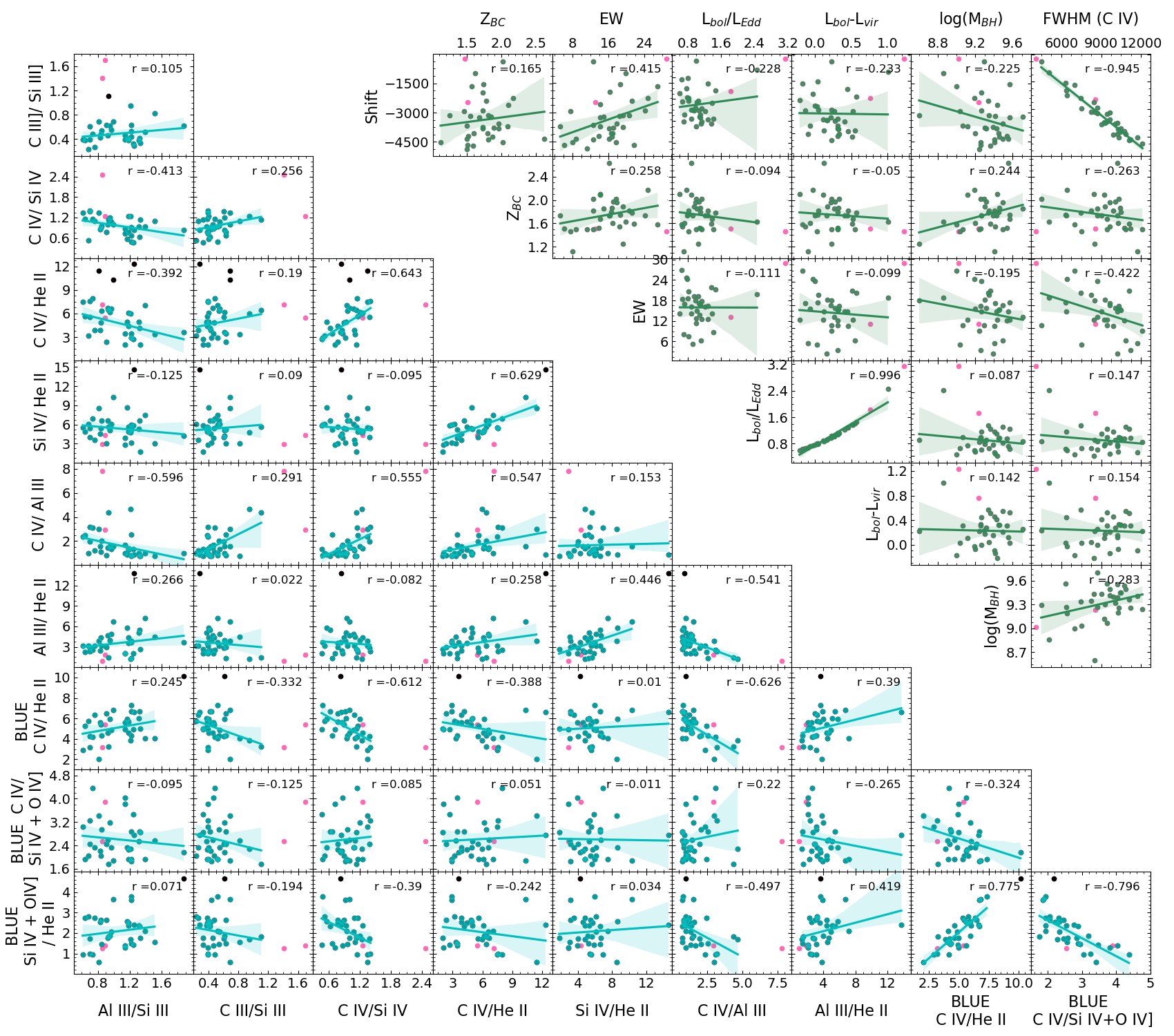}
    \caption{Correlations between diagnostic ratios for the broad and BLUE components (\textit{left lower matrix}), and between physical parameters (\textit{right upper matrix}). The pink dots are the outlier sources described in Sect. \ref{intruders} and were not considered in the statistical analysis. Additionally, for each panel we excluded, if necessary, the objects with the highest dispersion from the main relation (black dots). %For both matrixes, the 
    The blue and green dots are the values considered for each relation, the solid lines show the linear fit, the shaded areas represent the confidence interval of the lineal regression. For each relation the Spearman correlation coefficient ($r$) is shown in the right top corner. 
    }
    \label{fig:9x9m}
 \end{figure*}

Figure \ref{fig:F6_MP} shows the UV-plane selection criteria. The vast majority of our sample are xA quasars except for the two sources, J0103$-$1104 and J1231$+$0726 that were previously identified as outlying sources. They show a spectra not consistent with the xA spectral type and will not be considered for further statistic analysis. On the converse, the borderline object J0946$-$0124 is outside the xA part of the parameter plane. However, its error bars indicate that the criteria are still satisfied within 1 $\sigma$ confidence level. The overall appearance of the spectrum (Fig. \ref{fig:11}) supports the classification of J0946$-$0124 as an xA source. 

The lack of a clear trend shown in Figure \ref{fig:F6_MP}  may be due to the prior selection criteria for the xA bin in the 4DE1 parameter, because we are zooming into one bin of the 4DE1 that shows a small dispersion around well-defined parameter values.

%%%%%%%%%%%%%%%%%%%%%%%%%%%%%%%%%%%%%%%%%%%%%%%%%%%%%%%%%%%%%%
\subsection{Correlation between diagnostic ratios and physical parameters}
\label{sec:correlation}

Figure \ref{fig:9x9m} shows the correlations between all diagnostic ratios  employed in this work. Blue plots (left lower matrix) show the correlations between the diagnostic ratios from Table \ref{tab:ratios}, for both broad and blue components. Green plots (upper right matrix) represent the correlations between the physical parameters derived for each object. The pink dots correspond to the intruders described in Sect. \ref{intruders} and are not considered for the statistic analysis. In order to get a better correlation we also excluded  from the analysis  sources located very far from the main relation in the parameter plane due to measurement problems, the most frequent ones being absorptions and low S/N. These sources are shown as black dots.  The solid lines show the linear relation {for the} remaining sources. The Spearman correlation coefficient is reported at the  top right corner of each panel.

Most of the relations in Figure \ref{fig:9x9m} show a weak trend, in spite of the elimination of outlying sources. This  may be due to the prior selection criteria for the xA bin in the 4DE1 parameter, because we are zooming into one bin of the 4DE1 that shows a small dispersion around well-defined parameter values.

The panels involving BLUE \civ /(\siiv + \oiv) (second row from bottom) show no significant correlation as the flux ratio remains approximately constant. This constant value may imply that the wind emission in the HILs is active throughout the sample, and that the emitting gas might be in similar physical conditions. 

The relations that show the highest correlation coefficient (\textit{r}$>$0.77, last two panels of the last row) are between the BLUE components. However, these ratios are not statistically independent since they involve the same lines in abscissa and ordinate, and all of them are HILs and need similar conditions to occur.

Among all the ratios in the correlation matrix the most significant result is the relation between  BLUE \civ /\heiiuv{} and BC \civ / \aliii\ with a correlation coefficient of \textit{r} = 0.626,  showing a connection between the metallicity indicators for the BC and the BLUE. This correlation is consistent with the overall high-$Z$ scenario emerging from the present work: BLUE \civ /\heiiuv{} decreases with increasing \civ / \aliii, and both ratios increase with decreasing metallicity (Section \ref{ZfixednU} and following). However, also this anti-correlation should be viewed with care: most of the data points cluster around a typical value.  

\begin{figure*}[htbp]%4
     \centering
     \hspace{-1cm}
     %trim=left bottom right top
     \includegraphics[trim= 0. 0. 0. 0, clip, width=\textwidth]{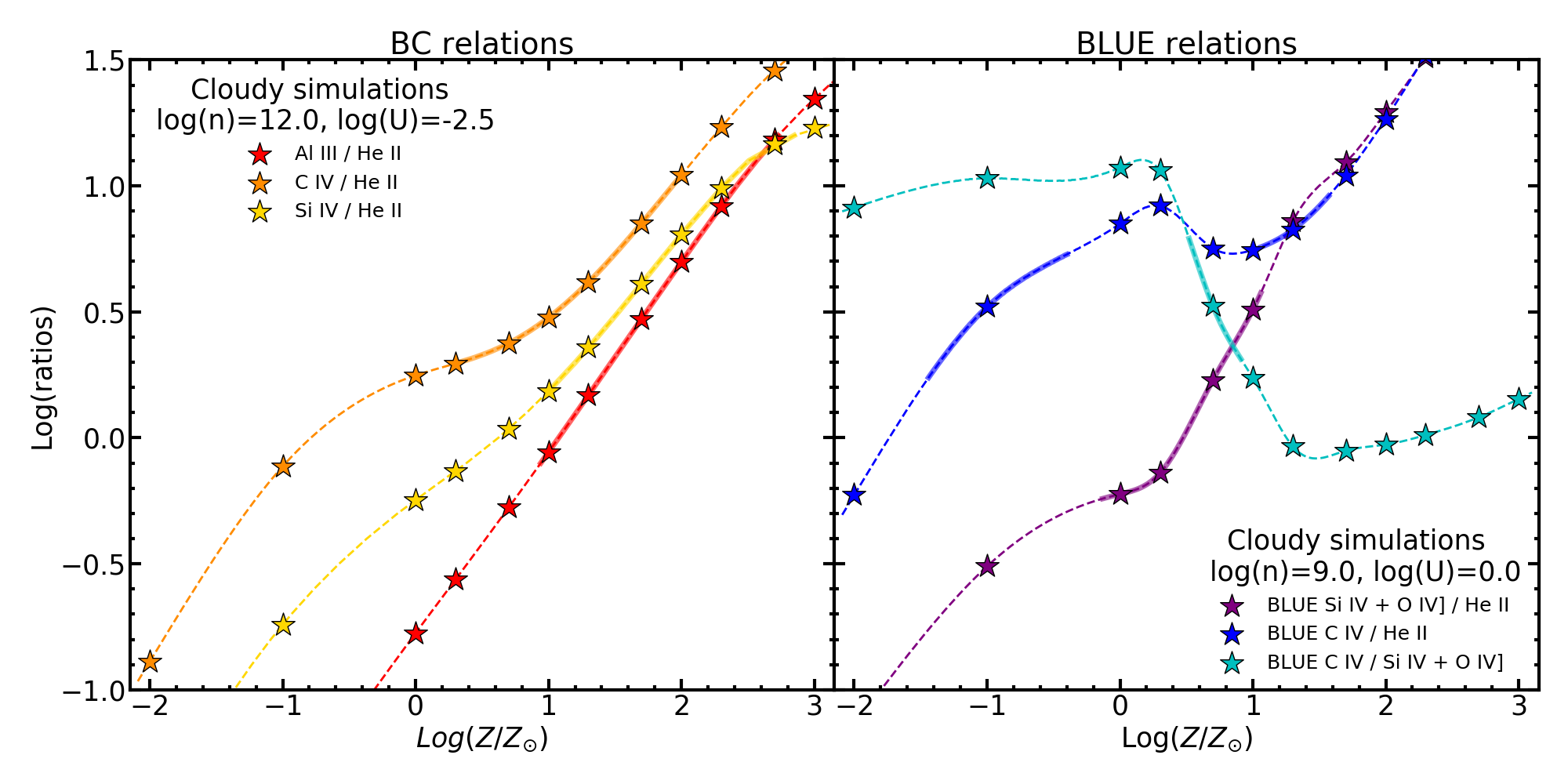}
    \caption{Intensity ratios trends predicted by {\tt CLOUDY} as a function of the logarithm of the metallicity, for fixed physical parameters $U$ and \nh. The \textit{left} panel shows the ratios of low/intermediate ionization potential  BC lines. The \textit{right} panel shows the ratios for high ionization potential emissions (BLUE components). In both panels, the stars represent the ratios predicted by {\tt CLOUDY}. The dashed lines are the interpolation between the {\tt CLOUDY} values obtained with a cubic spline, the thick solid line segments in each trend line show the range of line ratios covered by our sample, according to the measured flux ratios.
    }   
    \label{fig:trend}
 \end{figure*}

%%%%%%%%%%%%%%%%%%%%%%%%%%%%%%%%%%%%%%%%%%%%%%%%%%%%%%%%%%%%%%%%%%%%%%%%%%%%%%%
\begin{table*}
\centering
{\fontsize{9.5}{10}
\caption{Metallicity ($\log Z$) of the BC assuming fixed $U$, \nh.}
\begin{tabular}{ccccc | cccccccccccccc}
\hline\hline  & \multicolumn{4}{c|}{log(\nh)=12} & \multicolumn{4}{c}{log(\nh)=13} &  \\
{SDSS JCODE} & \aliii/\heiiuv & \civ/\heiiuv & \siiv/\heiiuv & Z$_{\mu_{1/2}}$ &\aliii/\heiiuv & \civ/\heiiuv & \siiv/\heiiuv & Z$_{\mu_{1/2}}$ \\
        (1) & (2) & (3) & (4)  & (5) & (6) & (7) & (8) & (9)  \\
\hline 
& & & & & & & & & \\
J0020+0740	&	$	1.73	\pm	0.17	$	&	$	1.62	\pm	0.15	$	&	$	2.02	\pm	0.15	$	&	$	1.73	\pm	0.15	$	&	$	2.00	\pm	0.17	$	&	$	2.02	\pm	0.15	$	&	$	2.26	\pm	0.15	$	&	$	2.02	\pm	0.15	$	\\
J0034$-$0326	&	$	1.49	\pm	0.17	$	&	$	0.42	\pm	0.18	$	&	$	1.46	\pm	0.18	$	&	$	1.46	\pm	0.18	$	&	$	1.76	\pm	0.17	$	&	$	1.30	\pm	0.18	$	&	$	1.79	\pm	0.18	$	&	$	1.76	\pm	0.18	$	\\
J0037$-$0238	&	$	1.22	\pm	0.17	$	&	$	1.58	\pm	0.15	$	&	$	1.80	\pm	0.15	$	&	$	1.58	\pm	0.15	$	&	$	1.46	\pm	0.17	$	&	$	1.99	\pm	0.15	$	&	$	2.09	\pm	0.15	$	&	$	1.99	\pm	0.15	$	\\
J0103$-$1104	&	$	1.43	\pm	0.17	$	&	$	1.51	\pm	0.15	$	&	$	1.74	\pm	0.15	$	&	$	1.51	\pm	0.15	$	&	$	1.69	\pm	0.17	$	&	$	1.94	\pm	0.15	$	&	$	2.04	\pm	0.15	$	&	$	1.94	\pm	0.15	$	\\
J0106$-$0855	&	$	1.74	\pm	0.17	$	&	$	1.06	\pm	0.15	$	&	$	1.96	\pm	0.15	$	&	$	1.74	\pm	0.15	$	&	$	2.00	\pm	0.17	$	&	$	1.57	\pm	0.15	$	&	$	2.22	\pm	0.15	$	&	$	2.00	\pm	0.15	$	\\
J0123+0329	&	$	1.81	\pm	0.16	$	&	$	0.97	\pm	0.14	$	&	$	1.95	\pm	0.14	$	&	$	1.81	\pm	0.14	$	&	$	2.07	\pm	0.16	$	&	$	1.51	\pm	0.14	$	&	$	2.21	\pm	0.14	$	&	$	2.07	\pm	0.14	$	\\
J0210$-$0823	&	$	1.94	\pm	0.18	$	&	$	1.11	\pm	0.17	$	&	$	1.50	\pm	0.17	$	&	$	1.50	\pm	0.17	$	&	$	2.19	\pm	0.18	$	&	$	1.62	\pm	0.17	$	&	$	1.82	\pm	0.17	$	&	$	1.82	\pm	0.17	$	\\
J0827+0306	&	$	1.50	\pm	0.18	$	&	$	1.14	\pm	0.17	$	&	$	1.68	\pm	0.17	$	&	$	1.50	\pm	0.17	$	&	$	1.77	\pm	0.18	$	&	$	1.65	\pm	0.17	$	&	$	1.98	\pm	0.17	$	&	$	1.77	\pm	0.17	$	\\
J0829+0801	&	$	2.01	\pm	0.16	$	&	$	1.44	\pm	0.14	$	&	$	1.87	\pm	0.14	$	&	$	1.87	\pm	0.14	$	&	$	2.25	\pm	0.16	$	&	$	1.89	\pm	0.14	$	&	$	2.14	\pm	0.14	$	&	$	2.14	\pm	0.14	$	\\
J0836+0548	&	$	2.21	\pm	0.17	$	&	$	1.69	\pm	0.15	$	&	$	2.11	\pm	0.15	$	&	$	2.11	\pm	0.15	$	&	$	2.41	\pm	0.17	$	&	$	2.07	\pm	0.15	$	&	$	2.33	\pm	0.15	$	&	$	2.33	\pm	0.15	$	\\
J0845+0722	&	$	1.80	\pm	0.17	$	&	$	1.35	\pm	0.14	$	&	$	1.92	\pm	0.14	$	&	$	1.80	\pm	0.14	$	&	$	2.06	\pm	0.17	$	&	$	1.82	\pm	0.14	$	&	$	2.18	\pm	0.14	$	&	$	2.06	\pm	0.14	$	\\
J0847+0943	&	$	1.92	\pm	0.17	$	&	$	1.59	\pm	0.15	$	&	$	1.81	\pm	0.15	$	&	$	1.81	\pm	0.15	$	&	$	2.17	\pm	0.17	$	&	$	2.00	\pm	0.15	$	&	$	2.10	\pm	0.15	$	&	$	2.10	\pm	0.15	$	\\
J0858+0152	&	$	1.60	\pm	0.15	$	&	$	1.25	\pm	0.13	$	&	$	1.78	\pm	0.13	$	&	$	1.60	\pm	0.13	$	&	$	1.87	\pm	0.15	$	&	$	1.74	\pm	0.13	$	&	$	2.07	\pm	0.13	$	&	$	1.87	\pm	0.13	$	\\
J0903+0708	&	$	1.65	\pm	0.18	$	&	$	0.36	\pm	0.18	$	&	$	1.12	\pm	0.18	$	&	$	1.12	\pm	0.18	$	&	$	1.92	\pm	0.18	$	&	$	1.29	\pm	0.18	$	&	$	1.49	\pm	0.18	$	&	$	1.49	\pm	0.18	$	\\
J0915$-$0202	&	$	1.25	\pm	0.18	$	&	$	1.14	\pm	0.15	$	&	$	1.86	\pm	0.15	$	&	$	1.25	\pm	0.15	$	&	$	1.50	\pm	0.18	$	&	$	1.64	\pm	0.15	$	&	$	2.13	\pm	0.15	$	&	$	1.64	\pm	0.15	$	\\
J0926+0135	&	$	1.55	\pm	0.16	$	&	$	1.53	\pm	0.14	$	&	$	1.81	\pm	0.14	$	&	$	1.55	\pm	0.14	$	&	$	1.82	\pm	0.16	$	&	$	1.96	\pm	0.14	$	&	$	2.09	\pm	0.14	$	&	$	1.96	\pm	0.14	$	\\
J0929+0333	&	$	2.17	\pm	0.17	$	&	$	1.95	\pm	0.16	$	&	$	2.34	\pm	0.16	$	&	$	2.17	\pm	0.16	$	&	$	2.38	\pm	0.17	$	&	$	2.27	\pm	0.16	$	&	$	2.53	\pm	0.16	$	&	$	2.38	\pm	0.16	$	\\
J0946$-$0124	&	$	1.29	\pm	0.16	$	&	$	1.63	\pm	0.13	$	&	$	1.91	\pm	0.13	$	&	$	1.63	\pm	0.13	$	&	$	1.55	\pm	0.16	$	&	$	2.03	\pm	0.13	$	&	$	2.18	\pm	0.13	$	&	$	2.03	\pm	0.13	$	\\
J1024+0245	&	$	1.99	\pm	0.17	$	&	$	1.65	\pm	0.14	$	&	$	1.97	\pm	0.14	$	&	$	1.97	\pm	0.14	$	&	$	2.23	\pm	0.17	$	&	$	2.04	\pm	0.14	$	&	$	2.22	\pm	0.14	$	&	$	2.22	\pm	0.14	$	\\
J1026+0114	&	$	2.05	\pm	0.17	$	&	$	1.30	\pm	0.15	$	&	$	2.02	\pm	0.15	$	&	$	2.02	\pm	0.15	$	&	$	2.28	\pm	0.17	$	&	$	1.78	\pm	0.15	$	&	$	2.26	\pm	0.15	$	&	$	2.26	\pm	0.15	$	\\
J1145+0800	&	$	2.09	\pm	0.18	$	&	$	1.43	\pm	0.17	$	&	$	2.21	\pm	0.17	$	&	$	2.09	\pm	0.17	$	&	$	2.31	\pm	0.18	$	&	$	1.88	\pm	0.17	$	&	$	2.42	\pm	0.17	$	&	$	2.31	\pm	0.17	$	\\
J1214+0242	&	$	1.87	\pm	0.16	$	&	$	1.60	\pm	0.14	$	&	$	1.85	\pm	0.14	$	&	$	1.85	\pm	0.14	$	&	$	2.13	\pm	0.16	$	&	$	2.01	\pm	0.14	$	&	$	2.13	\pm	0.14	$	&	$	2.13	\pm	0.14	$	\\
J1215+0326	&	$	1.62	\pm	0.17	$	&	$	1.78	\pm	0.14	$	&	$	2.05	\pm	0.14	$	&	$	1.78	\pm	0.14	$	&	$	1.89	\pm	0.17	$	&	$	2.14	\pm	0.14	$	&	$	2.29	\pm	0.14	$	&	$	2.14	\pm	0.14	$	\\
J1219+0254	&	$	1.57	\pm	0.17	$	&	$	1.74	\pm	0.14	$	&	$	1.88	\pm	0.14	$	&	$	1.74	\pm	0.14	$	&	$	1.84	\pm	0.17	$	&	$	2.11	\pm	0.14	$	&	$	2.15	\pm	0.14	$	&	$	2.11	\pm	0.14	$	\\
J1231+0725	&	$	1.02	\pm	0.15	$	&	$	1.70	\pm	0.13	$	&	$	1.47	\pm	0.13	$	&	$	1.47	\pm	0.13	$	&	$	1.23	\pm	0.15	$	&	$	2.08	\pm	0.13	$	&	$	1.80	\pm	0.13	$	&	$	1.80	\pm	0.13	$	\\
J1244+0821	&	$	1.96	\pm	0.16	$	&	$	1.66	\pm	0.14	$	&	$	2.03	\pm	0.14	$	&	$	1.96	\pm	0.14	$	&	$	2.20	\pm	0.16	$	&	$	2.05	\pm	0.14	$	&	$	2.27	\pm	0.14	$	&	$	2.20	\pm	0.14	$	\\
J1259+0752	&	$	1.50	\pm	0.18	$	&	$	0.89	\pm	0.18	$	&	$	1.94	\pm	0.18	$	&	$	1.50	\pm	0.18	$	&	$	1.77	\pm	0.18	$	&	$	1.45	\pm	0.18	$	&	$	2.20	\pm	0.18	$	&	$	1.77	\pm	0.18	$	\\
J1314+0927	&	$	1.69	\pm	0.18	$	&	$	0.70	\pm	0.18	$	&	$	1.51	\pm	0.18	$	&	$	1.51	\pm	0.18	$	&	$	1.96	\pm	0.18	$	&	$	1.36	\pm	0.18	$	&	$	1.84	\pm	0.18	$	&	$	1.84	\pm	0.18	$	\\
J1419+0749	&	$	1.68	\pm	0.15	$	&	$	1.53	\pm	0.14	$	&	$	1.69	\pm	0.14	$	&	$	1.68	\pm	0.14	$	&	$	1.95	\pm	0.15	$	&	$	1.95	\pm	0.14	$	&	$	1.99	\pm	0.14	$	&	$	1.95	\pm	0.14	$	\\
J1509+0744	&	$	1.74	\pm	0.17	$	&	$	1.24	\pm	0.15	$	&	$	1.60	\pm	0.15	$	&	$	1.60	\pm	0.15	$	&	$	2.01	\pm	0.17	$	&	$	1.73	\pm	0.15	$	&	$	1.92	\pm	0.15	$	&	$	1.92	\pm	0.15	$	\\
J1519+0723	&	$	1.92	\pm	0.17	$	&	$	1.43	\pm	0.15	$	&	$	1.89	\pm	0.15	$	&	$	1.89	\pm	0.15	$	&	$	2.17	\pm	0.17	$	&	$	1.88	\pm	0.15	$	&	$	2.16	\pm	0.15	$	&	$	2.16	\pm	0.15	$	\\
J1545+0156	&	$	1.86	\pm	0.17	$	&	$	0.96	\pm	0.15	$	&	$	1.52	\pm	0.15	$	&	$	1.52	\pm	0.15	$	&	$	2.11	\pm	0.17	$	&	$	1.50	\pm	0.15	$	&	$	1.85	\pm	0.15	$	&	$	1.85	\pm	0.15	$	\\
J1609+0654	&	$	2.63	\pm	0.18	$	&	$	2.07	\pm	0.14	$	&	$	2.70	\pm	0.14	$	&	$	2.63	\pm	0.14	$	&	$	2.72	\pm	0.18	$	&	$	2.36	\pm	0.14	$	&	$	2.75	\pm	0.14	$	&	$	2.72	\pm	0.14	$	\\
J1618+0704	&	$	1.81	\pm	0.18	$	&	$	1.17	\pm	0.18	$	&	$	1.73	\pm	0.18	$	&	$	1.73	\pm	0.18	$	&	$	2.07	\pm	0.18	$	&	$	1.67	\pm	0.18	$	&	$	2.03	\pm	0.18	$	&	$	2.03	\pm	0.18	$	\\
J2116+0441	&	$	1.85	\pm	0.15	$	&	$	2.02	\pm	0.14	$	&	$	2.20	\pm	0.14	$	&	$	2.02	\pm	0.14	$	&	$	2.10	\pm	0.15	$	&	$	2.32	\pm	0.14	$	&	$	2.41	\pm	0.14	$	&	$	2.32	\pm	0.14	$	\\
J2145$-$0758	&	$	1.73	\pm	0.17	$	&	$	1.73	\pm	0.16	$	&	$	1.90	\pm	0.16	$	&	$	1.73	\pm	0.16	$	&	$	2.00	\pm	0.17	$	&	$	2.10	\pm	0.16	$	&	$	2.17	\pm	0.16	$	&	$	2.10	\pm	0.16	$	\\
\\ \hline																																																	
$\mu$	&	$	1.74	\pm	0.17	$	&	$	1.47	\pm	0.15	$	&	$	1.88	\pm	0.15	$	&	$	1.73	\pm	0.15	$	&	$	2.00	\pm	0.17	$	&	$	1.91	\pm	0.15	$	&	$	2.15	\pm	0.15	$	&	$	2.03	\pm	0.15	$	\\
$\sigma$	&	$	0.31	\pm	0.01	$	&	$	0.40	\pm	0.01	$	&	$	0.28	\pm	0.01	$	&	$	0.28	\pm	0.01	$	&	$	0.29	\pm	0.01	$	&	$	0.28	\pm	0.01	$	&	$	0.23	\pm	0.01	$	&	$	0.23	\pm	0.01	$	\\
SIQR	&	$	0.18	\pm	0.01	$	&	$	0.25	\pm	0.01	$	&	$	0.13	\pm	0.01	$	&	$	0.18	\pm	0.01	$	&	$	0.17	\pm	0.01	$	&	$	0.20	\pm	0.01	$	&	$	0.11	\pm	0.01	$	&	$	0.14	\pm	0.01	$	\\
\hline
\end{tabular}
\tablefoot{(1) SDSS identification, (2-4) and (6-8) metallicity values estimated from \civ/\heii, \siiv+\oiv/\heii, and \aliii/\heii\ line ratios for fixed densities of log(\nh) = 12 and 13 {(in units of cm$^{-3}$)}, respectively. (5) and (9) median metallicity values for {the corresponding value of} fixed density.
%, (2), (3) and (4) metallicity values for \civ/\heii\, \siiv+\oiv/\heii\ and \aliii/\heii\ with uncertainties. 
The measurements of the six BALQ were excluded from this table.}
\label{tab:bc_z}}
\end{table*}

\begin{table*}[htbp]
\centering
{\fontsize{9.5}{10}
\caption{Metallicity ($\log Z$) of BLUE assuming fixed $U$, \nh\ }  \label{tab:blue_z}\tabcolsep=3pt
\begin{tabular}{ccrccc}
\hline\hline
& \multicolumn{4}{c}{log(\nh) = 9} \\
SDSS JCODE          & {\siiv+\oiv/\heiiuv} &  \multicolumn{1}{c}{\civ/\heiiuv} &  {\civ/\siiv+\oiv}   &  Z$_{\mu_{1/2}}$\\
(1) & (2) & \multicolumn{1}{c}{(3)} & (4) & (5) \\
\hline \\
J0020+0740	&	$	0.62	\pm	0.13	$	&	$	-0.73	\pm	^{	1.72	}	_{	0.35	}	$	&	$	0.73	\pm	0.07	$	&	$	0.62	\pm	^{	0.13	}	_{	0.13	}	$	\\
J0034$-$0326	&	$	0.57	\pm	0.15	$	&	$	-0.52	\pm	^{	1.81	}	_{	0.50	}	$	&	$	0.67	\pm	0.12	$	&	$	0.57	\pm	^{	0.15	}	_{	0.15	}	$	\\
J0037$-$0238	&	$	0.69	\pm	0.13	$	&	$	-0.86	\pm	^{	0.34	}	_{	0.30	}	$	&	$	0.86	\pm	0.07	$	&	$	0.69	\pm	^{	0.13	}	_{	0.13	}	$	\\
J0103$-$1104	&	$	0.62	\pm	0.14	$	&	$	0.66	\pm	^{	0.70	}	_{	1.58	}	$	&	$	0.65	\pm	0.09	$	&	$	0.65	\pm	^{	0.14	}	_{	0.14	}	$	\\
J0106$-$0855	&	$	0.94	\pm	0.14	$	&	$	0.75	\pm	^{	0.67	}	_{	1.59	}	$	&	$	0.90	\pm	0.09	$	&	$	0.90	\pm	^{	0.14	}	_{	0.14	}	$	\\
J0123+0329	&	$	1.07	\pm	0.13	$	&	$	0.98	\pm	^{	0.62	}	_{	1.47	}	$	&	$	0.93	\pm	0.07	$	&	$	0.98	\pm	^{	0.13	}	_{	0.13	}	$	\\
J0210$-$0823	&	$	0.66	\pm	0.17	$	&	$	-0.81	\pm	^{	1.82	}	_{	0.42	}	$	&	$	0.79	\pm	0.16	$	&	$	0.66	\pm	^{	0.17	}	_{	0.17	}	$	\\
J0827+0306	&	$	0.63	\pm	0.17	$	&	$	0.72	\pm	^{	0.73	}	_{	1.69	}	$	&	$	0.64	\pm	0.16	$	&	$	0.64	\pm	^{	0.17	}	_{	0.17	}	$	\\
J0829+0801	&	$	0.85	\pm	0.12	$	&	$	0.80	\pm	^{	0.63	}	_{	1.52	}	$	&	$	0.79	\pm	0.05	$	&	$	0.80	\pm	^{	0.12	}	_{	0.12	}	$	\\
J0836+0548	&	$	0.80	\pm	0.14	$	&	$	-0.79	\pm	^{	1.65	}	_{	0.35	}	$	&	$	0.93	\pm	0.09	$	&	$	0.80	\pm	^{	0.14	}	_{	0.14	}	$	\\
J0845+0722	&	$	0.89	\pm	0.13	$	&	$	0.72	\pm	^{	0.68	}	_{	1.55	}	$	&	$	0.86	\pm	0.07	$	&	$	0.86	\pm	^{	0.13	}	_{	0.13	}	$	\\
J0847+0943	&	$	0.86	\pm	0.13	$	&	$	-0.64	\pm	^{	1.78	}	_{	0.40	}	$	&	$	0.94	\pm	0.07	$	&	$	0.86	\pm	^{	0.13	}	_{	0.13	}	$	\\
J0858+0152	&	$	0.89	\pm	0.11	$	&	$	0.72	\pm	^{	0.63	}	_{	1.49	}	$	&	$	0.87	\pm	0.04	$	&	$	0.87	\pm	^{	0.11	}	_{	0.11	}	$	\\
J0903+0708	&	$	0.72	\pm	0.15	$	&	$	-0.52	\pm	^{	1.82	}	_{	0.50	}	$	&	$	0.77	\pm	0.12	$	&	$	0.72	\pm	^{	0.15	}	_{	0.15	}	$	\\
J0915$-$0202	&	$	0.63	\pm	0.14	$	&	$	-0.72	\pm	^{	1.76	}	_{	0.38	}	$	&	$	0.75	\pm	0.09	$	&	$	0.63	\pm	^{	0.14	}	_{	0.14	}	$	\\
J0926+0135	&	$	0.93	\pm	0.12	$	&	$	-0.42	\pm	^{	1.73	}	_{	0.45	}	$	&	$	0.94	\pm	0.05	$	&	$	0.93	\pm	^{	0.12	}	_{	0.12	}	$	\\
J0929+0333	&	$	1.05	\pm	0.14	$	&	$	0.92	\pm	^{	0.64	}	_{	1.57	}	$	&	$	0.96	\pm	0.09	$	&	$	0.96	\pm	^{	0.14	}	_{	0.14	}	$	\\
J0946$-$0124	&	$	0.74	\pm	0.13	$	&	$	-1.02	\pm	^{	0.24	}	_{	0.24	}	$	&	$	0.98	\pm	0.05	$	&	$	0.74	\pm	^{	0.13	}	_{	0.13	}	$	\\
J1024+0245	&	$	0.80	\pm	0.13	$	&	$	-0.56	\pm	^{	1.79	}	_{	0.43	}	$	&	$	0.83	\pm	0.07	$	&	$	0.80	\pm	^{	0.13	}	_{	0.13	}	$	\\
J1026+0114	&	$	0.85	\pm	0.13	$	&	$	0.91	\pm	^{	0.63	}	_{	1.52	}	$	&	$	0.74	\pm	0.07	$	&	$	0.85	\pm	^{	0.13	}	_{	0.13	}	$	\\
J1145+0800	&	$	0.90	\pm	0.17	$	&	$	0.80	\pm	^{	0.70	}	_{	1.71	}	$	&	$	0.83	\pm	0.16	$	&	$	0.83	\pm	^{	0.17	}	_{	0.17	}	$	\\
J1214+0242	&	$	0.75	\pm	0.12	$	&	$	0.79	\pm	^{	0.64	}	_{	1.53	}	$	&	$	0.73	\pm	0.05	$	&	$	0.75	\pm	^{	0.12	}	_{	0.12	}	$	\\
J1215+0326	&	$	0.38	\pm	0.12	$	&	$	-0.76	\pm	^{	1.62	}	_{	0.32	}	$	&	$	0.63	\pm	0.05	$	&	$	0.38	\pm	^{	0.12	}	_{	0.12	}	$	\\
J1219+0254	&	$	0.38	\pm	0.13	$	&	$	-1.04	\pm	^{	0.26	}	_{	0.27	}	$	&	$	0.71	\pm	0.07	$	&	$	0.38	\pm	^{	0.13	}	_{	0.13	}	$	\\
J1231+0725	&	$	0.55	\pm	0.14	$	&	$	-1.04	\pm	^{	0.25	}	_{	0.28	}	$	&	$	0.82	\pm	0.08	$	&	$	0.55	\pm	^{	0.14	}	_{	0.14	}	$	\\
J1244+0821	&	$	0.72	\pm	0.12	$	&	$	-0.64	\pm	^{	1.76	}	_{	0.37	}	$	&	$	0.81	\pm	0.05	$	&	$	0.72	\pm	^{	0.12	}	_{	0.12	}	$	\\
J1259+0752	&	$	0.63	\pm	0.15	$	&	$	-0.53	\pm	^{	1.82	}	_{	0.51	}	$	&	$	0.70	\pm	0.12	$	&	$	0.63	\pm	^{	0.15	}	_{	0.15	}	$	\\
J1314+0927	&	$	0.91	\pm	0.17	$	&	$	0.88	\pm	^{	0.69	}	_{	1.72	}	$	&	$	0.82	\pm	0.17	$	&	$	0.88	\pm	^{	0.17	}	_{	0.17	}	$	\\
J1419+0749	&	$	0.72	\pm	0.13	$	&	$	-0.75	\pm	^{	1.67	}	_{	0.34	}	$	&	$	0.82	\pm	0.05	$	&	$	0.72	\pm	^{	0.13	}	_{	0.13	}	$	\\
J1509+0744	&	$	0.92	\pm	0.14	$	&	$	0.86	\pm	^{	0.65	}	_{	1.58	}	$	&	$	0.86	\pm	0.09	$	&	$	0.86	\pm	^{	0.14	}	_{	0.14	}	$	\\
J1519+0723	&	$	0.91	\pm	0.13	$	&	$	0.65	\pm	^{	0.70	}	_{	1.53	}	$	&	$	0.91	\pm	0.07	$	&	$	0.91	\pm	^{	0.13	}	_{	0.13	}	$	\\
J1545+0156	&	$	1.05	\pm	0.14	$	&	$	0.91	\pm	^{	0.64	}	_{	1.56	}	$	&	$	0.95	\pm	0.09	$	&	$	0.95	\pm	^{	0.14	}	_{	0.14	}	$	\\
J1609+0654	&	$	0.87	\pm	0.13	$	&	$	0.89	\pm	^{	0.63	}	_{	1.54	}	$	&	$	0.77	\pm	0.07	$	&	$	0.87	\pm	^{	0.13	}	_{	0.13	}	$	\\
J1618+0704	&	$	1.14	\pm	0.15	$	&	$	1.64	\pm	^{	0.19	}	_{	0.36	}	$	&	$	0.87	\pm	0.12	$	&	$	1.14	\pm	^{	0.15	}	_{	0.15	}	$	\\
J2116+0441	&	$	-0.05	\pm	0.12	$ \ \	&	$	-1.36	\pm	^{	0.16	}	_{	0.21	}	$	&	$	0.69	\pm	0.05	$	&	$	-0.05	\pm	^{	0.12	}	_{	0.12	}	$ \ \ 	\\
J2145$-$0758	&	$	0.48	\pm	0.14	$	&	$	-1.13	\pm	^{	0.25	}	_{	0.23	}	$	&	$	0.73	\pm	0.09	$	&	$	0.48	\pm	^{	0.14	}	_{	0.14	}	$	\\
\\ \hline																																			
$\mu$	&	$	0.77	\pm	0.13	$	&	$	-0.47	\pm	^{	0.68	}	_{	0.50	}	$	&	$	0.82	\pm	0.07	$	&	$	0.77	\pm		0.13		$	\\
$\sigma$	&	$	0.23	\pm	0.01	$	&	$	0.86	\pm	^{	0.61	}	_{	0.62	}	$	&	$	0.10	\pm	0.03	$	&	$	0.21	\pm		0.01	$	\\
SIQR	&	$	0.14	\pm	0.01	$	&	$	0.78	\pm	^{	0.54	}	_{	0.60	}	$	&	$	0.08	\pm	0.01	$	&	$	0.11	\pm	0.01		$	\\

\hline
\end{tabular}
\tablefoot{(1) SDSS identification, (2-4) logarithm of metallicity values in solar units estimated from \siiv+\oiv/\heii, \civ/\siiv+\oiv, and \civ/\heii\ line ratios. (5) Logarithm of median metallicity values  for each object. The measurements of the six BALQ were excluded from this table.}
\label{tab:Z_blue}}
\end{table*}
 %%%%%%%%%%%%%%%%%%%%%%%%%%%%%%%%%%%%%%%%%%%%%%%%%%%%

\subsection{Metallicity inferences}
\label{fixed}

As mentioned in Sect.  \ref{photoion}, we use flux ratios as metallicy diagnostics for the BLR gas. The foundations of the method we apply are explained in \citetalias{sniegowskaetal21}. Here we briefly recall two main aspects.  

Among all the indicators proposed by different authors \citep[e.g.][]{nagaoetal06,shinetal13,marzianisulentic14} we give higher weight to the indicators involving \heii. The diagnostics ratios are \aliii/\heii, \civ/\heii, and \siiv+\oiv/\heii\ for the virialized BC, and \civ/\heii,  \siiv+\oiv/\heii,  \civ/\heii, and \siiv+\oiv/\civ\ for BLUE. We also considered the fitting complex for regions with $\lambda$ < 1300  \AA. However, even though \nivfull\ is present for most of our sample, we decided to avoid its fitting procedure because it appears to be weak, and in some cases it is severely affected by absorptions.

We  modeled the BLR assuming gas in much different physical conditions for the virialized and for the wind component:
%: a) virialized clouds for LIL and IIL descrived with log(\nh)= 12-13 and log($U$)= -2.5, and b) wind clouds for HIL descrived with log(\nh)= 9 and log($U$)= 0.
\begin{itemize}
    \item virialized: assumed with log(\nh) = 12-13 and log($U$ )= -2.5; 
    \item wind: assumed  with log(\nh) = 9 and log($U$) = 0.
\end{itemize}

The assumptions on density and ionization parameter for the virialized component stem from several studies suggesting low-ionization and high density \citep{baldwinetal96,baldwinetal04,marzianietal10}. Further work has set robust lower limit to the density of the low-ionization region, from the strength of IR Ca{\sc ii} triplet and of the \feii\ emission blend \citep{matsuokaetal07,martinez-aldamaetal15,pandaetal18,pandaetal19a,pandaetal20a} at $\log$\ \nh $\approx 11.5$. The low \civ/\hb\ ratios for the virialized component is best explained by a modest ionization parameter, $\log U \sim 2.5$ \citep{marzianietal10,pandaetal19a}. These values are in agreement with the most recent determination of \citet{sniegowskaetal21}.   The blueshifted component is by far less constrained. The higher \civ/\hb\ suggests higher ionization parameter than for the virialized component \citep{marzianietal10,sniegowskaetal21}. The absence of a strong blueshifted component in \aliii\ indicates moderate density, likely around $\log $ \nh\ $\sim $ 10. For this density the lack of any obvious \ciii\ blueshifted component (detection that may be however hampered by the severe blending with \siiii) implies very high ionization parameter $\log U \sim -0.5 - 0$.

\subsubsection{Estimates of $Z$ distributions at fixed  ($U$, \nh)}
\label{sec:ZfixednU}

Our six metallicity indicators are shown in Figure \ref{fig:trend}. Left panel shows the
{low ionization zone metallicity indicators,} and right panel shows the high ionization potential metallicity indicators or BLUE component indicators. For both panels the stars represent the flux ratios predicted by CLOUDY according to the specified physical conditions, we fitted each CLOUDY trend with a cubic spline (dashed lines) and derive the metallicity of our sample according to it. The thicker solid  lines show the ranges of metallicity derived from our sample.

In order to seek for a better description of the BLR, we decided to use two cases of density log(\nh) = 12 and log(\nh) = 13  for the BC ratios but keeping the ionization parameter fixed at log($U$)= -2.5. These high \nh\ values are  needed to explain \feii\ and the UV emission lines in the virialized component, as briefly summatized above.  Table \ref{tab:bc_z} lists the derived metallicities for the three broad component metallicity indicators and their median value {(Z$_{\mu_{1/2}}$)} for the log(\nh) = 12  and  the log(\nh) = 13 case.  Left panel of Figure \ref{fig:trend} shows only the case for log(\nh) = 12, it can be seen that our three BC indicators show a consistent trend. The derived median value is Z = 1.73 $\pm$ 0.15Z$_\odot$ \ for the virialized clouds as reported in table \ref{tab:bc_z}. The errors for the metallicity estimates, both BC and BLUE relations,  were estimated following the flux ratio error propagation but considering them as relative errors for their logarithmic display.

The wind clouds were described with a density of log(\nh) = 9 and a ionization parameter of log($U$) = 0, as shown in the right  panel of Fig \ref{fig:trend}. These ratios do not show consistent relations, especially for the BLUE \civ/\heiiuv \ ratio with a non-monotonic trend (also shown in Fig. 17 of \citetalias{sniegowskaetal21}).  The non-monotonic behavior occur in the interval from sub-solar to super solar metallicity, which means that  one value of the BLUE \civ/\heiiuv \ ratio may correspond to three values of the metallicity. Thus, for the sources that cross the trend (blue dashed line in Fig. \ref{fig:trend}) in three points we decided to take the average value. Table \ref{tab:blue_z} lists the derived metallicities for our three BLUE  metallicity indicators and their median value, with a general median value of Z = 0.77  $\pm$ 0.13 Z$_\odot$ (as reported in table \ref{tab:blue_z}), setting a lower metallicity limit for our sample.

\begin{figure*}
     \centering
     \hspace{-1cm}
     %trim=left bottom right to   
     \includegraphics[trim= 0.5 0.5 0.5 200.5, clip, width=0.5\textwidth]{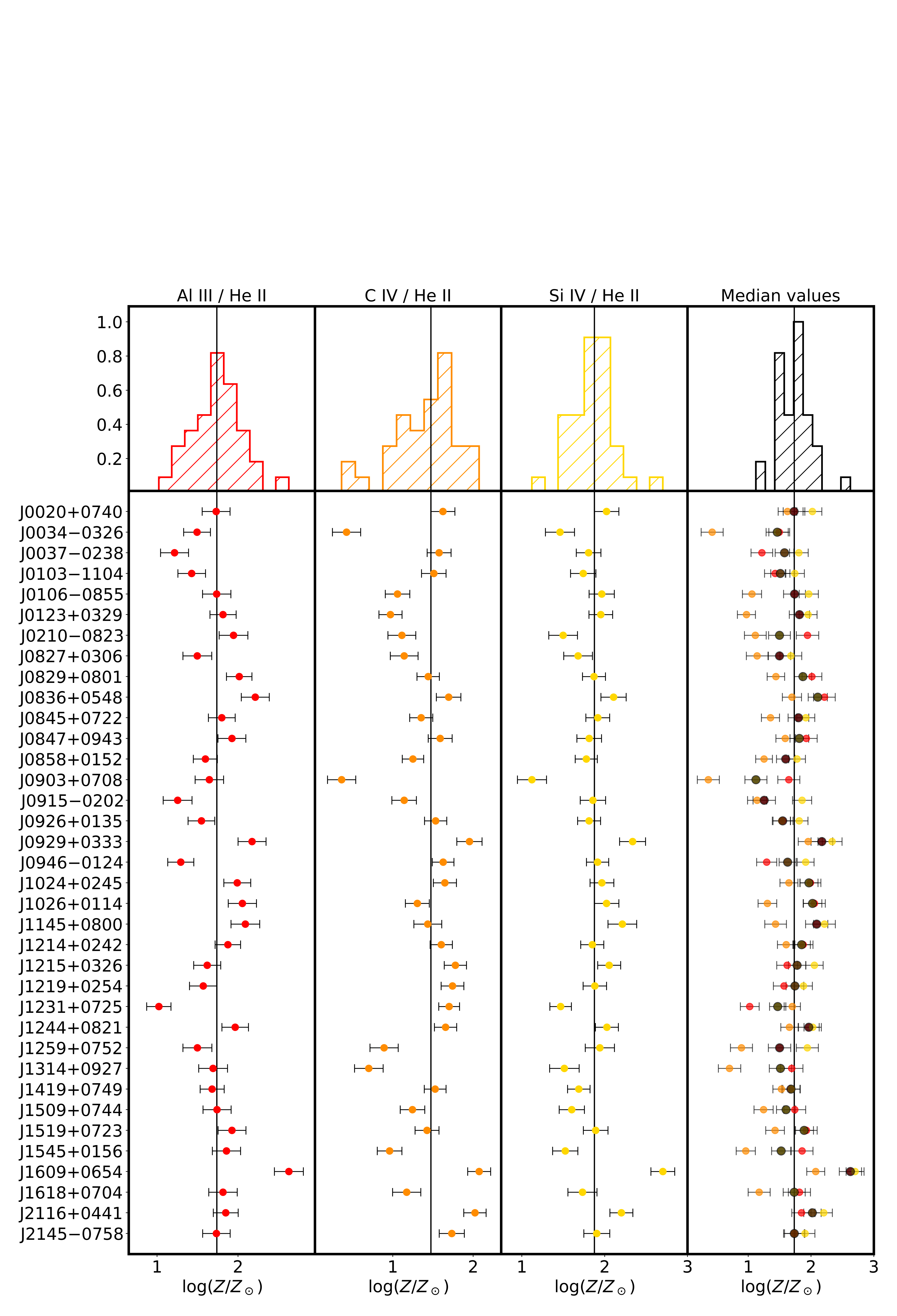}
    \includegraphics[trim= 0.5 0.5 0.5 200.5, clip, width=0.5\textwidth]{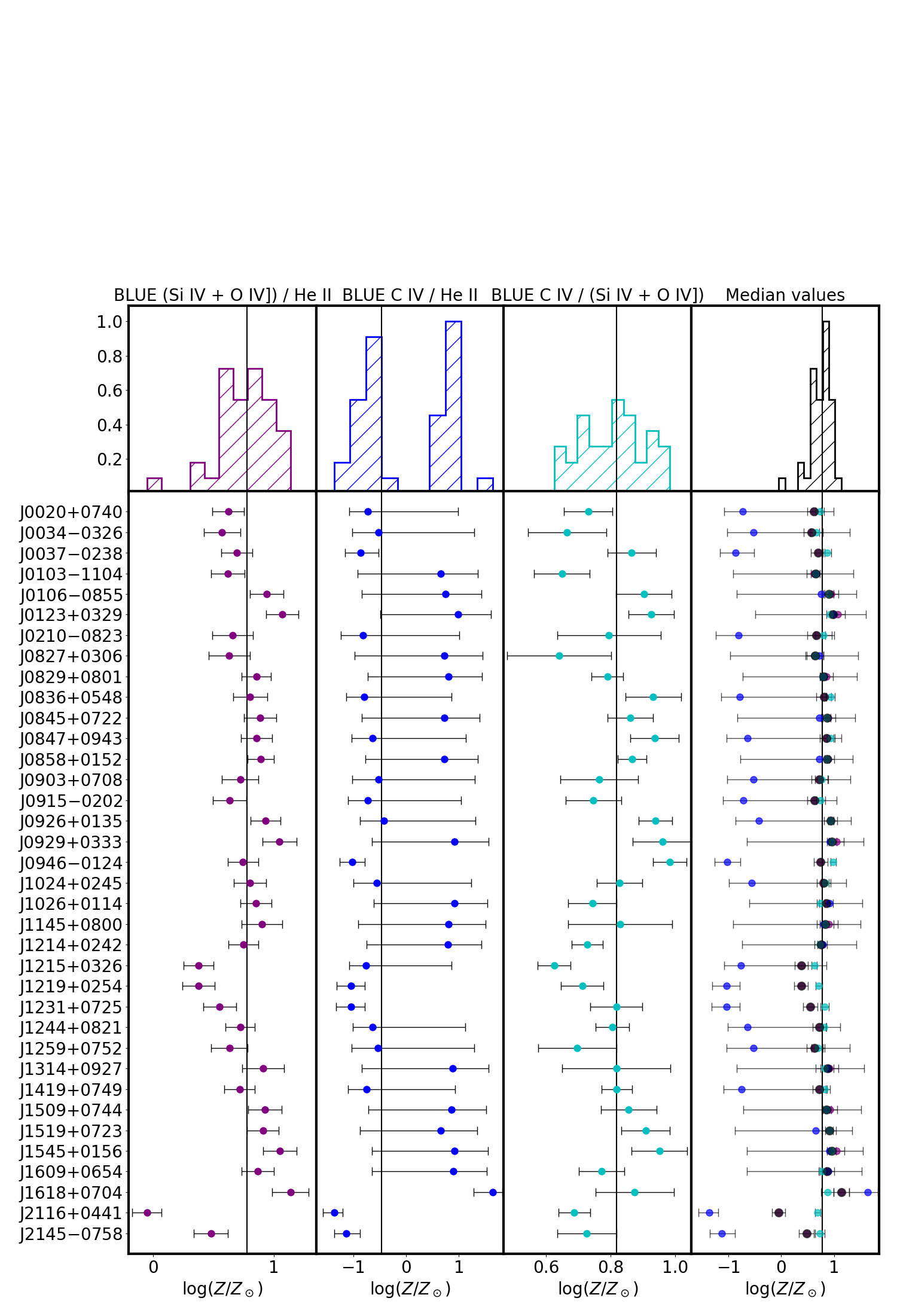}  \caption{Distribution (top panels) and individual (bottom panels) values of line ratio metallicity estimate involving \heii\ using broad components (left) and BLUE components (right). 
    }   
    \label{fig:z_bc_blue}
 \end{figure*}

Figure \ref{fig:z_bc_blue} left panel shows the  metallicity estimates  distributions  in log(Z/Z$_\odot$) for the BC intensity ratios shown in Fig. \ref{fig:redmix}. The right top and bottom panels show the superimposed estimates from the three ratios and their median values in black. The values derived from the BC \civ/\heiiuv \ relation are the ones with higher dispersion from the general median especially those with a strong \heiiuv \ feature such as J0034$-$0326 (Fig. \ref{fig:02}) and J0903+0708 (Fig. \ref{fig:16}).

Figure \ref{fig:z_bc_blue} right panel show the metallicity estimates distributions in log(Z/Z$_\odot$) for the BLUE intensity ratios shown in Fig. \ref{fig:bluemix}.  The right top and bottom panels show the superimposed estimates from the three ratios and their median values in black.  
What stands out the most is the bi-valuated behavior of the BLUE \civ/\heiiuv \ estimates, this is caused by the non-monotonic behavior of the respective CLOUDY relation shown in Fig. \ref{fig:trend}. For  the sources crossing the trend in three points, the dots are the mean of the three possible values and the error bars  are extended to lowest point where the source cross the trend (negative error bar) and the highest point where the source cross the trend (positive error bar). 

 %%%%%%%%%%%%%%%%%%%%%%%%%%%%%%%%%%%%%%%%%%%%%%%%%%%%%%%%%%%%%%

\subsubsection{Estimates of $Z$ relaxing the constraints on $U$\ and \nh}
\label{ZfreenU}
\begin{figure}[ht]
  \centering
\includegraphics[width=10cm]{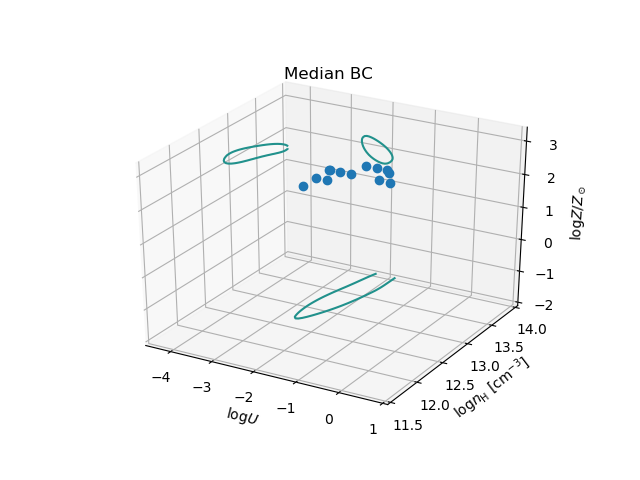}
\caption{The parameter space \nh, $U$, $Z$. Data points in 3D space are elements in the grid of the parameter space selected for not being different from $\chi^2_\mathrm{min}$\ by more than $\delta \chi^2 \approx 1$, also  satisfying the condition that the three intensity ratios used for the computation of the $\chi^2$ individually agree with the ratios predicted within the errors at  1$\sigma$\ confidence level. Data points were computed from the emission line ratios measured for the BC and referring to the median values in Table \ref{tab:ratios}. The individual contour line was smoothed with a Gaussian kernel.  
\label{fig:mediancr}}
\end{figure}

\begin{figure*}[htp!]
  %\centering
\includegraphics[width=10cm]{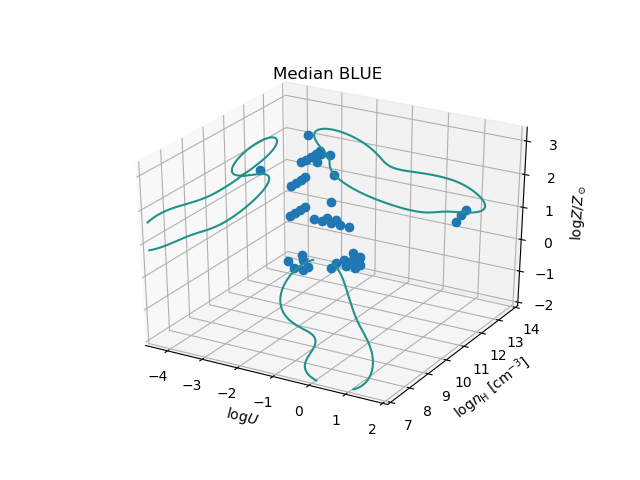}
\includegraphics[width=8cm]{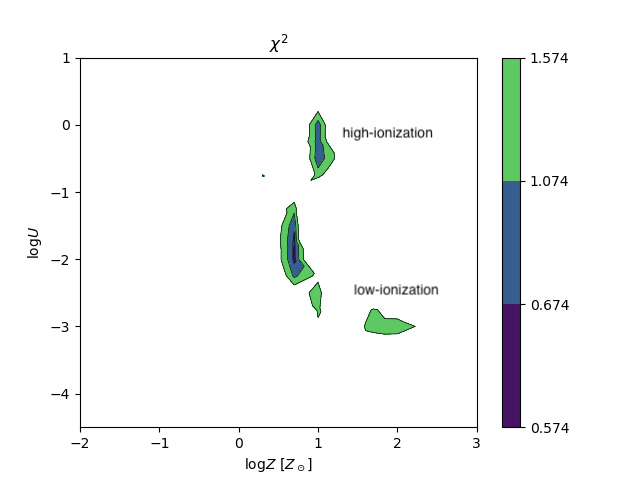}\\

\caption{Left: the parameter space \nh, $U$, $Z$, as in Figure \ref{fig:mediancr}, but for component BLUE. As in that case, data points were computed from the emission line ratios  referring to the median values in Table \ref{tab:ratios}. The individual contour line was smoothed with a Gaussian kernel.   Right:  the plane $\log U$ vs $\log Z$. The outer isophotal contours delimit the region $\chi_\mathrm{min} + 1$, where the $\chi_\mathrm{min}$\ is the value after smoothing with a Gaussian filter.   The low- and high-ionization solutions are marked.   
\label{fig:mediancr_blue}}
\end{figure*}

Table \ref{tab:zun} reports the results of the analysis relaxing the assumptions of fixed values of \nh, $U$. We used the complete range values of hydrogen density 7.00 $\leq \log(n_\mathrm{H}) \leq$ 14.00 and the ionization parameter $-4.5 \leq \log(U) \leq$ 1.00 described in Section \ref{photoion}. The $\log Z$, $\log U$, $\log$\nh\ and their associated uncertainties are reported for 33 sources of the sample (excluding BAL QSOs, and  sources that are not xA). The exclusion of sources that are not xA is due to the fact that their SED may not be consistent with the ones of the Population A sources: even if the intensity ratio can be measured with good precision, the present set of {\tt CLOUDY} simulation cannot be used to predict the metallicity of these sources. The last two rows list  the results for the median computed from the median of the ratios of the individual sources, $\mu_\frac{1}{2}$(Ratios), and the one computed from the medians of the individual object parameter values, $\mu_\frac{1}{2}$(Objects). The parameters reported in Table \ref{tab:zun} cluster around well-defined values, with medians  $\approx$ 50 $Z_\odot$, ionization parameter $\log U \approx -2$, and very high density $\log$\nh $\approx 13.75$. These values are  in good agreement with \citetalias{sniegowskaetal21}, implying that the additional sources that are now the majority of the sample and meet the selection criteria for being xA, also show ionization degrees and metallicity values in the same range. The result is largely a consequence of the inclusion of \aliii\ in the diagnostics. This line, involved also in the selection criteria, is very strong in xA with respect to quasars belonging to other spectral types along the main sequence \citep{aokiyoshida99,bachevetal04}, and \aliii\ strength is favored at high $Z$, high \nh. 

Two groups of sources stand out. The first one includes (1) sources that are extreme with $Z \sim \ 100 Z_\odot$. These sources have a large fraction of their gas mass made of metals \citepalias[][]{sniegowskaetal21}; (2) sources with relatively high $U \sim -1.5$. Five out of seven sources of the 1st group show the highest \siiv\ compared to \civ\ and are at the low end of the distribution of \civ\ and \aliii\ equivalent widths. However, J2116+0441 and J2145-0758 show relatively prominent \civ\ and lower \siiv, probably as an effect of a higher ionization parameter.  The second group has higher \civ/\aliii\ than average, although this has no strong implication for  $Z$: if $\log U \sim -1.75$, the median $\log Z \sim 1.7$ [$Z_\odot$]; only in two cases with  $\log U \sim -1.5$, the median $\log Z \sim 1.3$ [$Z_\odot$], lower than the full sample median. 

Table \ref{tab:zun_blue} provides the information on metallicity that is possible to extract from the BLUE intensity ratio. No \nh\ and $U$\ values are reported, as they are very poorly constrained (Fig. \ref{fig:mediancr_blue}). The bottom panel shows the  median case reported in Table \ref{tab:zun_blue} and reveals the existence of two main emitting regions in the parameter plane $U$\ vs $Z$. The range reported  in  Table \ref{tab:zun_blue} is the $Z$ range possible for the high-ionization and the intermediate ionization range ($U \sim -2$, right panel of Fig. \ref{tab:zun_blue}).

\begin{table*}
\centering
\caption{$Z$, $U$, \nh\ of individual sources and median derived from the BC}  \label{tab:zun}
\begin{tabular}{cccccccc}\hline\hline
{SDSS JCODE}      &{$\chi^2_\mathrm{min}$  }  & {$\log Z$\ [$Z_\odot$]} & {$\delta \log Z$\ [$Z_\odot$]} & {$\log U$} & {$\delta \log U$} 
 & {$\log $ \nh} & {$\delta \log $ \nh}  \\  
(1) & (2) & (3) & (4)  & (5) & (6) & (7)\\ 
\hline
J0020+0740 & 0.094 & 2.00 & 2.00 -- 2.00 & -2.00 & -2.00 -- -1.75 & 13.25 & 13.00 -- 13.50 \\ 
J0034$-$0326 & 2.128 & 1.30 & 0.00 -- 3.00 & -2.25 & -3.75 -- -1.50 & 13.50 & 7.25 -- 14.00 \\
J0037$-$0238 & 0.192 & 1.30 & 1.30 -- 1.30 & -1.50 & -1.75 -- -1.00 & 11.75 & 11.50 -- 12.00 \\
J0103$-$1104 & 0.046 & 1.70 & 1.70 -- 1.70 & -2.00 & -2.00 -- -2.00 & 12.75 & 12.75 -- 12.75 \\
J0106$-$0855 & 2.725 & 1.30 & 1.70 -- 1.70 & -1.75 & -3.50 -- 0.00 & 14.00 & 11.25 -- 14.00 \\
J0123+0329 & 4.549 & 0.70 & 0.00 -- 3.00 & -3.75 & -3.75 -- 1.00 & 14.00 & 7.00 -- 14.00 \\
J0210$-$0823 & 0.837 & 0.70 & 0.70 -- 1.70 & -3.75 & -3.75 -- -3.75 & 14.00 & 13.75 -- 14.00 \\
J0827+0306 & 0.346 & 1.30 & 1.30 -- 1.70 & -2.00 & -2.25 -- -1.75 & 13.75 & 13.00 -- 13.75 \\
J0829+0801 & 1.881 & 1.70 & 0.70 -- 2.00 & -2.00 & -3.75 -- -1.50 & 13.75 & 12.00 -- 14.00 \\
J0836+0548 & 1.610 & 1.70 & 1.00 -- 2.30 & -2.00 &-3.75 -- -1.50 & 14.00 & 12.00 -- 14.00 \\ 
J0845+0722 & 1.156 & 1.70 & 1.00 -- 2.00 & -1.75 & -2.50 -- -1.50 & 13.75 & 12.75 -- 14.00 \\
J0847+0943 & 0.494 & 2.00 & 1.00 -- 2.00 & -2.50 & -2.75 -- -2.25 & 12.75 & 12.25 -- 14.00 \\
J0858+0152 & 0.645 & 1.30 & 1.30 -- 1.70 & -2.25 & -2.50 -- -2.00 & 13.75 & 12.75 -- 13.75 \\
J0903+0708 & 0.979 & 0.30 & 0.00 -- 1.30 & -3.75 & -3.75 -- -3.50 & 14.00 & 13.75 -- 14.00 \\
J0915$-$0202 & 0.249 & 1.30 & 1.30 -- 1.30 & -1.50 & -1.75 -- -1.25 & 13.50 & 13.50 -- 14.00 \\
J0926+0135 & 0.146 & 1.70 & 1.70 -- 1.70 & -2.25 & -2.25 -- -2.00 & 12.50 & 12.50 -- 12.75 \\
J0929+0333 & 0.334 & 1.70 & 1.70 -- 2.30 & -2.25 & -2.25 -- -1.75 & 14.00 & 12.75 -- 14.00 \\
J0946$-$0124 & 0.109 & 1.30 & 1.30 -- 1.30 & -0.75 & -1.00 -- -0.75 & 11.00 & 11.00 -- 11.25 \\
J1024+0245 & 0.578 & 2.00 & 1.00 -- 2.00 & -2.25 & -2.50 -- -2.00 & 13.50 & 12.50 -- 14.00 \\
J1026+0114 & 2.638 & 1.70 & 0.30 -- 3.00 & -1.75 & -3.75 -- -0.25 & 13.75 & 10.25 -- 14.00 \\
J1145+0800 & 1.997 & 1.70 & 0.70 -- 2.70 & -1.75 & -3.75 -- -0.50 & 14.00 & 12.50 -- 14.00 \\
J1214+0242 & 0.271 & 1.70 & 1.70 -- 2.00 & -2.25 & -2.50 -- -2.25 & 13.75 & 12.75 -- 13.75 \\
J1215+0326 & 0.116 & 1.70 & 1.70 -- 1.70 & -1.75 & -2.00 -- -1.75 & 12.25 & 12.25 -- 12.25 \\
J1219+0254 & 0.139 & 1.70 & 1.70 -- 1.70 & -2.25 & -2.25 -- -2.00 & 12.25 & 12.25 -- 12.50 \\
J1231+0725 & 0.037 & 1.00 & 1.00 -- 1.00 & -2.00 & -2.00 -- -2.00 & 11.25 & 11.25 -- 11.25 \\
J1244+0821 & 0.596 & 2.00 & 1.30 -- 2.00 & -2.25 & -2.50 -- -1.75 & 13.50 & 12.50 -- 14.00 \\
J1259+0752 & 1.952 & 1.30 & 0.00 -- 1.70 & -1.75 & -3.50 -- -0.50 & 13.75 & 12.00 -- 14.00 \\
J1314+0927 & 2.378 & 0.00 & 0.00 -- 3.00 & -3.50 & -3.75 -- -0.50 & 14.00 & 7.25 -- 14.00 \\
J1419+0749 & 0.146 & 2.00 & 1.00 -- 2.00 & -2.50 & -2.50 -- -2.25 & 13.00 & 12.25 -- 14.00 \\
J1509+0744 & 0.733 & 1.70 & 1.30 -- 1.70 & -2.50 & -2.75 -- -2.00 & 13.50 & 12.50 -- 13.75 \\
J1519+0723 & 1.309 & 1.70 & 0.70 -- 2.00 & -2.00 & -2.75 -- -1.50 & 13.75 & 12.25 -- 14.00 \\
J1545+0156 & 1.648 & 0.70 & 0.30 -- 1.70 & -3.75 & -3.75 -- -1.75 & 14.00 & 12.75 -- 14.00 \\
J1609+0654 & 1.673 & 2.00 & 1.70 -- 3.00 & -2.00 & -3.75 -- -0.50 & 14.00 & 10.00 -- 14.00 \\
J1618+0704 & 1.472 & 1.70 & 0.30 -- 2.00 & -2.25 & -3.75 -- -1.25 & 13.50 & 12.00 -- 14.00 \\
J2116+0441 & 0.063 & 2.00 & 2.00 -- 2.00 & -2.00 & -2.00 -- -2.00 & 12.50 & 12.50 -- 12.50 \\
J2145$-$0758 & 0.127 & 2.00 & 1.30 -- 2.00 & -2.25 & -2.25 -- -2.00 & 13.00 & 12.00 -- 14.00 \\            
\hline                                                                              $\mu_\frac{1}{2}$(Ratios) &   0.277   &          1.70   &  1.30 --  2.00		&       -2.00 &  -2.25  --   -2.00     &    13.75 & 13.25 -- 14.00       \\       
 $\mu_\frac{1}{2}$(Objects) &  \ldots & 1.70  &    1.50  --   1.90 &  -2.00   & -2.25  --  -1.75 &  13.75   &  13.25  -- 14.00 \\       
 \hline
 \end{tabular}
\tablefoot{The measurements of the six BALQ were excluded from this table.}
\end{table*}

\begin{table*}
%\tabletypesize{\scriptsize}
%\tabcolsep=2pt
\centering
\caption{$Z$, $U$, \nh\ of individual sources and median derived from BLUE}  \label{tab:zun_blue}
\begin{tabular}{ccccl}
\hline\hline
{SDSS JCODE}    & $\chi^2$      & {$\log Z$\ [$Z_\odot$]} &  $\delta \log Z$\ [$Z_\odot$]  & Notes \\  
(1) & (2) & (3) & (4) & (5)\\
\hline
J0020+0740 & 0.001845 & 0.3 -- 1.0 & 0.2 -- 1.1 & \\          
J0034$-$0326 & 0.00657 & 0.7 & 0.0 -- 1.3 & 1400 blend absorbed \\         
J0037$-$0238 & 0.02968 & 0.7 -- 1.0 & 0.3 -- 1.3 & \\          
J0106$-$0855 & 0.00667 & 1.0 & 0.7 -- 1.7 & \\            
J0123+0329 & 0.04137 & 1.0 & 0.7 -- 1.7 & \\            
J0210$-$0823 & 0.00469 & 1.0 & 0.3 -- 1.7 & low EW  low S/N \\       
J0216+0115* & \ldots & \ldots & \ldots  & no 1400 blend \\         
J0252$-$0420* & \ldots & \ldots & \ldots  & only 1900 blend available \\        
J0827+0306 & 0.00756 & 0.7 -- 1.0 & 0.0 -- 1.3 & low S/N \\        
J0829+0801 & 0.02020 & 0.7 & 0.6 -- 1.1 & \\            
J0836+0548 & 0.00302 & 0.7 & 0.5 -- 1.7 & \\            
J0845+0722 & 0.00926 & 1.0 & 0.7 -- 1.7 & \\            
J0847+0943 & 0.0173 & 1.3 & 0.65 -- 1.75 & \\            
J0858+0152 & 0.0206 & 0.7 & 0.0 -- 1.5 & 1400 blend problematic \\         
J0903+0708 & 0.0027 & 1.0 & 0.3 -- 1.7 & \\            
J0915$-$0202 & 0.00128 & 1.0 & 0.1 -- 1.3 & \\            
J0926+0135 & 0.03320 & 1.3 & 0.6 -- 1.7 & \\            
J0929+0333 & 0.01158 & 1.0 & 0.7 -- 2.0 & \\            
J0932+0237* & \ldots & \ldots & \ldots  & no 1400 blend \\         
J0946$-$0124 & 0.0235 & 0.7 & 0.5 -- 1 & borderline \\           
J1013+0851* & \ldots & \ldots & \ldots & no 1400 blend \\         
J1024+0245 & 0.0055 & 0.7 -- 1.0 & 0.6 -- 1.5 & \\          
J1026+0114 & 0.0036 & 0.7 & 0.6 -- 1.3 & \\            
J1145+0800 & 0.0109 & 1.0 & 0.3 -- 1.7 & \\            
J1205+0201* & \ldots & \ldots & \ldots & no 1400 blend \\         
J1214+0242 & 0.0154 & 0.7 & 0.2 -- 0.8 & \\            
J1215+0326 & 0.0509 & 1.0 & 0 -- 1.3 & \\            
J1219+0254 & 0.0136 & 0.0 & -0.7 -- 1.0 & \\            
J1244+0821 & 0.0246 & 1.0 & 0.5 -- 1.2 & \\            
J1259+0752 & 0.0228 & 0.7 & 0.0 -- 1.3 & \\            
J1314+0927 & 0.0078 & 0.7 & 0.4 -- 1.7 & low S/N \\          
J1419+0749 & 0.0171 & 1.0 & 0.5 -- 1.3 & \\            
J1509+0744 & 0.0175 & 1.0 & 0.7 -- 1.7 & \\            
J1516+0029* & \ldots & \ldots & \ldots  & only 1900 blend available \\
J1519+0723 & 0.0043 & 1.0 & 0.6 -- 1.7 & \\            
J1545+0156 & 0.0082 & 1.0 & 0.7 -- 1.7 & Abs. \\           
J1609+0654 & 0.0163 & 0.7 & 0.6 -- 1.0 & \\            
J1618+0704 & 0.0123 & 1.0 & 0.7 -- 2 & \\            
J2116+0441 & 0.0356 & \ldots & \ldots & Abs. at 1400  no high ion. solution \\ %-1.00 -1.1 -- 2.3
J2145$-$0758 & 0.0214 & \ldots & \ldots & Abs. at 1400  no high ion. solution \\%0 -1   -1. -- 1.2  J214502.56$-$075805.6               \\
 \hline
$\mu_\frac{1}{2}$(Ratios) & 0.7 -- 1.0 & 0.6 -- 1.2 &   \\       
 $\mu_\frac{1}{2}$(Objects) & 1.0 & 0.85 -- 1.15  \\
 \hline 
\end{tabular}
\tablefoot{BALQ are marked with an asterisk}(*). 
\end{table*}

\section{Discussion}
\label{Discussion}

We focus the discussion on three main topics: (1) the adequacy of the fitting methods, comparing the {\tt specfit} and the profile ratios as well as the results on diagnostic ratios and $Z$ obtained in this paper and the ones of \citetalias{sniegowskaetal21}; (2) the correlation between $Z$ and relevant physical and observational parameters. Of them, the \civ\ integrated properties  appear to be especially relevant; (3) the possibility of elemental pollution i.e., that the relative abundances of the elements deviate from the ones of solar chemical composition.

\subsection{Fitting methods}

\subsubsection{{\tt specfit} vs profile ratios}
\label{specpr}

\begin{figure*}
    \centering
     %trim=left bottom right top
     \includegraphics[trim= 20. 15. 70. 50, clip, width=0.77\textwidth]{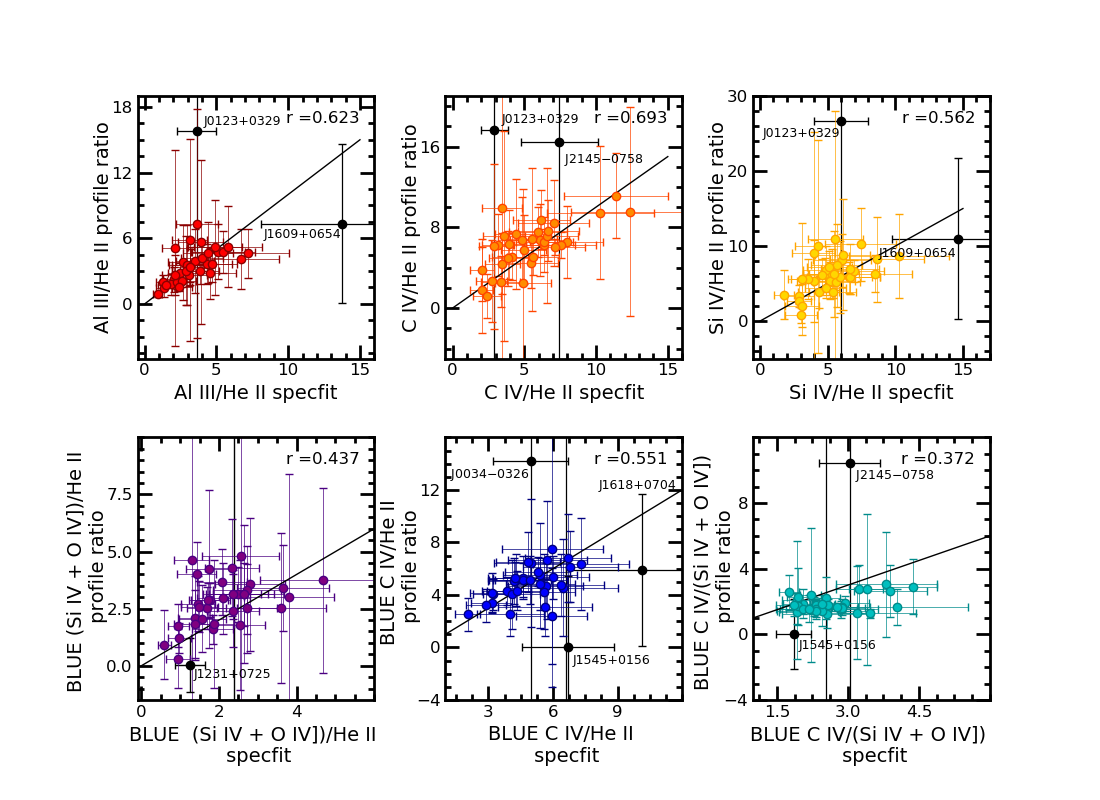}
    \caption{Top, from left to right: relation between intensity ratios \civ/\heii,
    \aliii/\heii\ and \civ/\heii\ BC computed with {\tt specfit} and with profile ratio technique described in Section \ref{specpr}. Bottom: comparison of intensity ratios for (\siiv+\oiv])/\heii\ for the BLUE component. Sources with the largest discrepancy from the equality line are identified.
    }   
    \label{fig:sppr}
 \end{figure*}

Fig. \ref{fig:sppr} compares the measurements of our six metallicity indicators for the two methods employee in this work, the \texttt{specfit} analysis (Sec. \ref{fitting}) and  the ratios of profile normalized intensities  (Sec. \ref{sec:profile}). Each panel shows  the equality line (black solid line) and the errors bars according to the fitting method. {Top panels correspond to the BC ratios and bottom panels to the BLUE ratios. The identified black dots are sources that  stray out from the rest of the sample, the cause of this behavior is different for BC and BLUE ratios:}

\begin{itemize}
    \item {BC ratios black dots (e.g. J0123+0329, J1545+0156, J1609+0654): this is due to deficient measurements of the \heiiuv\ emission which tend to be underestimated beside the \civ{} \ feature. The  {\tt specfit} modeling can compensate the surrounding narrow or broad absorptions while the profile measurements take the median intensity over a selected range directly from the spectra and can return close to zero or negative  results for low S/N spectra.
    \item BLUE ratios black dots (e.g J1231+0725, J0034$-$0326, J2145$-$0758): what causes the discrepancies for the BLUE ratios is the strong absorptions in these three features. Similar to the discussed for BC ratios, the \texttt{specfit} modeling can compensates the missing information from the original signal while the profile measurements only takes the median of the existence intensity within a given range and for the case of absorptions the resulting median intensity is diminished, as briefly  explained in Sec. \ref{sec:profile}, compare to the \texttt{specfit} measurements. We could modify the routine to skip the negative intensities (for the case of strong absorptions) but for profiles with less intense absorptions the result will still be the same. }
\end{itemize}
 The two measurements agree within the uncertainties in the wide majority of cases, {being the BLUE \civ/(\siiv+\oiv) ratio the one showing the lower degree of correlation with a trend mainly constant}. As reference we  consider r$\gtrsim$0.4 as a lower limit for  statistical significant correlation (with $\lesssim$ 1\%  of probability of not being correlated) for our $\sim$36 sources data set. If the BC \civ/\heiiuv\ ratios are compared, disagreements occur in the sense that the \civ/\heiiuv\ ratio measured with {\tt specfit} is much lower than the one from the profile ratio. This is due to the fact that the {\tt specfit} modeling compensates for narrow or broad absoprtion and is probably less affected by low S/N. The reverse case J1609+0654  is associated with  absorption and low S/N at the \heiiuv\ rest frame that effects both the {\tt specfit} results and the profile ratios. Similar considerations apply to the BLUE radial velocity domain: the outlier J1026+0114 is affected by an absorption that is interpolated across in the {\tt specfit} model. 

The comparison between the two methods of measurements show that, even in the fairly homogeneous sample of the present work, there is a range of values in the $Z$-sensitive diagnostic ratios: $2 \lesssim$ \civ/\heii\ $\lesssim 10$, $1 \lesssim$ \aliii/\heii $\lesssim 6$, $3 \lesssim $ \civ/\heii\ BLUE $\lesssim 10$. Outside of these range, extraordinarily large or small values are more likely due to measurement issues than to physical differences.

\subsubsection{Comparison with the results of \citetalias{sniegowskaetal21}}
\label{s21}

\begin{figure*}
    \centering
     %trim=left bottom right top
    \includegraphics[trim= 20. 15. 70. 50, clip, width=\textwidth]{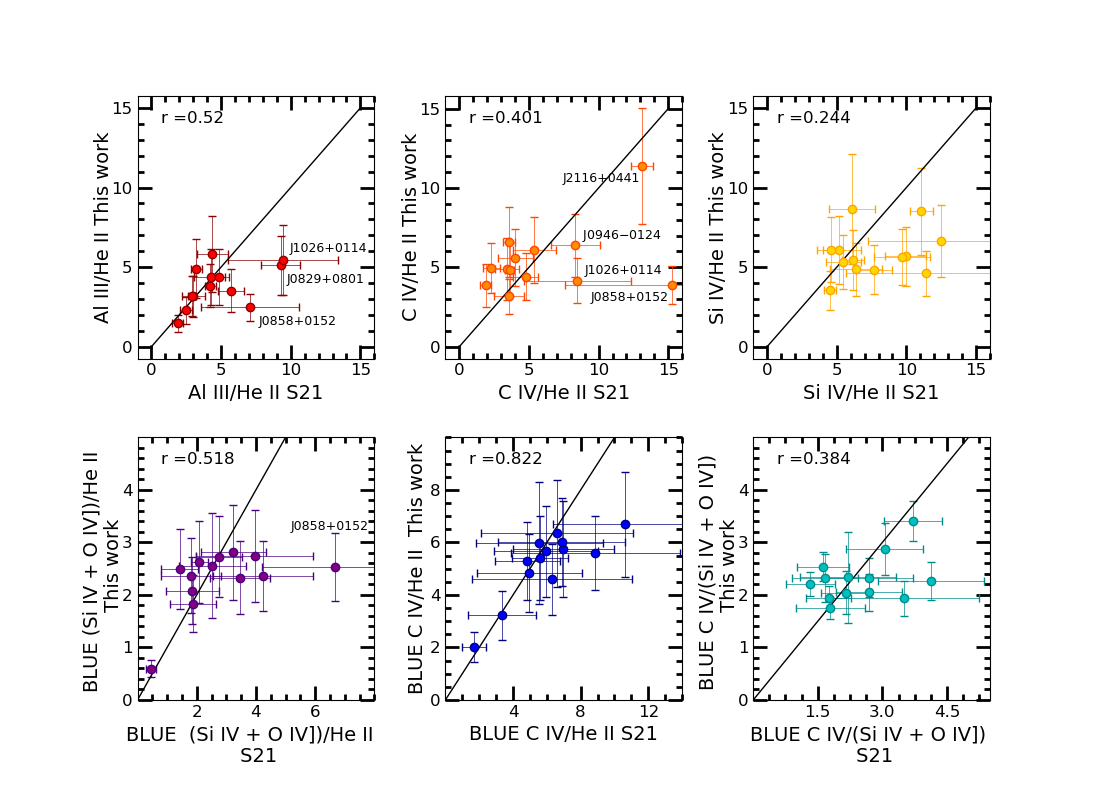}
    \caption{From left to right: Relation between intensity ratios \civ/\heii,
    \aliii/\heii, \civ/\heii\ BLUE, (\siiv+\oiv/)\heii, and (\siiv+\oiv/)\heii\ BLUE for the 13 objects in common with \citetalias{sniegowskaetal21}. The labels identify sources with largest deviation from the 1:1 lines. 
    }   
    \label{fig:kgms}
 \end{figure*}
The agreement between the measurements in the present thesis and the ones in common with  \citetalias{sniegowskaetal21} is fair for a 13 sources data set, as shown in Fig. \ref{fig:kgms}. For this subsample we are considering r$\gtrsim$0.7 as a lower limit for  statistical significant correlation (with $\lesssim$ 1\%  of probability of not being correlated). The main reason for this difference is most likely to a subestimation of the \heiiuv \ intensity in \citetalias{sniegowskaetal21}, yielding unrealistic values for the ratios involving \heiiuv. The difference between the medians for the two sets of measurements for the ratio \civ/\heii\ BC is $\approx 0.05 \pm 0.11$ \ dex, i.e., within the observational uncertainty. The source that is deviating most in Fig. \ref{fig:kgms}, J0858+0152 %\object{SDSS J085856.00+015219.4},
is associated with the difficult continuum placement around 1640  \AA: the value of \heii\ intensity reported by \citetalias{sniegowskaetal21} is most likely an underestimate. The discrepancy brings attention to the need to consider high S/N for the measurement of the \heii\ line.  However, the measurements remain statistically consistent with  a median deviation of $\approx -0.04 \pm 0.12$ for \aliii/\heii. The BLUE \civ/\heii\ ratio measurements appear to be  fully consistent  ($\mu \approx -0.02 \pm 0.04$\ dex).

The derived $Z$\ values from the \civ/\heii\  BC for fixed physical conditions are also in very good agreement: the median difference for the 13 objects in common is $\approx 0.11 \pm 0.22$, with the SIQR scatter less than a factor $\approx 2$, correspondent to  the uncertainties associated with the method. 

Regarding the BLUE component, a comparison is possible only for 8 objects. The free $\log Z$ estimates are $\approx 1.3$\ and $1.0$ for \citetalias{sniegowskaetal21} and the present work, suggesting metallicities typically in between 10 and 20 times the solar value.

\subsection{Analysis of $Z$ distributions at fixed  ($U$,\nh) for {\tt specfit} and profile ratio measurements}
\label{ZfixednU}

\begin{figure*}[htbp]%4
     \centering
     \hspace{-1cm}
     %trim=left bottom right top
     \includegraphics[trim= 50. 35. 70. 50, clip, width=0.95\textwidth]{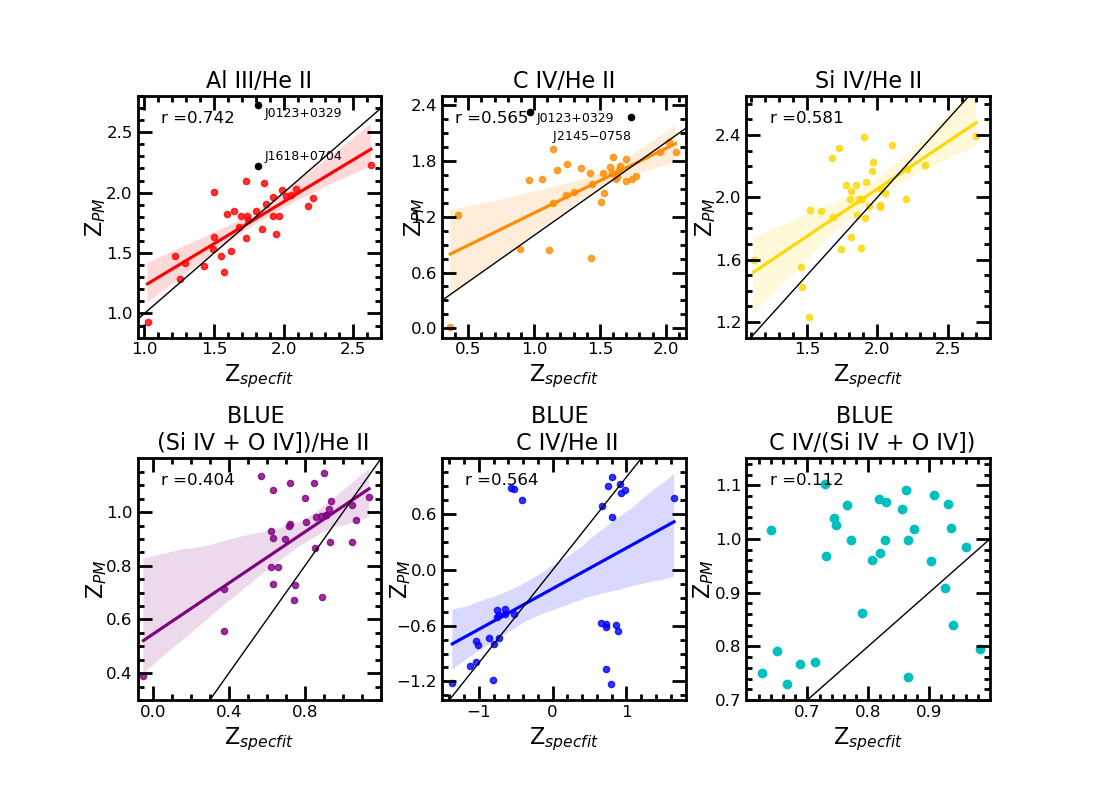}
    \caption{Metallicity comparison of our two fitting methods. Abscissas corresponds to {\tt specfit} results and the ordinates to profile ratio measurements results . Each panel shows the values per metallicity relation, top left corner indicates the Spearman correlation coefficient, the black solid line shows the equality  line.  Sources  with  the  largest  discrepancy  from  the equality line are discarded and identified as black dots.
    }   
    \label{fig:trend_pm}
 \end{figure*}

 Table \ref{tab:metal_nomr} lists the $Z$ results for the normalized profile  measurements, and we compare the metallicity results obtained with our two fitting methods in Figure \ref{fig:trend_pm}.  
The three upper panels showing the comparison for \aliii/\heiiuv, \civ/\heiiuv, \siiv/\heii\ demonstrate a high degree of correlation between the two methods, with the slope of the best fitting line close to the 1:1 relation. For the BLUE component, the situation is less clear. The \civ/(\siiv +\oiv) ratio is basically a scatter-plot: all sources are apparently with typical metallicities in the range $\log Z/Z_\odot \sim 0.6 - 1.1$. Some of our sources show a low  S/N mainly around the 1400 \AA \ and 1550 \AA \ regions. This affects our measurements and metallicity estimates giving a ratio that can not find a solution with the {\tt CLOUDY} metallicity relations. The case of \civ/\heii\ is especially cumbersome: due to the non-linear relation between ratio and $Z$, for some of the sources the two methods produce highly discordant values.

\begin{table*}[h]
\centering
\fontsize{8.2}{10}\selectfont %\tabcolsep=3pt
%\tabletypesize{\scriptsize\tabcolsep=2pt}
\tabcolsep=3pt
\caption{Metallicity of normalized intensities measurements}
\begin{tabular}{ccccccccccccccccccc}
\hline\hline
 &   {BC} & {BC} & {BC} & {BLUE} & {BLUE} & {BLUE} \\
{SDSS JCODE}  & {\aliii/\heiiuv} & {\civ{}/\heiiuv} & {\siiv{}/\heiiuv} &
 {(\siiv{}+\oiv)/\heiiuv} & {\civ{}/\heiiuv} & {\civ{}/(\siiv{}+\oiv)}\\
(1) & (2) & (3) & (4)  & (5)  & (6) & (7) \\
\hline
J0020+0740	&	$	1.63	\pm	0.17	$	&	$	1.61	\pm	0.19	$	&	$	1.94	\pm	0.15	$	&	$	0.93	\pm	0.19	$	&	$	-0.48	\pm	^{	1.36	}	_{	-1.05	}	$	&	$	0.97	\pm	0.11	$	\\
J0034$-$0326	&	$	1.53	\pm	0.22	$	&	$	1.22	\pm	0.35	$	&	$	1.55	\pm	0.20	$	&	$	1.13	\pm	2.98	$	&	$	1.85	\pm	^{	2.84	}	_{	0.92	}	$	&	$	0.73	\pm	0.46	$	\\
J0037$-$0238	&	$	1.48	\pm	0.16	$	&	$	1.74	\pm	0.14	$	&	$	1.99	\pm	0.14	$	&	$	0.90	\pm	0.16	$	&	$	-0.73	\pm	^{	0.86	}	_{	-1.03	}	$	&	$	1.00	\pm	0.15	$	\\
J0103$-$1104	&	$	1.39	\pm	0.24	$	&	$	1.35	\pm	0.24	$	&	$	1.67	\pm	0.24	$	&	$	0.80	\pm	0.39	$	&	$	0.69	\pm	^{	1.60	}	_{	-1.51	}	$	&	$	0.79	\pm	0.26	$	\\
J0106$-$0855	&	$	1.81	\pm	0.19	$	&	$	1.61	\pm	0.21	$	&	$	2.17	\pm	0.19	$	&	$	1.04	\pm	0.35	$	&	$	0.90	\pm	^{	1.71	}	_{	-1.32	}	$	&	$	0.96	\pm	0.21	$	\\
J0123+0329	&	$	2.73	\pm	1.81	$	&	$	2.32	\pm	1.81	$	&	$	\ldots	$	&	$	0.97	\pm	0.77	$	&	$	0.86	\pm	^{	1.63	}	_{	-1.13	}	$	&	$	0.91	\pm	0.74	$	\\
J0210$-$0823	&	$	1.66	\pm	0.36	$	&	$	0.85	\pm	0.42	$	&	$	0.33	\pm	0.59	$	&	$	0.80	\pm	0.31	$	&	$	-1.19	\pm	^{	-0.72	}	_{	-1.86	}	$	&	$	1.19	\pm	0.30	$	\\
J0827+0306	&	$	2.01	\pm	0.77	$	&	$	1.92	\pm	0.76	$	&	$	2.25	\pm	0.76	$	&	$	0.73	\pm	0.35	$	&	$	-1.07	\pm	^{	-0.45	}	_{	-1.88	}	$	&	$	1.02	\pm	0.28	$	\\
J0829+0801	&	$	1.97	\pm	0.22	$	&	$	1.56	\pm	0.26	$	&	$	1.99	\pm	0.23	$	&	$	0.87	\pm	0.28	$	&	$	0.57	\pm	^{	1.34	}	_{	-0.92	}	$	&	$	0.86	\pm	0.26	$	\\
J0836+0548	&	$	1.96	\pm	0.21	$	&	$	1.82	\pm	0.22	$	&	$	2.34	\pm	0.21	$	&	$	0.96	\pm	0.23	$	&	$	-0.80	\pm	^{	1.30	}	_{	-1.53	}	$	&	$	1.06	\pm	0.24	$	\\
J0845+0722	&	$	1.85	\pm	0.32	$	&	$	1.73	\pm	0.32	$	&	$	2.10	\pm	0.33	$	&	$	0.99	\pm	0.17	$	&	$	-0.61	\pm	^{	1.25	}	_{	-1.11	}	$	&	$	1.09	\pm	0.08	$	\\
J0847+0943	&	$	1.81	\pm	0.22	$	&	$	1.67	\pm	0.23	$	&	$	2.04	\pm	0.21	$	&	$	0.98	\pm	0.15	$	&	$	-0.42	\pm	^{	1.34	}	_{	-0.90	}	$	&	$	1.02	\pm	0.08	$	\\
J0858+0152	&	$	1.82	\pm	0.19	$	&	$	1.77	\pm	0.27	$	&	$	2.08	\pm	0.20	$	&	$	0.68	\pm	0.90	$	&	$	-0.58	\pm	^{	1.48	}	_{	-1.54	}	$	&	$	0.74	\pm	0.94	$	\\
J0903+0708	&	$	1.85	\pm	0.39	$	&	$	0.01	\pm	1.04	$	&	$	1.60	\pm	0.40	$	&	$	1.11	\pm	0.35	$	&	$	0.87	\pm	^{	1.72	}	_{	-1.50	}	$	&	$	1.06	\pm	0.15	$	\\
J0915$-$0202	&	$	1.29	\pm	0.24	$	&	$	1.35	\pm	0.30	$	&	$	1.89	\pm	0.20	$	&	$	0.90	\pm	0.19	$	&	$	-0.73	\pm	^{	1.08	}	_{	-1.13	}	$	&	$	1.03	\pm	0.13	$	\\
J0926+0135	&	$	1.47	\pm	0.14	$	&	$	1.45	\pm	0.15	$	&	$	1.75	\pm	0.14	$	&	$	0.89	\pm	0.33	$	&	$	0.75	\pm	^{	1.33	}	_{	-0.63	}	$	&	$	0.84	\pm	0.33	$	\\
J0929+0333	&	$	1.89	\pm	0.29	$	&	$	1.90	\pm	0.30	$	&	$	2.21	\pm	0.28	$	&	$	1.03	\pm	0.24	$	&	$	0.82	\pm	^{	1.55	}	_{	-0.97	}	$	&	$	0.99	\pm	0.17	$	\\
J0946$-$0124	&	$	1.42	\pm	0.09	$	&	$	1.64	\pm	0.08	$	&	$	1.86	\pm	0.09	$	&	$	0.67	\pm	0.17	$	&	$	-0.80	\pm	^{	-0.44	}	_{	-1.13	}	$	&	$	0.80	\pm	0.18	$	\\
J1024+0245	&	$	2.02	\pm	0.36	$	&	$	1.70	\pm	0.40	$	&	$	2.23	\pm	0.36	$	&	$	1.05	\pm	0.17	$	&	$	0.88	\pm	^{	1.55	}	_{	-0.76	}	$	&	$	1.00	\pm	0.09	$	\\
J1026+0114	&	$	1.98	\pm	0.17	$	&	$	1.46	\pm	0.20	$	&	$	1.95	\pm	0.17	$	&	$	1.11	\pm	0.21	$	&	$	0.92	\pm	^{	1.64	}	_{	-0.97	}	$	&	$	1.04	\pm	0.07	$	\\
J1145+0800	&	$	2.03	\pm	0.30	$	&	$	0.76	\pm	1.64	$	&	$	2.18	\pm	0.30	$	&	$	1.14	\pm	0.61	$	&	$	1.00	\pm	^{	1.99	}	_{	-2.00	}	$	&	$	1.07	\pm	0.14	$	\\
J1214+0242	&	$	1.90	\pm	0.13	$	&	$	1.84	\pm	0.14	$	&	$	2.08	\pm	0.13	$	&	$	0.73	\pm	0.66	$	&	$	-1.23	\pm	^{	1.17	}	_{	-2.00	}	$	&	$	1.10	\pm	0.92	$	\\
J1215+0326	&	$	1.52	\pm	0.11	$	&	$	1.64	\pm	0.24	$	&	$	2.03	\pm	0.10	$	&	$	0.71	\pm	0.23	$	&	$	-0.50	\pm	^{	1.31	}	_{	-1.00	}	$	&	$	0.75	\pm	0.25	$	\\
J1219+0254	&	$	1.34	\pm	0.10	$	&	$	1.61	\pm	0.09	$	&	$	1.67	\pm	0.09	$	&	$	0.56	\pm	0.24	$	&	$	-0.99	\pm	^{	-0.79	}	_{	-1.19	}	$	&	$	0.77	\pm	0.23	$	\\
J1231+0725	&	$	0.93	\pm	0.11	$	&	$	1.59	\pm	0.06	$	&	$	1.42	\pm	0.06	$	&	$		\ldots		$	&	$	-0.76	\pm	^{	1.03	}	_{	-1.16	}	$	& \ldots	$				$	\\
J1244+0821	&	$	1.81	\pm	0.19	$	&	$	1.75	\pm	0.17	$	&	$	2.05	\pm	0.21	$	&	$	0.96	\pm	0.11	$	&	$	-0.47	\pm	^{	1.25	}	_{	-0.88	}	$	&	$	0.96	\pm	0.09	$	\\
J1259+0752	&	$	1.63	\pm	0.30	$	&	$	0.85	\pm	0.66	$	&	$	1.94	\pm	0.36	$	&	$	1.08	\pm	0.15	$	&	$	-0.47	\pm	^{	1.34	}	_{	-1.02	}	$	&	$	1.16	\pm	0.10	$	\\
J1314+0927	&	$	1.81	\pm	0.44	$	&	$	-0.64	\pm	0.80	$	&	$	1.23	\pm	0.59	$	&	$	0.99	\pm	0.42	$	&	$	-0.65	\pm	^{	1.50	}	_{	-1.76	}	$	&	$	1.07	\pm	0.27	$	\\
J1419+0749	&	$	1.71	\pm	0.45	$	&	$	1.67	\pm	0.45	$	&	$	1.88	\pm	0.45	$	&	$	0.95	\pm	0.12	$	&	$	-0.43	\pm	^{	1.26	}	_{	-0.80	}	$	&	$	0.97	\pm	0.09	$	\\
J1509+0744	&	$	1.77	\pm	0.21	$	&	$	1.44	\pm	0.24	$	&	$	1.91	\pm	0.21	$	&	$	1.01	\pm	0.25	$	&	$	-0.59	\pm	^{	1.36	}	_{	-1.24	}	$	&	$	1.06	\pm	0.15	$	\\
J1519+0723	&	$	1.97	\pm	0.26	$	&	$	1.67	\pm	0.27	$	&	$	1.99	\pm	0.40	$	&	$	0.99	\pm	0.19	$	&	$	-0.57	\pm	^{	1.32	}	_{	-1.13	}	$	&	$	1.08	\pm	0.09	$	\\
J1545+0156	&	$	2.08	\pm	0.57	$	&	$	1.59	\pm	0.58	$	&	$	1.92	\pm	0.57	$	&	$	0.89	\pm	0.57	$	&	\ldots	&	\ldots	\\
J1609+0654	&	$	2.23	\pm	0.43	$	&	$	1.90	\pm	0.47	$	&	$	2.39	\pm	0.42	$	&	$	2.92	\pm	15.93	$	&	$									$ \ldots 	&	$	1.00	\pm	0.06	$	\\
J1618+0704	&	$	2.22	\pm	0.62	$	&	$	1.70	\pm	0.63	$	&	$	2.32	\pm	0.62	$	&	$	1.06	\pm	0.47	$	&	$	0.78	\pm	^{	1.74	}	_{	-2.00	}	$	&	$	1.02	\pm	0.22	$	\\
J2116+0441	&	$	1.70	\pm	0.18	$	&	$	2.01	\pm	0.16	$	&	$	1.99	\pm	0.17	$	&	$	0.39	\pm	0.71	$	&	$	-1.21	\pm	^{	-0.88	}	_{	-1.63	}	$	&	$	0.77	\pm	0.73	$	\\
J2145$-$0758	&	$	2.10	\pm	0.69	$	&	$	2.27	\pm	0.68	$	&	$	2.39	\pm	0.68	$	&	$	-1.01	\pm	1.83	$	&	$	-1.03	\pm	^{	-0.68	}	_{	-1.40	}	$	&	$	-1.30	\pm	1.83	$	\\
\\ \hline																																										
$\mu_{1/2}$	&	$	1.81	\pm	0.23	$	&	$	1.64	\pm	0.27	$	&	$	1.99	\pm	0.22	$	&	$	0.96	\pm	0.28	$	&	$	-0.49	\pm	^{	1.33	}	_{	-1.13	}	$	&	$	1.00	\pm	0.19	$	\\
$\sigma$	&	$	0.32	\pm	0.30	$	&	$	0.56	\pm	0.39	$	&	$	0.38	\pm	0.31	$	&	$	0.51	\pm	2.67	$	&	$	0.85	\pm	^{	0.89	}	_{	0.54	}	$	&	$	0.41	\pm	0.36	$	\\
SIQR	&	$	0.18	\pm	0.10	$	&	$	0.15	\pm	0.15	$	&	$	0.13	\pm	0.12	$	&	$	0.12	\pm	0.16	$	&	$	0.76	\pm	^{	0.22	}	_{	0.27	}	$	&	$	0.11	\pm	0.09	$	\\
\hline
\end{tabular}
\tablefoot{Columns are as follows: (1) SDSS short name, (2-7) {ratios from normalized intensities measurements.} The measurements of the six BALQ were excluded from this table. \\ 
$^{\ast \ast}$ The ellipsis {(...)} correspond to deficient measurements.}
\label{tab:metal_nomr}
\end{table*}

\subsection{{Correlation between $Z$\ and AGN physical properties}}
\label{acc}

As we have previously mentioned in Sec. \ref{sec:correlation}, due to the prior selection criteria for the xA bin  and also to our narrow $z$ range, we are not   expecting clear trends between metallicity $Z$, luminosity, \mbh\ and \lledd: we selected sources that are extreme in the 4DE1 parameter space.  Instead, most of the results are yielding well-defined parameters that corroborate  the efficiency of the 4DE1 selection criteria. For instance,  the metallicity and the Eddington ratio  are found not to be  correlated ($r\approx -0.094$; Fig. \ref{fig:9x9m}).  Both parameters however show a small dispersion around  well-defined values  log(Z$_\mathrm{BC}$)= 1.73 $\pm$0.15 and \lledd\ = 1.01 $\pm$0.16.  The lack of significant correlations can  be  seen in the upper right panels of  Fig. \ref{fig:9x9m}. The only exception is the full profile of the \civ \ emission (BC + BLUE component) FWHM and centroid at   half intensity  (indicated as shift in Fig. \ref{fig:9x9m}) : the Pearson correlation coefficient is $r\approx -0.97$, and  Spearman's  $ \approx -0.95$. These values have extremely high significance  for a sample  of 36 objects. This result is however not immediately related to $Z$ trends (the $r$\ values suggest no significant correlations of shift and FWHM of \civ\ with $Z$ (see Fig. \ref{fig:9x9m}), and will be presented and discussed elsewhere.

\subsection{Highly-supersolar abundances}
\label{super}

Accepted at face value, the $Z$\ derived for the virialized component of the BC implies metallicities 20 -- 50 times the solar values. These determinations are consistent with the results of \citetalias{sniegowskaetal21} who also found median values of 50, 20, 80\ for the $Z$\ estimated from the \aliii/\heiiuv{}, \civ/\heiiuv, \siiv/\heiiuv\ ratios, respectively. Similarly in both studies a median $Z \approx 50 Z_\odot$\ is found if the condition of fixed \nh\ and $U$\ is relaxed. The consistency with \citetalias{sniegowskaetal21} is in a way hardly surprising, since the method of measurement and analysis is the same, and the quasars belong to the same spectral class. 

Changes in metallicity that are associated to changes in the physical conditions ($U$, \nh) for the BC are modest. A small change in the $Z$ estimates are possible if the $U$ is increased by $\sim 0.5$ to 1 dex, up to $\log U \sim -1.5$. In this case, the median $Z$\ derived for our sample would decrease to $\log Z/Z_\odot \approx 1.2 $. Increase of \nh\ or decrease in $U$\ would increase the $Z$ estimates. The last 3 columns of Table \ref{tab:bc_z} report the values if the density is increased to \nh $=10^{13}$\ cm$^{-3}$: in this case the values remain very high $\sim 100 Z_\odot$, and the difference between the estimation from \civ\ and \aliii\ is reduced.  The parameter space (\nh, $U$, $Z$) shown in Fig. \ref{fig:mediancr} indicates that the $Z$ is well-constrained, and that a change in physical conditions still satisfying the observed ratios would imply a change in $Z$, i.e.  $\delta \log Z \lesssim 0.5$.    

The values we obtain for the BC are therefore consistent with $Z \gtrsim   20 Z_\odot$. This value should be taken as a reference for further discussion, and  is not much higher from the estimate obtained for high-$z$\ quasars in several previous studies \citep{baldwinetal03,warneretal04,nagaoetal06b,shinetal13,sulenticetal14} that indicate $Z \sim 10 Z_\odot$\ in sources that are not as extreme as the one considered here, with lower \aliii\ and \siiv\ with respect to \civ. These conditions can be interpreted as associated with lower $Z$.   

The $Z$\ values derived for the BLR of quasars are instead very high in the context of their host galaxies: the highest $Z$\ value measured in molecular cloud is around $5 Z_\odot$\ \citep{maiolinomannucci19}.  However, the nuclear and circumnuclear regions of quasars differ markedly from a normal interstellar environment. The   passage of stars through the disk involves the formation of accretion modified object that eventually reach high mass and explode, after a short evolution, as core collapse supernovae \citep{collinzahn99,chengwang99}. Stars in the nuclear region  can rapidly become very massive ($M \gtrsim 100 M_\odot$). These stars undergo core collapse, and contribute to polluting the disk with heavy elements through the high metal yields of supernova ejects \citep{cantielloetal21}. The compact remnants may continue accretion sweeping the accretion disk and giving rise to repeated supernova events \citep{lin97}. The metallicity enhancement is predicted to lead to $\sim 10 -  20 Z_\odot$\ \citep{wangetal11}, consistent with the value obtained for the xA sources. 

\subsection{Abundance pollution}
\label{pollu}

\begin{figure*}[ht]
  \centering
\includegraphics[width=9.cm]{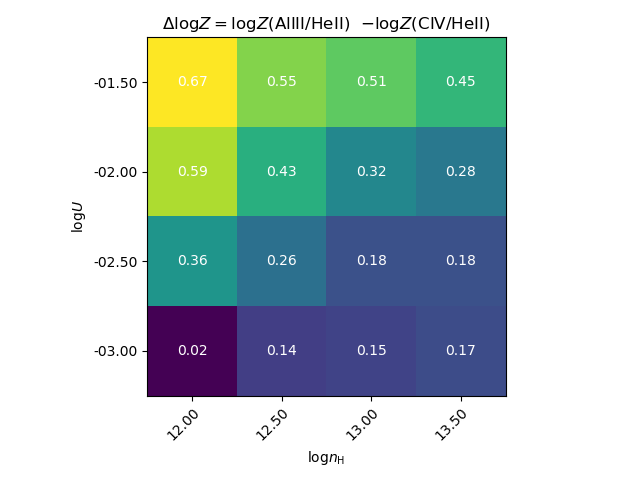}
\includegraphics[width=9.cm]{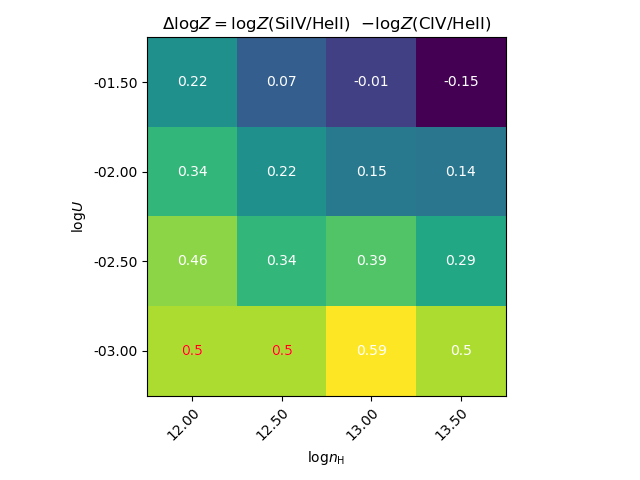}
\caption{Left panel: difference in the metallicity estimates from the \aliii/\heii\ intensity ratio and from the \civ/\heii\ ratio of the broad component, as a function of ionization parameter and density. The difference remain significant at a $2 \sigma$\ confidence level down to $\Delta \log Z = \log Z$(\aliii/\heii) - $\log Z$(\civ/\heii) $\approx 0.3$. Right: same, for difference   in the metallicity estimates from the (\siiv+\oiv)/\heii\ ratio ($\approx$ \siiv/\heii) and from the \civ/\heii\ ratio. The two cases with red values are lower limits.}   
\label{fig:pollution}
\end{figure*}

xA sources were selected to have strong \aliii\ by definition, and the line is strongly sensitive to $Z$, for fixed physical condition. High \nh\ and moderate or relatively high $U$\ contribute to make the line stronger \citep{marzianietal20}. Figure \ref{fig:pollution} shows the distribution of the metallicity differences $\delta \log Z$ = $\log Z$(\aliii) $- \log Z$(\civ) and    $\log Z$(\siiv) $- \log Z$(\civ). For the most likely values of the ionization parameter, {-2 $\lesssim \log U \lesssim$ -2.5}, the value of $Z$ is consistently higher if derived from the \aliii/\heii\ ratio or \siiv/\heii\ ratio than from the one derived from \civ/\heii, by a factor of $\approx 1.5 -2$. The very high $Z$\ values derived for the BC are due to the inclusion of the \aliii\ lines. While values $\approx 10 - 20 Z_\odot$ seems to be possible for the xA sources, higher values $\sim 50 - 100 Z_\odot$, might be more associated with deviation from the  relative solar elemental abundances.

\subsection{Metal segregation}

The comparison between the $Z$\ derived from \civ\ BC fixed, and the  $Z$ for BLUE free can be carried out for both \citetalias{sniegowskaetal21}\ and the present paper. The BC fixed medians are 1.7 \citepalias{sniegowskaetal21}\ and 1.3 (present paper), vs.  1.3 and 1.0 for the respective blue components. Considering that the uncertainty is $\delta \log Z \approx 0.3$\ dex, the comparison suggests an overall consistency in the $Z$ determinations. However, in both works there is  a small discrepancy between the $Z$\ estimates for the BC and BLUE. Selective enrichment of BLUE and BC is in principle possible. Radiation driven winds are primarily accelerated via resonant scattering of metal lines \citep[e.g., \citealt{progaetal00,progakallman04}, ][and references therein]{giustiniproga19}. However, the metal ions are coupled to hydrogen and helium  through Coulomb collisions, and physical conditions needed to lead a decoupling are extreme and outside the range considered for the BLR \citep{baskinlaor12}. A second mechanism is associated with the interaction of compact objects crossing the disk and the disk itself \citep{wangetal11}: star formation in self-gravitating disks can give rise to  metallicity gradients in broad-line regions, with the higher $Z$ in the innermost regions where a radiation driven wind is expected to develop.  Also in this case, the wind component should be the more metal rich.  The expected condition is that BLUE might be more affected by metal segregation (if at all), so that $Z_\mathrm{BLUE} \gtrsim Z_\mathrm{BC}$ and not the converse, nor $Z_\mathrm{BLUE} \sim Z_\mathrm{BC}$.  

The very high $Z$\ estimates, reaching $Z\sim 100 Z_\odot$, are associated with the inclusion of the \aliii\ and \siiv\ lines. $Z$\ estimates   from  \civ/\heii\ may be consistent for BLUE and BC within the uncertainties, and suggest $Z \sim 10 Z_\odot$. Therefore, confirmatory data are needed for any discrepancy between BLUE and BC $Z$\ estimates, also considering that the diagnostics of the BLUE is poor (in addition to the difficulties in the measurement of the blue wings, the \civ/\heii\ has a non-monotonic behaviour with $Z$\ at high $U$\ (as shown in the Appendix of \citetalias{sniegowskaetal21}). Our preliminary conclusion is that  BLUE- and BC-derived $Z$ \ values are consistent, if we exclude estimators that could be affected by elemental overabundance, as briefly discussed above.

\section{Conclusion}

We studied  the UV spectroscopic properties of a sample of xA quasars focusing on {their} metal content. The sample sources are an extension of the sample of \citetalias{sniegowskaetal21} which included only 13 objects, and was mainly aimed to assess the  applicability of a method based on the intensity of the strong emission lines \civ, \aliii, \siiv+\oiv\ relative to \heii, with the heuristic separation between a virialized (symmetric and unshifted, BC), and a {blue-shifted} component (BLUE). 

The main result of the present investigation is the confirmation of very high metallicity values, in excess of 5 times solar, and more likely in the range 10 - 50 $Z_\odot$, as found by  \citetalias{sniegowskaetal21}. 

\begin{itemize}
   \item For the BC, and the most likely fixed physical condition (\nh $\sim 10^{12}$ \cmc; $U \sim 10^{-2.5}$) suggest $Z \approx 50 Z_\odot$ , with a 40\%\ scatter. We remark that this high value is a consequence of the inclusion {of the \aliii/\heii\ and \siiv+\oiv/\heii\ ratios in the $Z$\ estimate}.
   \item For the BLUE, again at fixed physical conditions, the most likely value is $Z \approx 6 Z_\odot$. The discrepancy with the BC needs to be accounted for. 
   \item Relaxing the assumption of fixed physical conditions, the three indepenent ratios allow for an independent estimate of \nh, $U$, and $Z$. The results on $Z$\ are consistent with the case for fixed physical conditions ($Z \approx 50 Z_\odot$), the ionization parameter is $\log U \sim - 2$, and the derived \nh\ is however significantly larger.  
    \item Without fixed physical conditions,   $Z \sim 10 Z_\odot$\ is found for BLUE, with $U$\ and \nh\ largely unconstrained.   
    \item We noted that the $Z$\ derived from \civ/\heii\ ($Z\approx 30 Z_\odot$) is lower than the ones from \aliii\ and \siiv+\oiv/\heii. We explored the ratios between  \aliii/\heii\ and \civ/\heii\ derived $Z$  as a function of \nh\ and $U$, and found that the difference is preserved within  factors 1.5 -- 5 in the range $-2.5 < \log U < -1.5$, and $12 \lesssim \log$\nh $\lesssim 13$ (\cmc).  
    \item We tentatively explained this result by abundance pollution, and suggest that the elemental abundances of the BLR gas might be significantly different from solar, especially for elements like Silicon and Aluminium. Our suggestion is supported by the models that attempts to account for the complexity of the active nucleus, that encompasses star forming regions and even accretion-modified stars. The conditions of the xA sample are most likely very different from the ones expected in a passively evolving, elliptical host. 
    \item Considering only the \civ/\heii\ as a $Z$ estimator, we are left in the uncomfortable situation that the relation between  this ratio and $Z$\ is not monotonic at high $U$. This effect involves the highest $U$\ down to $\log U \sim -1$. Therefore we consider that the best estimator for solar-metallicity estimates is the BC \civ/\heii\ ratio, that yields values around $Z \approx 30 Z_\odot$ with a sample standard deviation of 0.15 dex, implying $20 Z_\odot \lesssim Z \lesssim 40 Z_\odot$. 
    \item The main accretion parameters (\mbh, $L$, \lledd) of the present sample cluster around a well defined value, with modest scatter comparable to the individual error of measurements. The spectral similarity also suggest that we are observing almost "clones" of the same AGN. The \lledd\ value $\sim 1$\ confirms that these sources are high radiators. 
    \item The present sample shows that the \civ\ full profile parameters are highly correlated among themselves, as seen in previous studies. Especially striking is the correlation between FWHM and centroid at half maximum. Since all relevant parameters are constrained in a small range, we suggest that the correlation is mainly due to orientation effects. 
    \item The {\tt specfit} analysis requires a multicomponent decomposition that may not be strictly unique. It is actually a model imposed to the data on the basis of trends observed along the quasar main sequence. To test the validity of this model, we also considered a model independent measure, based on the profile ratios in fixed wavelength range, where BC and BLUE give the maximum emission, but otherwise without profile decomposition. 
\end{itemize}

The {profile} ratio method is in principle applicable to large samples of quasars that are highly accreting. However, it relies heavily on the \heii\ emission line which is weak and with a very broad profile (as true also for the optical He {\sc ii}$\lambda$4686 emission, \citealt{marzianisulentic93}). Dedicated observations yielding high-S/N are needed to improve the precision of the $Z$\ estimates.  {This is not a standard refrain: the potential of the method exploiting the \heii\ profile measurements can be unleashed only with precise \heiiuvfull\ measurements.} \vfill

\begin{acknowledgements} 
DD and CAN acknowledge support form grant IN113719 PAPIIT UNAM. CAN acknowledges support from CONACyT project Paradigmas y Controversias de la Ciencia 2022-320020. MS and SP acknowledge the financial support from the Polish Funding Agency National Science Centre, project 2017/26/A/ST9/-00756 (MAESTRO 9). SP acknowledges partial financial support from the Conselho Nacional de Desenvolvimento Científico e Tecnológico (CNPq) Fellowship (164753/2020-6).
\end{acknowledgements} 
\vfill

%\pagebreak
%\eject\newpage

%\vfill
%\end{document} 

%\pagebreak\pagebreak
%\clearpage\clearpage\eject
%\pagebreak\pagebreak
%\newpage
%%%%%%%%%%%%%%%%%%%%%%%%%%%%%%%%%%%%%%%%%%%%%%%%
\onecolumn
\begin{appendix}
\section{Rest-frame spectra and fits}
\label{app:spec}

The following Figures show the results of the {\tt specfit} analysis. 

\begin{figure*}[hp!]
    \centering
     %trim=left bottom right top
     \includegraphics[trim= 0.0 0. 0. 0., clip, width=0.92\textwidth]{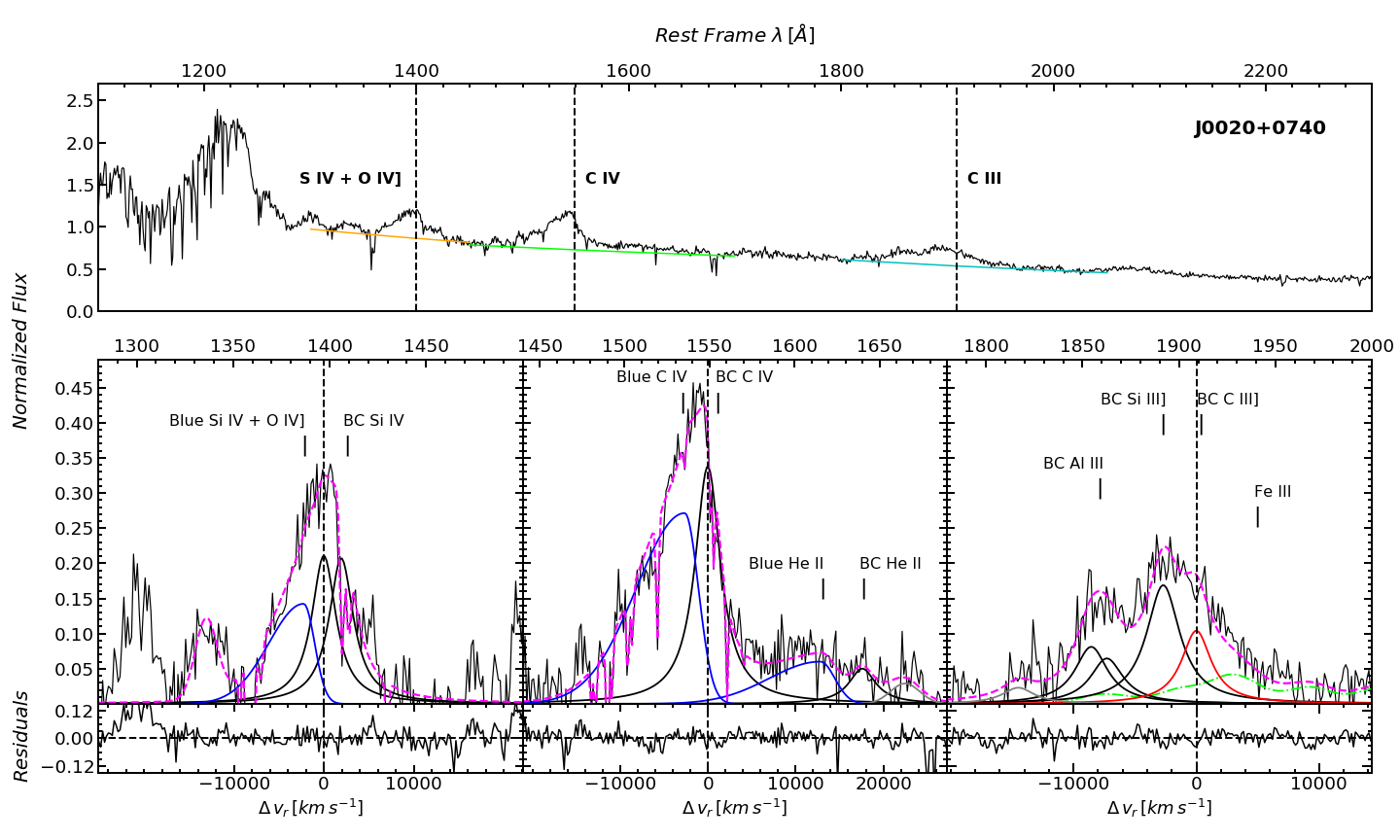}
   \includegraphics[trim= 0.0 0. 0. 0., clip, width=0.92\textwidth]{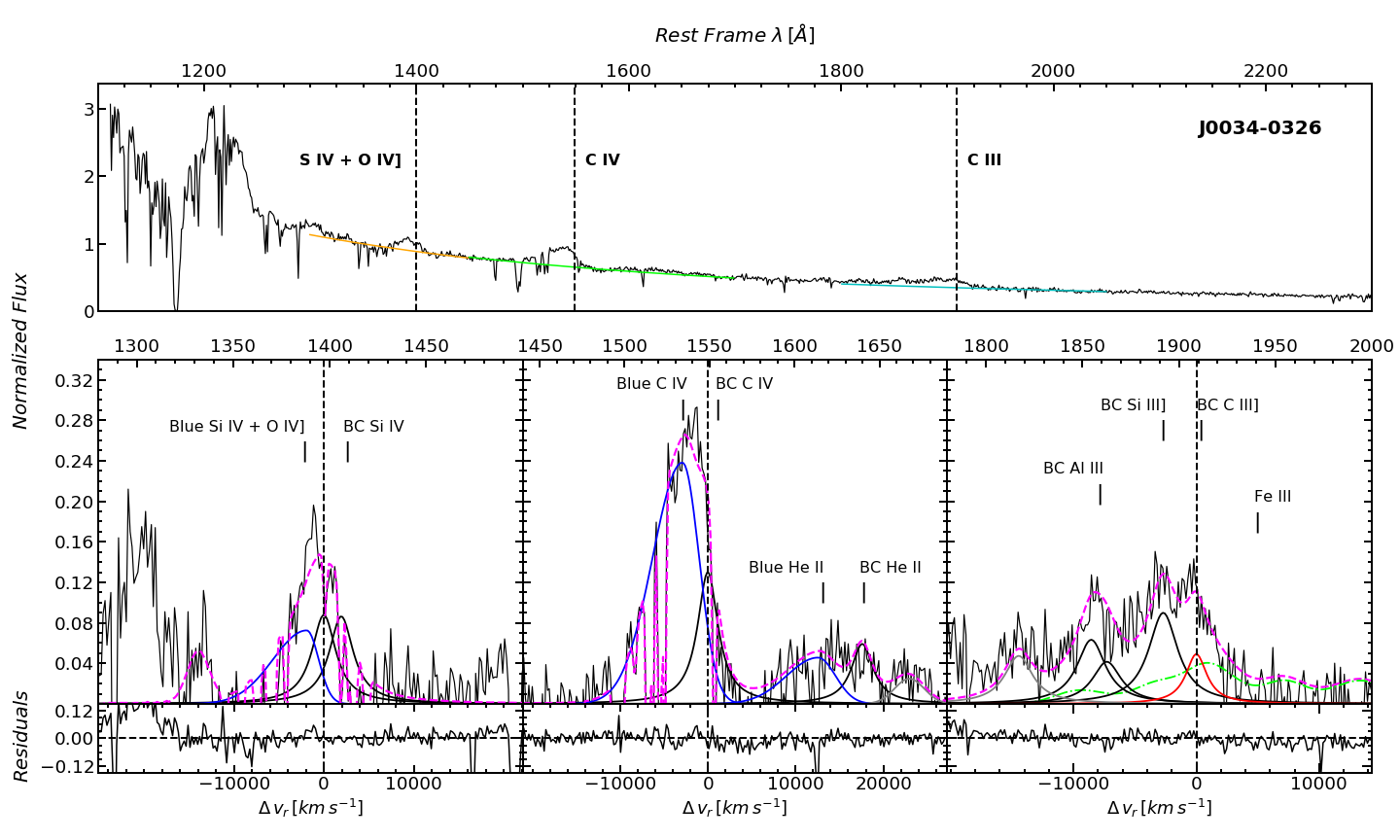} 
   \caption{Spectra of \object{SDSSJ002023.12+074041.1} and \object{SDSSJ003411.37-032618.2}, with {\tt specfit} analysis results. For each spectrum, the top panel shows the full  spectrum  as a function of rest-frame  wavelength, after normalization at 1350 \AA. The actual continuum ranges employed in the fitting of the three spectral regions discussed in the paper are marked. The bottom panels show the decomposition of the blends. Black lines: observed spectrum under continuum subtraction. Magenta dashed lines: full model of the observed spectrum.  Thick black lines: BC of \siiv, \civ, \heii, \aliii\ and \siiii. Blue lines: {blue-shifted} components of \siiv+\oiv, \civ\ and \heii. Grey lines: faint lines affecting the blends or \aliii\ BLUE component.      }   
    \label{fig:01}
 \end{figure*}

 \begin{figure*}
    \ContinuedFloat
    \centering
     %trim=left bottom right top
     \includegraphics[trim= 0.0 0. 0. 0., clip, width=\textwidth]{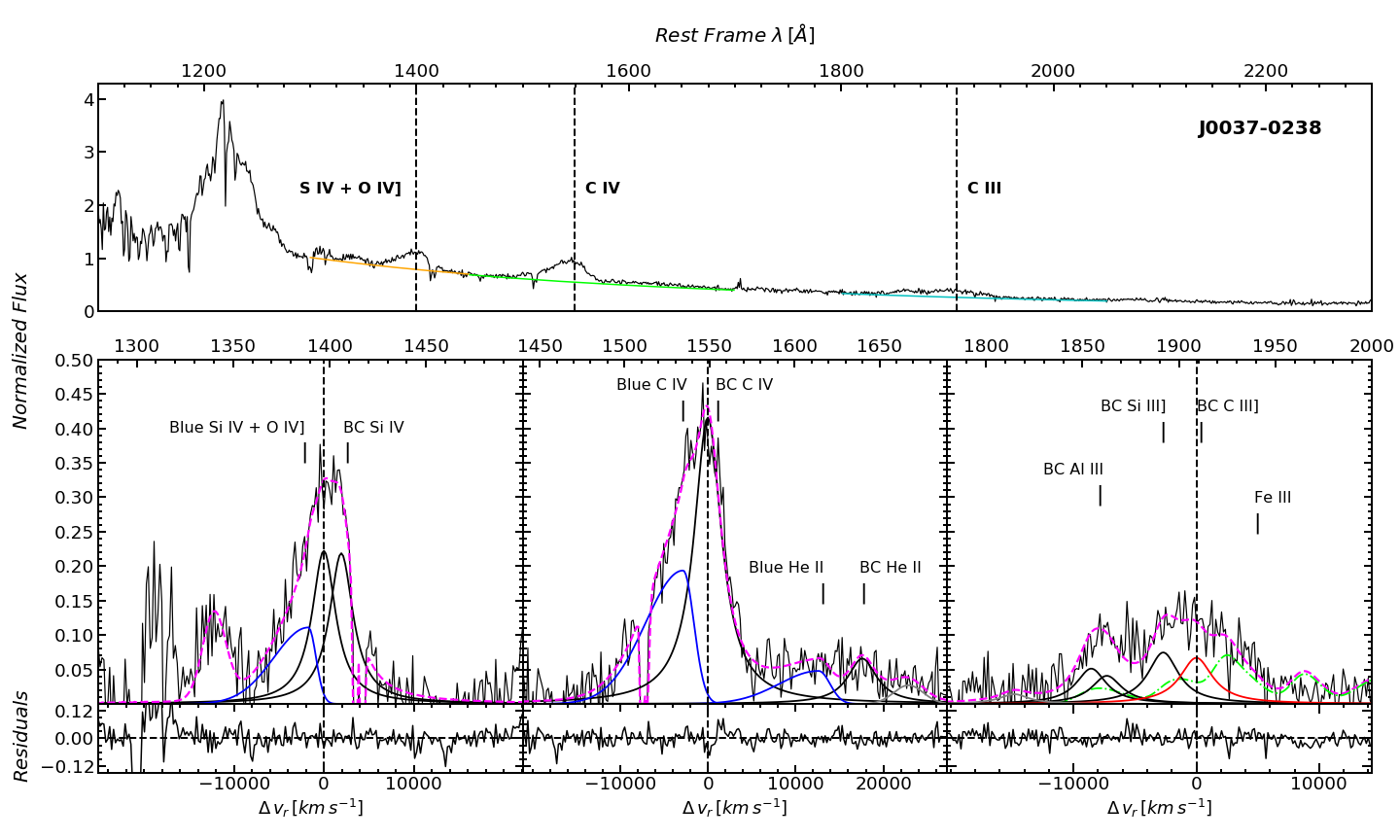}
     \includegraphics[trim= 0.0 0. 0. 0., clip, width=\textwidth]{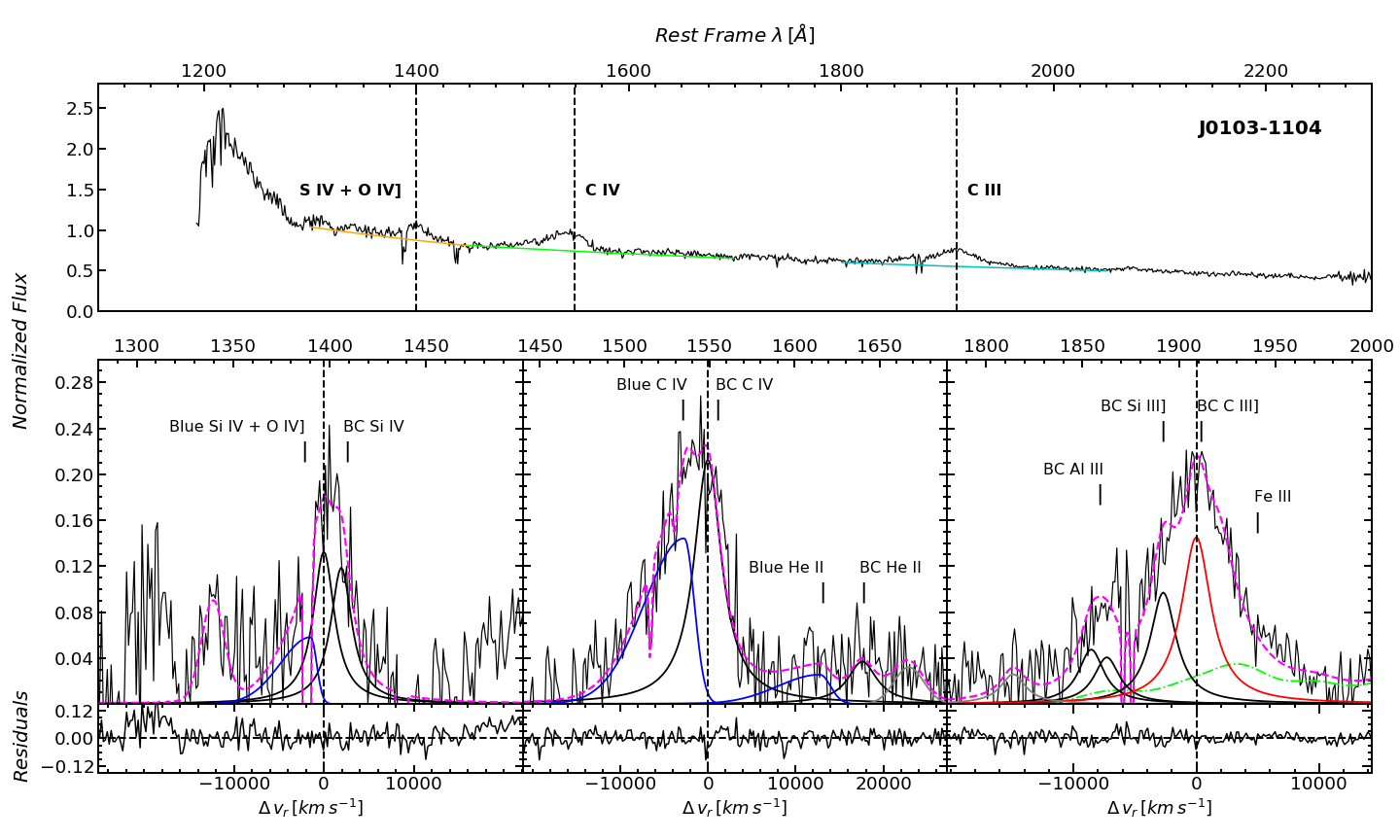}    \caption{continued.}   
    \label{fig:02}
 \end{figure*}

\begin{figure*}
    \ContinuedFloat
    \centering
     %trim=left bottom right top
     \includegraphics[trim= 0.0 0. 0. 0., clip, width=\textwidth]{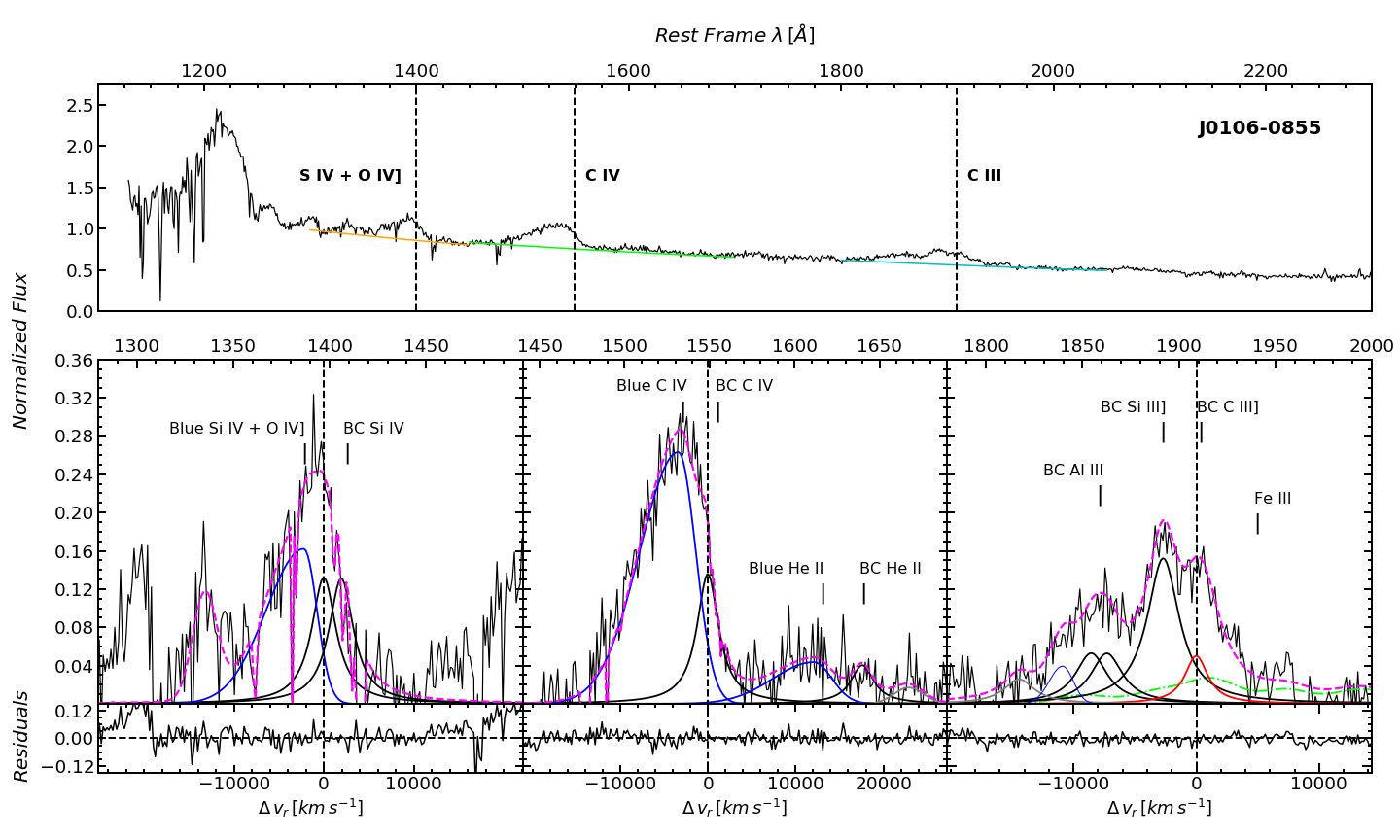}
    \includegraphics[trim= 0.0 0. 0. 0., clip, width=\textwidth]{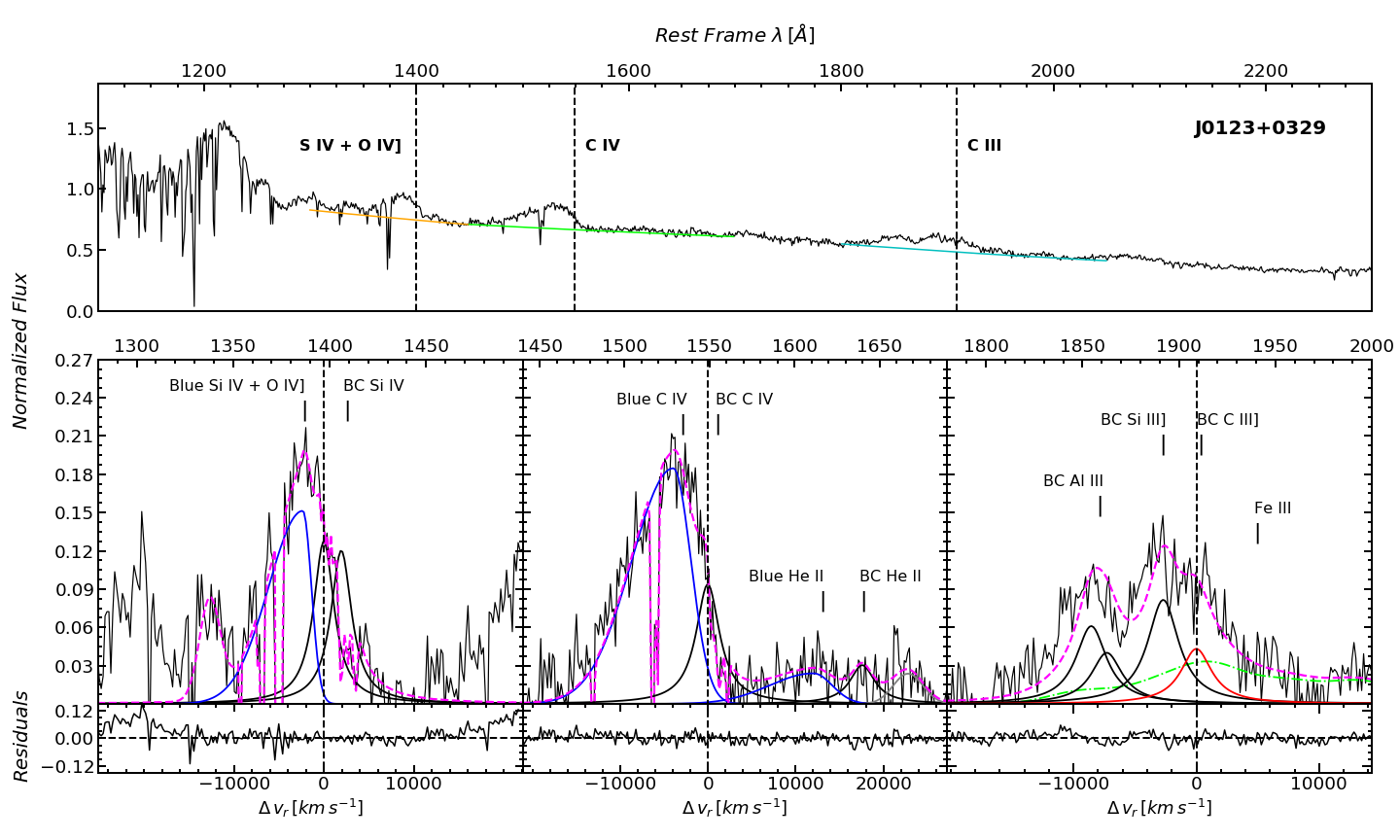}
    \caption{continued.}   
    \label{fig:03a}
 \end{figure*}

 \begin{figure*}
    \ContinuedFloat
    \centering
     %trim=left bottom right top
     \includegraphics[trim= 0.0 0. 0. 0., clip, width=\textwidth]{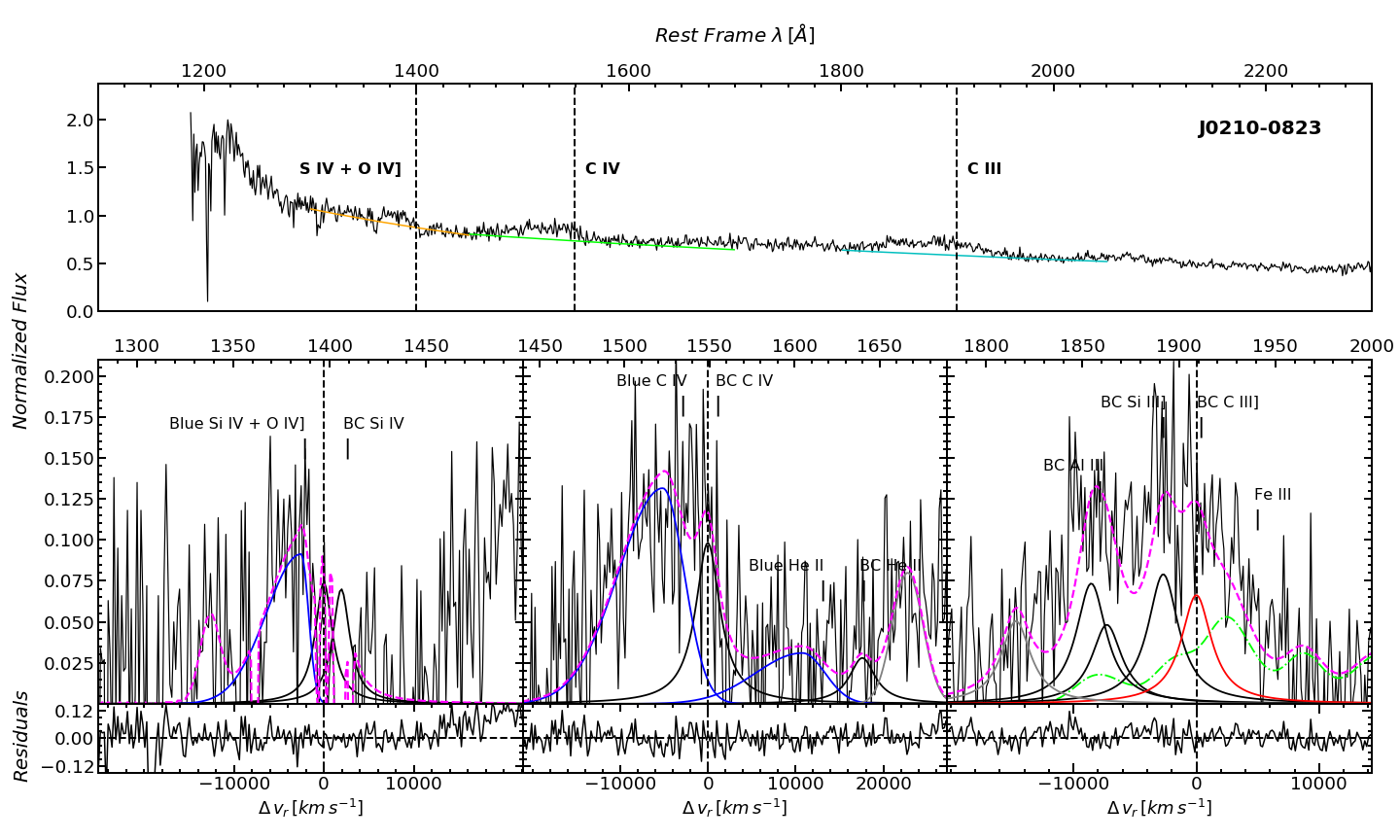}
     \includegraphics[trim= 0.0 0. 0. 0., clip, width=\textwidth]{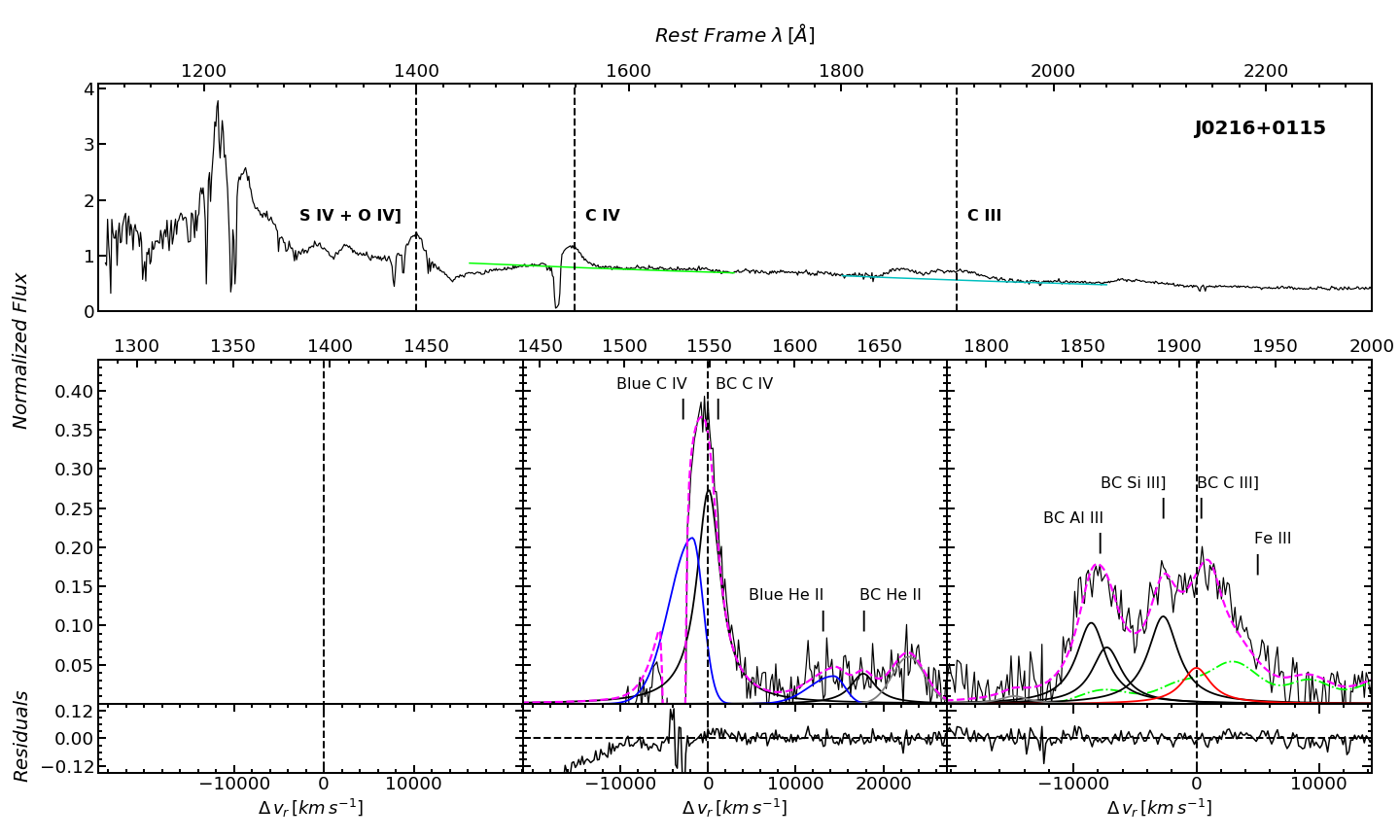}
    \caption{continued.}   
    \label{fig:04}
 \end{figure*}

  \begin{figure*}
    \ContinuedFloat
    \centering
     %trim=left bottom right top
     \includegraphics[trim= 0.0 0. 0. 0., clip, width=\textwidth]{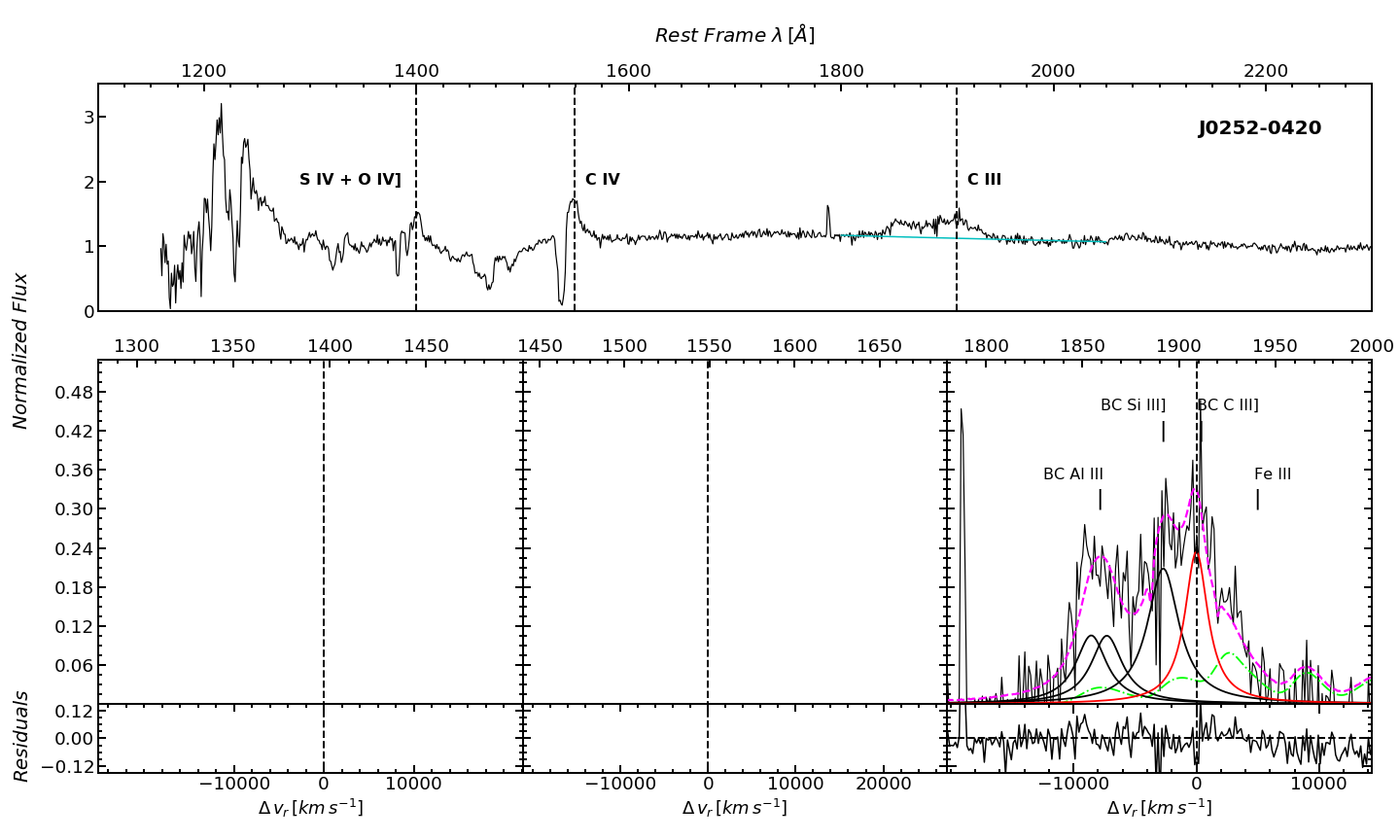}
     \includegraphics[trim= 0.0 0. 0. 0., clip, width=\textwidth]{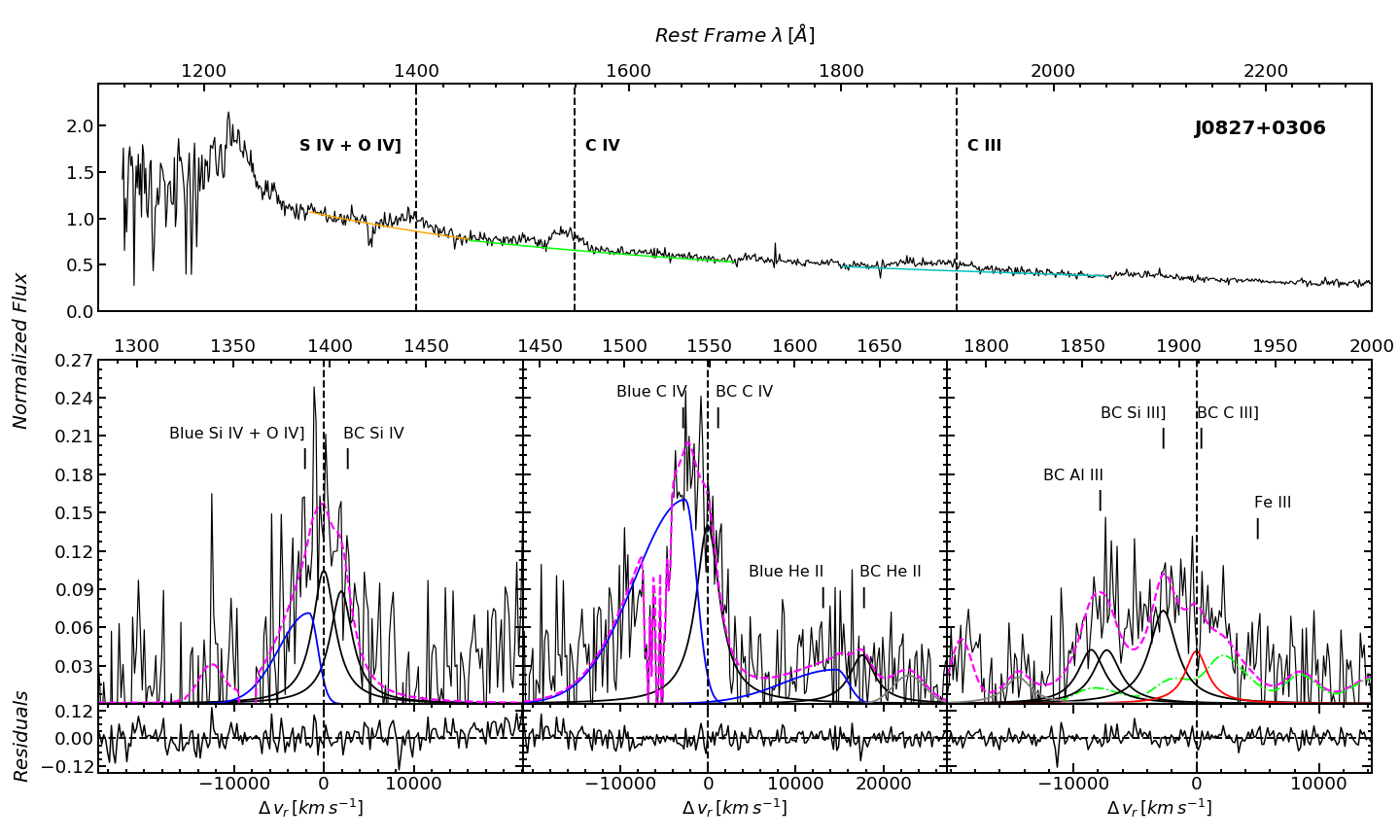}
   \caption{continued.}   
    \label{fig:05}
 \end{figure*}

 \begin{figure*}
    \ContinuedFloat
    \centering
     %trim=left bottom right top
     \includegraphics[trim= 0.0 0. 0. 0., clip, width=\textwidth]{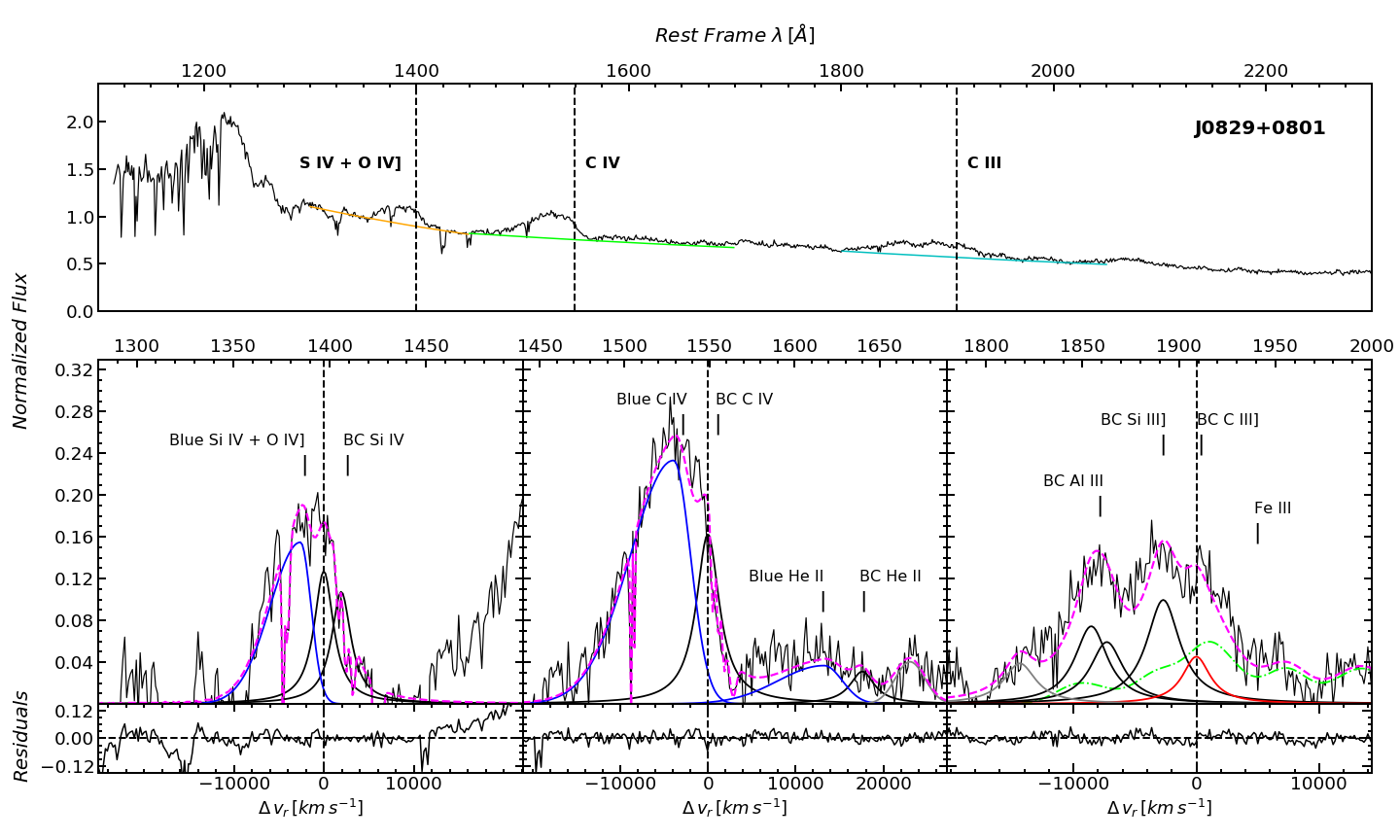}
      \includegraphics[trim= 0.0 0. 0. 0., clip, width=\textwidth]{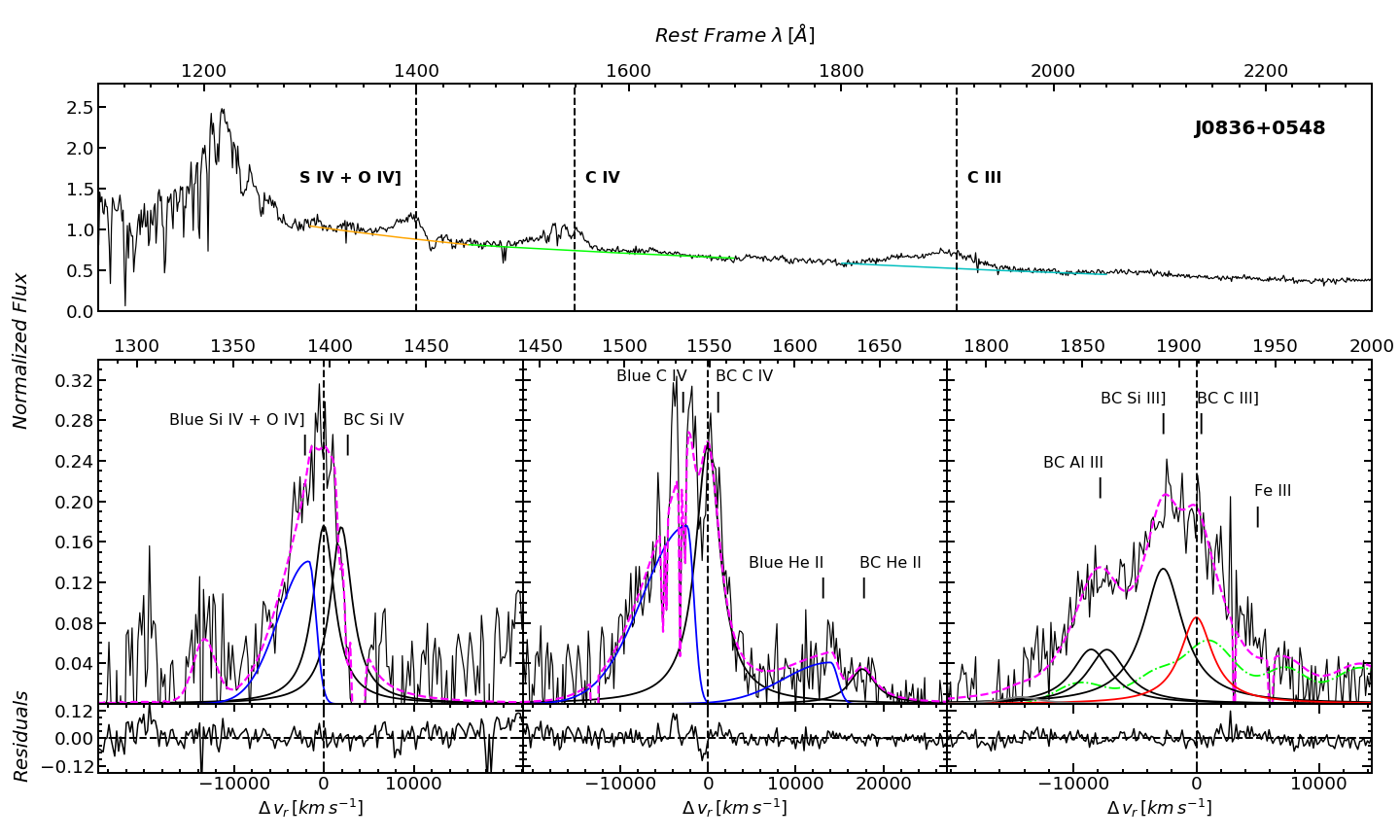}
   \caption{continued.}   
    \label{fig:6}
 \end{figure*}

  \begin{figure*}
    \ContinuedFloat
    \centering
     %trim=left bottom right top
     \includegraphics[trim= 0.0 0. 0. 0., clip, width=\textwidth]{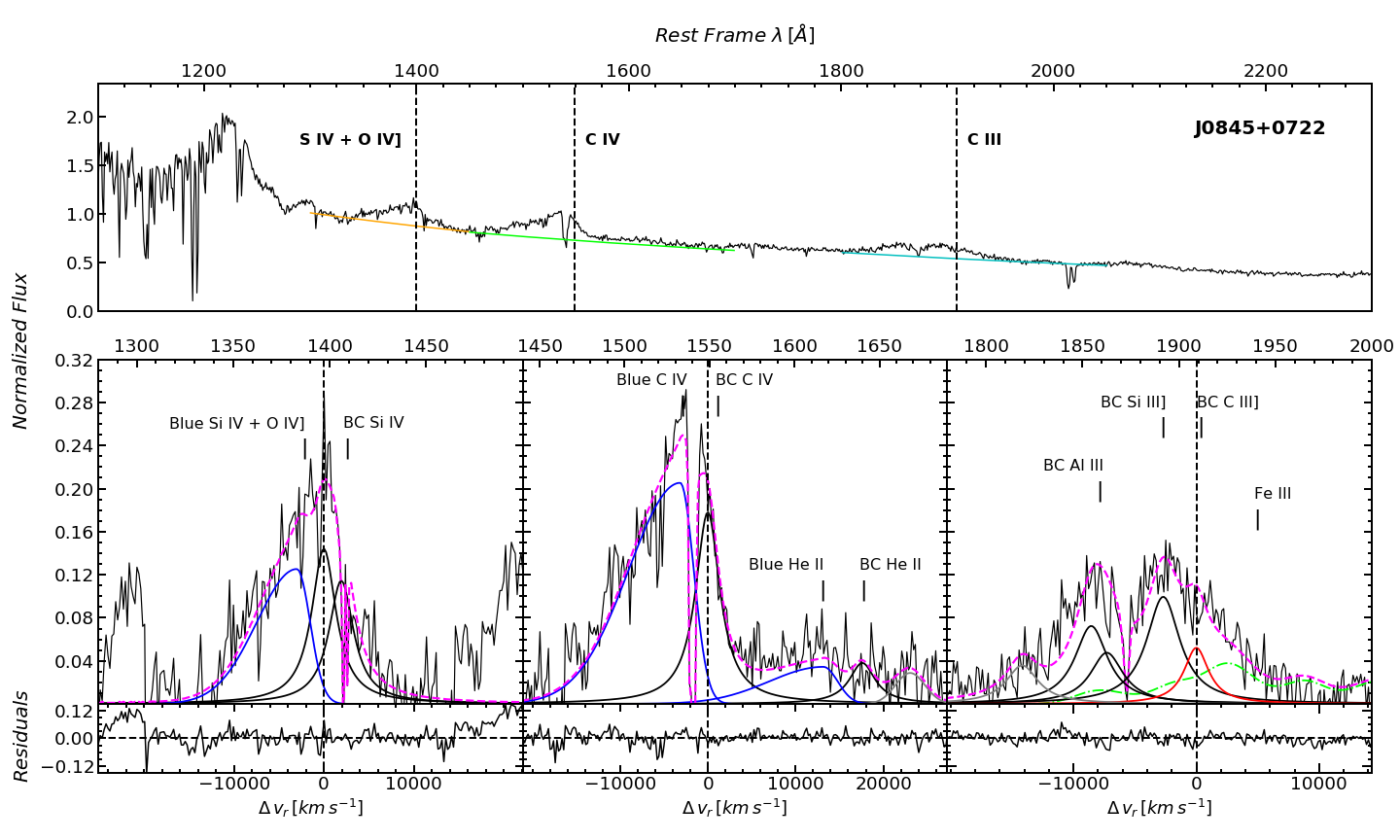}
    \includegraphics[trim= 0.0 0. 0. 0., clip, width=\textwidth]{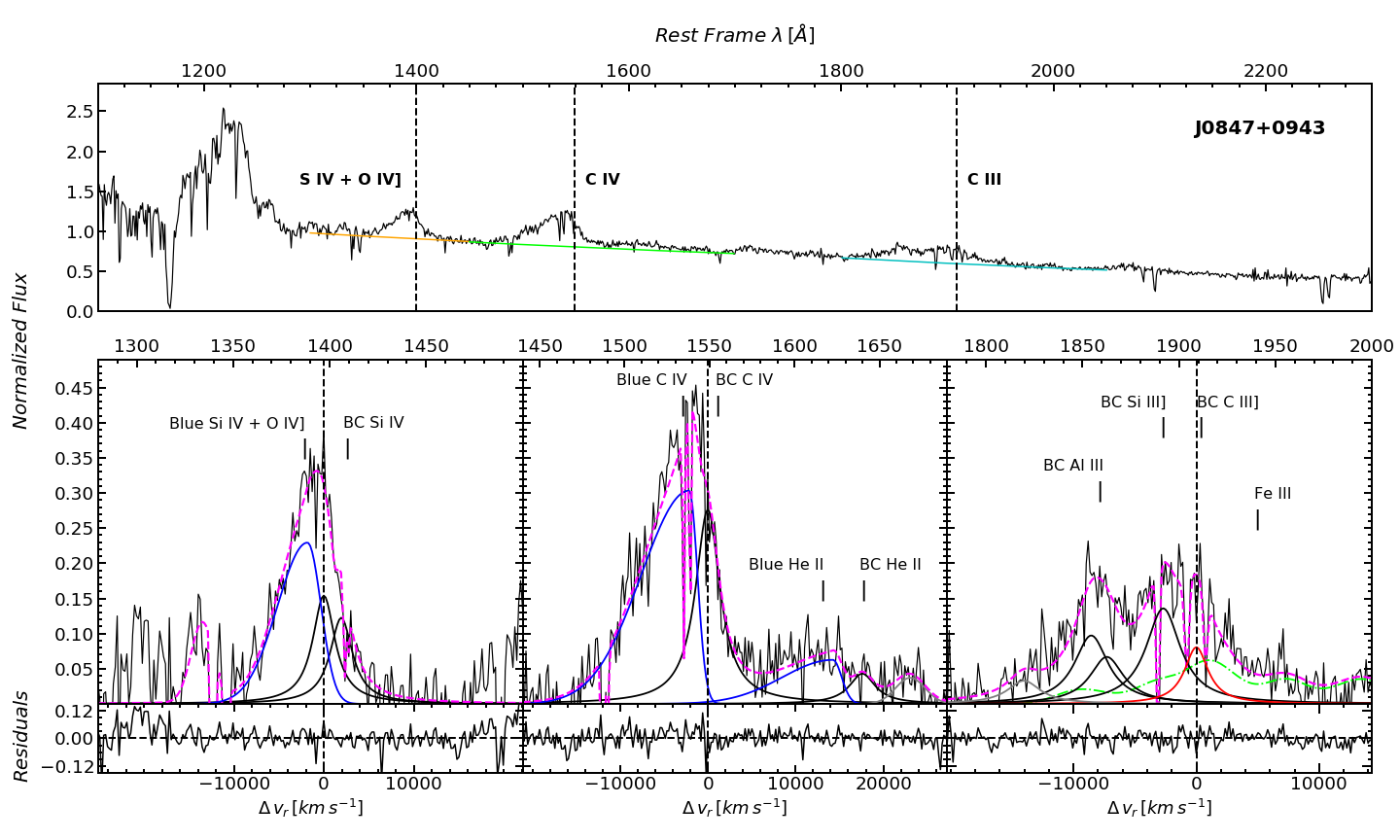}
    \caption{continued.}   
    \label{fig:7}
 \end{figure*}

  \begin{figure*}
    \ContinuedFloat
    \centering
     %trim=left bottom right top
     \includegraphics[trim= 0.0 0. 0. 0., clip, width=\textwidth]{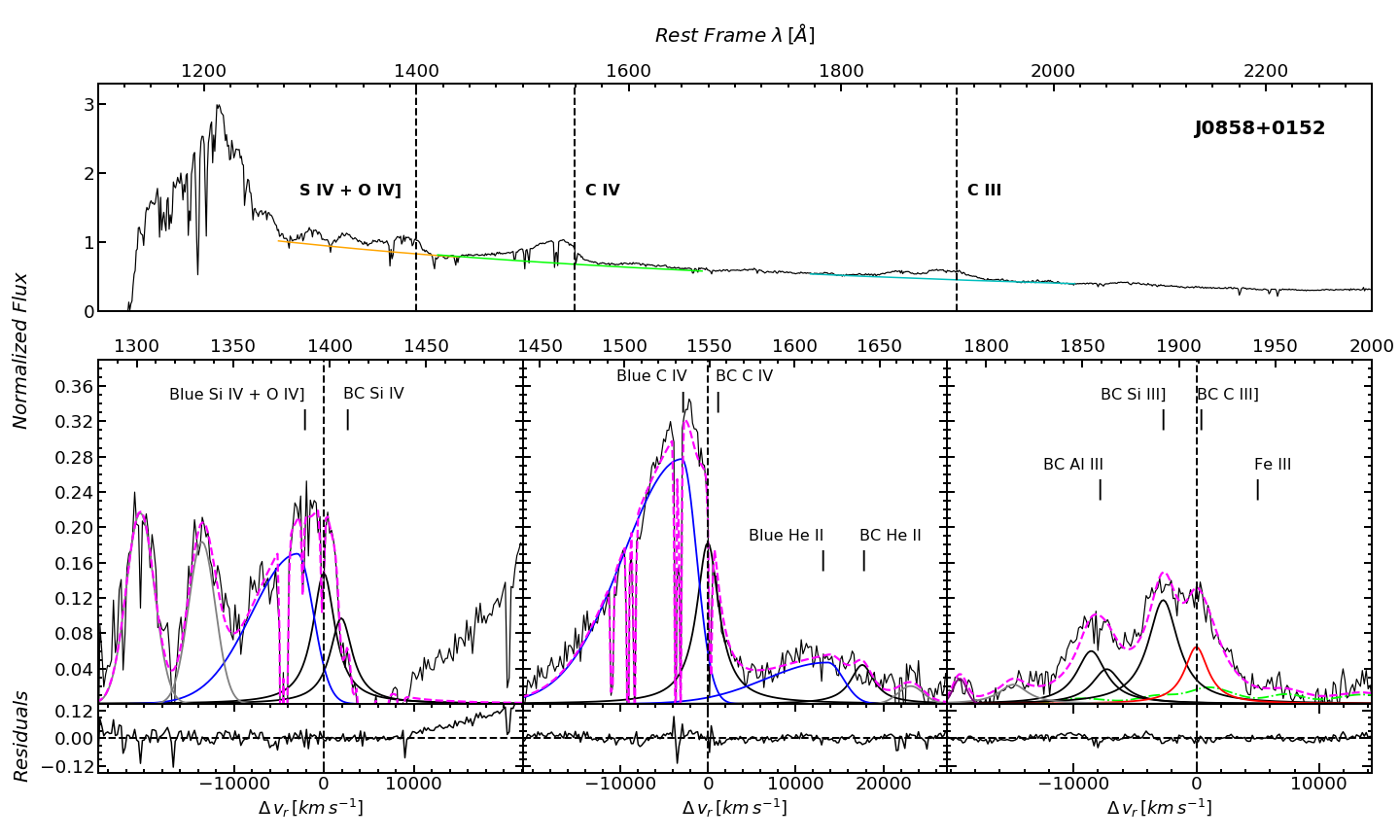}
    \includegraphics[trim= 0.0 0. 0. 0., clip, width=\textwidth]{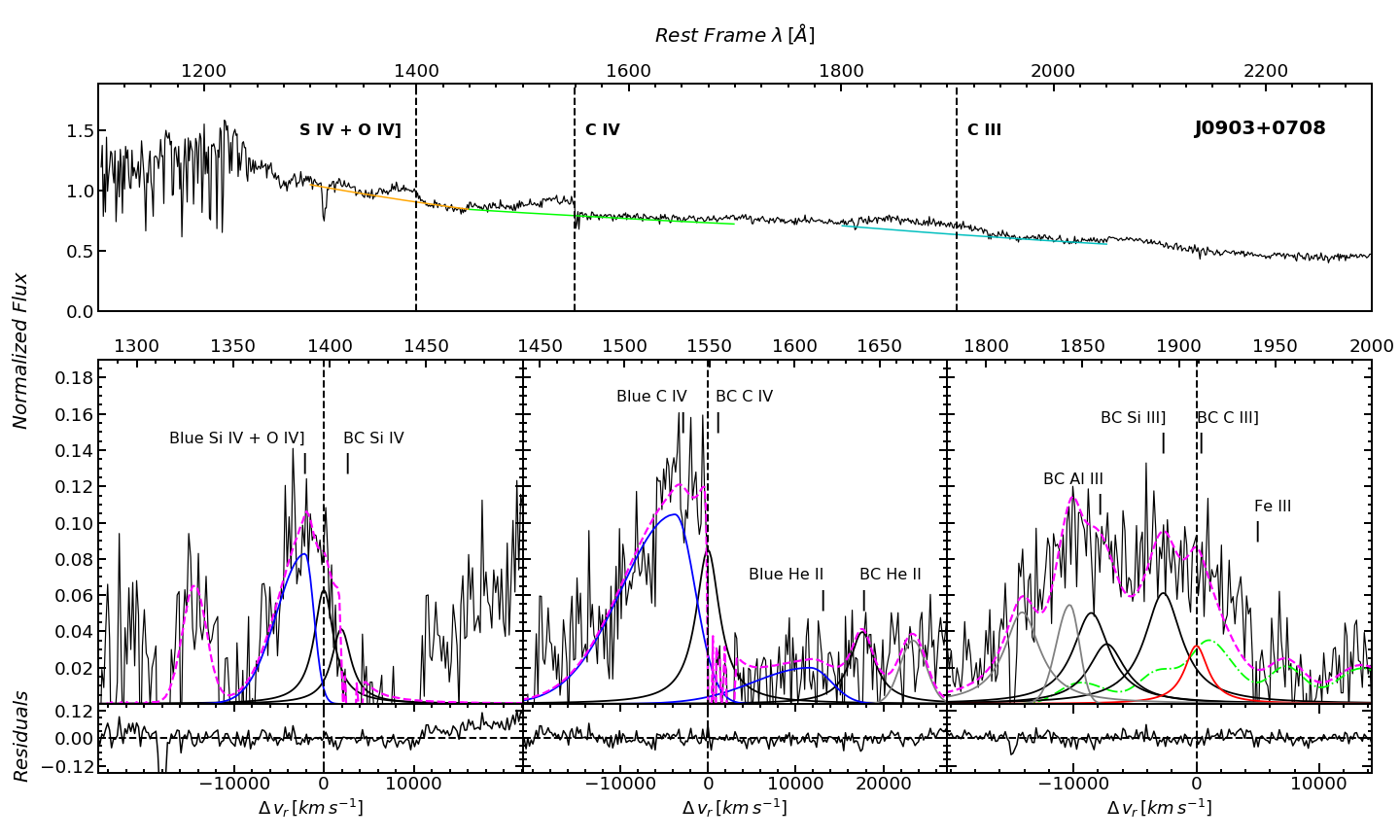}
    \caption{continued.}   
    \label{fig:8}
 \end{figure*}
 
  \begin{figure*}
    \ContinuedFloat
    \centering
     %trim=left bottom right top
     \includegraphics[trim= 0.0 0. 0. 0., clip, width=\textwidth]{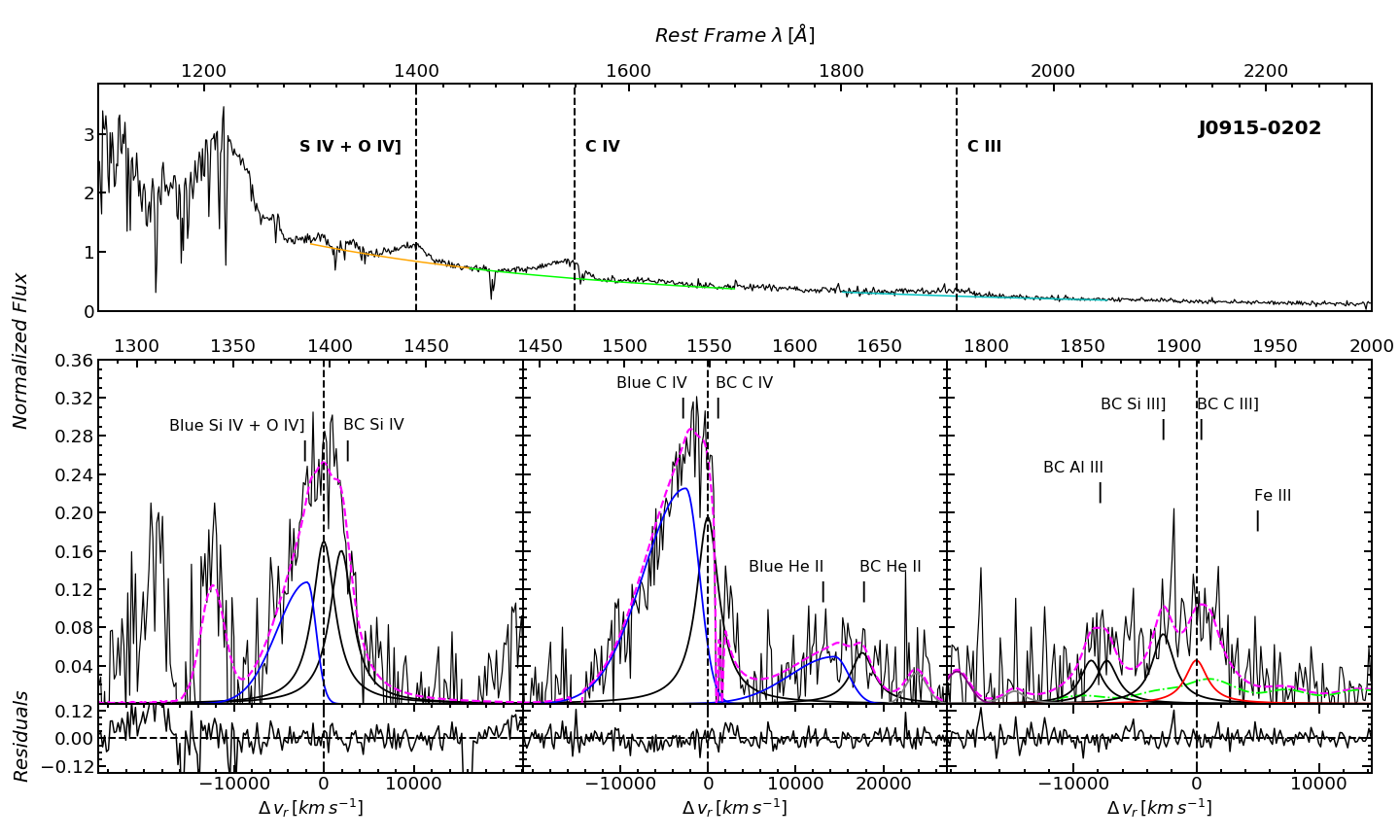}
     \includegraphics[trim= 0.0 0. 0. 0., clip, width=\textwidth]{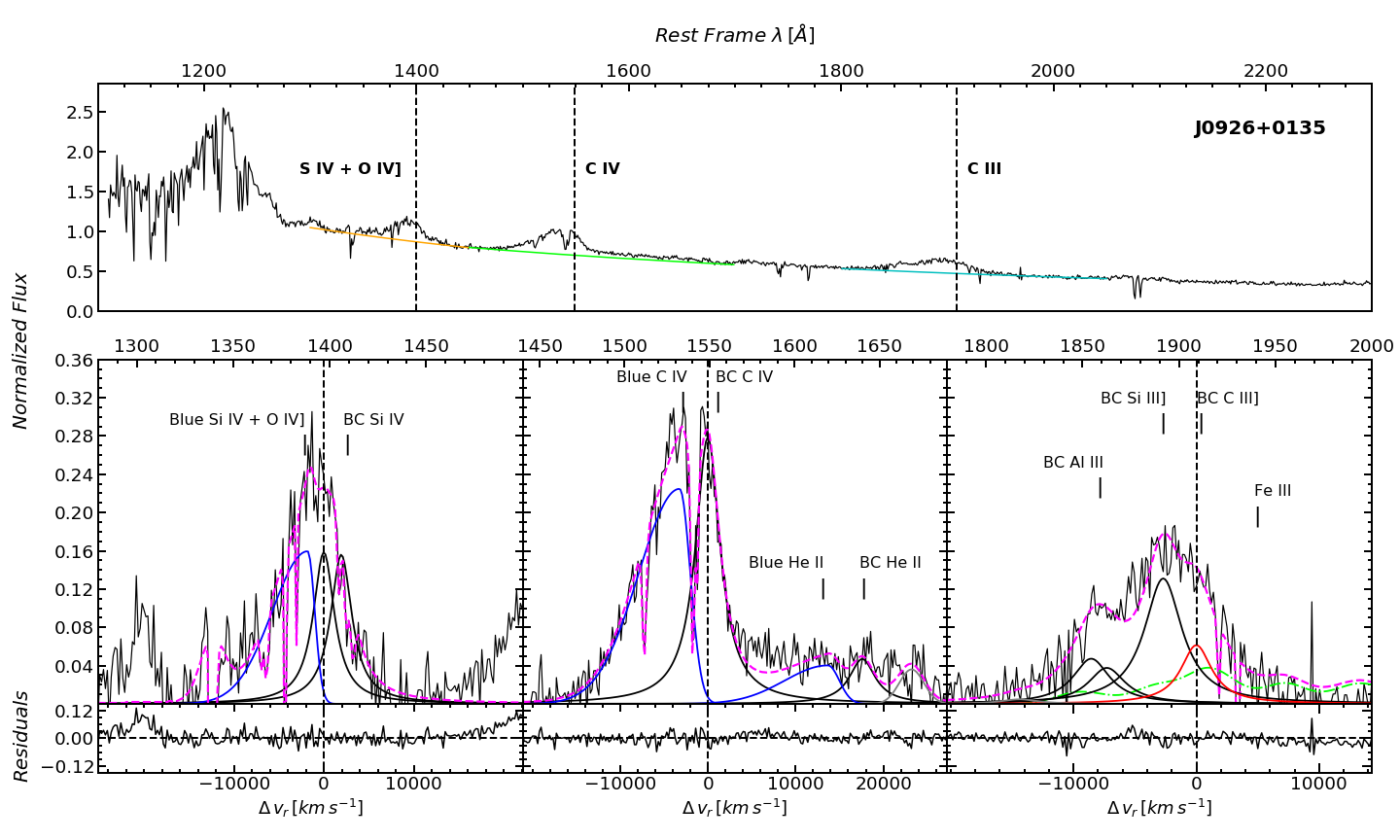}
         \caption{continued.}   
    \label{fig:9}
 \end{figure*}

   \begin{figure*}
    \ContinuedFloat
    \centering
     %trim=left bottom right top
     \includegraphics[trim= 0.0 0. 0. 0., clip, width=\textwidth]{_19--SDSSJ092919.45+033303.4.png}
    \includegraphics[trim= 0.0 0. 0. 0., clip, width=\textwidth]{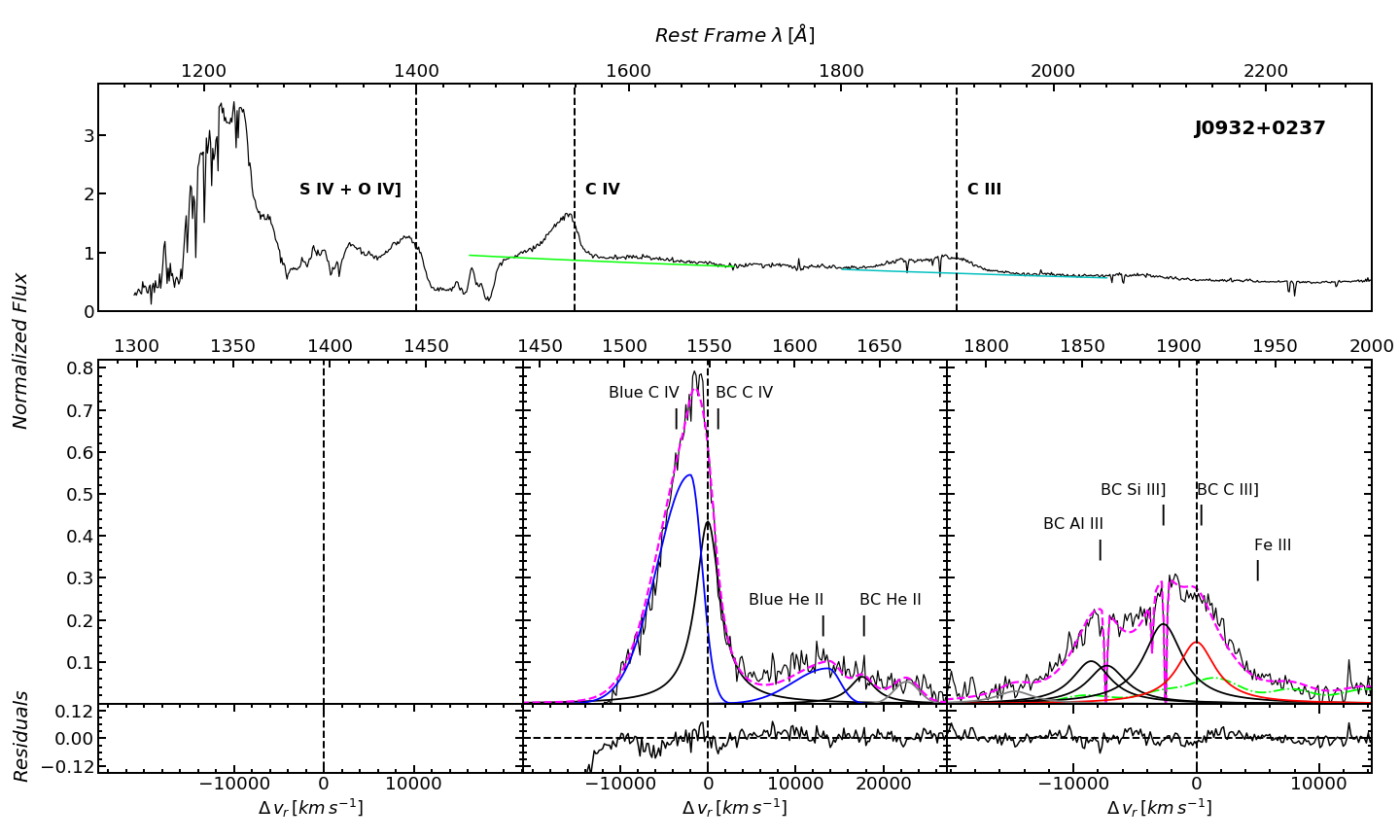}
    \caption{continued.}   
    \label{fig:10}
 \end{figure*}

   \begin{figure*}
    \ContinuedFloat
    \centering
     %trim=left bottom right top
    \includegraphics[trim= 0.0 0. 0. 0., clip, width=\textwidth]{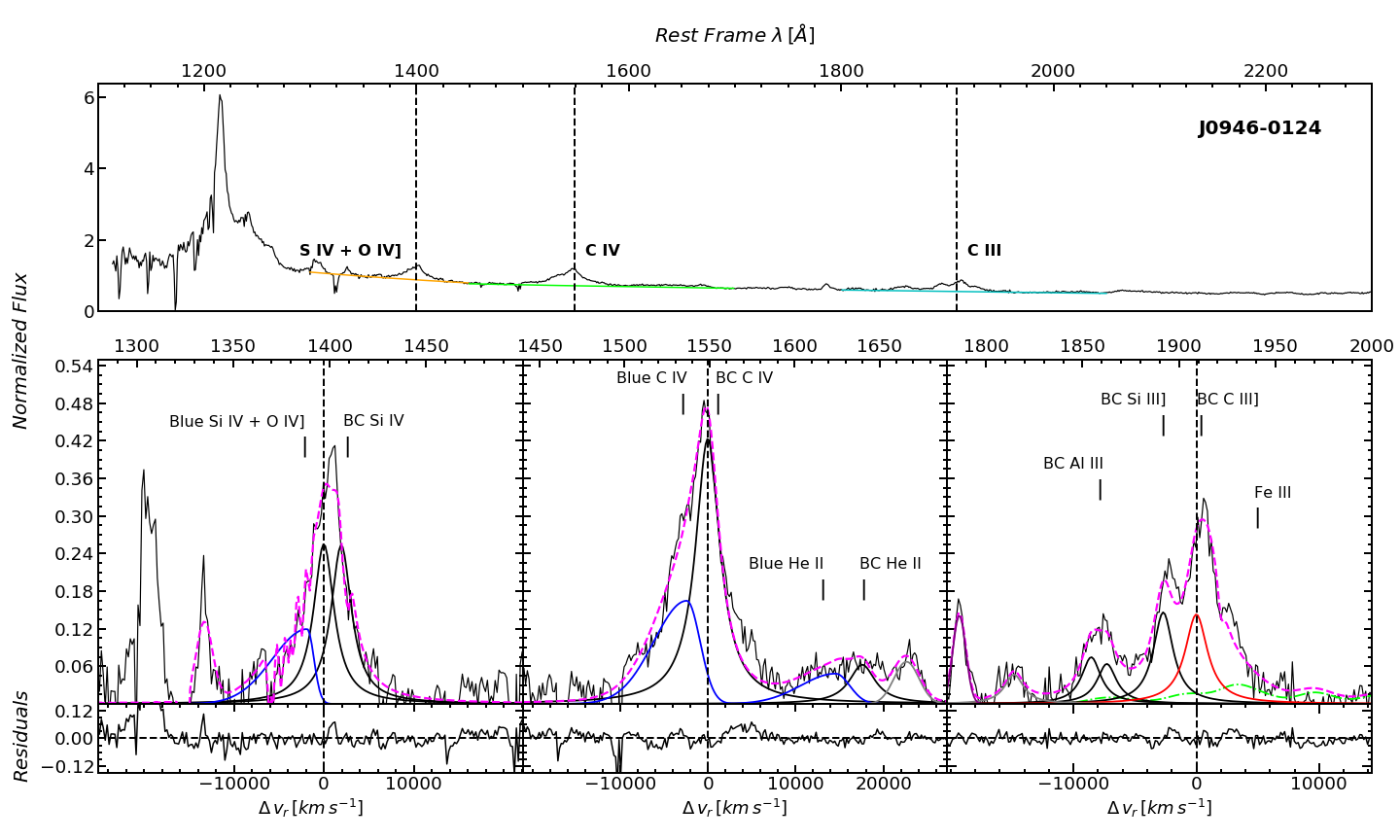}     \includegraphics[trim= 0.0 0. 0. 0., clip, width=\textwidth]{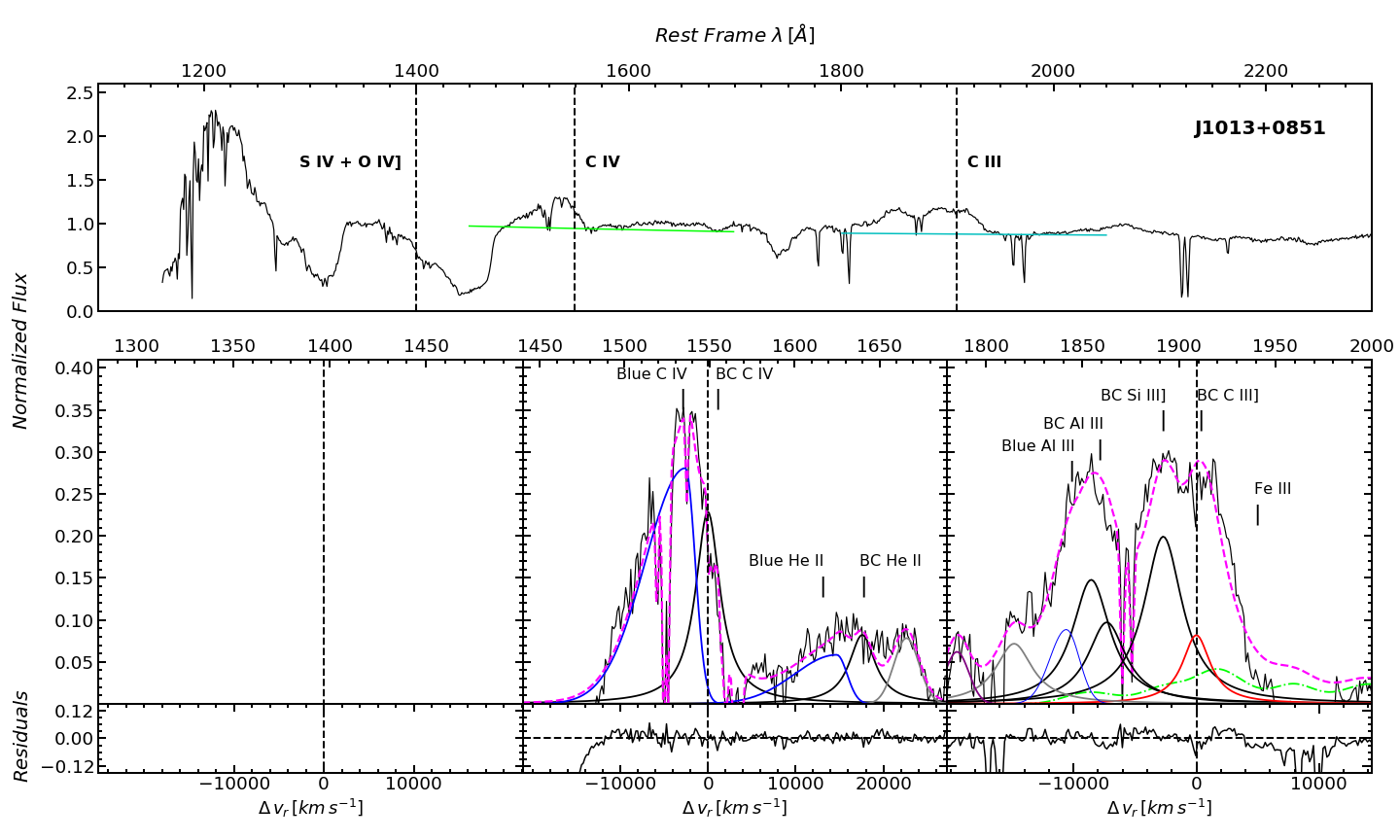}

    \    \caption{continued.}   
    \label{fig:11}
 \end{figure*}

  \begin{figure*}
    \ContinuedFloat
    \centering
     %trim=left bottom right top
     \includegraphics[trim= 0.0 0. 0. 0., clip, width=\textwidth]{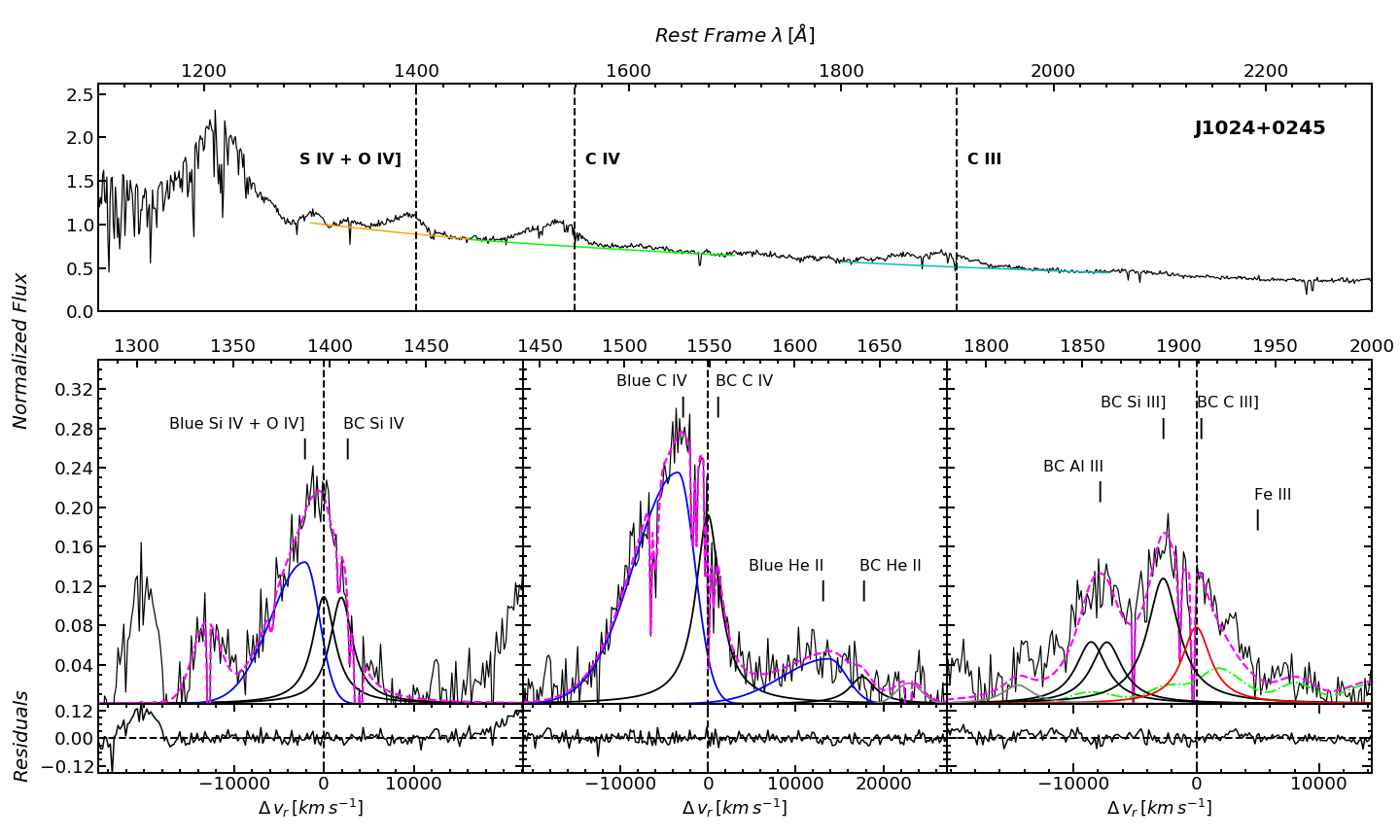}     \includegraphics[trim= 0.0 0. 0. 0., clip, width=\textwidth]{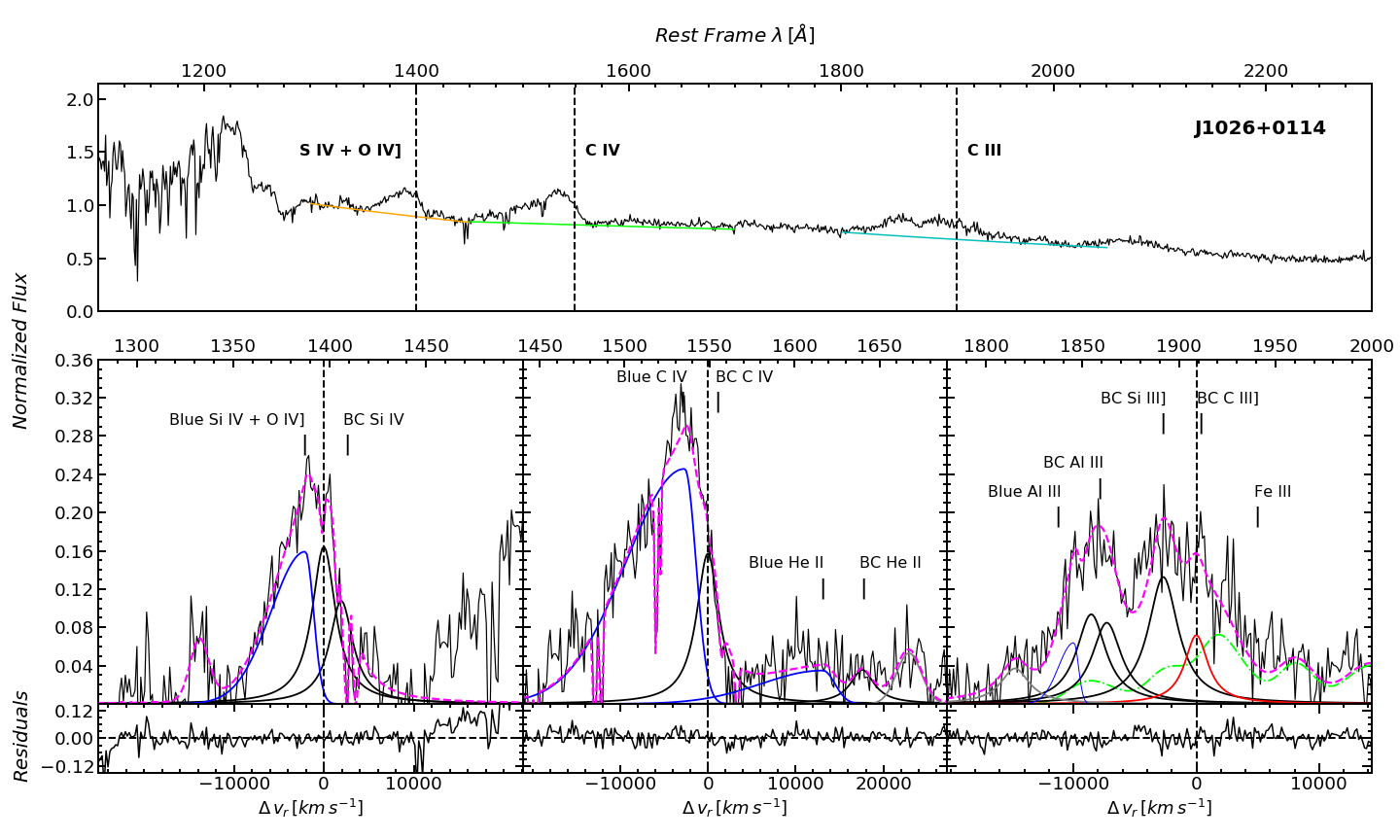}
   \caption{continued.}   
    \label{fig:12}
 \end{figure*}

   \begin{figure*}
    \ContinuedFloat
    \centering
     %trim=left bottom right top
     \includegraphics[trim= 0.0 0. 0. 0., clip, width=\textwidth]{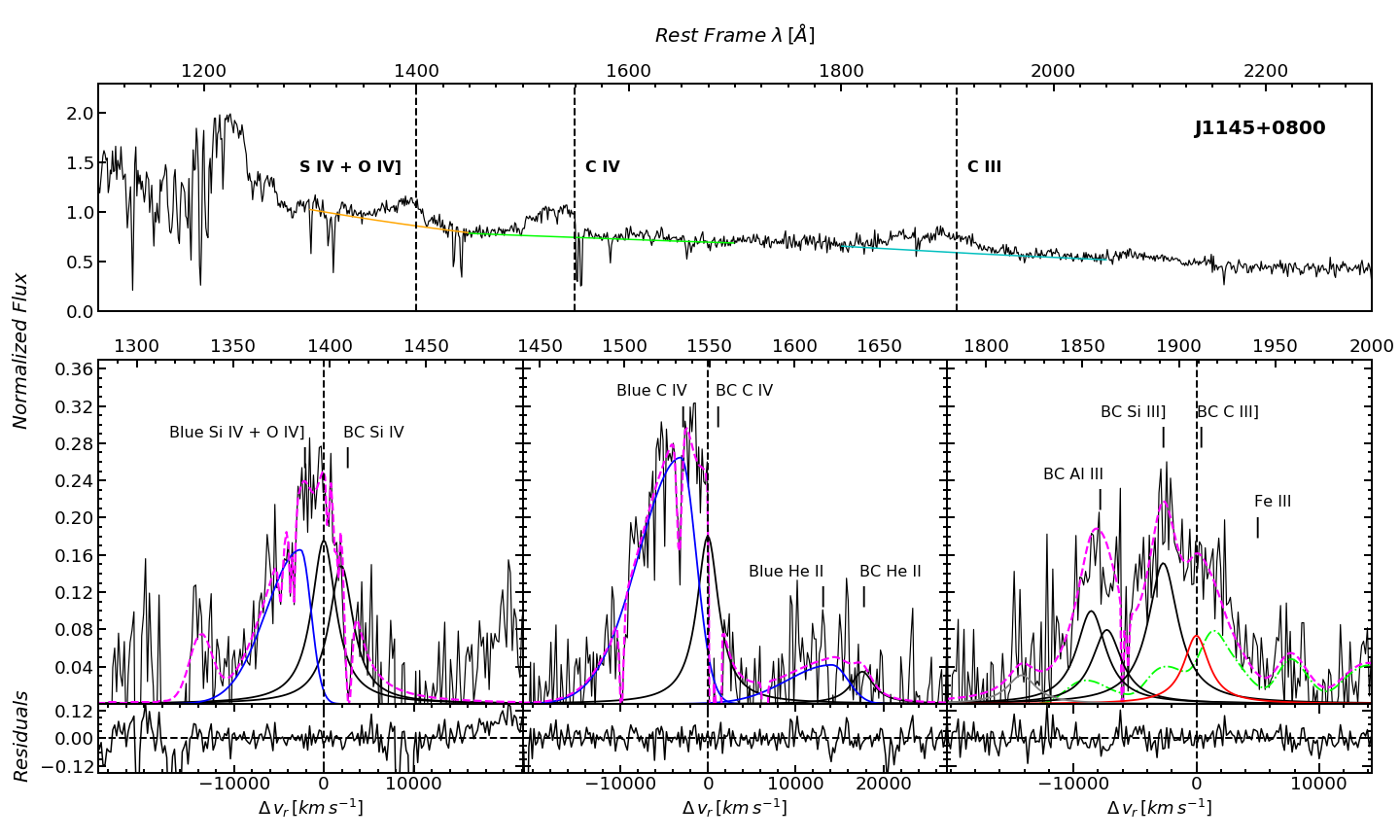}
          \includegraphics[trim= 0.0 0. 0. 0., clip, width=\textwidth]{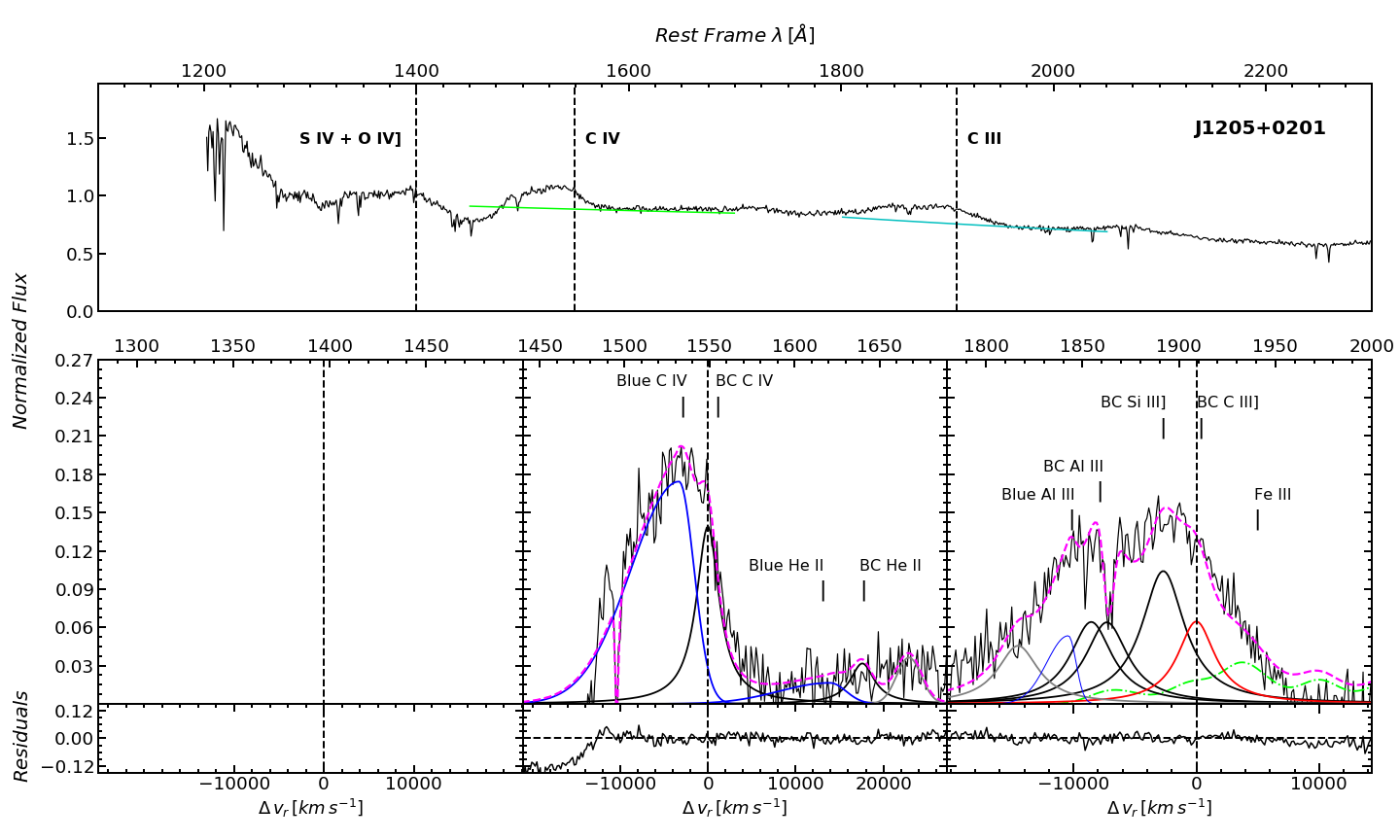}
    \caption{continued.}   
    \label{fig:13}
 \end{figure*}

  \begin{figure*}
    \ContinuedFloat
    \centering
     %trim=left bottom right top
       \includegraphics[trim= 0.0 0. 0. 0., clip, width=\textwidth]{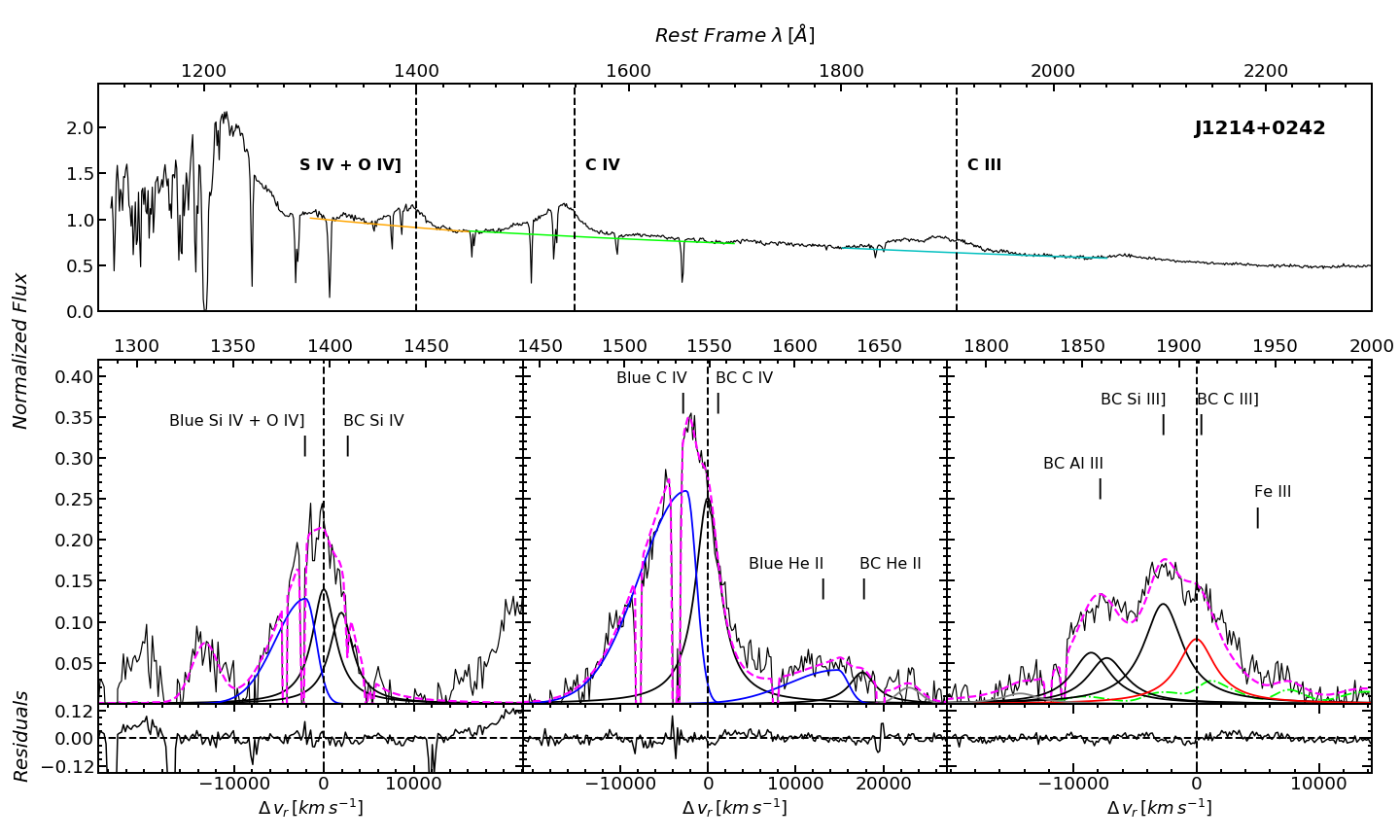}
    \includegraphics[trim= 0.0 0. 0. 0., clip, width=\textwidth]{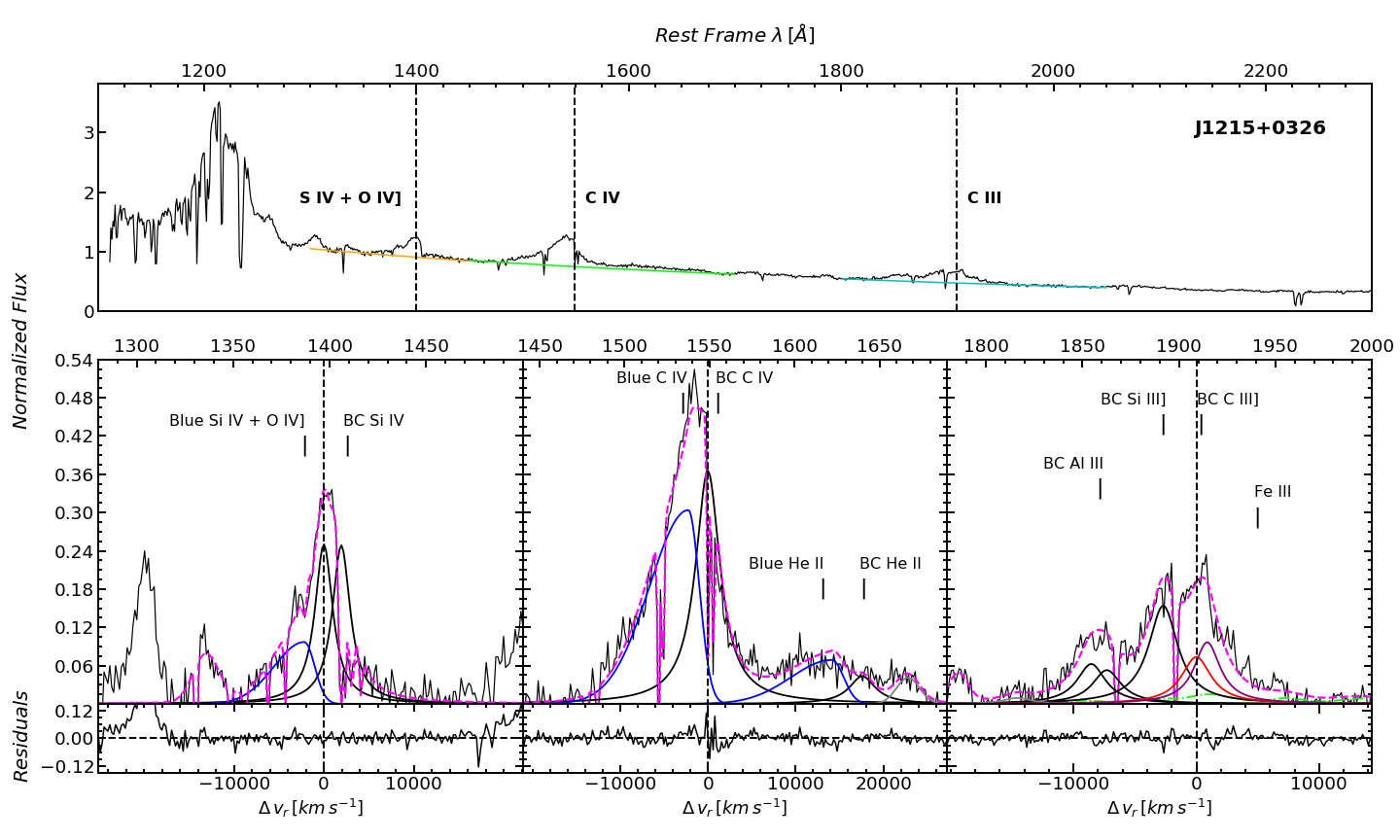}
     \caption{continued.}   
    \label{fig:14}
 \end{figure*}

  \begin{figure*}
    \ContinuedFloat
    \centering
     %trim=left bottom right top
     \includegraphics[trim= 0.0 0. 0. 0., clip, width=\textwidth]{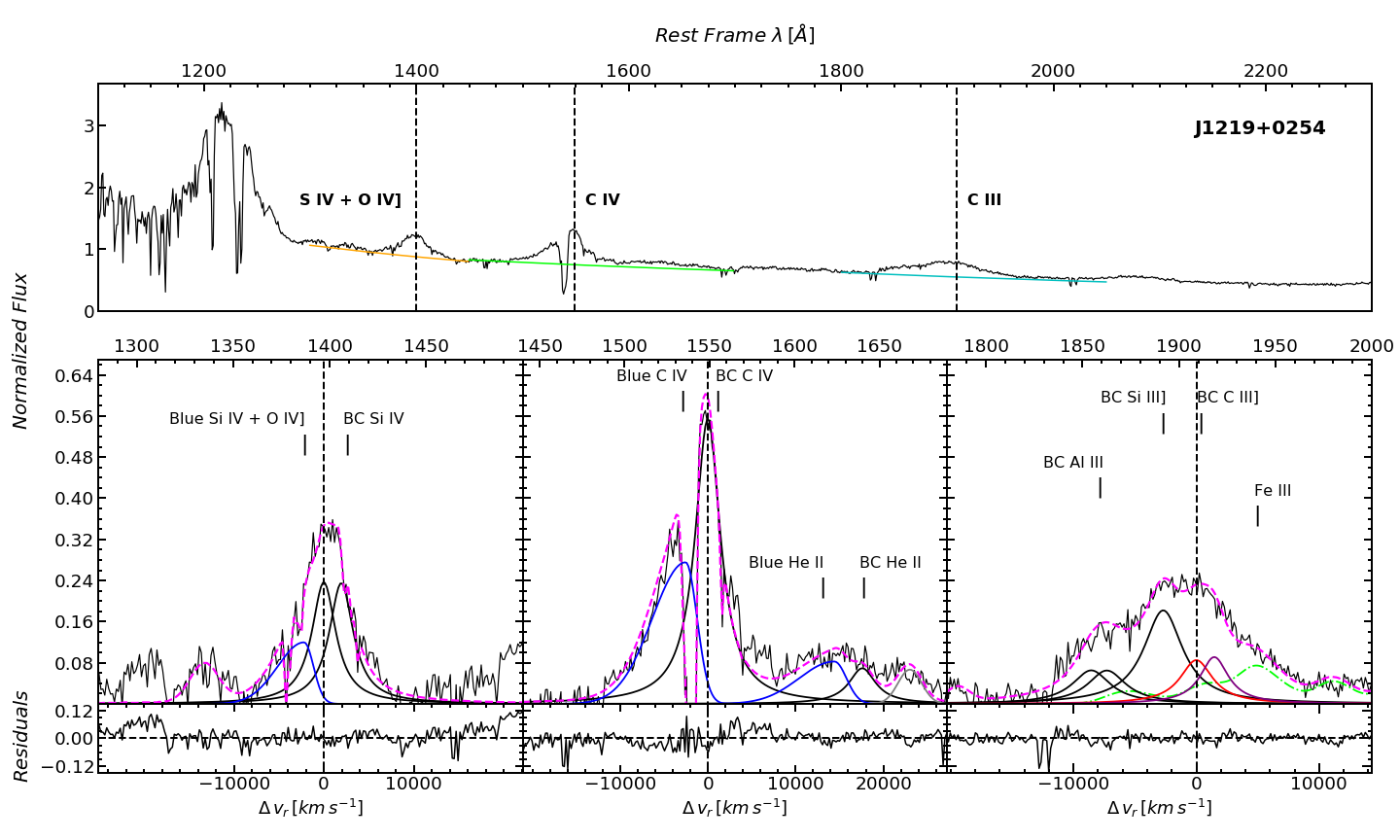}
   \includegraphics[trim= 0.0 0. 0. 0., clip, width=\textwidth]{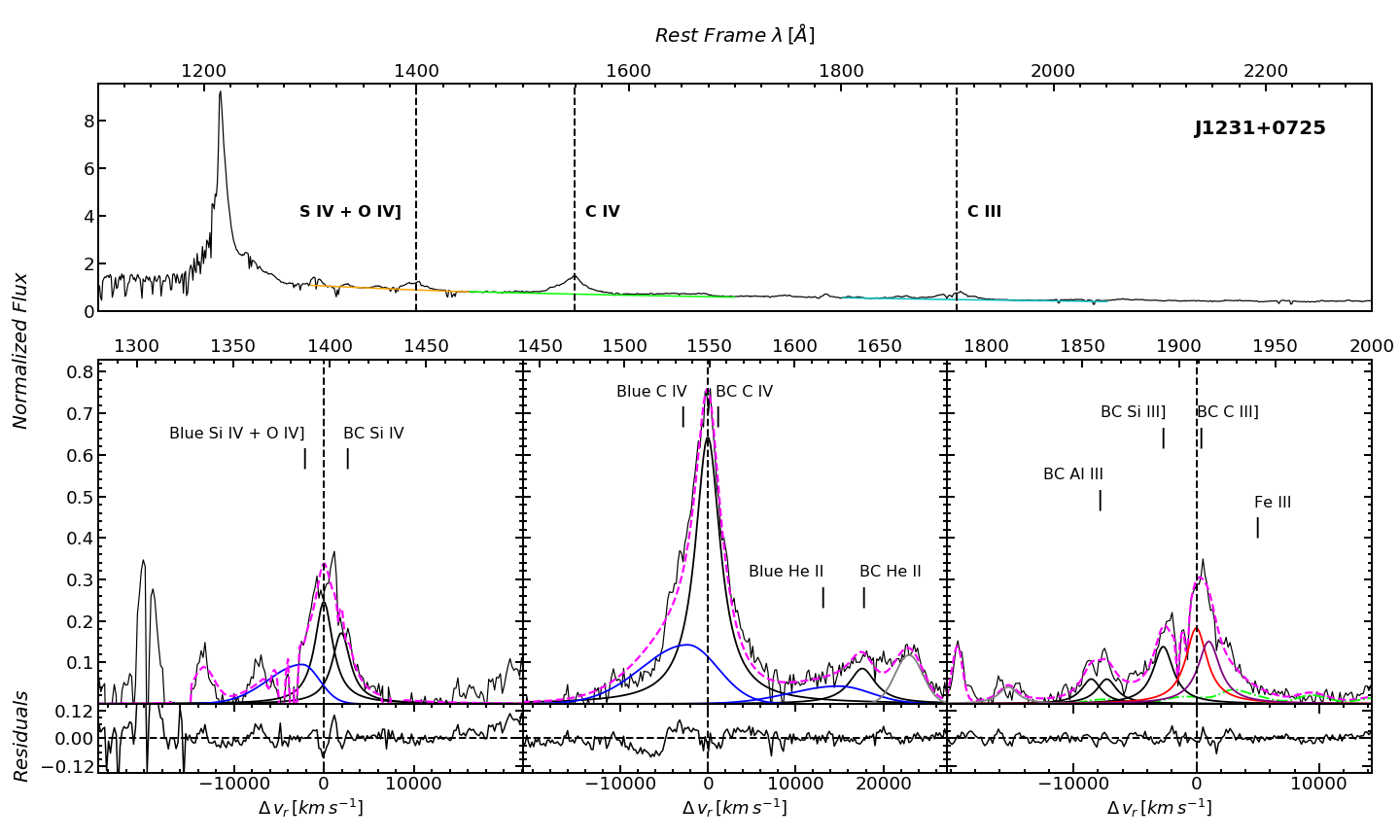}
  \caption{continued.}   
    \label{fig:15}
 \end{figure*}
 
  \begin{figure*}
    \ContinuedFloat
    \centering
    \includegraphics[trim= 0.0 0. 0. 0., clip, width=\textwidth]{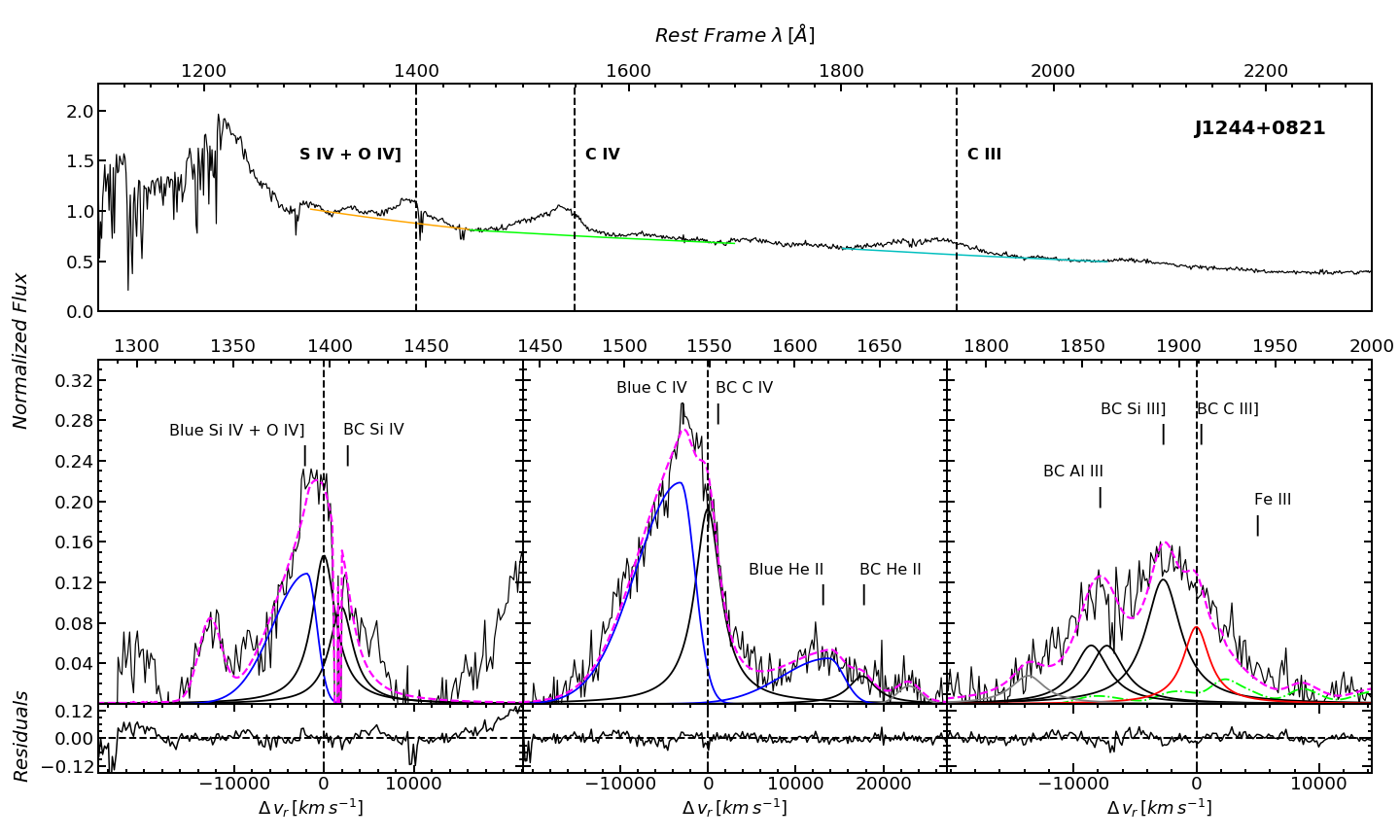}
     \includegraphics[trim= 0.0 0. 0. 0., clip, width=\textwidth]{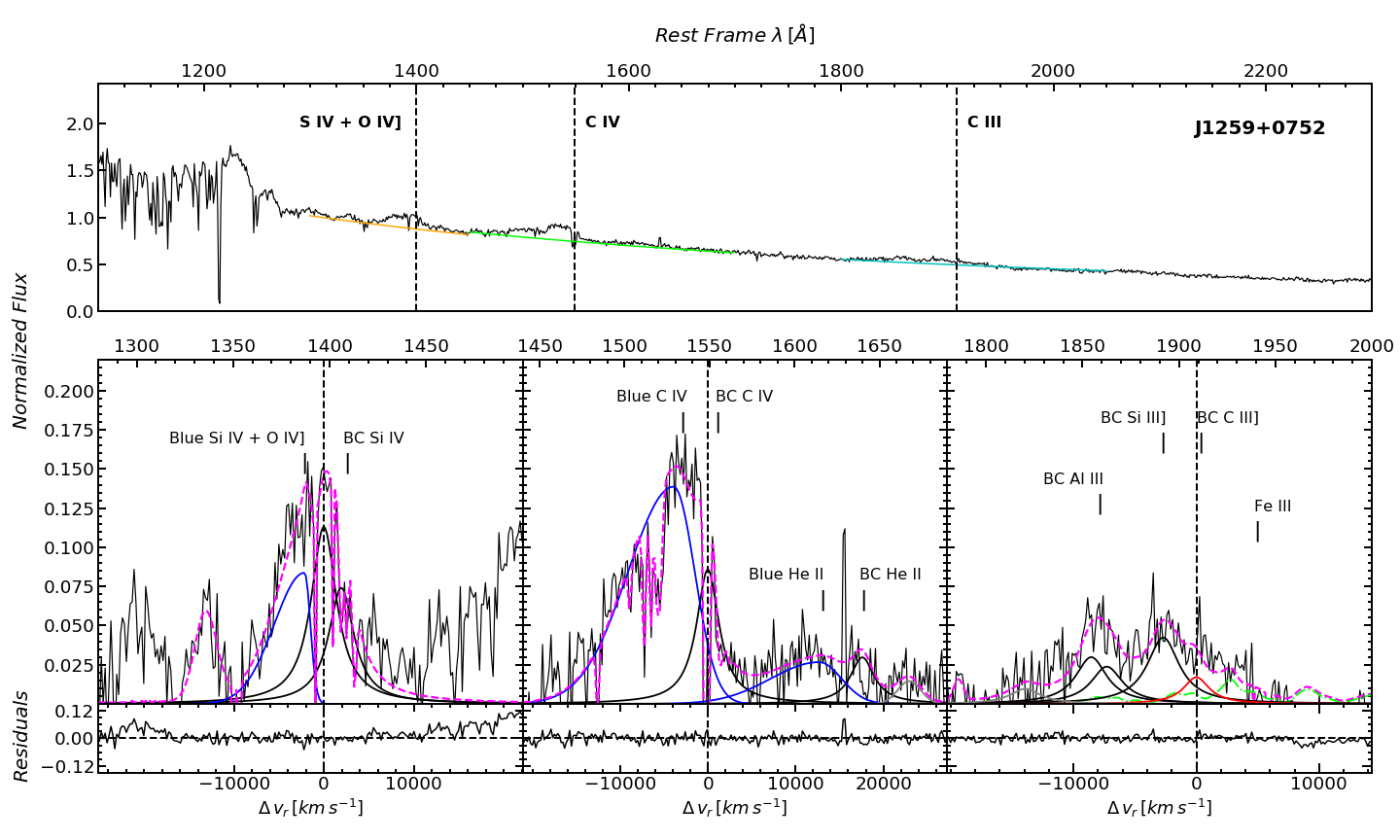}
   \caption{continued.}   
    \label{fig:16}
 \end{figure*}

  \begin{figure*}
    \ContinuedFloat
    \centering
     %trim=left bottom right top
     \includegraphics[trim= 0.0 0. 0. 0., clip, width=\textwidth]{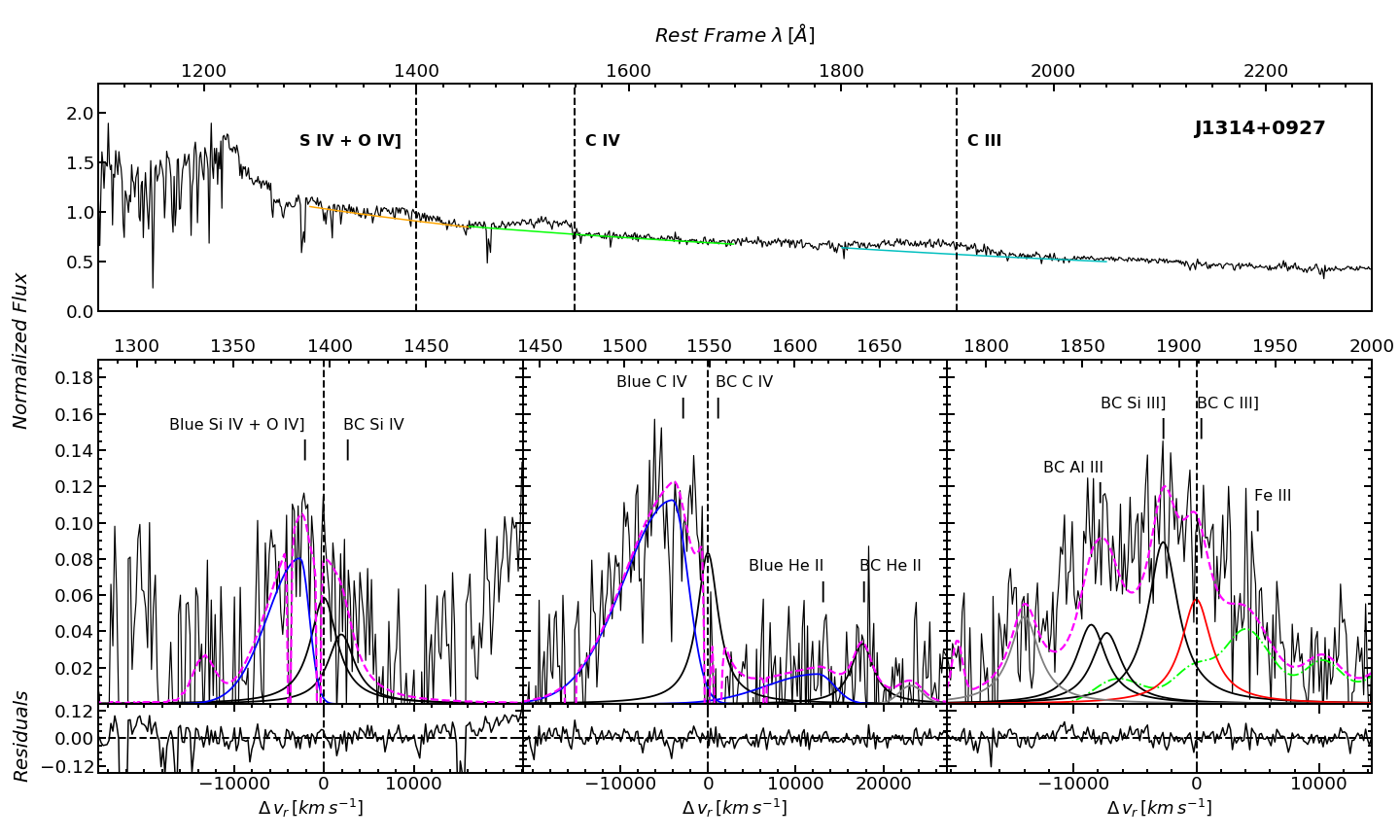}
     \includegraphics[trim= 0.0 0. 0. 0., clip, width=\textwidth]{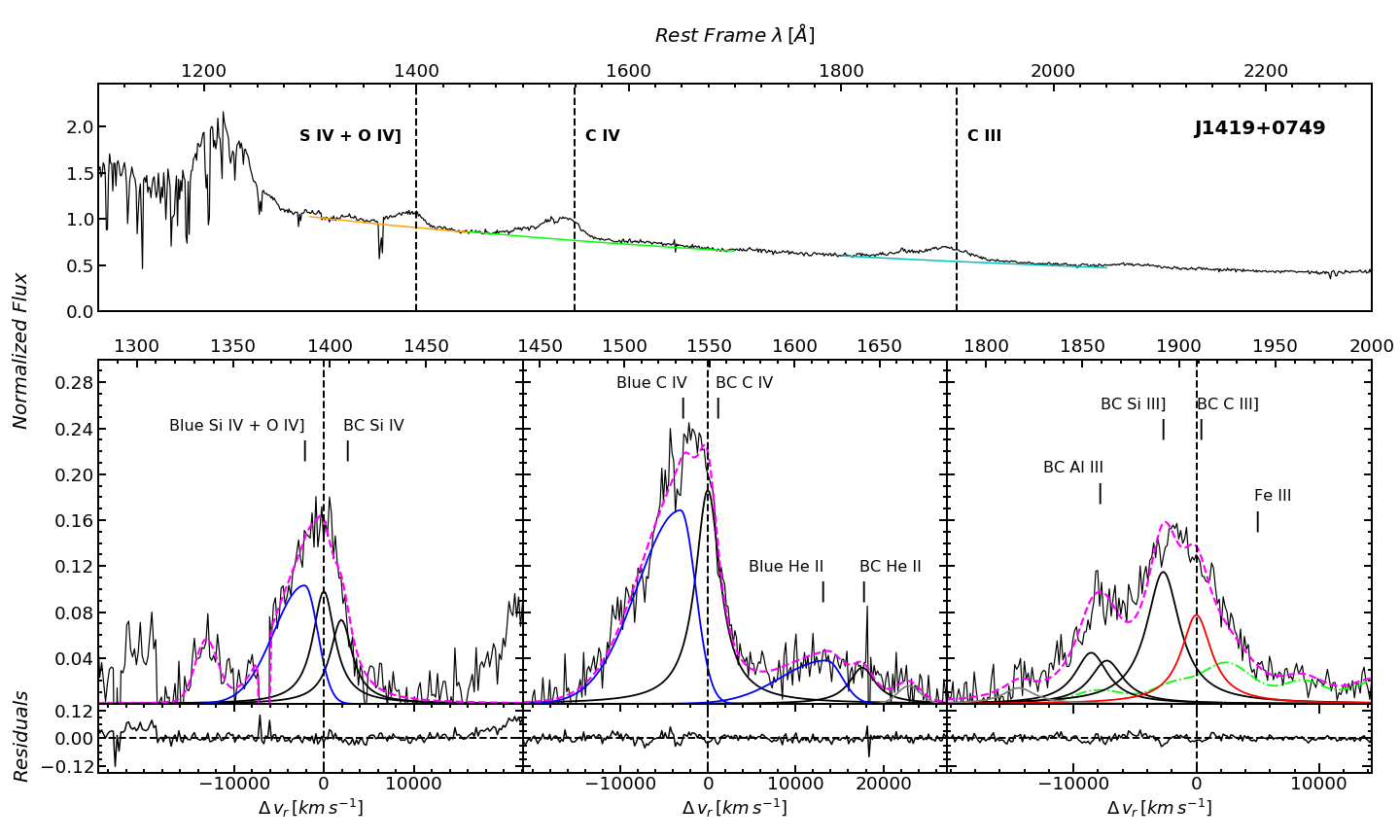}
   \caption{continued.}   
    \label{fig:17}
 \end{figure*}

  \begin{figure*}
    \ContinuedFloat
    \centering
     %trim=left bottom right top
     \includegraphics[trim= 0.0 0. 0. 0., clip, width=\textwidth]{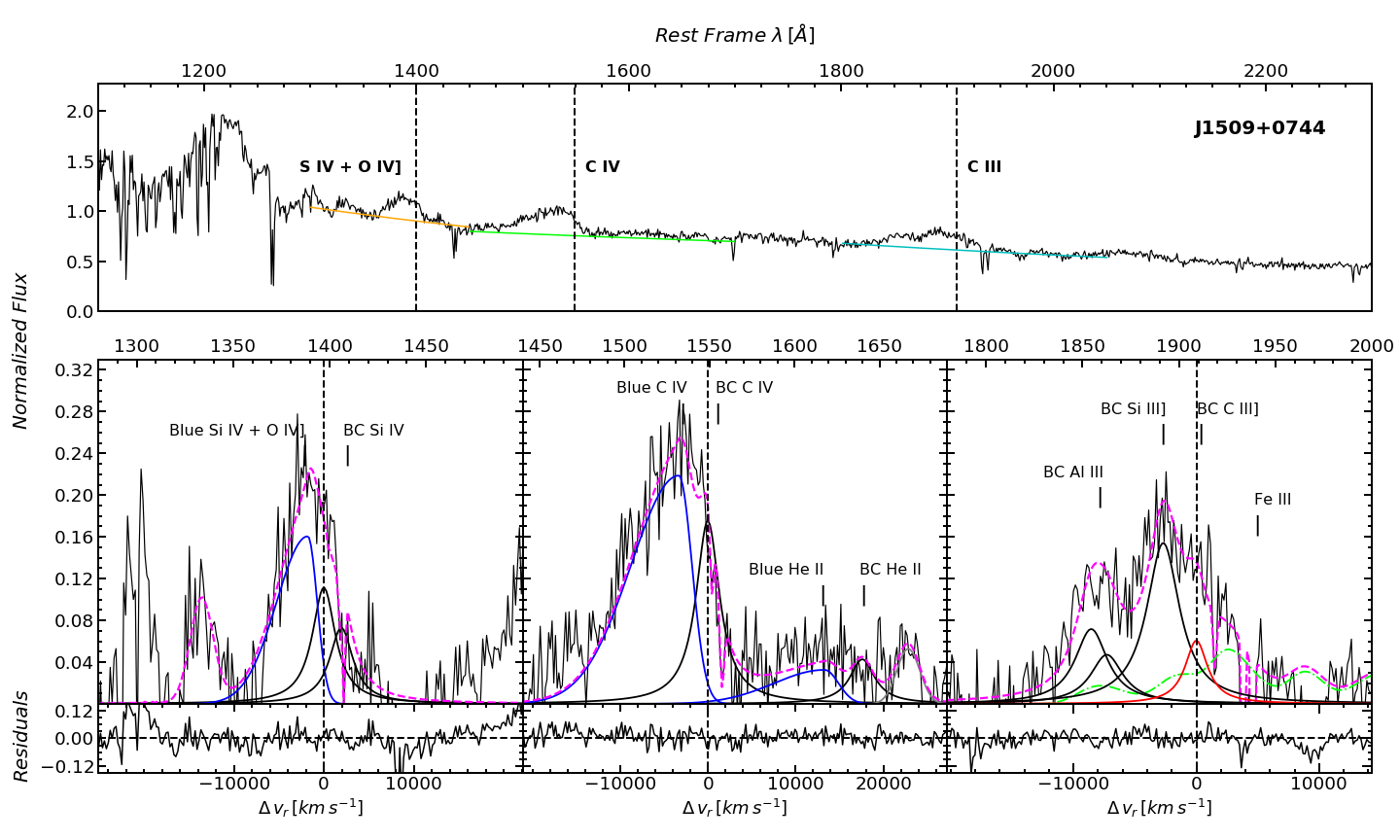}
    \includegraphics[trim= 0.0 0. 0. 0., clip, width=\textwidth]{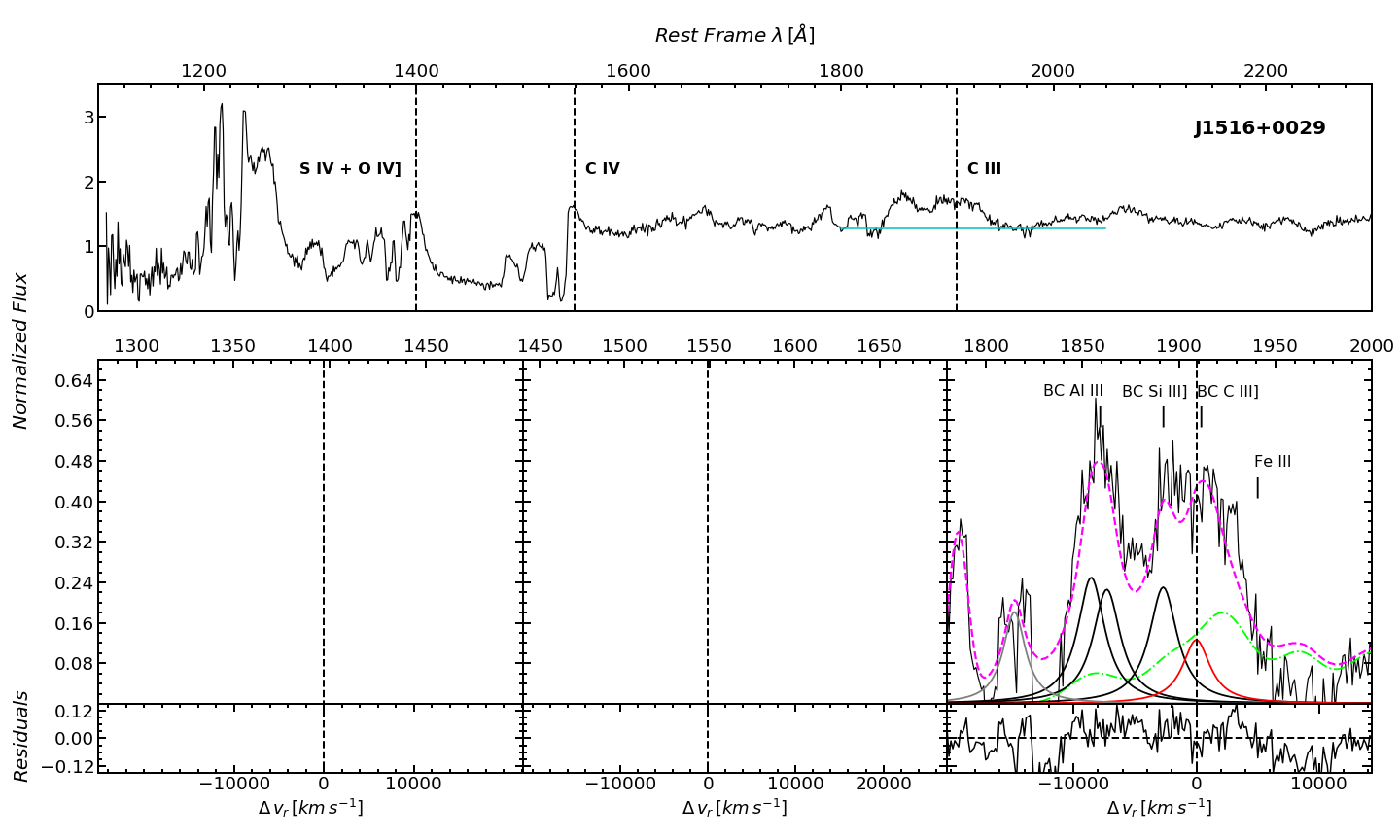}
    \caption{continued.}   
    \label{fig:18}
 \end{figure*}

  \begin{figure*}
    \ContinuedFloat
    \centering
     %trim=left bottom right top
     \includegraphics[trim= 0.0 0. 0. 0., clip, width=\textwidth]{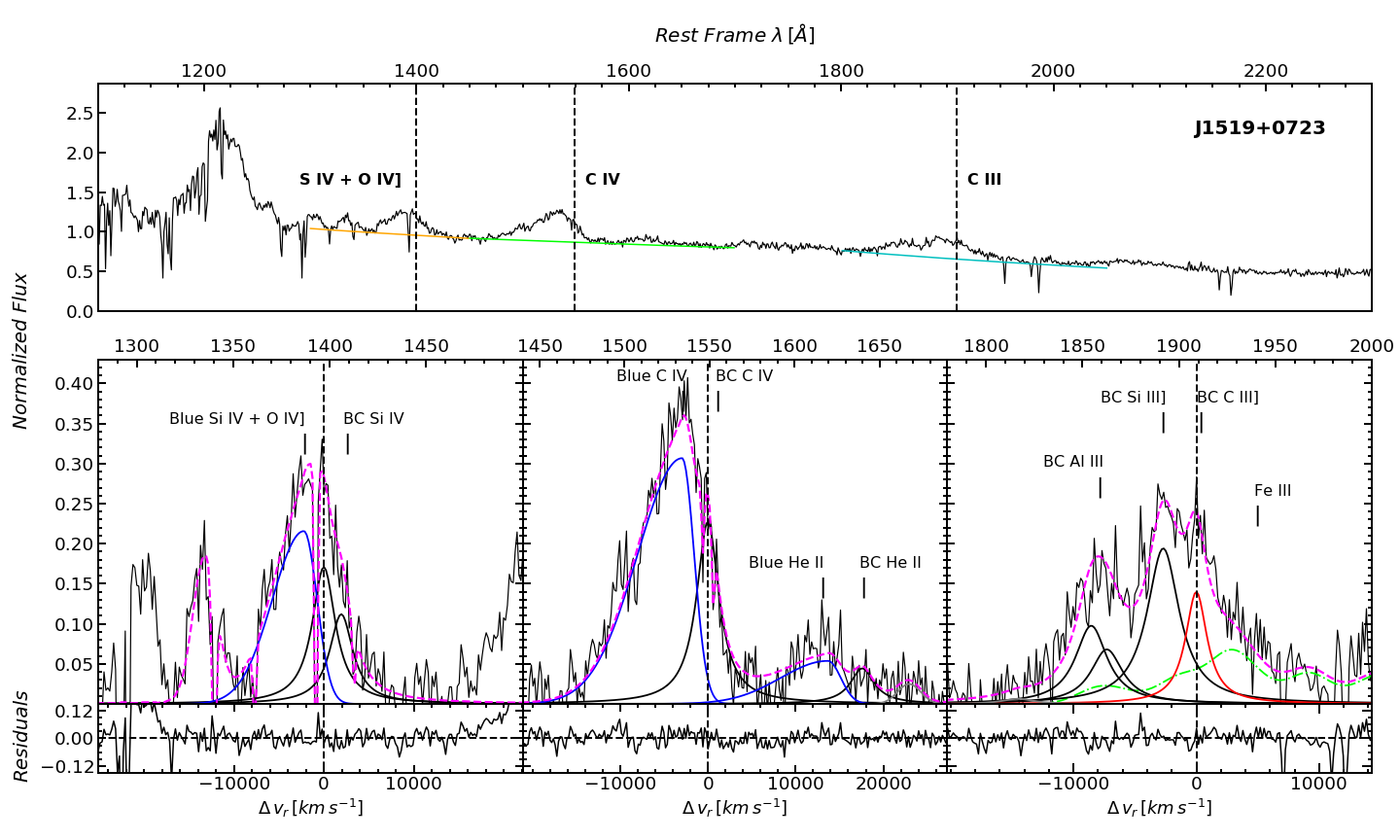}
     \includegraphics[trim= 0.0 0. 0. 0., clip, width=\textwidth]{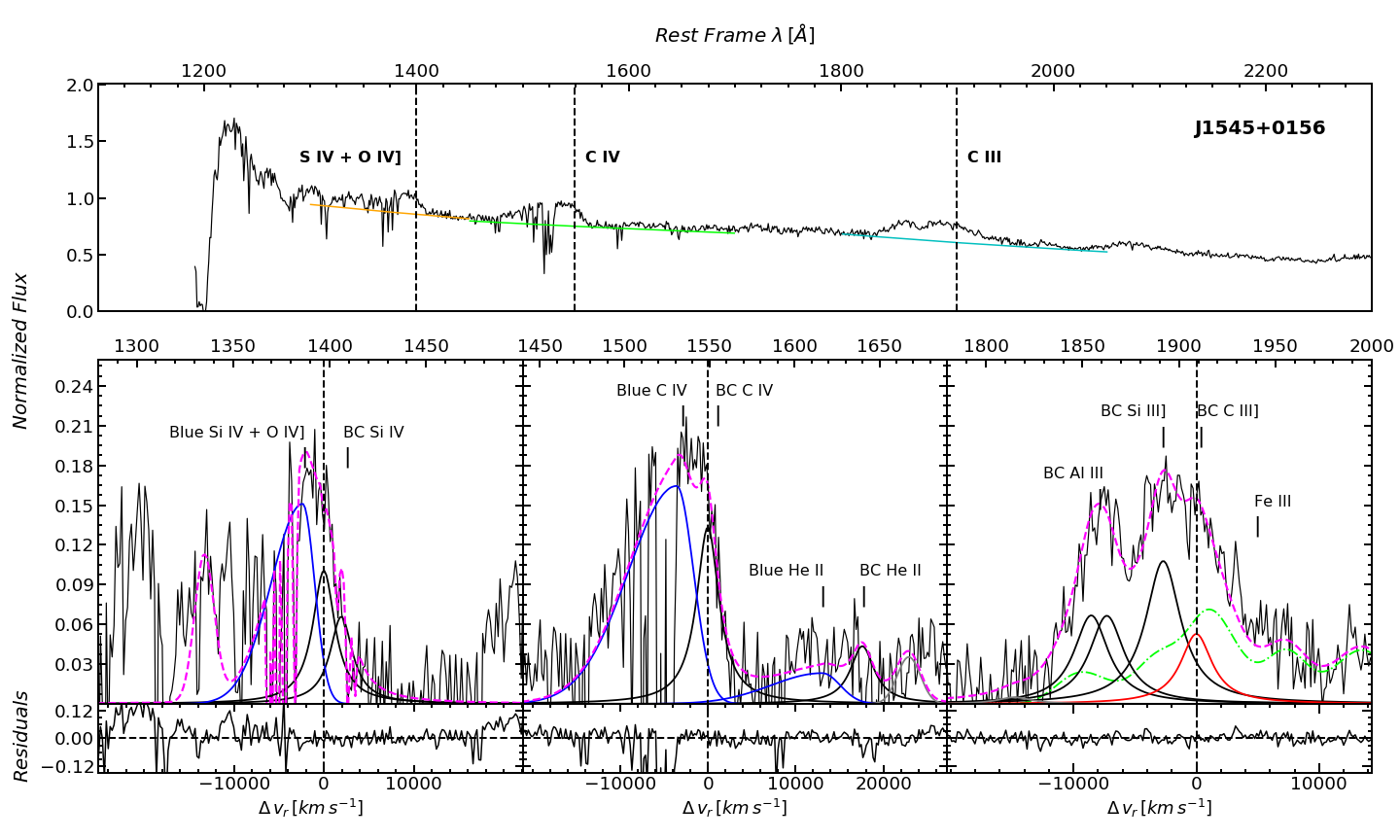}
   \caption{continued.}   
    \label{fig:19}
 \end{figure*}

  \begin{figure*}    \centering
    \ContinuedFloat
     %trim=left bottom right top
     \includegraphics[trim= 0.0 0. 0. 0., clip, width=\textwidth]{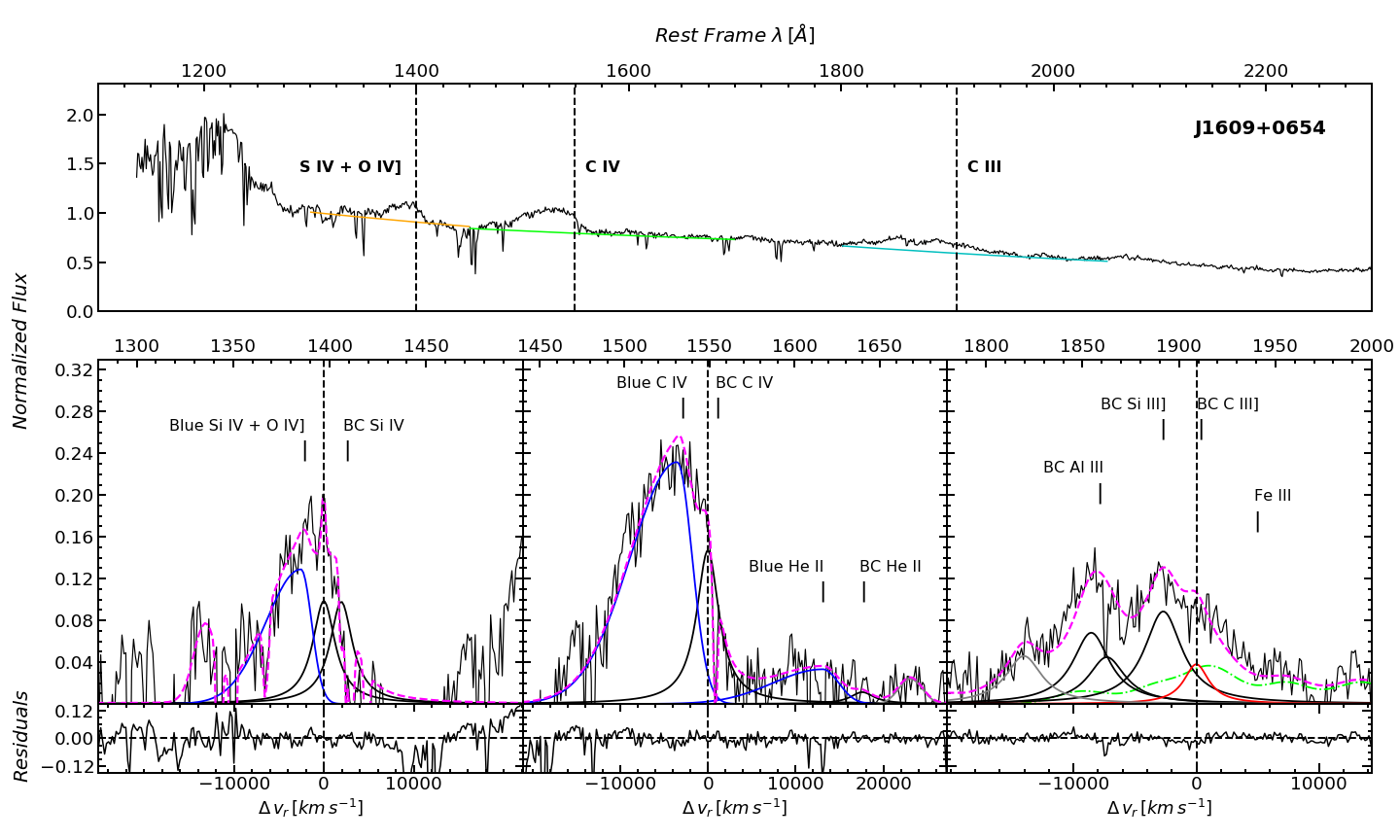}
       
     \includegraphics[trim= 0.0 0. 0. 0., clip, width=\textwidth]{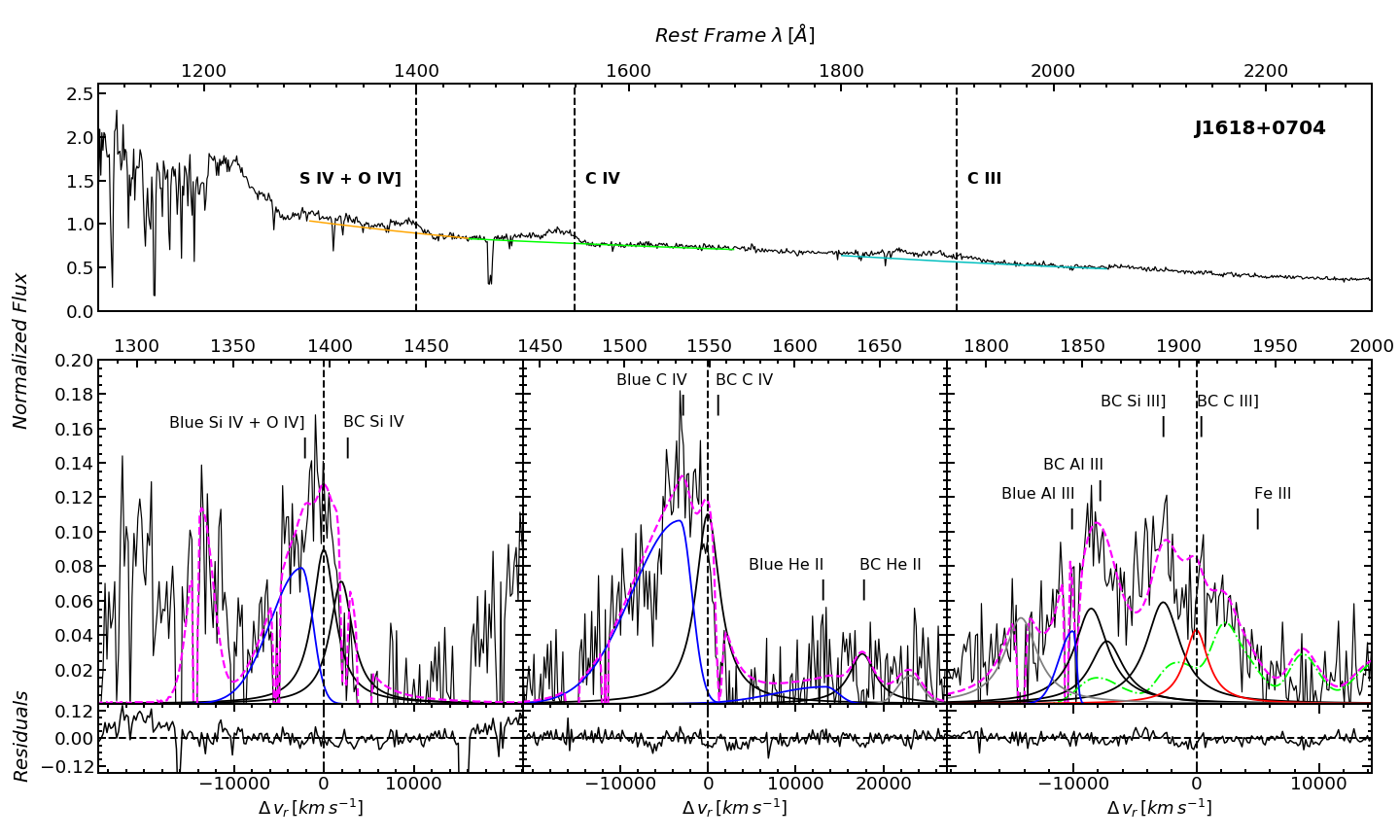}
    \caption{continued.} \label{fig:20} 
 \end{figure*}

    \begin{figure*}
    \ContinuedFloat
        \centering
        \includegraphics[trim= 0.0 0. 0. 0., clip, width=\textwidth]{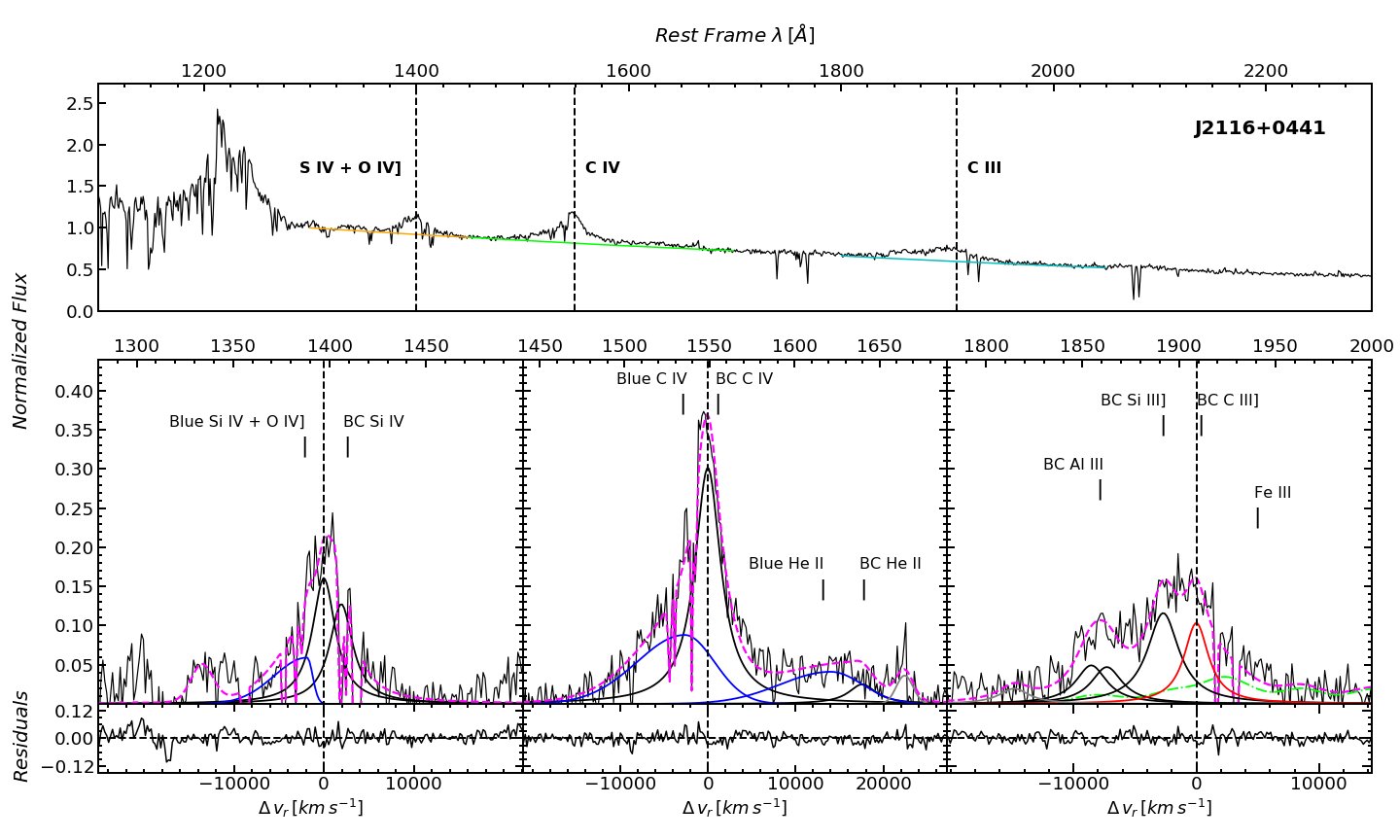}
     \includegraphics[trim= 0.0 0. 0. 0., clip, width=\textwidth]{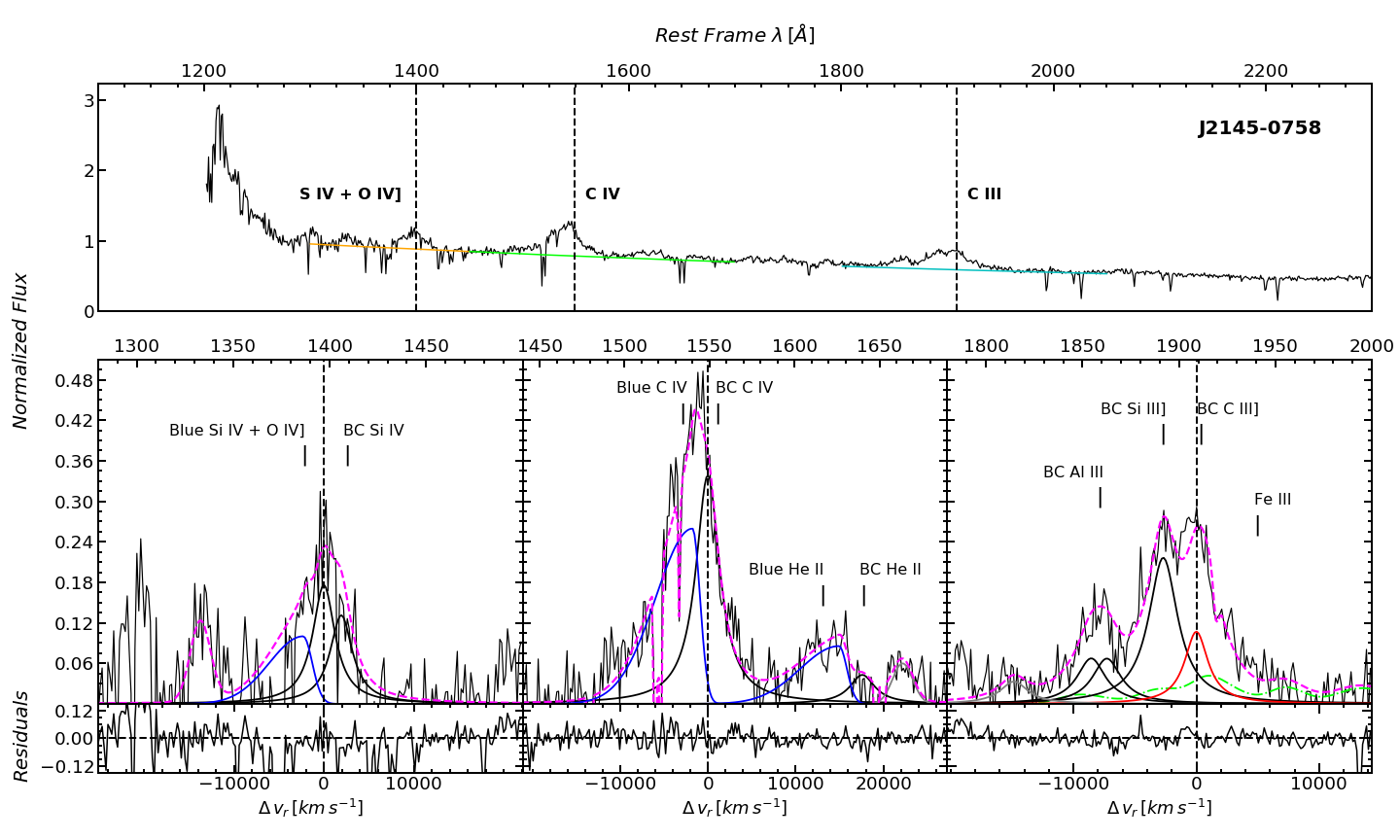}
    \caption{Cont.} \label{fig:21}
    \end{figure*}

%%%%%%%%%%%%%%%%%%%%%%%%%%%%%%%%%%%%%%%%%%%%%%%%%%%%

\vfill\eject\newpage
\pagebreak\pagebreak
\clearpage
\twocolumn
\section{Consideration on the spectral appearance of sample sources}
\label{ind}

\citet{martinez-aldamaetal18} carried out a  very thorough descriptive analysis of the xA spectra. The spectra of the present sample resemble closely the composite spectrum shown by those authors.  We will therefore limit ourselves to a few remarks.  

\paragraph{ BAL QSOs.} J0252$-$0420, an example of this type of sources, presents strong and prominent absorptions mainly between 1400  \AA \ and 1550  \AA. The absorption {wipes out} the emissions in these regions and inhibits their interpretation. We have 6 sources with this characteristic in our sample (as specified in Col. 9 of Table \ref{tab:general}). We measured the regions that seem less affected but BAL QSOs are not considered for further analysis, save for the general comment reported below. \\ The FWHM(\aliii) of the BAL sources is predominantly above the sample mean with the lowest value 2788 \kms \ for J1516+0029 and the highest 4272 \kms \ for J1205+0201 (also the highest value of the sample), although the EW(\aliii) values are consistent with the average {for} the non-BAL sample. If we compare our available \aliii\ flux ratios this would lead us to super solar values $\sim$15 Z$_\odot$, as already derived by \citet{baldwinetal96}.

\paragraph{"mini-BALs".} In addition to the broad, strong absorptions in sources classified as   BAL QSOs, we also found the presence of strong narrow absorptions in 8 of our spectra. These absorptions are mainly in the 1400  \AA \ and 1550  \AA\ regions and blur out just a section of the emission, so that we can still retrieve the emission profile. J0034$-$0326 is an example where absorptions  blur out small radial velocity sections in both  the aforementioned spectral regions. The metallicity of these sources remains consistent with the median value of the sample.    

\paragraph{Strong BC/weak lined quasars.} The values of the equivalent width of {\em all} lines save \lya\ are  small, including the \civ\ BLUE, with a distribution of the full \civ\ profile (BC+BLUE) peaked around 10 \AA, with a small dispersion. Half of the sample sources qualify as "weak lined quasars" \citep{diamond-stanicetal09,luoetal15,marzianietal16a}. Sources that have prominent \civ\ BC with respect to CIV BLUE are present in our sample but are labeled as outliers or borderline sources. Two cases in point are J1231+0725 and J2116+0441. 

\paragraph{J0858+0152} is measured as  one of the most metal rich sources in the sample of \citetalias{sniegowskaetal21}. It visually meets all the indicators signaling high $Z$: ratio \siiv/\civ $\sim 1$, extreme \aliii/\siiii, low $W$ \civ. The result concerning the metallicity relies, however, on the continuum placement in the region at $\approx$1350  \AA, where severe blending with \oi\ and \siii\ and ultimately with the very extended \lya\ red wing makes it difficult to identify the continuum level properly.

\end{appendix}

\end{document}